\documentclass{pinchcrn}   
                                             \usepackage{graphicx}
\usepackage{dcolumn}
\usepackage{bm}
\usepackage{amsmath}
\usepackage{amssymb}
\usepackage{mathrsfs}
\usepackage{bbm}
\usepackage{epsfig}
\usepackage[OT2,OT1]{fontenc}
\newcommand\cyr{%
\renewcommand\rmdefault{wncyr}%
\renewcommand\sfdefault{wncyss}%
\renewcommand\encodingdefault{OT2}%
\normalfont
\selectfont}
\DeclareTextFontCommand{\textcyr}{\cyr}

\usepackage{wrapfig}
\usepackage{eurosym}

\title{WHAT IS LIFE? \\ ~ \\ {\LARGE \bf Sub-cellular Physics of Live Matter}}

\author{Antti J. Niemi}
\affiliation{Department of Physics and Astronomy, Uppsala University,
P.O. Box 803, S-75108, Uppsala, Sweden \\ Laboratoire de Mathematiques et Physique Theorique
CNRS UMR 6083, F\'ed\'eration Denis Poisson, Universit\'e de Tours,
Parc de Grandmont, F37200, Tours, France \\
Department of Physics, Beijing Institute of Technology, Haidian District, Beijing 100081, China}

\begin{document}

\maketitle

\dedication{Dedicated to the 70$^{th}$  anniversary of Schr\"odinger's 1944 Lectures in Dublin.}

\preface

\begin{quote}
\noindent
"{\it There are at present fundamental problems in theoretical 
physics awaiting solution, e.g. the relativistic formulation of quantum mechanics and the nature of 
atomic nuclei (to be followed by more difficult ones such as the problem of life).}"  (P.A.M. Dirac 1931)
\end{quote} 

\vskip 0.3cm

70 years ago Erwin Schr\"odinger presented a series of lectures at the Dublin Institute for Advanced Studies
with the title  {\it What is Life? The Physical Aspect of the Living Cell.} 
The lectures were subsequently published in the form of a small 
book  \cite{Schrodinger-1944}.
According to Google Scholar this book has been
cited even more frequently than Schr\"odinger's  articles on quantum mechanics. In particular, it was
credited by Crick and Watson as the source of inspiration that led them to reveal the double helix structure
of DNA.

My lectures at les Houches were a celebration of the anniversary of Schr\"odinger's lectures, and for that
reason I decided to share a title.

Besides  Crick and Watson, Schr\"odinger's  book has been, and continues to be, a source of 
inspiration to generations of physicists. But it seems to me that 
its full dimensionality might  not yet have been fully comprehended, nor appreciated. In particular, 
the book has many parallels
to Philip Anderson's highly inspiring 1972 article {\it More is different. Broken symmetry and the nature of the 
hierarchical structure of science.} \cite{Anderson-1972}.   
Schr\"odinger clearly  
realised that {\it Bol'she}
({\cyr{bol}}{\tiny b}{\cyr{she}}) makes a difference when he wrote  that   

\vskip 0.3cm

\begin{quote}
{\it ... living matter, while not eluding the 'laws of physics' as established up to date, is likely to 
involve 'other laws of physics' 
hitherto unknown, which, however, once they have been revealed, will form just as integral a part of this science as the  
former.
}
\end{quote}

\vskip 0.3cm

Time should be ripe to accept a universal formal definition
of life in terms of proteins and their dynamics:
Proteins are the workhorses 
of all living organisms,   proteins are true nano-machines that participate in 
all the metabolic activities which constitutes life as we know it. 
From this perspective life is indeed something that 
can be modelled  and understood
using  both known and still to be revealed 
laws that govern sub-cellular physics of live matter.  For a physicist like me 
this  is an exciting  way to try and answer Schr\"odinger's question.

\vskip 0.3cm
The underlying theme in my lectures is to view proteins as an exciting
example of a physical system where much of Anderson's  {\it Bol'she} can be found. Indeed, proteins 
seem to bring together  {\it most}
of the contemporary lines of research in modern theoretical physics: Geometry of 
string-like structures, topological solitons, spin chains, integrable models, equilibrium and
non-equilibrium statistical physics, quest for entropy, emergent phenomena, ... and much, much more. Moreover,
the tools that are needed to fully understand proteins range from 
highly formal to extensively numerical, and for a theorist there are almost endless opportunities to
address questions with direct and important experimental relevance; physical, chemical, biological, medical, ... 
The amount of data is {\it overwhelming} and it is readily accessible. Experiments can be done and directly compared with 
theoretical calculations and numerical computations; the subject  continues to grow at a highly exponential rate.

\vskip 0.2cm
\noindent
These lectures  were prepared for students in condensed matter physics, both theoretical and experimental.
I assumed the students did not really have any prior knowledge of proteins, that they did not even
know how protein looks like
at the atomic level. Thus I begin  with a {\it protein minimum}. It explains the basic facts that I think one needs
to know, to get started in research on physics of proteins. The rest of the lectures address proteins from the point of view
of a physicist, from a perspective that I hope appeals to the way a physicist thinks. As such the presentation 
could be somewhat intimidating to chemists and biologists who might find the concepts and techniques that I
introduce as foreign, something they have not seen and are not accustomed to, in the context of a problem which
is traditionally viewed as theirs.  However,
I assure you that everything I describe is very simple. A good command of basic algebra is all that it really takes
to follow these lectures. Indeed, proteins brings 
physics together in a unique fashion
with biology, chemistry, 
applied mathematics, even medical research.  Theory and experiments.
With the goal to understand matter that is alive. I hope you {\it catch the bug}, too!

\vskip 0.5cm
\noindent
These lectures are available on-line at
\[
{\tt http://topo-houches.pks.mpg.de/?page_id=255
}
\]

\acknowledgements

I wish to thank my many collaborators, postdocs and students  from whom I have learned so much on the subject of these 
lectures. They include  Si Chen, Alireza Chenani, Maxim Chernodub, Jin Dai,   Ulf Danielsson, Jan Davidsson, Thomas Garandel,  Ivan Gordeliy, Valentin Gordeliy, 
Jianfeng He, Yanzhen Hou, Konrad Hinsen,  Shuangwei Hu, Theodora Ioannidou, Ying Jiang, Gerald Kneller, Andrei 
Krokhotine, Nevena Litova, Jiaojia Liu,  Adam Liwo,  Martin Lundgren,   Gia Maisuradze,   Nora Molkenthin, Alexander 
Molochov, Alexander Nasedkin, Daniel Neiss, Xuan Nguyen, Stam Nicolis, 
Xubiao Peng, Fan Sha,  Harold Scheraga, Adam Sieradzan, Ann Sinelnikova,  Maxim Ulybushev,  Yifan Zhou, and many 
many others (to whom I apologise if name is not mentioned above).  I also thank Frank Wilczek for making me
curious about Anderson's {\cyr{bol}}{\tiny b}{\cyr{she}}.

\vskip 0.2cm

\noindent
My research has been supported by CNRS PEPS grant, Region Centre Recherche d'Initiative Academique grant, 
Sino-French Cai Yuanpei Exchange Program (Partenariat Hubert Curien), Vetenskapsr\aa det, Carl Trygger's Stiftelse 
f\"or vetenskaplig forsk\-ning, and Qian Ren Grant. I thank 
the organisers of the {\it Topological Aspects in Condensed Matter Physics} Claudio Chamon, Mark Goerbig and 
Roderich Moessner  for giving me the opportunity 
to present these lectures at Les Houches. I thank International Institute of Physics in Natal, Brazil 
for hospitality during the writing of  these notes.  Last but not least, thanks to the great audience at Les Houches!

\tableofcontents

\maintext

\chapter{A Protein Minimum}

\section{Why proteins}

Proteins are nano-scale machines that control and operate all metabolic processes in all living organisms. 
Proteins  often have to function with extreme precision: 
Like most  machines, those made of proteins need to have their parts and pieces 
in the right place, in a good shape and finely tuned.
How else could these self-producing nano-machines work in such a great harmony,  
cooperate over an enormous range of scales and 
uphold something as complex as life? 
Indeed, it is widely understood that 
the biological function of a protein depends critically on its three dimensional geometry. 
From this perspective 
the so called  {\it 
protein folding problem} that aims to explain and derive 
the shape of a biologically active protein using laws of physics,
addresses the origin of life itself \cite{Dill-2007,Dill-2012}.

Furthermore, a wrong fold is  a common cause for a protein to lose its function. A 
wrongly folded protein can be dangerous, even fatal, to a biological organism. It is now widely understood that diverse 
neurodegenerative diseases including various forms of dementia such as Alzheimer's and Parkinson's, 
type-II diabetes, and about half of all cancers, 
are caused by wrong folds in certain proteins \cite{Chiti-2006}. At the same time, bacteria are on the rampage  and
emergent resistance through evolutionary processes 
renders existing antibiotics ineffective at a rapid pace \cite{davies-1996,walsh-2000,fischbach-2009}. 
No effective methods 
and treatments have been found to prevent or cure viral maladies like HIV, Ebola, or 
respiratory syndromes such as SARS and MERS. Our future protection against these and various other
harmful and deadly pathogens depends on our 
skills and knowledge to develop conceptually new, protein-level  approaches to fight and eliminate our enemies.
Research on proteins is really about 'Saving the Planet' as much as in any video game or movie 
ever made. But it is for real: By doing research  on proteins you have a change to become 
a real-life  {\it Gordon Freeman}.
  
For all these and many other reasons the ability to accurately describe the physics of proteins, their  structure and 
dynamics, would have an enormous impact  on biology, pharmacy, and health sciences. 
It would provide huge benefits to the society by paving ways to prevent and cure  many tormenting diseases.  
In particular, it would provide us an answer to  {\it what is life} along the lines foreseen by Schr\"odinger.

\vskip 0.3cm
In the following I will give a short introduction to proteins, what you need to get your research 
started as a physicist. For those
who are really seriously interested in
the biological aspects, I recommend the textbook {\it Molecular Biology of the Cell} \cite{Alberts-2014}.

\section{Protein chemistry and the genetic code}

Proteins are one dimensional linear polymers. Proteins are composed of twenty different amino acids
that share a number of structural properties: There are the  {\it backbone} atoms which are common to
all amino acids. There are the {\it residues} or {\it side-chains} which are different for each of the twenty
amino acid.

%
%
%
%
%
%
%
%
%
%
%
\begin{figure}[h]         
\begin{center}            
  \resizebox{6cm}{!}{\includegraphics[]{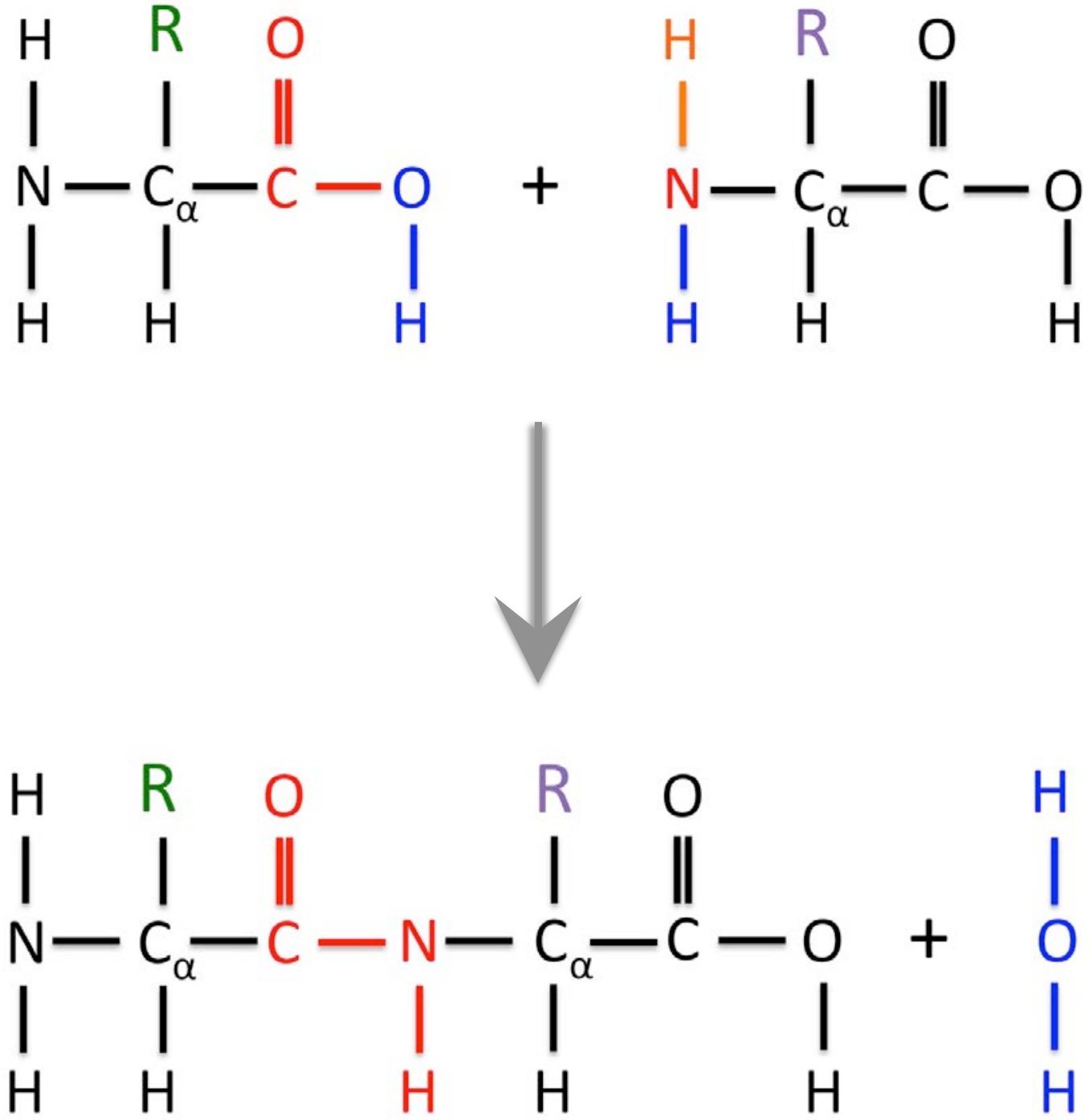}}
\end{center}
\caption{Amino acids have a common backbone with heavy atom pattern $-\mathrm N-\mathrm C-
\mathrm C-\mathrm O-$ but there are also 
twenty different residues (side-chains) which we denote $\mathrm R$. 
When two  amino acids combine together we obtain a
di-peptide, in addition of a water molecule. }   
\label{fig-1}    
\end{figure}
In Figure \ref{fig-1}  we show the chemical composition of a generic amino acid. When two amino acids meet, a chemical process can
take place  that joins them together into a di-peptide plus water, as shown in the Figure. 
When this process repeats itself  we eventually arrive at a long 
polypeptide chain {\it a.k.a.} protein 
as shown in Figure \ref{fig-2}. Once the protein attains the correct shape, it becomes ready for biological action.
%
%
%
%
%
%
%
%
%
%
%
{
\footnotesize
\begin{figure}[h]         
\begin{center}            
  \resizebox{8.5cm}{!}{\includegraphics[]{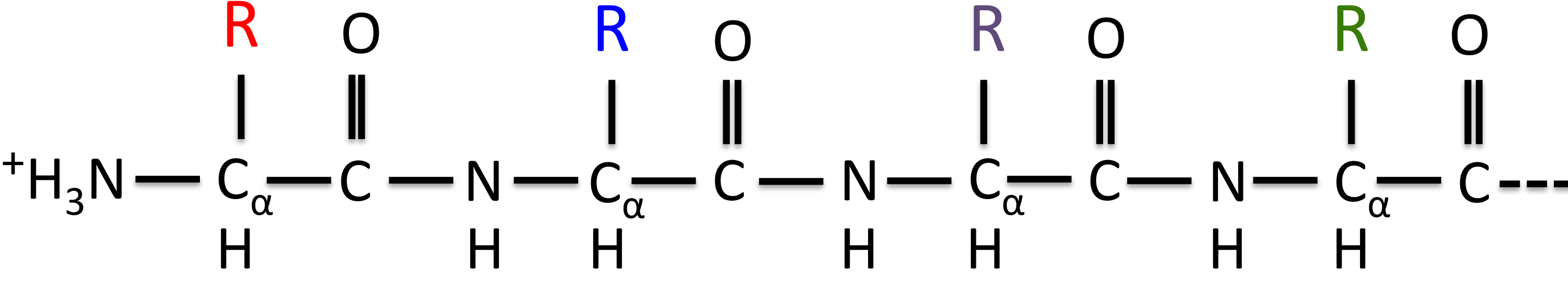}}
\end{center}
\caption{Proteins are long linear chains of amino acids,  each with a self-similar backbone 
structure but twenty different residue structures (R). }   
\label{fig-2}    
\end{figure}
}

Note the carbon atoms that are denoted C$\alpha$  in Figures \ref{fig-1} and \ref{fig-2}. 
These are called the $\alpha$-carbon, and they have
a central r\^ole in  protein structure. As shown in the Figures 
the C$\alpha$ carbons connect the residues to the backbone; the C$\alpha$  
forms the center of a $sp3$-hybridised  tetrahedron which subjects it
to strong steric constraints, and  holds it rigidly in place in relation  to the other atoms.
As we shall find 
the C$\alpha$ carbons largely determine the shape of the
protein. Thus much of our subsequent analysis of protein structure and dynamics 
is based on the central r\^ole of the C$\alpha$, for reasons that become increasingly apparent as we proceed.

In a living organism like you and me, the instructions for making proteins are stored in our {\it genome}.
At the level of DNA the {\it genetic code} consists of a sequence of nucleobases, that connect 
the two strands of DNA.  A group of three nucleobases corresponds to a single amino acid; there is a
segment of DNA, for each protein.
The genetic code is copied from DNA  to RNA in a process  called {\it transcription}. This process, like
all other processed in our body 
is driven by various proteins. Particular proteins  called  enzymes act as 
catalysts to help and control complex biological reactions.  

Our DNA consists of four different  nucleobases. Hence 
there are $4\times 4 \times 4 = 64$ different combinations. But 
two of them   are instructions to {\it stop} the process 
of transcription. Thus we have a total of 62 combinations of nucleobases that encode the 
20 amino acids, the genetic code is degenerate.

\begin{quote}
{\it Research project: From the point of view of physics, we have an appetising
similarity between genetic code where a group of three nucleobases  corresponds to an amino acid,
and the Standard Model where baryons are made of three quarks. 
Can you find a symmetry principle akin the Eightfold Way that relates the $62$ codons to the 20 amino acids?
Hint: A good way to start trying is to follow  \cite{Frappat-1998}.}
\end{quote}

Once formed, the RNA has the mission to
carry the genetic code to a
ribosome. A ribosome is essentially a nano-scale 3D printer. It is made of proteins, and it has the duty to produce
new proteins according to the
instructions given to it by  RNA. The process where ribosome combines amino acids into 
a protein chain is called {\it translation}.

\section{Data banks and experiments}

The amount of data and information which is
available in internet is abundant, both on the genetic code  and on proteins.
Ther are various open-access  libraries both on the sequences and on the structures of proteins; the
amount of data is already more than any single person can possibly ever analyse, and it
continues to increase at an exponential rate.

For those who are mainly  interested 
in biology and related bioinformatics, an excellent resource on protein 
sequences and their biological function is
\begin{equation}
{\tt http://www.uniprot.org/ }
\label{uniprot}
\end{equation}
This data bank contains presently almost 90 million different protein sequences.  As shown in Figure \ref{fig-3} (left)
the number of known sequences grows at a very high,  exponential rate.
%
%
%
%
%
%
%
%
%
%
%
{
\footnotesize
\begin{figure}[h]         
\begin{center}            
  \resizebox{13cm}{!}{\includegraphics[]{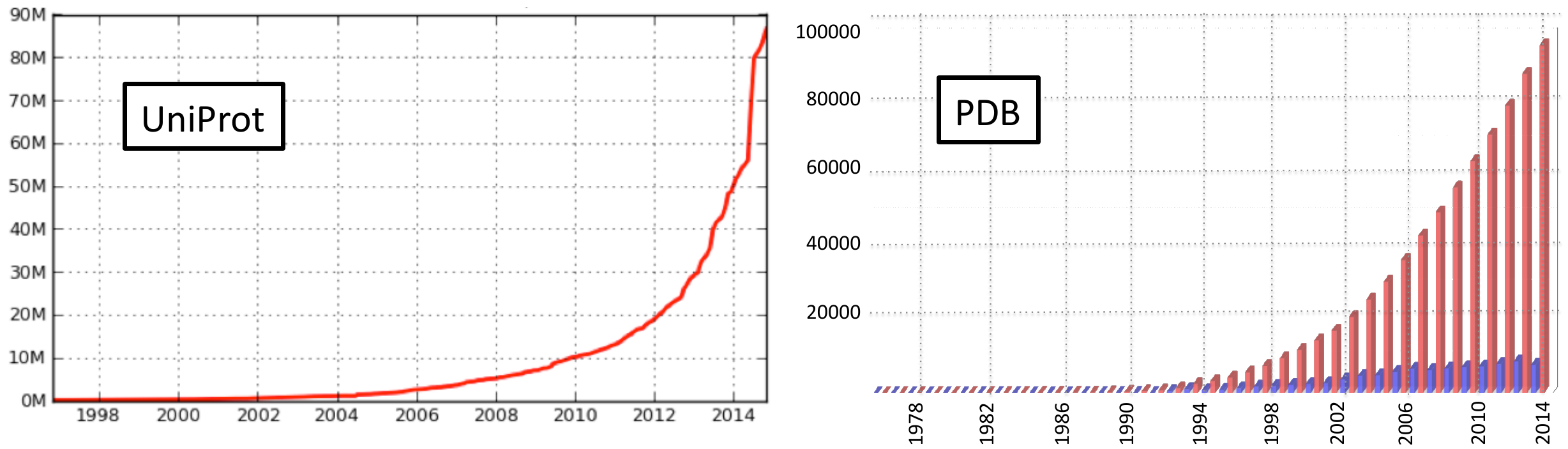}}
\end{center}
\caption{{\it Left}: The increase in the number of sequences in {\it Uniprot}, as a function of year, taken from
(\ref{uniprot}). {\it Right}: The increase in the number of structures in PDB as a function of year. Both annual 
increase and accumulated total are shown.   
Taken from  (\ref{pdb}).}   
\label{fig-3}    
\end{figure}
}

For those who are mainly interested in physics of proteins, 
{\it Protein Data Bank} (PDB) is an excellent resource 
\begin{equation}
{\tt http://www.pdb.org/ }
\label{pdb}
\end{equation}
As shown in Figure \ref{fig-3} (right) the number of known protein structures in PDB is 
around 100.000 and growing. But not at all as fast as the number of known sequences, 
only about $0.1\%$
of known protein sequences have a known structure.

Numerous other good sources of information exist and can be found in internet. For example
{\it PSI-Nature Structural Biology Knowledgebase}  is a comprehensive database
for various structural aspects of proteins. It can be found  at
\begin{equation}
{\tt http://sbkb.org/}
\label{sbkb}
\end{equation}

\vskip 0.2cm
Most of the 100.000  structures in PDB
have been resolved
using x-ray crystallography. But other techniques are also  being used. In particular, the number of  
NMR structures is increasing.  Until now it has been very difficult to resolve 
long protein sequences using NMR techniques. Most NMR structures are quite short, those 
with more than 100 amino acids are rare.  
The advantage of NMR where no crystallisation is needed over x-ray crystallography is,
that NMR can more easily provide dynamical information. It is possible to follow proteins in motion using 
NMR, while crystallised structures have problems moving. But even x-ray techniques such as small 
angle x-ray scattering (SAXS) and wide angle x-ray scattering (WAXS) are now being developed, to observe proteins
in motion. In the near future, equipments including free electron lasers can  provide detailed structural
and dynamical information, at very short time scales. Various other methods are also in use and under development: 
The experimental study of protein structure and  dynamics is still very much in its infancy. This makes the
study of proteins into an exciting  field to enter, also for those who are experimentally minded.  New techniques
are developed and introduced, all the time.

The protein crystals in PDB are {\it ordered}, they are
commonly presumed to display a crystallised conformation which is close
to the biologically active one.  But  {\it most} proteins are apparently  intrinsically 
unstructured. Such proteins can not be crystallised into any kind of  biologically unique conformation.
When these proteins are biologically active, they  do not have any single conformation.  
Instead they change their form, perpetually. Most proteins in our body are like this, intricate
nano-machines that are in a constant action. 
Very little, in fact next-to-nothing, is known on the structural aspects of intrinsically 
unstructured proteins. In these lectures  we shall look at examples of both ordered  
and intrinsically disordered proteins.

Our experimental considerations  will mainly make use of
a subset of crystallographic PDB structures which
have been measured with a ultra-high precision, with a resolution better than 1.0 \AA. 
The reason why we prefer to use these ultra-high resolution structures is due to a process called
{\it refinement} that commonly takes place during  experimental data
validation and model building \cite{Read-2011}. 
During refinement and validation, one iteratively improves the parameters of an 
approximate trial structure of the experimental observations, until one obtains some kind of 
a best fit between the trial structure and the observed diffraction 
pattern. As  an {\it Ansatz}, and as reference,   the process utilises  
known experimental crystallographic structures. Widely used experimentally determined,
highly accurate template libraries of small molecules 
include  the one by Engh and Huber \cite{Engh-2001}. 
Thus, the process of refinement 
{\it might} introduce a bias towards structures that are already known. In particular, it is not clear
to what extent the structure of a small molecule persists in the complex, highly interactive  
environment of a large protein.

Indeed, it is  important to recognise and keep in mind 
that the  PDB data files  are prone to all kinds of errors
\cite{Read-2011,Joosten-2009,Nature-2009,Dauter-2014}.  The data should be 
used with care. {\it Molprobity} 
is an example of a web-server that can be used to analyse the quality of an experimental 
protein structure. It can be found at
\begin{equation}
{\tt http://molprobity.biochem.duke.edu/}
\label{molprobity}
\end{equation}

One might presume that in the case of ultra-high resolution 
structures, those that have been measured with better than 1.0 \AA ~resolution, 
the quality of observed diffraction pattern should be very good. These structures should have  much less
need for refinement, during the model building.  Thus, they should  be   much less biased 
towards known structures. The number of misplaced atoms should be relatively low.

%
%
%
%
%
%
%
%
%
%
%
{
\footnotesize
\begin{figure}[h]         
\begin{center}            
  \resizebox{7cm}{!}{\includegraphics[]{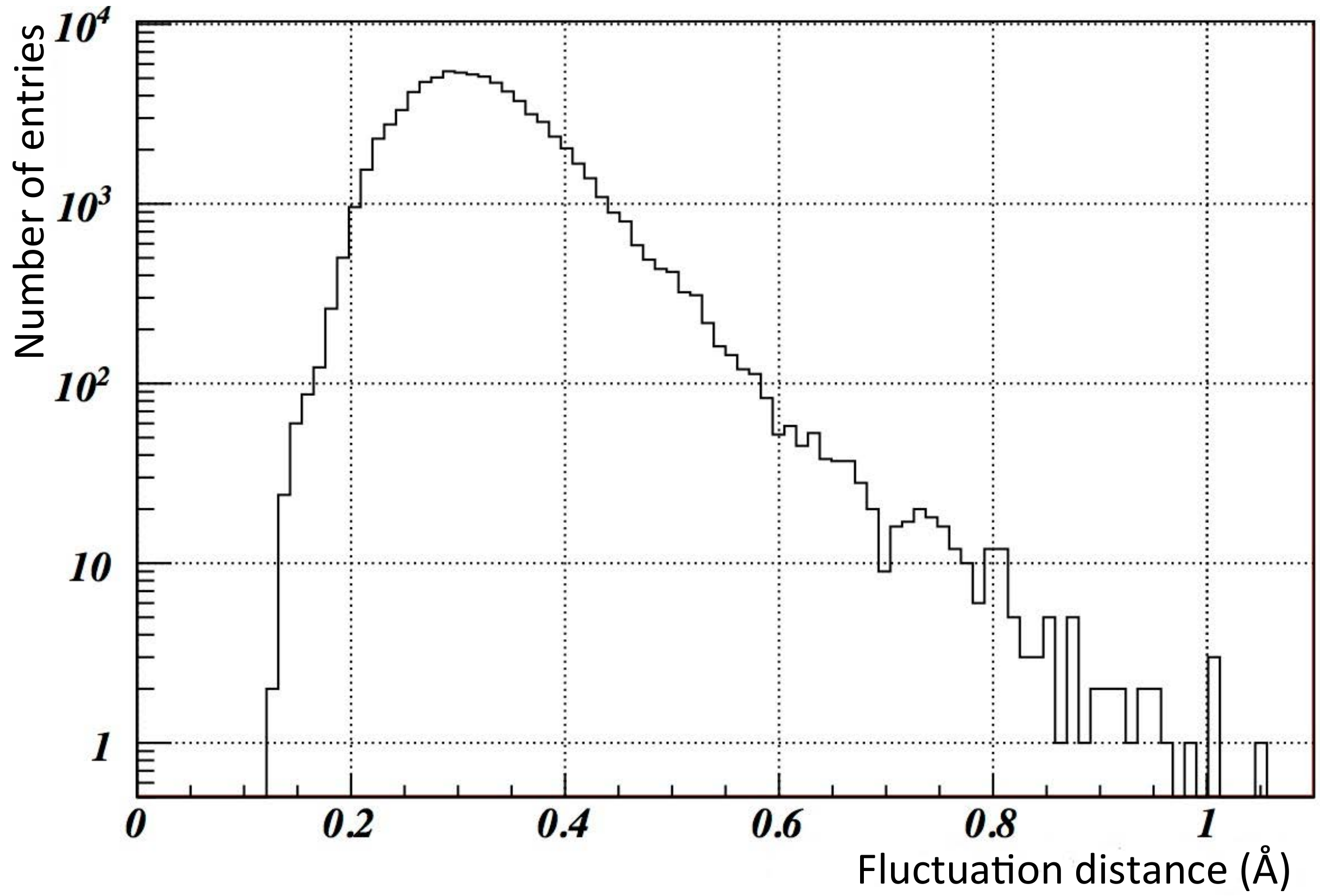}}
\end{center}
\caption{The distribution of the Debye-Waller fluctuation distance for the C$\alpha$ atoms, among those crystallographic PDB structures that have been measured with better than 1.0 \AA~ resolution.}
  \label{fig-4}    
\end{figure}
}

In order to minimise radiation damage, 
crystallographic structures are often measured at temperatures which are near that of 
liquid nitrogen {\it i.e.} around 80-90 Kelvin. Thus the thermal fluctuations in the atomic coordinates should 
be small. In  the PDB data, the experimental uncertainty 
in the atomic coordinates is estimated by  the (temperature) B-factors. Besides the thermal fluctuations,
these  B-factors summarise also all the other uncertainties that the experimentalist thinks affects the precision. 
In Figure \ref{fig-4} we show the distribution of the Debye-Waller fluctuation distance in our subset of 
the ultra-high precision structures, for the  C$\alpha$ atom coordinates. 
The fluctuation distance can be estimated using the Debye-Waller relation,
\begin{equation}
\sqrt{<\! \mathbf x^2 \!> } \ \approx \ \sqrt{ \frac{B}{8\pi^2} }
\label{dw}  
\end{equation}
This corresponds roughly  
to the one standard deviation uncertainty in the experimentally measured coordinate values.  
In Figure \ref{fig-5}  is an example of a generic PDB entry 
that shows how the B-factors are listed   
together with the atomic coordinates. 
%
%
%
%
%
%
%
%
%
%
%
{
\footnotesize
\begin{figure}[h]         
\begin{center}            
  \resizebox{12cm}{!}{\includegraphics[]{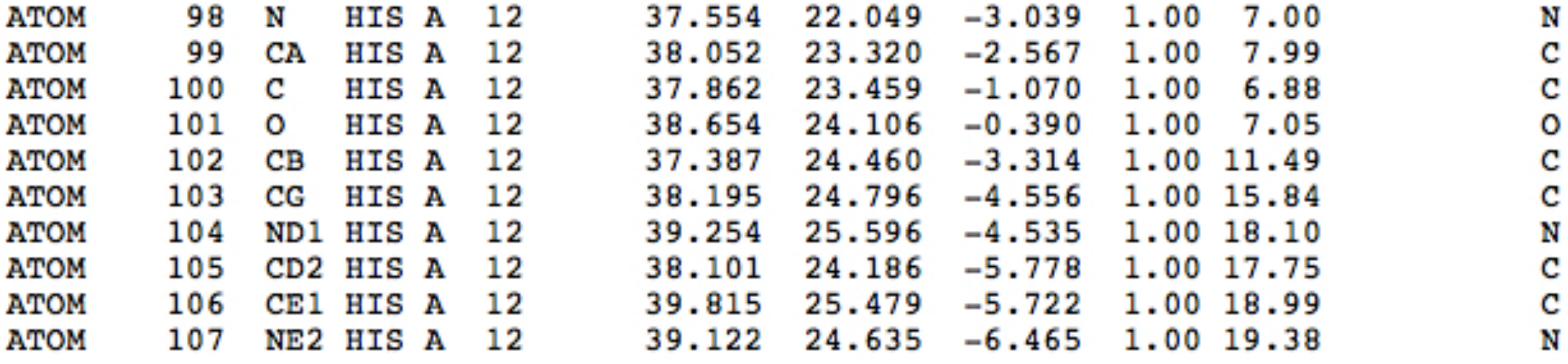}}
\end{center}
\caption{
An example of PDB file, in this case the amino acid histidine (HIS). The second column
lists the atom number (98-107) along the backbone. The third column lists the type of the 
atom; CA stands for C$\alpha$, and 
the entries  98-101 are the backbone
N-C$\alpha$-C-O atoms of HIS. The entries 102-107 are side chain atoms. In this case, HIS is the 12$^{th}$
amino acid along the backbone. The ($x,y,z$) coordinates are listed in the following three columns. The 
B-factors are listed in the last column, before a list of the chemical symbols. Note that in this list, the hydrogen atoms 
are absent. Hydrogens can be difficult to observe.}  
\label{fig-5}    
\end{figure}
}

According to Figure \ref{fig-4}, among our ultra-high resolution PDB structures 
the one standard deviation error distance in the C$\alpha$ atomic positions peaks in the range of 0.3 - 0.5 \AA ngstr\"om.
The lower bound is around 0.15 --  0.2 \AA ngstr\"om  and in these lectures we 
shall adopt $\sim 0.15$ \AA~ 
{\it i.e.} around 20$\%$ of the radius of the carbon atom, as the lower bound 
estimate for the size of
quantum mechanical zero point fluctuations in the C$\alpha$ positions.   Note that historically $\sim$ 0.2 \AA ~ has been  considered as the boundary between
x-rays and $\gamma$-rays.

\section{Phases of proteins}

Like most linear polymers, proteins  have a highly  complex phase structure that 
can depend on a multitude of factors, including
the chemical structure  of a polymer and its solvent, temperature, pressure, changes in solvent's acidity  
and many other environmental factors  \cite{Degennes-1979,Schafer-1999}.  
In a good solvent environment,  the interactions between 
a polymer segment and the solvent  molecules usually cause the polymer to expand, and the polymer behaves 
like a self-avoiding random walk. In a poor solvent environment,  such as water that surrounds proteins 
in our cells,
the polymer-polymer self-interactions dominate and  the polymer tends to collapse 
into a space filling conformation. These two phases are separated by a  $\theta$-regime, 
where the repulsive and attractive interactions cancel each other 
and the polymer has the geometric character of a random walk (Brownian motion).

In the limit where the number $N$ of  monomers  is very large, many aspects of the phase structure
become {\it universal} \cite{Kadanoff-1966,Wilson-1971,Wilson-1974}. 
An example of a universal quantity in the case of a linear polymer such as a protein, 
is the compactness index  $\nu$. It is defined in terms  of 
the radius of gyration $R_g$ \cite{Degennes-1979,Schafer-1999,Huggins-1941,Flory-1942,Li-1995}
\begin{equation}
R^2_g \ = \ \frac{1}{2N^2}  \sum_{i,j} ( {\bf r}_i  
- {\bf r}_j )^2
\label{Rg0}
\end{equation}
Here $ {\bf r}_i $ are the coordinates of all the atoms in the polymer. In the case of a protein, with no loss of
generality we may 
restrict  $\mathbf r_i$ to  the coordinates of the  backbone C$\alpha$ atoms only.
The compactness index $\nu$ governs  the large-$N$ asymptotic form of (\ref{Rg0}). 
When the number $N$ of monomers becomes very large, we have \cite{Li-1995}
\begin{equation}
R^2_g \ \buildrel{N \ {\rm large}}\over{\longrightarrow}  \
R_0^2 N^{2\nu} ( 1 + R_1 N^{-\delta_1} + ... ) 
\label{R}
\end{equation}
It should be obvious that  $\nu$  coincides with the inverse Hausdorff dimension of the structure.   
Besides the compactness index $\nu$, the critical exponents  $\delta_1$ {\it etc.}
are also universal quantities.
But the form factor $R_0$ that characterises the effective distance between 
the monomers in the large $N$  limit, and  the subsequent amplitudes 
$R_1 \ etc. $ that parametrise  the finite size corrections, are not universal \cite{Li-1995}.
These parameters can in principle be computed from the chemical structure of the 
polymer and solvent, in terms of  
environmental factors such as temperature and pressure.

As a universal quantity, $\nu$ is independent of the detailed atomic structure.  
Different values of $\nu$ correspond  to  different phases of polymer.
The four commonly accepted  mean-field values  of $\nu$ are
\begin{equation}
\nu \ =  \ \left\{  \ \, \begin{matrix} 1/3 \\ 1/2  \\ 3/5  \\ 1 \end{matrix} \right.
\label{nuval}
\end{equation}

Under poor solvent conditions such as in the case of proteins in our cells, a linear single chain polymer 
collapses into the space filling 
conformation and we have the mean field exponent $\nu = 1/3$. 
For a fully flexible ideal chain the mean field
value is $\nu = 1/2$. This phase takes place at the $\theta$-regime that 
separates the collapsed phase from the high temperature self-avoiding random walk phase, for which 
we have the mean field Flory 
value $\nu =  3/5$.  Finally,  when $\nu =1$ the polymer is 
like a rigid stick. Some proteins are like this, some collagens for example.

\begin{quote}
{\it Research project: Three dimensional dynamical systems such as the Lorenz equation provide numerous
examples of space curves with attractors that have all kind of Hausdorff dimensions. Can 
you find physical examples of polymers (proteins)
where  $\nu$ takes values that correspond to phases which are different from the four listed in
(\ref{nuval})?} 
\end{quote}

The mean-field values of the critical exponents $\nu$, $\delta_1$ {\it etc.} in equation (\ref{R})
may be corrected by fluctuations. In particular,  in the universality class of the self-avoiding random walk
the improved values are  \cite{Degennes-1972,Leguillou-1980}
\begin{equation}
\begin{matrix} \nu & = & 0.5880 \pm 0.0015 \\
\delta_1 & = & 0.47 \pm 0.03 \end{matrix}
\label{zinn1}
\end{equation}
The computation of (\ref{zinn1})  in  \cite{Degennes-1972,Leguillou-1980}  utilises the concept of
universality  to
argue that the three dimensional self-avoiding random walk  is in same universality class with the 
$O(n)$ symmetric  scalar field theory with a quartic self-interaction, in the limit where the number of 
components $n \to 0$. A subsequent numerical Monte Carlo 
evaluation of the critical exponents (\ref{zinn1}), computed using a self-avoiding random walk model on 
a square lattice \cite{Li-1995} gave very similar 
values 
\begin{equation}
\begin{matrix} \nu &  =  & 0.5877 \pm 0.0006 \\  
\delta_1 & = & 0.56 \pm 0.03 \end{matrix}
\label{zinn2}
\end{equation}


In  the case of crystallographic PDB protein structures,  we  may evaluate the dependence of 
the radius of gyration  on the number of residues using  the coordinates of the C$\alpha$
atoms. The result   is shown in  Figure \ref{fig-6}. 
%
%
%
%
%
%
%
%
%
%
%
{
%
%
\begin{figure}[h]         
\begin{center}            
  \resizebox{6.5cm}{!}{\includegraphics[]{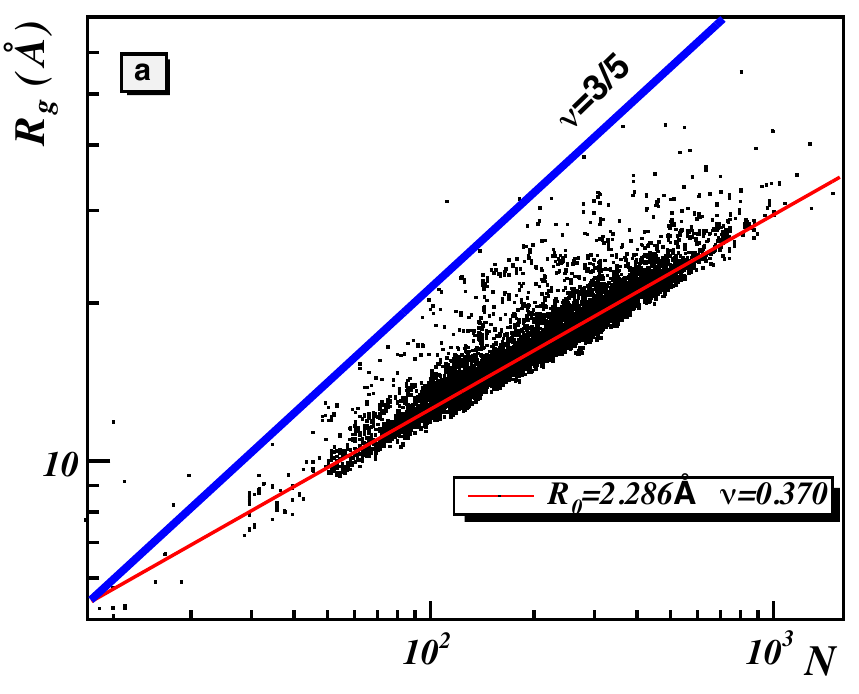}}
\end{center}
\caption{The C$\alpha$ radius of gyration as a function of residues, in the case of those monomeric 
crystallographic PDB proteins  that have been measured with better than 2.0 \AA~ resolution.  The red line is the
least-square linear fit, and the blue line denotes the Flory value $\nu = 3/5$.}   
\label{fig-6}    
\end{figure}
}
A least-square fit to the data gives 
\begin{equation}
R_g \ \approx \ R_0 N^{\nu} \ \approx \  2.280 \, N^{0.375} \ \ \ {\rm \AA}
\label{Rg}
\end{equation}
Note that proteins are not homopolymers. But when $N$ increases,  the detailed amino acid structure 
of a protein should become increasingly irrelevant in determining the relation between the 
radius of gyration and the number of residues. For long protein chains, the 
inhomogeneity due to amino acids should be treated as a  finite size 
correction in (\ref{R}), when $N$ becomes very large.  Indeed, in the limit of very large
number of residues a generic protein is like a chain along which the 20 residues have been quite
randomly distributed. It should be  like a spin-chain embedded in $\mathbb R^3$
where each residue is a spin variable, with 20 different (random) values. Thus, when the 
ratio $20/N$ becomes very small the effect of an
individual residue  becomes small in an average, statistical sense. 
The protein approaches a homopolymer that
is equipped with an "averaged" residue.

\begin{quote}
{\it Research project: Develop theory of spin chains embedded in $\mathbb R^3$. } 
\end{quote}

\section{Backbone geometry}

According to Figure \ref{fig-6}, those proteins that can be crystallised are in the collapsed $\nu \approx 1/3$
phase. To describe the properties of their 
thermodynamical phase state, we need to identify a proper set of
{\it order parameters} in the sense of Landau, Ginzburg  and Wilson;
the concept of order parameter is 
described in numerous textbooks.\footnotemark\footnotetext{In the context of protein research,
order parameters are sometimes called {\it reaction coordinates}.}
A local order parameter is a systematically constructed effective dynamical 
variable, that describes collectively a set of elemental degrees of freedom such as atoms and molecules,
in a system that is subject to the laws of statistical physics. Examples of an order 
parameter include magnetisation in the case of a ferromagnet, director in a nematic
crystal, condensate wave function in superfluid $He^4$, and Cooper pair in a BCS superconductor. 
The concept of an order parameter is often intimately related to the concept of 
symmetry breaking and emergent phenomena. For example,  
in each of the cases that we mentioned, we have a symmetry that
becomes broken, and this symmetry breaking
gives rise to emergent structures. In particular, 
the breaking of the symmetry is described in terms of the properties of the
pertinent order parameter, in each case.

In the case of a protein, we have already concluded that the phase structure relates to the aspects of 
protein geometry. The different
phases of a protein are characterised by different Hausdorff dimensions (\ref{nuval}). 
Moreover, proteins have an  apparent
symmetry that has  become broken: 

\vskip 0.2cm
Amino acids are chiral molecules. An amino acid can be either left-handed ({\tt L}) or right-handed ({\tt D}).
The only exception is glycine which has no chirality. For two amino acids to form a di-peptide as shown
in Figure \ref{fig-1},  they must have the same chirality; you can't easily shake someones left hand with your
own right.  For some reason the symmetry
between {\tt L} and {\tt D} is broken in Nature, practically all amino acids that appear in proteins of living
organisms from prokaryotes such as bacteria to eukaryotes
like us, are left-handed chiral.  
This symmetry breaking  apparently reflects itself to higher level geometric structures of proteins:  As a 
polymer chain, the proteins that are found in living organisms are more often  twisted in a {\it right-handed} 
manner, that the opposite. Thus, any local order parameter that describes the phase 
properties of proteins in living organism,
should somehow capture the helical aspects of protein geometry.
\vskip 0.2cm

%
%
%
%
%
%
%
%
%
%
%
{
\footnotesize
\begin{figure}[h]         
\begin{center}            
  \resizebox{10cm}{!}{\includegraphics[]{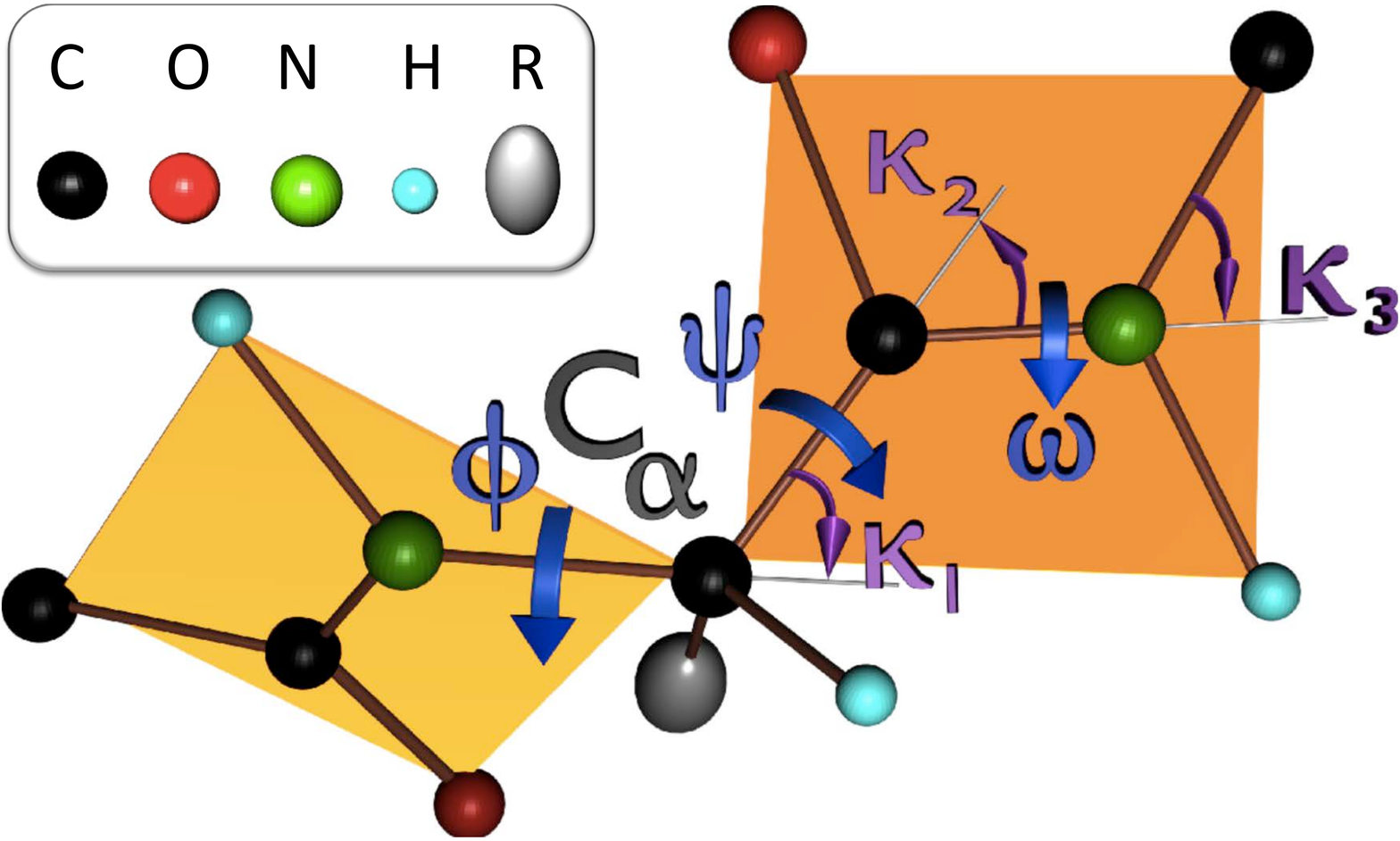}}
\end{center}
\caption{The atoms along the protein backbone form peptide planes; the two covalent bonds C=O and N-H are 
anti-parallel. The figure also shows the definition of the various bond and torsion angles, between the covalent
bonds. The two torsion angles ($\phi, \psi$) are the Ramachandran angles.  
}   
\label{fig-7}    
\end{figure}
}
Figure \ref{fig-7}  details the local geometry  of a protein. In this Figure we identify a  C$\alpha$ atom
together with its covalently bonded N, C, H atoms, and the residue R that starts with a covalently 
bonded carbon atom called C$\beta$. 
The covalent bonds between these 
five atoms form a $sp3$ hybridised  tetrahedron, with C$\alpha$ at the centre. Take the C atom
to be the top of the tetrahedron, and N, H, R  as the three bottom vertices. Consider the axis of the tetrahedron
that runs along the covalent bond from the C to C$\alpha$ and look down this axis from C towards C$\alpha$.
If the H atom is in the clockwise direction from residue R, the amino acid is left-handed; this
is the case in Figure \ref{fig-7}. 
Otherwise, the amino acid  is right-handed.

In Figure \ref{fig-7} we have also
two  {\it peptide planes}, one which is prior to the C$\alpha$ atom and another 
which is after the C$\alpha$ atom. The C$\alpha$ atom 
is located at the joint vertex of the two adjacent peptide planes. 
We proceed to analyse in detail the properties of the peptide plane geometry.

It turns out that  the  geometry of the peptide planes is  indeed {\it very} rigidly planar. The planarity
is measured by the angle $\omega$ which is shown in Figure \ref{fig-7}; it 
is the angle between the C=O covalent bond, and the N-C$\alpha$ covalent bond (or N-H covalent bond)
The values of $\omega$ are found to fluctuate {\it very} little around $\omega = \pi$,
which corresponds to the {\it trans}-conformation of the backbone and is shown in the Figure; there are a few entries, mainly 
involving the amino acid {\it proline}, where the backbone is in the {\it cis}-conformation where
$\omega$ vanishes. 
The  {\it cis}-conformation is  equally planar, 
the fluctuations around   $\omega = 0$ are minimal. 
Figure \ref{fig-8} shows the distribution of the $\omega$ angles in our ultra high resolution subset of
crystallographic
protein structures, those  that have been measured with better than 1.0 \AA ~ resolution. 
%
%
%
%
%
%
%
%
%
%
%
{
\footnotesize
\begin{figure}[h]         
\begin{center}            
  \resizebox{12cm}{!}{\includegraphics[]{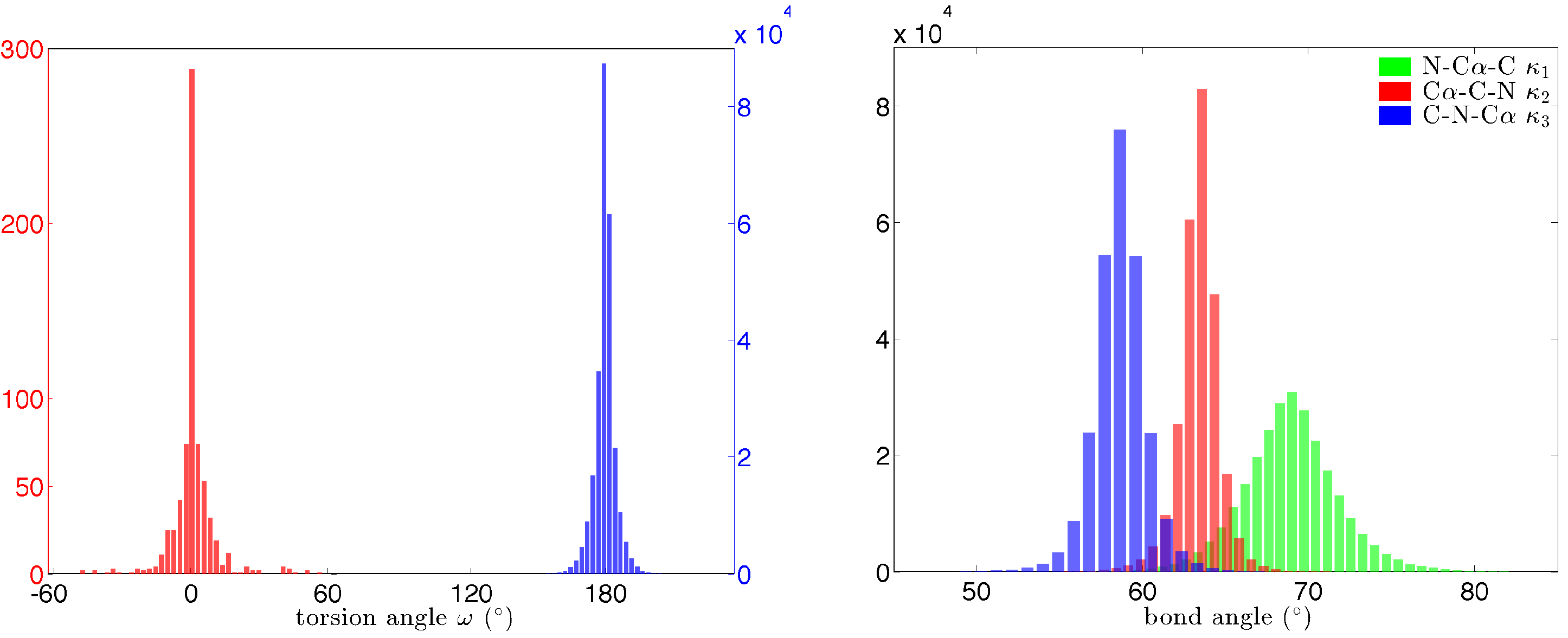}}
\end{center}
\caption{Left: Distribution of the torsion angle $\omega$ between the covalent bonds C=O and N-H in Figure \ref{fig-6},
in our ultra high 1.0 \AA~ resolution PDB subset. The result shows that the geometry of the peptide planes
is indeed  planar, with very high precision. For {\it trans} we have $\omega \approx \pi$ and
for {\it cis} we have $\omega \approx 0$.  Right: The distribution of the three bond angles ($\kappa_1, \kappa_2, \kappa_3$)
defined in Figure \ref{fig-7} in our data set. The variation around the average values is relatively small.}   
\label{fig-8}    
\end{figure}
}
In addition, the values of the three covalent bond angles ($\kappa_1, \kappa_2, \kappa_3$) that are defined in Figure \ref{fig-7} are shown in this Figure. Their values are likewise  subject to relatively small variations.

The various covalent bond lengths along the protein backbone have also values that fluctuate very little
around their average values. Figure  \ref{fig-9} shows the various bond lengths between the heavy atoms along
the backbone. We also show the distance between two consecutive C$\alpha$ atoms, which is also subject to
very small fluctuations.
%
%
%
%
%
%
%
%
%
%
%
{
\footnotesize
\begin{figure}[h]         
\begin{center}            
  \resizebox{12cm}{!}{\includegraphics[]{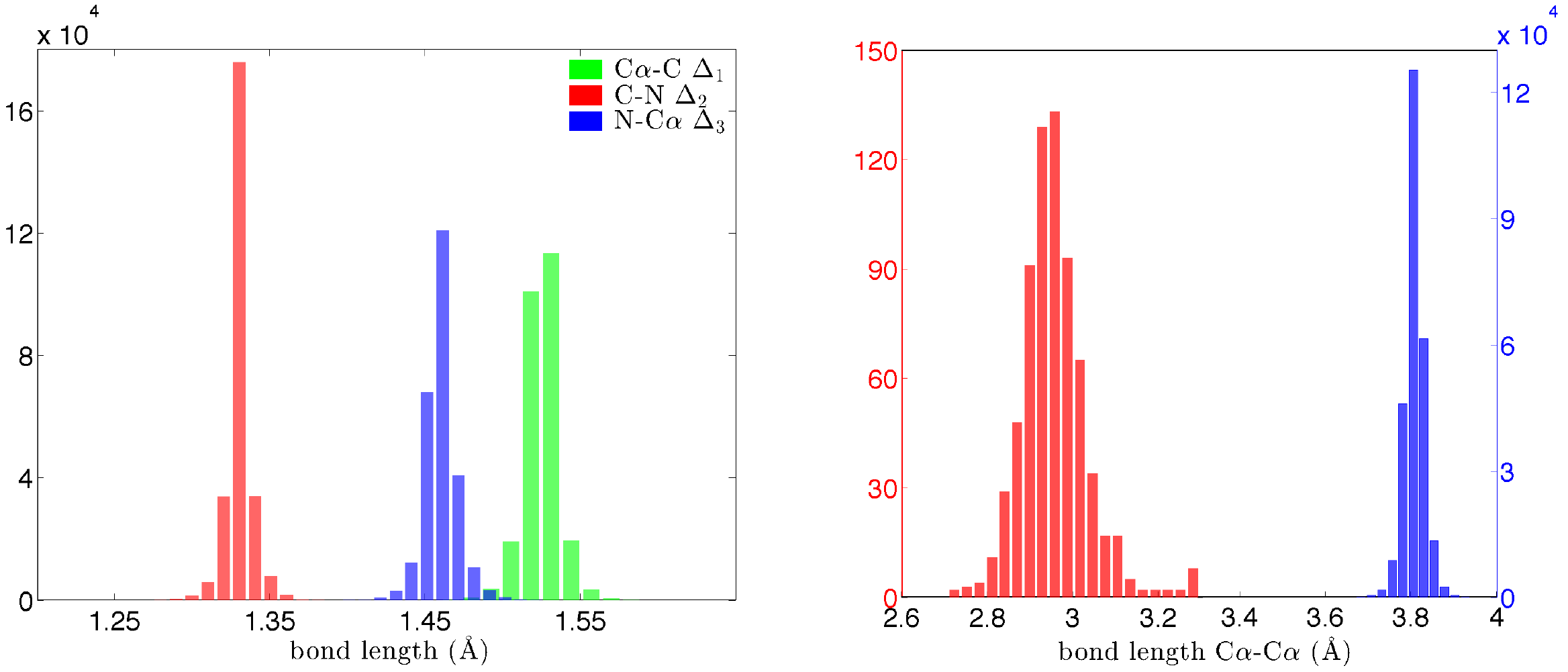}}
\end{center}
\caption{Left: Distribution of the three covalent bond lengths C$\alpha$-C, C-N and N-C$\alpha$ shown
in Figure \ref{fig-7}, along the protein backbone.  Right: Distribution of the length between  neighboring 
C$\alpha$ atoms, along the protein backbone. The smaller value $\sim 3.0$ \AA~ corresponds to {\it cis},
and the larger value  $\sim 3.8$ \AA~ is for {\it trans}. 
}   
\label{fig-9}    
\end{figure}
}

\section{Ramachandran angles}

The Figures \ref{fig-8} and \ref{fig-9} show that the three covalent bond angles ($\kappa_1, \kappa_2, \kappa_3$),
the torsion angle $\omega$ and the various covalent bond lengths reveal no dependence on local
geometry. Each of these variables have fairly uniform distributions, which is apparently quite 
insensitive to variations in local backbone geometry.   
Thus we are left  with only the two Ramachandran angles ($\phi,\psi$) in Figure \ref{fig-7}
as  the potential local order parameters, to characterise local geometry along the backbone. 
Indeed,  it turns  out that  the variations in their values are {\it not} small, and in particular
appears to depend on backbone geometry.
This is shown by the Ramachandran map in Figure \ref{fig-10}, that displays the 
($\phi,\psi$) distribution in PDB structures which have been measured with better than 2.0 \AA ~ resolution. 
%
%
%
%
%
%
%
%
%
%
%
{
\footnotesize
\begin{figure}[h]         
\begin{center}            
  \resizebox{6cm}{!}{\includegraphics[]{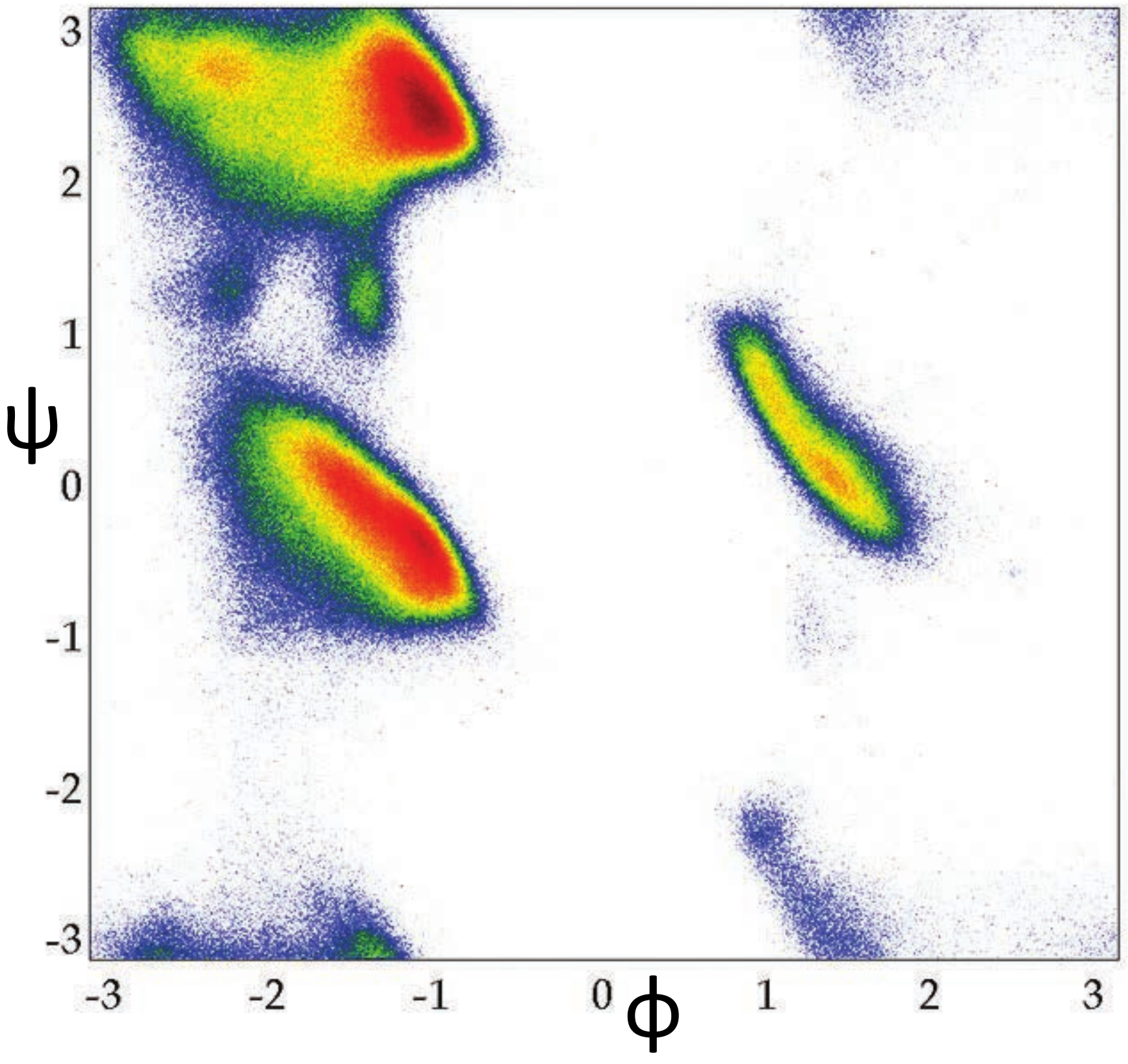}}
\end{center}
\caption{Distribution of Ramachandran angles ($\phi,\psi$) defined in Figure \ref{fig-7} in radians. 
}   
\label{fig-10}    
\end{figure}
}
Note the asymmetry, both in $\phi$ and in $\psi$; this asymmetry translates into helicity of the protein backbone. 
Regular right-handed helical structures (right-handed $\alpha$-helices) are quite  
common, while left-handed helical structures are very rare.

Since the phase structure of a protein relates to its geometry, we may expect that 
the set of the two Ramachandran angles could be utilised as the local order parameters to
describe the phase structure of  proteins. 
However, it turns out that this is not the case \cite{Hinsen-2013}: The Ramachandran angles form an incomplete set of
local order parameters. To show this, we consider all the PDB structures in 
our data set of ultra-high resolution structures, those that have been measured with better than 1.0 \AA~ resolution. 
We do the following {\it reconstruction}:

From the PDB coordinates of the atoms we first compute the numerical values of all the 
Ramachandran angles, for each and every peptide plane. Then we continue and do the inverse: We start from 
the Ramachandran angles that we have computed, and we assume that all the other angles 
and bond lengths have their average values. We then try to reconstruct the original protein structure, by computing
the C$\alpha$ coordinates. 

It turns out that this reconstruction of the coordinates, going from C$\alpha$ to Ramachandran and the back, usually fails. It is not possible 
to do such a reconstruction, 
in the case of a generic protein.
%
%
%
%
%
%
%
%
%
%
%
{
\footnotesize
\begin{figure}[h]         
\begin{center}            
  \resizebox{6cm}{!}{\includegraphics[]{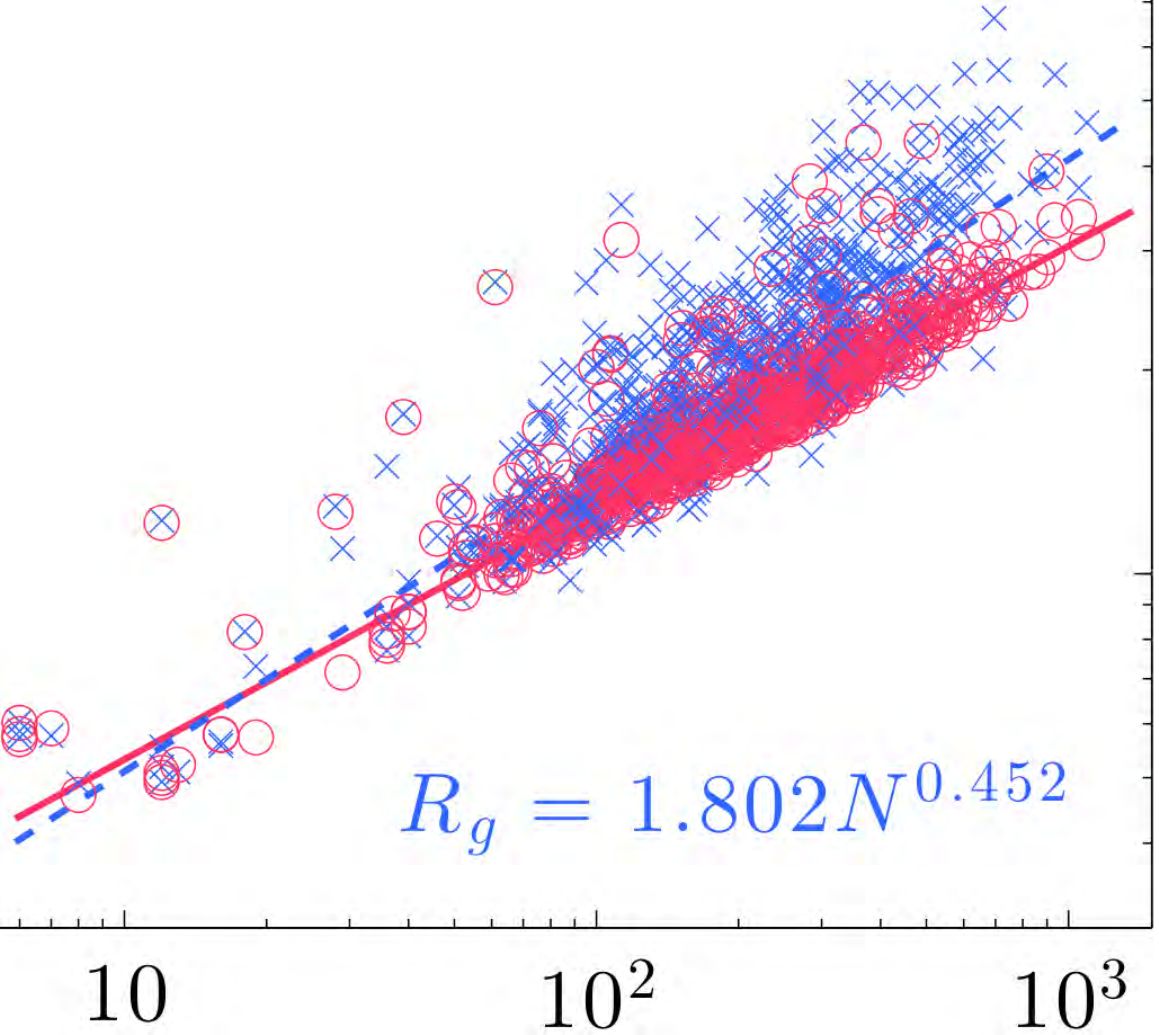}}
\end{center}
\caption{Comparison of $R_g$ between the PDB structures, and the structures obtained by
a reconstruction which is based on Ramachandran angles, with all other backbone angles and
bond lengths set to their ideal values. Red circles are the original PDB structures, blue crosses  
are the reconstructed structures.
}   
\label{fig-11}    
\end{figure}
}

In Figure \ref{fig-11}  we show how the reconstructed proteins fail to reproduce even
the correct fractal geometry of folded proteins  \cite{Hinsen-2013}. Instead of (\ref{Rg}), 
we obtain the asymptotic relation
\[
R_g \ \approx \ 1.8 \, N^{0.45} \ \ {\rm \AA}
\]
for the reconstructed protein structures:  The inverse Hausdorff dimension $\nu = 0.45$ is incorrect.
In order to reconstruct the correct fractal geometry, it turns out that we need to include {\it all} the angular
variables in Figure \ref{fig-7}, as variables. Only for the bond lengths, can the average PDB values  be used.

Eventually, in a subsequent lecture, we shall construct a {\it different}  set of local 
order parameters and we show their completeness. However, we first address the modelling of
proteins in terms of their (primary) full atom level description.

\section{Homology modelling}

As shown in Figure \ref{fig-3} (left),
the number of sequences in {\it Uniprot} grows at a rate  which is much higher  than that of structures in PDB,  
shown in Figure \ref{fig-3} (right). 
The gap  is  enormous and keep on growing: Sequencing is now fast, cheap and routine while 
crystallographic protein structure determination is difficult, time consuming, and often very
expensive; apparently the average cost of a PDB structure is around 100 kUSD. 
Thus it is  impossible to close the
gap between sequences and structures by purely experimental methods. To close this gap 
we need to develop efficient and accurate computational methods.

Homology\footnotemark\footnotetext{Homology between two proteins is commonly measured on the basis of amino acid
sequence similarity. High sequence similarity is a sign of  shared ancestry, but for short
proteins it can also occur due to chance.} 
 modelling \cite{Marti-2000,Schwede-2003,Zhang-2009} 
together with other comparative modelling techniques, is presently  
the most reliable and effective approach to generate a three-dimensional model of 
a protein structure from the knowledge of its amino acid sequence. These  
methods are consistently the top performers in the bi-annual 
{\it Critical Assessment of protein Structure Prediction} (CASP) tests, see 
\[
{\tt http://predictioncenter.org/}
\]
Homology modelling techniques aim to construct the atomic level structure of a given 
protein (target), by comparing its amino acid sequence to various 
libraries of known, homologically related
protein structures (templates). Apparently the reason why this kind of methods work is as follows:
It seems to be the case that  the 
number of possible protein folds in nature is limited,   and that the
three-dimensional protein structures seem  to be better conserved than the amino acid 
sequences \cite{Chothia-1992}. However, the quality of a model obtained for the target  is largely dictated 
by the evolutionary distance between the target  and the available template structures. 
If no closely homologous template can be found, these approaches typically  fail. 

\vskip 0.2cm 

From the point of view of physics, any template based approach has the disadvantage that 
there is no energy function. Thus, no energetic analyses of structure and dynamics can
be performed. For this, other techniques need to be introduced:  To understand
the physical properties of a protein we need to know the energy function.

\begin{quote}
{\it Research project: Try to develop a structure prediction approach that uses the 
best of both worlds: One that finds an  initial Ansatz structure using templates, and then 
develops it using techniques of physics. For hints, continue reading ... } 
\end{quote}

\section{All-atom models}

\begin{quote}
{\it  ...  if we were organisms so
sensitive that a single atom, or even a few atoms,
could make a perceptible impression on our
senses -Heavens, what would life be like! \rm (E. Schr\"odinger)}
\end{quote}

We present a short overview of all-atom molecular dynamics, where impressive progress is being made.
The aim of an all-atom approach  is to model the way a protein folds, 
by following the time evolution of each and
every atom involved including those of the surrounding solvent (water). But
despite impressive progress, the subject remains very much under development and provides enormous
challenges  to those brave enough to face them:
Both conceptual level and technical level problems remain to be resolved, involving 
issues relevant to  physics, chemistry, computer science, and also problems for 
those interested in optimisation and efficient algorithm development.

There are many examples of all-atom energy functions, called force fields in this context. 
The most widely used are {\it Charmm} \cite{Charmm} and {\it Amber} \cite{Amber}; we also 
mention {\it Gromacs} \cite{Gromacs} which is a 
popular platform for performing molecular dynamics (MD) simulations using different force fields. 

A typical all atom force field that describes  a protein chain with $N$ atoms has the following generic form 
\begin{equation}
E(r^N) = \sum\limits_{bonds} \frac{k_b}{2} ( l - l_{0})^2 + 
\sum\limits_{angles} \frac{k_a}{2} ( \kappa - \kappa_{0})^2
+ \sum\limits_{torsions} \frac{V_n}{2} \left[  1 + \cos( n \omega - \gamma )\right)]  
\label{hos}
\end{equation}
\begin{equation}
+ \sum\limits_{j=1}^{N-1}\sum\limits_{i=j+1}^{N} \left\{ \epsilon_{i,j} \left[ \left( \frac{r_{0ij}}{r_{ij}}\right)^2 -
2\left( \frac{ r_{0ij}}{r_{ij}}\right)^6 \right] + \frac{q_i q_j}{4\pi \epsilon_0 r_{ij}}\right\}
\label{ff}
\end{equation}
The first two contributions in (\ref{hos}) describe harmonic oscillations  of the bond lengths and bond angles
around certain {\it ideal values} ($l_0, \kappa_0$). The third contribution involves torsion angles and evokes 
a mathematical pendulum, similarly with {\it ideal value} ground state(s) given by $\gamma$. 
Torsion angles $\omega$ 
are often  found to be much more flexible than bond angles $\kappa$, thus 
the numerical values of $V_n$ are  commonly orders of magnitude smaller than those of $k_a$. Moreover, 
a mathematical pendulum which is used for the torsion angles in lieu of a harmonic oscillator, allows for 
larger amplitude  motions and multiple equilibrium states, which is consistent with their more flexible character. 

The second contribution (\ref{ff}) involves the 6-12 Lennard-Jones potential that  
approximates the interaction between a pair of neutral atoms; the form is chosen for computational efficiency. 
Finally, we have
the Coulomb interaction. In practice, long-range interactions (\ref{ff}) 
are cut off beyond a distance around 10 \AA ngstr\"om, again for computational efficiency.
The "ideal" values of the various parameters are usually determined by a process of optimisation, using 
comparative simulations. For  parameter fine tuning, one may use
very short peptides that have accurately known experimental structures. Such structures  can be  
found for example in the Engh-Huber library \cite{Engh-2001}.

In a full all-atom MD simulation one solves the Newton's equation of motion with (\ref{hos}), (\ref{ff}) in an environment
of water which in a full all-atom approach is also treated explicitly. We note that {\it e.g.} the $r^{-12}$ 
term in (\ref{ff}) models short range Pauli repulsion due to overlapping 
electron orbitals, and the  $r^{-6}$ term emerges from  long range van der Waals interactions. Thus these terms
have a quantum mechanical origin, and  the proper interpretation of the ensuing Newton's equation is
not in terms of a classical equation but in terms of a semi-classical one.

A MD simulation solves the Newton's equation iteratively, with a time-step $\Delta \tau$. This time-step
should be short in comparison to the shortest time scale $\Delta t_{min}$, 
that characterises the fastest atomic level oscillations.
The ratio of the two defines a dimensionless 
expansion parameter. For good convergence of the iterative, discretised Newton's 
equation we should have
\begin{equation}
\frac{\Delta \tau}{\Delta t_{min}} << 1
\label{coup}
\end{equation}
This implies that $\Delta \tau$ should be no more than a few femto-seconds: 
The modelling of protein folding  in an all-atom MD is conceptually  a weak-coupling  expansion
in terms of the dimensionless ratio (\ref{coup}).

\begin{quote}
{ \it Research project: The Newton's equation for mathematical pendulum
\[
\ddot \omega = V \sin\omega
\]
is integrable. Its naive discretisation 
\[
\ddot \omega \ \to \frac{1}{\epsilon^2} (\omega_{n+1} - 2\omega_n + \omega_{n-1} ) 
\]
yields the so-called standard map \cite{Ott-2002}
\[
 \omega_{n+1} - 2 \omega_n + \omega_{n-1} = \epsilon^2 V \sin\omega
\]
which is not integrable.  However, an  integrable discretisation of
the mathematical pendulum is known. See for example  Chapter VIII of ref. \cite{Korepin-1997}.  
Find which one describes protein folding more accurately.
 }
 \end{quote}

\section{All-atom simulations}

Enormous  computational resources have been developed and dedicated to solve 
the {\it protein folding problem} \cite{Dill-2007,Dill-2012}. 
From the point of view of molecular dynamics this
amounts to a numerical simulation
of the all-atom time evolution in a protein, from a random chain initial condition
to the natively folded conformation
using a force field such as (\ref{hos}), (\ref{ff}).
For example,
the {\it Blue Gene} family of supercomputers was originally designed by IBM to address
the problem of protein folding and gene development, which explains  the name. Subsequently special purpose
computers have been constructed to address the folding problem, at all-atom level of scrutiny. 
Thus far the most powerful is the
{\it Anton} supercomputer, built by D.E. Shaw Research \cite{Shaw-2008,Shaw-2009,Lindorff-2011}. In 
the case of relatively short proteins, {\it Anton}
can produce a few microseconds of {\it in vitro} folding trajectory  per day {\it in silico} \cite{Shaw-2009}; this is
about 3-4 
orders of magnitude more than a Blue Gene can achieve.
In a series of MD simulations of 12 fast-folding proteins, from chignon with 10 residues to genetically modified
$\lambda$-repressor with 80 residues, and
with each protein capable of folding within a number of micro seconds {\it in vitro}, {\it Anton} was able to produce 
dynamical trajectories that reproduced the experimentally observed folded  structures, in some examples 
with an impressive precision \cite{Lindorff-2011}.  At the moment, 
these results set the benchmark for all-atom protein folding simulations.
But despite the encouraging results obtained by {\it Anton}, several issues remain to be
overcome before proteins can be routinely folded at all-atom level, starting from a knowledge 
of the amino acid sequence only. We name a few, as a  challenge for future research:

\vskip 0.2cm
$\bullet~$ For the majority of proteins it takes much longer than a millisecond or so   to fold. For example
myoglobin, which is probably the most widely studied protein and that we shall fold in the sequel, 
needs about 2.5 seconds {\it in vitro} to reach its native state when it starts from 
a random chain initial condition. Thus, it would take at least 1.000 years to 
simulate the folding of myoglobin  at the level of all-atoms, using presently available computers.

\vskip 0.2cm
$\bullet~$ In an all-atom MD simulation with explicit water, an increase in the number of water molecules 
quickly exhausts the capacity of presently available computers. For example, in the case of the 80 residue
$\lambda$-repressor mutant simulation using  {\it Anton}, only around 11.000 explicit water molecules could be included.
This should be contrasted to the physiological conditions: The normal pH of blood plasma is around 7.4. 
Since  pH is defined as the $\log_{10}$  of the reciprocal of the hydrogen ion activity  
in a solution, this translates to an average of one proton per $10^{7.4}$  water molecules. Protein folding
is strongly affected by pH; proteins have different natively folded states, at different pH values. Thus, it remains 
a formidable task to describe physiologically relevant pH environments in a truly all-atom manner. 

\vskip 0.2cm
$\bullet$ All-atom force fields utilise a quadratic, harmonic oscillator approximation around
the ideal values of the bond lengths and angles (\ref{hos}); mathematical pendulum in the case of torsional angles.
As long as the atomic fluctuations around the 
ideal values remain {\it very} small, this is a decent approximation.  But whenever the atoms deviate from 
their ideal positions more than "just a little",  higher order non-linear corrections are inevitably present and can not
be ignored.  The existing all-atom force fields are not designed to account for this. The force fields
are not built to describe protein conformations in a realistic manner,
whenever the lengths and angles are not {\it very} close to their ideal values.

\section{Thermostats}

\begin{quote}
{\it This section describes a technical aspect, that is not needed in the rest of the
lectures. We add this section as we feel it addresses a highly important yet 
still open theoretical issue that needs to be addressed by any all-atom modelling approach.}
\end{quote}

Finally, we have the theoretically highly interesting problem of {\it thermostatting}:
An all-atom simulation solves the Newton's equation. As a consequence 
energy is conserved and we have a microcanonical ensemble. 
But in a living cell the energy is not conserved, instead  temperature is fixed.
Accordingly proteins in living organisms should be described in terms of a canonical ensemble. 
One needs to convert the microcanonical ensemble which is described by the all-atom Newton's equation,
into a canonical one. To achieve a conversion, one  adds {\it thermostats} to the system.  
A thermostat models an environment which maintains its own 
temperature constant while interacting with the system of interest:
We refer to \cite{Gallavotti-2014} for a detailed description of thermostats.

The Langevin equation is an example of a thermostat which is well grounded in physical principles.
In the case of uniform damping we write it as follows,
\begin{equation}
\ddot{ \mathbf x_i} \ = \ -\nabla_i E(\mathbf x)  - \lambda \dot{\mathbf x}_i + \mathbf F_i(t)
\label{lang1}
\end{equation}
\begin{equation}
< \! \mathbf F_i(t) \!> = 0 \ \ \ \ \ \& \ \ \ \ \ <\! \mathbf F_i(t)\cdot  \mathbf F_j(s) \!> \ = \ \lambda k_B T \delta_{ij}(t-s)
\label{lang2}
\end{equation}
The derivation of the Langevin equation assumes 
two sets of variables, {\it a.k.a.} particles: There are light, small  particles that are fast.  And there are
heavy  particles
that are slow. The Langevin equation 
describes the dynamics of the latter,
in the limit where the ensuing particles are much slower and heavier than the small and fast ones.
The derivation is based 
on standard arguments of statistical physics. Thus, the Langevin equation
is {\it a priori} a well grounded and rigorous method to introduce temperature into a Newtonian system. 

In the case of all-atom MD the Langevin equation 
can not be used.  There are no small and fast variables around.
The oscillating atoms are themselves the small and fast variables.
Moreover, from a conceptual point of view
the presence of a white noise (\ref{lang2}) {\it de facto} converts the ensuing numerical algorithm into a
Monte Carlo process, albeit an elaborated one.

Many alternatives to  Langevin thermostat have been introduced. An example is the Gau\ss ian thermostat \cite{Gallavotti-2014}.
Unlike Langevin's, it is {\it deterministic}. 
Instead of small and fast background variables, one couples the variables 
$\mathbf x_i$ of interest to explicit  thermostat 
variables $\mathbf X_k$, with equations of motion 
\begin{equation}
m_i \ddot{\mathbf x}_i \ = \ - \nabla_i E(\mathbf x) - \nabla_i \mathcal E(\mathbf x, \mathbf X) 
\label{gauss1}
\end{equation}
\begin{equation}
M_{k} \ddot{\mathbf X}_k \ = \ -\nabla_k \mathcal U (\mathbf X) - \nabla_k \mathcal E(\mathbf x, \mathbf X) - 
\alpha_k \dot{\mathbf X}_{k}
\label{gauss2}
\end{equation}
The  thermostatting effect is modelled by the last term in (\ref{gauss2}); the $\alpha_k$  is determined
by subjecting the auxiliary variables to the non-holonomic constraint
\[
\frac{M_k}{2} \dot{\mathbf X}_{k}^2 = \frac{3}{2} k_B T 
\]
\[
\Rightarrow \ \ \alpha_k \ = \ ( \frac{3}{2} k_B T)^{-1}
\left[ \frac{M_k}{2} \dot{\mathbf X}_{k}^2 + 
\dot{\mathbf X}_{k} \left[  \nabla_k \mathcal U(\mathbf X) + 
\nabla_k \mathcal E(\mathbf x,\mathbf X) 
\right]\right]
\]
for each of the thermostat variable. 
The disadvantage of a Gau\ss ian thermostat is in the lack of a Hamiltonian character  
in the equations of motion (\ref{gauss1}), (\ref{gauss2}).  

A Hamiltonian, albeit singular,  thermostatting has been proposed by 
Nos\'e and Hoover \cite{Nose-1984,Hoover-1985,Martyna-1992,Liu-2000}. 
Their thermostat 
is probably the most widely used  in the context of all atom protein folding simulations.  
In the simplest variant the all-atom phase space is extended by a single {\it ghost} 
particle with a singular, logarithmically divergent potential that  provides the temperature for all the 
rest\footnotemark\footnotetext{ We remark that this ghost particle is a little like a Higgs particle that
gives the mass for other particles in sub-atomic physics. Except that instead of mass it gives 
temperature, and it can not be observed.}.  The singular character of the potential  introduces 
an inexhaustible heat reservoir, in essence.

Following \cite{Niemi-2014a} we consider the toy-model case of a single canonical degree of freedom ($p,q$) in
the presence of a single Nos\'e-Hoover thermostat degree of freedom ($P,Q$). The classical action is
\cite{Nose-1984,Hoover-1985,Martyna-1992,Liu-2000}
\begin{equation}
S \ = \ \int\limits_{-\mathrm T}^{\mathrm T} \! dt \, \left\{  p \dot q + P \dot Q -  \frac{ p^2 }{ 2mQ^2} - V(q) - \frac{P^2}{2M}  - \frac{1}{\beta_0} 
\ln Q 
\right\}
\label{nose1}
\end{equation}
We may assume  that $V[q]$ has the double-well  profile 
\begin{equation}
V[q] = \lambda(q^2 - m^2)^2
\label{V}
\end{equation}
We are interested in the tunnelling amplitude between the two minimum energy configurations
$q = \pm m$,
\begin{equation}
<\! +m, \mathrm T  | \!-m,-\mathrm T  \!> = \int \hskip -0.8cm  \int\limits_{q(-\mathrm T) = -m}^{q(+\mathrm 
T) = +m} \hskip -0.4cm 
[dp][dq] e^{i S}
\label{m+-}
\end{equation}
The integration over $p$ is Gau\ss ian but yields a Jacobian factor that depends on $Q$. This Jacobian has the 
same functional form as the last term in (\ref{nose1}), thus it can be absorbed by a redefinition (renormalisation)
of $\beta_0 \to \beta$.   As usual, we evaluate the transition amplitude using Euclidean (imaginary) time formalism, obtained
by sending $t \to i t$.  The Euclidean
action is 
\begin{equation}
S \ = \ \int\limits_{-\mathrm T}^{\mathrm T} \! dt \, \left\{  \frac{1}{2} Q^2 q_t^2 + V[q] + \frac{1}{2} Q_t^2 + \frac{1}{\beta}
\ln  Q \right\}
\label{nose2}
\end{equation}
and we search for a finite action instanton  trajectory that connects the two states $q=\pm m$; note that
the continuation  to imaginary time "inverts" the potential term, as shown in Figure \ref{fig-12}.
%
%
%
%
%
%
%
%
%
%
%
{
\footnotesize
\begin{figure}[h]         
\begin{center}            
  \resizebox{8cm}{!}{\includegraphics[]{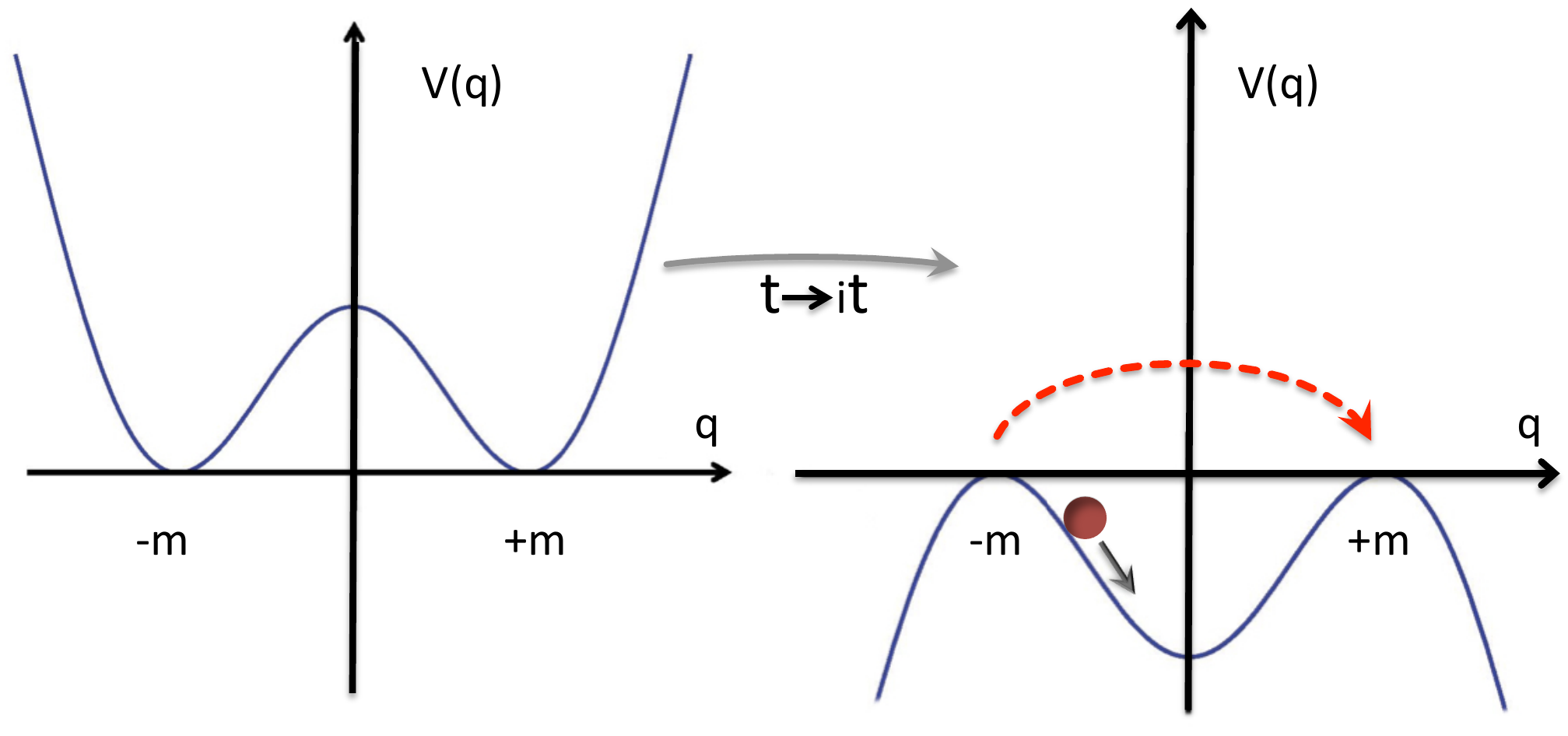}}
\end{center}
\caption{Analytic continuation to Euclidean time has the effect of inverting the potential $V(q)$. The instanton
is a trajectory which starts at (Euclidean) time $-\mathrm T$ from the local maximum at $q=-m$ and reaches the
local maximum at $q=-m$ at time $+\mathrm T$, as shown in the Figure on right.
}   
\label{fig-12}    
\end{figure}
}
The instanton is a solution to the  equations of motion
\begin{eqnarray}
Q^2 q_{tt}  & = &  V_q - 2 Q \, Q_t q_t \  \simeq  V_q - \gamma q_t
\label{nose2a}
\\
Q\, Q_{tt} & = & Q^2 q_t^2  + \frac{1}{\beta} 
\label{nose2b}
\end{eqnarray}
Note how the coupling between $q$ and the thermostat variable $Q$  gives rise to an 
effective friction-like coefficient $\gamma(t)$. 

We first consider the case where the thermostat field $Q$ is absent. This amounts to setting $Q\equiv 1$ 
in (\ref{nose2}),   (\ref{nose2a}) and removing (\ref{nose2b}), leaving
us with the equation
\[
q_{tt} = V_q \ = \ 4\lambda q (q^2 - m^2)
\]
By adjusting the initial velocity we conclude that a solution exist
which starts from $q=-m$ at time $-\mathrm T$, and ends at $q=+m$ at time $+\mathrm 
T$. This solution is the instanton
that gives rise to a finite tunneling amplitude between $\pm m$; in the limit $\mathrm T \to \infty$ the instanton
has  the familiar double-well topological soliton profile
\begin{equation}
q(t) = m \tanh \! \left( \! \sqrt{2\lambda}\, m (t-a) \! \right)
\label{topsol} 
\end{equation}

Now suppose the thermostat field is present. We first consider a scenario, where 
at $\pm \mathrm T$ the system reaches a stationary 
state where $q = \pm q_0 \not= \pm m$.  Since the action (\ref{nose2})  should remain finite as $\mathrm T \to \infty$,
we arrive at the Gibbsian relation
\begin{equation}
Q(\mathrm T) \ \buildrel{\mathrm T\to\pm\infty}\over{\longrightarrow} \ Q_\pm \ = \ e^{-\beta V[q_\pm]}
\label{gibbsi}
\end{equation}
This  proposes that $\beta$ is indeed the Bolzmannian temperature factor, when positive.

Next, we integrate (\ref{nose2b}) and then take the limit $\mathrm T\to \infty$; since the Euclidean action should be
finite, we may assume  that  $Q_t$  should vanish as $\mathrm T \to \infty$ which removes the surface term.  
We find
\begin{equation}
\int\limits_{-\infty}^\infty \! dt \, \left\{  Q_t^2 + Q^2 q_t^2 \right\} \ = \ -  
\int\limits_{-\infty}^\infty \! dt \,  \frac{1}{\beta}
\label{nhoo}
\end{equation}
For a non-trivial tunnelling configuration and with a finite  Euclidean 
action (\ref{nose2})  the integral on the left-hand-side of (\ref{nhoo})  
should be  finite and non-vanishing.   
But since the  {\it l.h.s.} is manifestly non-negative, 
the Bolzmann temperature factor $\beta$ can not be positive as it should. 
Thus  we conclude that the presence of the thermostat field suppresses tunnelling. We note that a
suppression of tunneling amplitude by Nos\'e-Hoover thermostat  
in the case of double-well potential has  been observed in numerical simulations \cite{Liu-2000}.

Proteins regularly need to tunnel over various  different potential barriers in their presumably highly rugged 
energy landscape, when proceeding from a random initial configuration to the natively folded state. Thus 
we suspect that simulations using Nos\'e-Hoover thermostats can  cause  complications whenever we have 
a protein for which we can expect that the folding pathway goes thru various intermediates and  molten globules.

\begin{quote}
{\it Research project: Investigate how the suppression of tunneling  amplitudes in the case of a properly modified
Nos\'e-Hoover thermostat could be avoided. }
\end{quote}

Other kind of thermostats have also been introduced, in particular we mention the Berendsen 
thermostat \cite{Berendsen-1984}.
These thermostats that are often convenient in numerical simulations, are designed to approach
canonical ensembles in a limit. But they commonly lack a Hamiltonian interpretation
{\it i.e.} are non-physical,
and thus the physical interpretation of the results is not there.  
While this might not be an issue when
the goal is {\it simply} to find a local energy minimum of an all-atom force field such as (\ref{hos}), (\ref{ff}), 
a non-physical thermostat can not be used to model the dynamics of  proteins.

\section{Other physics-based approaches:} 

The all-atom approaches where the discretised 
Newton's equation is solved iterately,  are conceptually weak coupling 
expansions in the  dimensionless parameter (\ref{coup}). To describe the folding 
of most proteins, one needs to be able to extend this expansion and ensure its convergence, over some
fifteen orders of magnitude or even more. From the 
perspective of sub-atomic physics, this is like extending perturbative Standard Model calculations
all the way to Planck scale. 

Several approaches have been developed, with the goal to introduce an expansion parameter
that corresponds to a time scale which is clearly larger than the  femtosecond scale. Such  {\it coarse-grained}  force fields 
average over the very short time scale atomic fluctuations: 
If the denominator in (\ref{coup}) can be increased,  so can the numerator, and it  becomes 
possible to develop expansions which reach longer {\it in vivo} 
time scales with no increase in the {\it in silico} time.
In practice, a  carefully crafted coarse grained force field can cover up to around three orders of 
magnitude longer folding trajectories than all-atom approaches, while still 
maintaining a good overall quality.  Here we mention in particular  UNRES as an example
of such a carefully crafted coarse grained force field \cite{Liwo-1997,Scheraga-2007,Liwo-2014}. See  the homepage 
\[
{\tt http://www.unres.pl/}
\]

Finally, we  comment on the various versions of the G\~o model  
\cite{Go-1983}  and the closely related elastic (Gau\ss ian) network models  \cite{Haliloglu-1997}.
These approaches  have played a very important r\^ole, to gain insight to protein folding 
in particular when the 
power of computers is insufficient for any kind of serious all-atom folding simulations.
In these models the folded configuration is presumed to be known; the individual 
atomic coordinates of the folded
protein chain
appear as an input.
A simple energy function is then introduced, tailored to ensure  that the known folded configuration  
is a minimum energy ground state. In the G\~o model the energy could be as simple as a square well
potential which is centered at the native conformation. In elastic network models the atoms 
are connected by harmonic oscillators, with energy minima that correspond to the natively folded state.

Since the positions of all the relevant atoms appear as parameters
in these models,  they contain more parameters than 
unknown and thus no predictions can be made. Only a description is possible.
From the point of view of a
system of equations, these models are over-determined. In any {\it predictive} energy function the number of adjustable 
parameters must remain  {\it smaller} than the number of independent atomic coordinates.  
Otherwise, no predictions can be made, no physical principles can  be tested. Even though a description is possible.

%
%
%
%
%
%
%
%
%
%
%
\chapter{Bol'she }
%
%

%
%

%
%
%

\vskip 0.3cm
In 1972 Anderson wrote an article \cite{Anderson-1972} entitled {\it More is different} that has been inspirational to many. 
In particular, he argued for the importance of emergent phenomena.  But  the call for {\it Bol'she}  is already
present in  Schr\"odinger's  1944 book:

\begin{quote}
{\it  ... living matter, while not eluding the 'laws of physics' as established up to date, is 
likely to involve 'other laws of physics' hitherto unknown, which, however, once
they have been revealed, will form just as integral a part of this science as the former.}
\end{quote}

\noindent
However, Anderson makes  a {\it crucially} important point that can not be found in Schr\"odinger's  book.
Anderson's article has this point even as a  subtitle:  {\it Broken symmetry and the nature of the hierarchical structure of science}.
Anderson realised that in order for emergent phenomena to give 
rise to structural self-organisation,  one needs  a  symmetry 
which becomes broken.  

We have already pointed out that in the case of proteins  
a broken symmetry is present.  Individual amino acids can be either
left-handed chiral or right-handed chiral, equally.  But for some reason, living matter  is built almost 
exclusively from amino acids  that are left-handed chiral. We note that, apparently as a consequence, protein 
chains are predominantly  chiral with right-handed helicity. 

%
%
%
%
%
%

\section{The importance of symmetry breaking}

Consider a fluid dynamical system such as water, the atmosphere, or any other scenario 
that can be described by the Navier-Stokes or Euler equation or their many descendants.  
These are fundamentally atomic systems, but with an enormous number of  constituents. 
Their macroscopic properties are governed  and often with a very high precision,  
by the properties of a local order parameter 
which computes  the fluid velocity.
Structures such as vortices and  tornadoes, and solitonic waves 
like the one that emerges from the Korteweg - de Vries equation,  
are all described by a solution that  breaks an underlying symmetry. 

A fluid dynamical vortex line is a familiar example of a  highly regular collective state of individual atoms, with 
a topological character.  It is an example of an emergent structure:
At the atomic level  of scrutiny, the individual atoms and molecules 
that constitute the vortex are subject to 
random, brownian thermal  motion. By following a single individual water molecule 
you can not really conclude whether
a vortex is present.  The vortex materialises only at the macroscopic 
level, when the  individual haphazard atomic motions become collectively 
self-organised  into a regular pattern.  

It would be incomprehensible to construct a macroscopic vortex line such 
as a tornado in atmosphere from purely atomic level considerations, even in principle:
 A vortex is an example of a soliton, and  solitons 
can not be constructed simply by adiabatically building up individual atomic level 
interactions as small perturbations around a  ground state which consists of free 
individual atoms.  A (topological) soliton emerges when non-linear interactions combine elementary 
constituents into a localised collective  excitation that is stable against small 
perturbations and cannot decay, unwrap or disentangle.

\section{An optical illusion}

We start to describe the importance of symmetry breaking at an intuitive level. We consider
a simple  optical illusion, {\it not} a physical example. But it 
nevertheless demonstrates how {\it Bol'she} gives rise to a symmetry that becomes broken,
and the illusion of breaking symmetry  leads to the formation of structure, in our eyes. 

We then proceed to consider two physically relevant examples, where a very similar broken symmetry is 
present. But now in a physically relevant fashion. Most importantly, each of the two examples we present describes
a simple physical scenario  that shares many features with proteins. 

In Figure \ref{fig-13} we have two sets of cubes. On the left we have two individual cubes, 
and on the right there are more cubes. If one looks at the cubes on 
the right, one  should be able to visualise some order. For example  light grey steps that 
come down, from the left. Now  rotate the Figure, slowly. When one  keeps ones eyes focused 
on the two individual  cubes on the left, nothing really happens  besides an overall rotation of the two cubes. 
But if one instead focuses on the set of cubes on right, one should observe a rapid
transition:  There is a point where the direction of the steps changes, abruptly, so that
after a rotation by 180 degrees we still have the same light grey steps, 
still coming down from the left.  

The system on the right is {\it Bol'she}, with  a  discrete
$\mathbb Z_2$ symmetry under a
rotation by 180 degrees. There are two ground states, and our mind chooses one, thus
breaking the symmetry. In fact, the scenario is very much like 
that in Figure \ref{fig-12}, there are two identical ground states. 
In this sense Anderson's {\it Bol'she} is  present in Figure \ref{fig-13}. 
No similar optical illusory effect is observed 
when the two individual  cubes on the left are rotated. 
%
%
%
%
%
%
%
%
%
%
%
{
\footnotesize
\begin{figure}[h]         
\begin{center}            
  \resizebox{10.5cm}{!}{\includegraphics[]{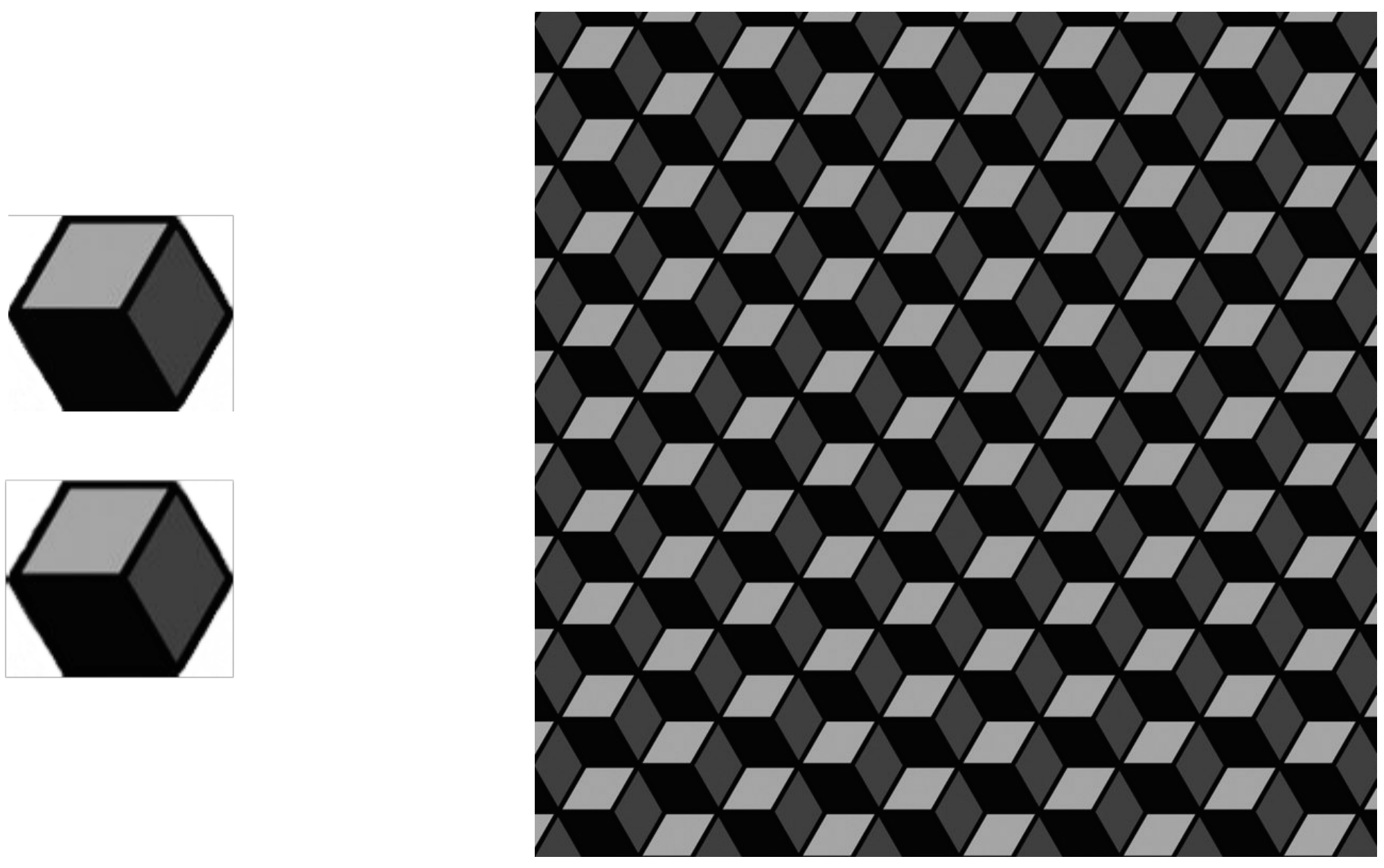}}
\end{center}
\caption{Rotate the figure, slowly, by 180 degrees. When you focus your eyes on the 
two individual cubes on the left, nothing
much happens. However, if you focus on the set of cubes on the right, you see an abrupt 
effect like a phase transition between two different but identical ground states; 
we have a $\mathbb Z_2$ symmetry that has become broken by visual 
perception.
}   
\label{fig-13}    
\end{figure}
}

We now proceed to describe two physical examples where a similar kind of $\mathbb Z_2$ symmetry becomes broken,
with equally dramatic {\it physical} - not illusory - consequences.

\section{Fractional charge}

Polyacetylene in $\it trans$ conformation is like a simplified  
protein.  Figure \ref{fig-14} a) shows the structure.
%
%
%
%
%
%
%
%
%
%
%
{
\footnotesize
\begin{figure}[h]         
\begin{center}            
  \resizebox{8.5cm}{!}{\includegraphics[]{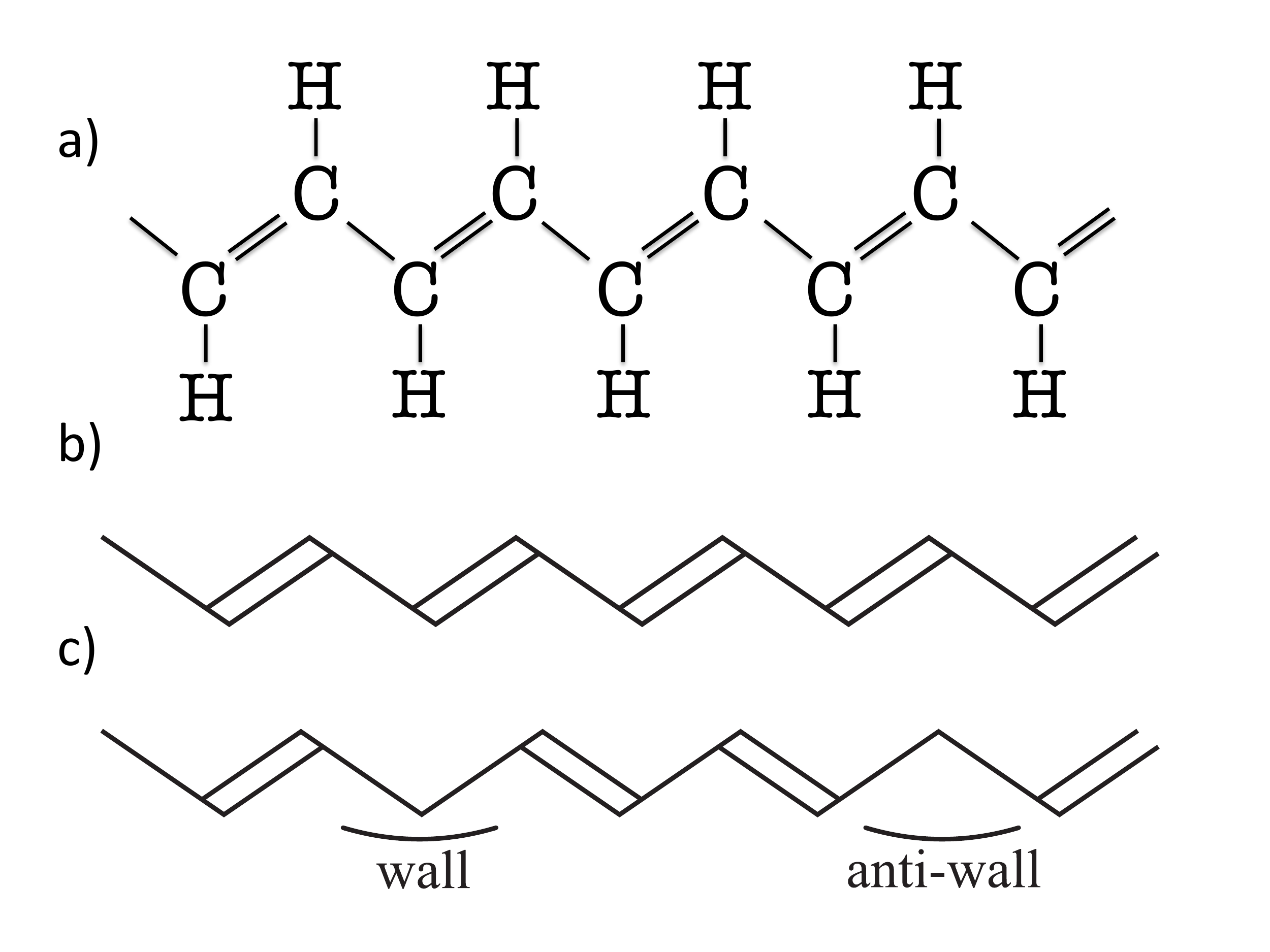}}
\end{center}
\caption{a) The {\it trans} polyacetylene. b) One of the two degenerate
ground states in a {\it trans}-polyacetylene chain. The other ground state
is obtained by a reflection that interchanges the double bonds and the
single bonds. c) A state of a {\it trans}-polyacetylene chain with one of the
double bonds converted into a single bond and then transported to form
two domain walls carrying fractional electric charges.
}   
\label{fig-14}    
\end{figure}
}
There is a "backbone" that consists entirely of carbon atoms, and at each carbon atom
there is a "side chain" with a single hydrogen. Much like in a protein, but simpler. 
%
%
In Figure~\ref{fig-14} b)
we depict the ({\it trans}) polyacetylene chain by combining each (CH) unit into a single vertex. The
double line describes the $\sigma$-bond and the single line is the  $\pi$-bond, between two
consecutive C atoms.
Due to a Peierls instability the asymmetry of the chain
causes the ground state to be doubly degenerate. The free energy acquires the 
double-well profile that we have depicted in  Figure {\ref{fig-12}. The two
ground states are related to each other by a $\mathbb Z_2$ reflection (parity)
symmetry of the polyacetylene chain about a (CH) site, which exchanges  
the $\sigma$-bonds  and the  $\pi$-bonds.

When we choose one of the ground states, we break the symmetry. But  we may introduce  domain
walls, that interpolate between two different ground states. 
In Figure~\ref{fig-14} c) we show an example. In this Figure we have two domain walls.  Each 
interpolates between two different ground states;  between the two domain walls we have a region where the 
the $\sigma$-bonds and $\pi$-bonds have become interchanged.  Quantitatively, in terms of the double well 
potential shown in Figure \ref{fig-12}, each of the two domain walls correspond to a topological soliton
(instanton) profile (\ref{topsol}).

We now argue that we have {\it Bol'she} which makes things different: The chain in Figure~\ref{fig-14} c)
is obtained from the  chain in Figure~\ref{fig-14} b) by  removal of a single electron. There are 15 bonds in the 
Figure~\ref{fig-14} b) and 14 in the Figure~\ref{fig-14} c). The removal of a single electron converts 
a double bond into a single bond, and we have simply moved  the bonds around to make the
two identical domain walls.  Since the structures in Figure~\ref{fig-14} b) and in
Figure~\ref{fig-14} c) differ from each
other only by the removal of a single bond and since the
two domain walls are identical, related to each other by parity, the two domain walls
must share equally the quantum numbers of the missing bond: The two 
domain walls have electric charge half, each.

This phenomenon of fermion number  (charge)  
fractionalisation demonstrates how Anderson's  {\it Bol'she} really makes a difference:
Such exotic  quantum number assignments could never
be obtained simply by linearly superposing an integer number of initially non-interacting
electrons and holes adiabatically, in a continuous manner, into a weakly interacting system:
{\it A charge half state simply can not be made by combining together any finite number 
of particles with an integer charge. } Unless something {\it Bol'she} is  involved.

\vskip0.2cm

{\it Fine points:} A bond line in a polyacetylene corresponds to two electrons, one
with spin up and the other with spin down. Thus an isolated domain wall must
have a net electric charge which is equal in magnitude but opposite
in sign to that of a single electron. But since the
spins of the electrons that have been removed are paired,
the domain wall carries no spin. This unusual spin-charge
assignment has been observed experimentally and
it constitutes the essence of fermion number fractionalisation \cite{Shirakawa-1977,Jackiw-1981,Semenoff-1986}
that gives rise  to electric conductivity along the polyacetylene.
In the absence of the spin doubling we would observe a
domain wall  that carries half the electric charge of one electron.
Note that if we add a single electron to a domain wall,
we obtain a state which is charge neutral but carries the
spin of the electron. Alternatively, if we remove a single electron
from a domain wall we obtain a state with spin one-half
and a charge that equals (minus)
twice that of the electron. Neither of these states are possible, without  {\it Bol'she}.

\section{Spin-charge separation} 

\vskip 0.6cm
We shall eventually argue that proteins are {\it very} much like one dimensional spin-chains; the side chains 
are akin spin variables along the backbone. Thus, our second  example is a 
one dimensional spin chain.  For conceptual clarity 
we may assume that the spin variables 
are single electrons (or maybe protons like H${+}$). We assume that the background
lattice prefers a ground state which is an antiferromagnetic
N\'eel state where all the spins point into an opposite
direction from their nearest neighbours.  Furthermore, we assume that in the 
ground state all the sites have single occupancy, and that
there is a very strong repulsive force between the electrons which 
prevents a double occupancy. Such a  ground state is a Mott insulator,
and we have depicted the structure in Figure~\ref{fig-15}.
%
%
%
%
%
%
%
%
%
%
%
{
\footnotesize
\begin{figure}[h]         
\begin{center}            
  \resizebox{8.5cm}{!}{\includegraphics[]{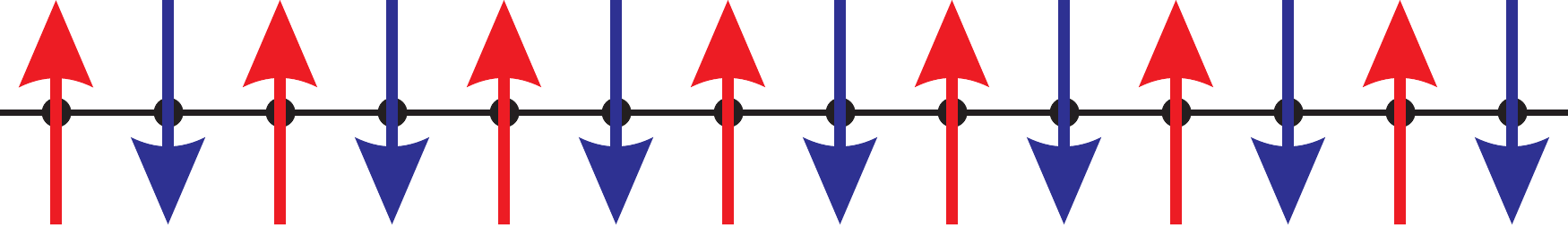}}
\end{center}
\caption{One the two degenerate ground states in a
N\'eel antiferromagnet, with
alternating spin directions along a one dimensional lattice of
electrons. The $\mathbb Z_2$ symmetric ground state
is obtained by reversing the direction of every spin.
}   
\label{fig-15}    
\end{figure}
}
As in the case of a polyacetylene, the ground state is doubly degenerate:
The $\mathbb Z_2$ symmetry transformation operates by reversing the direction of the
spin at every single lattice site. By choosing one of the two
ground states, we break the $\mathbb Z_2$  symmetry.

If we reverse the direction of a single spin along the chain,
we form a localized configuration with three parallel neighboring spins; see Figure \ref{fig-16} a).
%
%
%
%
%
%
%
%
%
%
%
{
\footnotesize
\begin{figure}[h]         
\begin{center}            
  \resizebox{8.5cm}{!}{\includegraphics[]{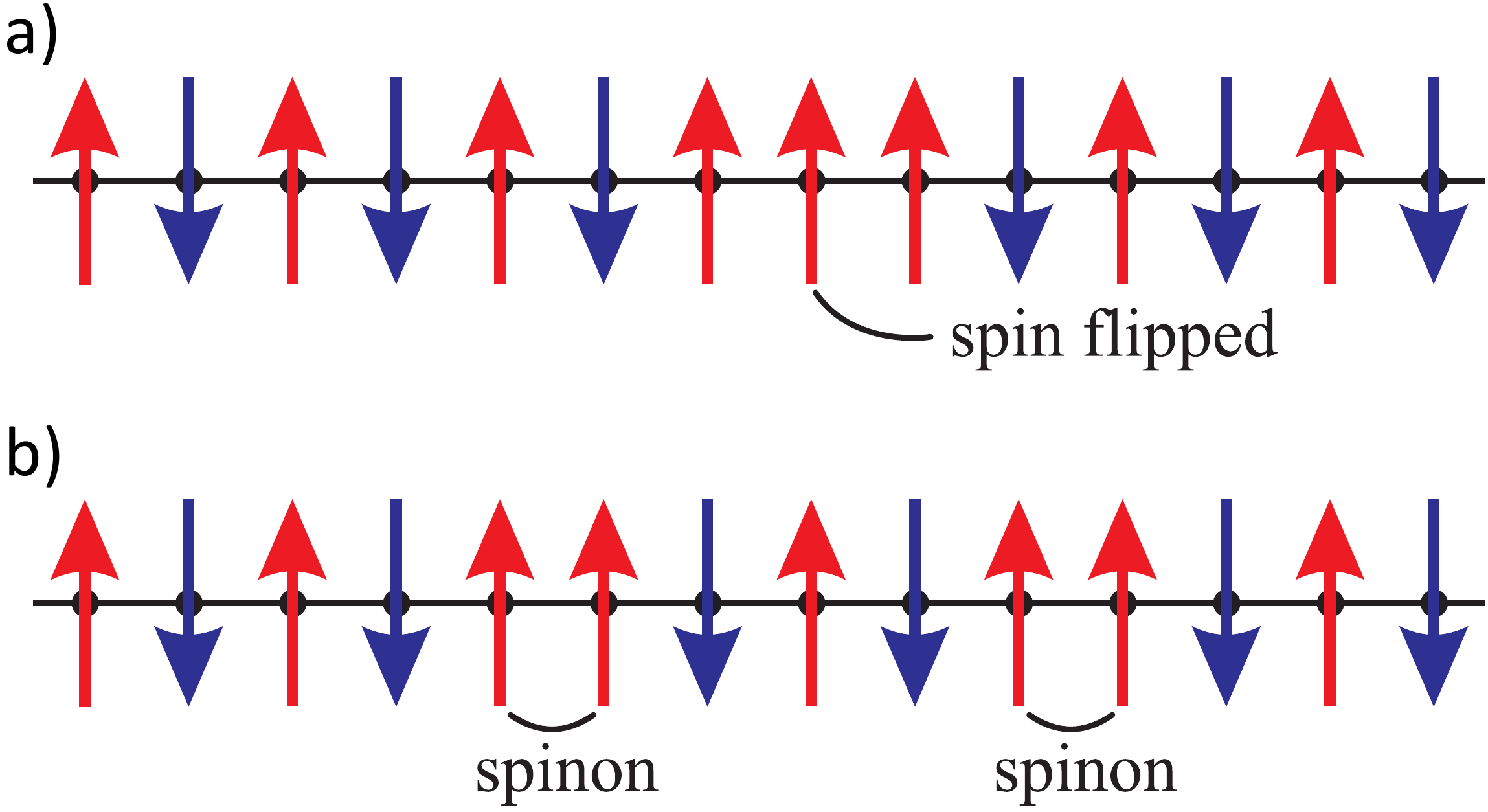}}
\end{center}
\caption{a) The same as in Figure~\ref{fig-15} but with the spin of
only one electron reversed. b) The state of Figure a)  is decomposed
into two domain walls representing the spinons.}   
\label{fig-16}    
\end{figure}
}
By successively reversing the direction of neighboring spins
but without changing
the total spin we can decompose this configuration into two
separate domain walls, each consisting of two parallel spins. These
domain walls interpolate between the
two ground states of the spin chain, as shown in Figure~\ref{fig-16} b).
Since the lattices in Figure~\ref{fig-16} a) and b)
differ from each other only by the flip of a single
spin, the total change in the spin between the two lattices
is one. The two domain walls in Figure~\ref{fig-16} b) are
also identical. Thus each must have a spin equal to one-half. 

Since we have made the two domain walls without adding or removing
any electrons, each of them must be charge neutral.  We conclude
that the domain walls are {\it spinons}, they describe the same
spin degree of freedom as a single electron but with no charge \cite{Baskaran-1987}.
The two  domain walls interpolate between the
two different ground states of the antiferromagnetic chain,  very much 
like the domain walls in the case of polyacetylene.

\vskip 0.2cm

Now we consider the same antiferromagnetic lattice but with one of the
electrons removed as shown in Figure~\ref{fig-17}.
%
%
%
%
%
%
%
%
%
%
%
{
\footnotesize
\begin{figure}[h]         
\begin{center}            
  \resizebox{8.5cm}{!}{\includegraphics[]{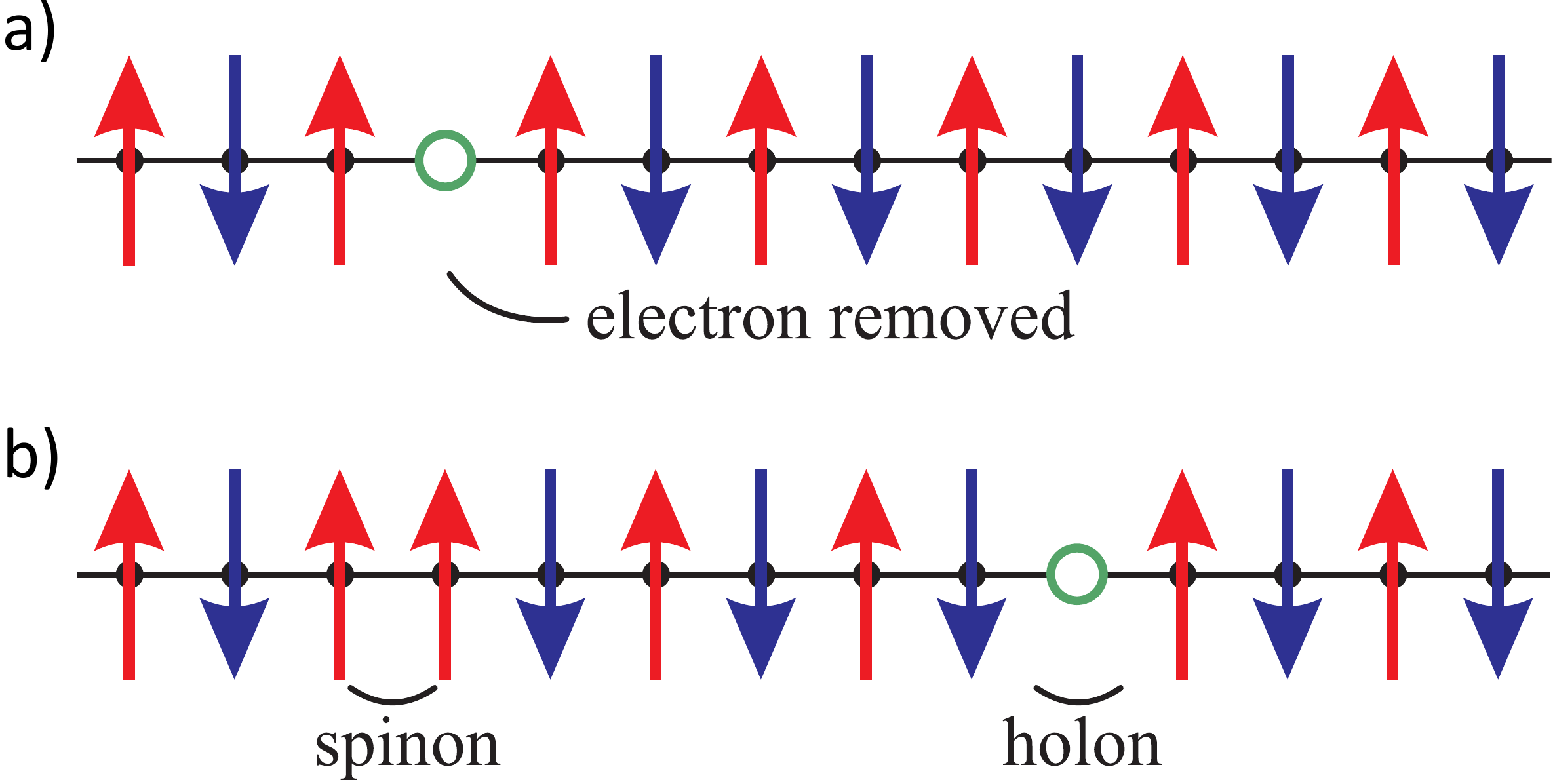}}
\end{center}
\caption{a) Spinon (left) and chargon (right) states in the underdoped state. 
b) The state of Figure a)  is decomposed
into two domain walls representing the spinons.}   
\label{fig-17}    
\end{figure}
}
This corresponds to an underdoped case,we have a hole in 
the spin chain. When we move the hole to the right 
we arrive at the situation
depicted in Figure~\ref{fig-17} b). 
In addition of the  hole we have here another
domain wall which is similar to the two domain walls that
we have in Figure~\ref{fig-16} b). This domain wall is a spinon,
it has spin one-half but carries no charge. Since we have removed 
one electron and since the spinor does not carry charge, 
the hole must carry a
charge which is opposite to that of one electron.
But no spin is available for
this hole, it describes a spinless {\it holon}.

The ARPES experiment at Lawrence Berkeley Laboratory has confirmed
that such spinons and holons do exist,  they have been observed
in a one dimensional copper oxide (SrCuO${}_2$) wire \cite{Kim-2006}.

Note that in Figure~\ref{fig-17} a) the two spins on the left and on the right
of the hole are parallel with each other but in Figure~\ref{fig-17} b) the
two spins are opposite to each other. This confirms that like the spinon,
the spinless holon is a domain wall that interpolates between the
two different ground states of the chain.

%
%
%
%
%
%
%
%
%
%
%
{
\footnotesize
\begin{figure}[h]         
\begin{center}            
  \resizebox{8.5cm}{!}{\includegraphics[]{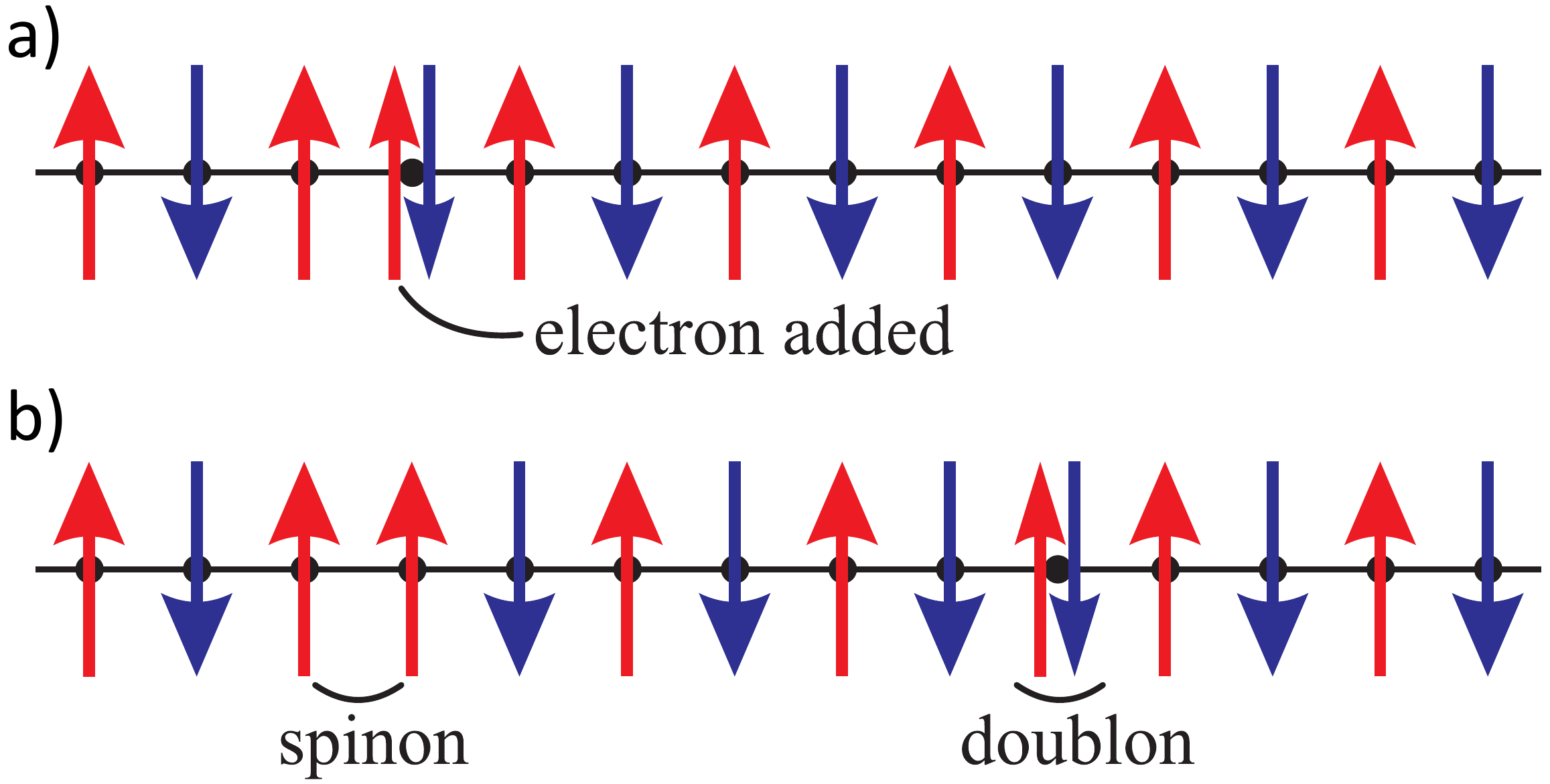}}
\end{center}
\caption{a) The N\'eel state with one
electron added (overdoped case).
b) Spinon (left) and doublon (right) states in the overdoped state.}   
\label{fig-18}    
\end{figure}
}
Finally, in Figure~\ref{fig-18} a) we have added a single electron to the lattice.
This example is of particular conceptual interest since it allows 
us to directly address what happens to a (pointlike) electron when 
it enters the antiferromagnetic environment: 
The presence of the electron introduces a single site with a 
double bond $(\uparrow\downarrow)$. The chain is now 
over-doped.  As before we transport the
double filled state,  {\it e.g.}
to the right so that we  arrive at the situation depicted in Figure~\ref{fig-18} b).
Note that due to Pauli exclusion the transport occurs so that
we move alternatively either a spin-up or a spin-down state one step
to the right. The final configuration shown in Figure~\ref{fig-18} b) describes
two separate domain walls that both interpolate between the two 
distinct ground
states of the spin chain. One of these domain walls is again
a spinon. The other one is a {\it doublon}. Since one electron has been added and
since spinor has spin but no charge, we conclude that the
doublon does not have any spin but it carries the entire charge
of one electron. The charge of the doublon 
is opposite to that of the holon.

The two examples  we have discussed, fermion number fractionalisation and spin charge separation,
make it plain and clear how much difference {\it Bol'she} can do:  It would be impossible to
make states with the spin of an electron but no charge, or states with the
charge of an electron but no spin, simply by superposing an integer number of
non-interacting electrons and then adiabatically switching on their mutual interactions.
For states with such exotic quantum numbers we need to have an environment 
with a symmetry that has become broken.

\section{All-atom is Landau liquid}

The Landau (Fermi) liquid is a paradigm on which much of our understanding of many-body systems 
like metals is based. This paradigm states, that in a physical system with a large number of atoms, 
each atom retains its individual integrity. The properties of a material system which is described by
a Landau liquid,  
can be understood by superposing its individual constituents in a weak 
coupling expansion around a given ground state.
In particular a Landau liquid which is made of
electrons and protons can only have  spin and charge assignments which are obtained by 
superposing the individual spins and charges. This is the case when
the properties of the system can be understood using the notion of adiabaticity: We imagine that we
start from an initial condition where the elemental  constituents have no 
mutual interactions. The interactions are then turned on,  adiabatically, in a continuous manner.  
Accordingly  the ground state of the original non-interacting system 
becomes continuously deformed 
into the ground state of the interacting system.  
%
%

The two examples that  we have described, polyacetylene and antiferromagnetic spin chain,
show that the Landau liquid paradigm is not a universal one. In a Landau liquid system it would
be impossible to have states with exotic quantum numbers such as an electric charge 
which is half of that of a single electron.
The Landau liquid paradigm can fail  whenever we have emergent 
structures that display symmetries which become broken.
In such scenarios there are often collective excitations like topological solitons which 
can not be built simply by adding together small adiabatic perturbations around a 
ground state of non-interacting  elemental constituents.

\vskip 0.2cm
The all-atom description (\ref{hos}), (\ref{ff}) of protein force field 
implicitly assumes the Landau liquid paradigm. 
According to (\ref{hos}), (\ref{ff}) the  individual atoms oscillate around their ideal
values, under the influence of  a potential which is either a harmonic oscillator
or a mathematical pendulum. The Lennard-Jones and Coulomb potentials 
introduce continuously evolving deformations around  the  
ideal atomic positions,  in a manner that can be modelled by a weak coupling expansion  
of the iterative Newton's equation in powers of (\ref{coup}); in practical simulations these long range interactions
are tuned off, beyond a distance around 10 \AA ngstr\"om.
 
It remains to be seen whether an all-atom  Landau liquid description of proteins 
breaks down. But the basal ingredient, that of a broken $\mathbb Z_2$ symmetry
which also appears in our examples of polyacetylene and antiferromagnetic spin chain,
is certainly present: The amino acids are left-handed chiral, and as a consequence
proteins that constitute live matter prefer  
right-handed helicity along their backbone.

%
%
%
%
%
%
%
%
%
%
%
\chapter{Strings in three space dimensions}
%
%
%

\begin{quote}
{ \it  Thus we have come to the conclusion that an
organism and all the biologically relevant
processes that it experiences must have an
extremely 'many-atomic' structure and must be
safeguarded against haphazard, 'single-atomic'
events attaining too great importance. \rm (E. Schr\"odinger)
 }
\end{quote}

\noindent
We start our search of broken symmetry and the ensuing {\it Bol'she}  that makes us alive, by considering 
differentiable (class $\mathcal C^3$) strings in $\mathbb R^3$.  
\section{Abelian Higgs Model and the limit of slow spatial variations}  
\label{sect2}
%

The Abelian Higgs Model (AHM) is the paradigm 
framework to describe vortices as solitons. Solitonic vortices are important to 
many physical phenomena, from cosmic strings in Early Universe to 
type-II superconductors. In particular,  the Weinberg-Salam model of electroweak interactions with its
Higgs boson is a non-Abelian extension of AHM.

\vskip 0.2cm

The  AHM involves a single complex scalar (Higgs) field $\phi$ and a vector field  $A_i$. 
These fields are
subject to the U(1) gauge transformation
\begin{equation}
\begin{array}{lcl}
\phi \  & \to  & \ e^ {ie \, \vartheta } \phi\\
A_i \ & \to & A_i + \partial_i \vartheta
\end{array}
\label{u1}
\end{equation}
where $\vartheta$ is  a function, and $e$ is a parameter. The standard AHM Hamiltonian is
\begin{equation}
{\mathcal H} = \frac{1}{4} G^2_{ij} + |(\partial_i - i e A_i)
\phi|^2 + \lambda \left(|\phi|^2 -
v^2
\right)^2
\label{H1}
\end{equation}
where
\[
G_{ij} = \partial_i A_j - \partial_j A_i
\]
When the space dimension $D$ is odd, a Chern-Simons term ($ChS$) can be added to (\ref{H1}). Explicitly, 
\begin{equation}
\begin{matrix} D = 1: & \ \ \ ChS & \sim & A & \\
D=3: & \ \ \ ChS & \sim &  AdA & \\
D=5: & \ \ \  ChS & \sim & AdAdA & \\ 
& & & & etc.
\end{matrix}
\label{chs}
\end{equation}
The Chern-Simons term  is the paradigm way to break parity.

In a material system (\ref{H1}), (\ref{chs}) is the Kadanoff-Wilson  
energy  in the limit where the fields have slow spatial variations 
\cite{Kadanoff-1966,Wilson-1971}. To describe this limit,
we start from the  full free energy of a  material 
system which is  based on the AHM field multiplet; we denote it
\[
\mathcal F ( \phi, A_i)
\]
This free energy is in general a non-local functional of the field variables.   But it must be gauge invariant.   
Thus, it  can only depend on manifestly gauge 
invariant combinations of the fields such as
\[
|\phi|^2 , \ \ \  |(\partial_i - i e A_i) \phi |^2 \  , \ \dots 
\]
Consider the limit where the length scale that is associated to spatial variations of the 
field variables becomes very large in comparison to other characteristic length scales of the system. 
In this limit we can expand the free energy in powers of the gauge covariant derivatives of the fields. 
The expansion looks like this \cite{Coleman-1973}:
\begin{equation}
 \mathcal F ( \phi, A_i) \ = \ V(|\phi |) +  Z(|\phi |)\,  |(\partial_i - i e A_i) \phi |^2   + ChS(A) + 
 W(|\phi |) \,  G_{ij}^2 
 + \ \dots
\label{weinberg}
\end{equation}
The leading term is called the effective potential.
The higher derivative terms are multiplied by functions $Z(|\phi |)$, $W(|\phi |) $  {\it etc}.  

\vskip 0.2cm

\noindent
The AHM energy 
(\ref{H1}), (\ref{chs})  constitutes the leading order non-trivial contribution to (\ref{weinberg}), 
in powers of fields and their covariant
derivatives. 
%
\vskip 0.2cm

We introduce a set of new variables $(J_i, \rho, \theta)$, obtained from
($A_i,\phi$) by the following change of variables
\begin{equation}
\begin{array}{lcl}
\phi \  & = & \ \rho \cdot e^ {i \theta } \\
A_i \ & \to & J_i = \frac{i}{2 e
|\phi|^2}\left[ \phi^* ( \partial_i -  i e A_i ) \phi - c.c. \right]
\end{array}
\label{asu}
\end{equation}
We can introduce these new variables whenever $\rho \not= 0$. Note that both  
$\rho$ and $J_i$ are gauge invariant under 
the gauge transformation (\ref{u1}). But
\[
\theta \ \to \ \theta + \vartheta
\]
When we  write (\ref{H1}), (\ref{chs})   in terms of 
these new variables (\ref{asu}), we have
\begin{equation}
{\mathcal H} = \frac{1}{4} \left(
J_{ij} + \frac{2\pi}{e} \sigma_{ij}
\right)^2
\!\!
+ (\partial_i \rho)^2 + \rho^2 J_i^2
+ \lambda \left(\rho^2 - \eta^2 \right)^2 \ + \ ChS
\label{H2}
\end{equation}
where
\[
J_{ij} = \partial_i J_j - \partial_j J_i
\]
and 
\begin{equation}
\sigma_{ij} =
\frac{1}{2\pi}\, [\partial_i , \partial_j ] \theta
\label{ds1}
\end{equation}
We observe  that  (\ref{H2}) involves only variables
that are manifestly $U(1)$ gauge invariant. In particular, unlike in the case of
(\ref{H1}), in (\ref{H2}) the local gauge invariance is entirely removed,
by a change of variables \cite{Chernodub-2008}.

The term $\sigma_{ij}$ is a string current. It has  a
Dirac $\delta$-like support which
coincides with the world-sheet of the cores of 
vortices. 
When (\ref{H2}) describes  a finite energy vortex, (\ref{ds1}) subtracts a singular
string contribution that appears in $J_{ij}$. Since $J_i$ is singular in the presence of a vortex line,
it makes  a divergent  contribution to the third term in the {\it r.h.s.} of (\ref{H2}). But the divergence 
becomes removed, provided the density $\rho$ vanishes on the world-sheet of the vortex core.
Thus the vanishing of $\rho$ along a string-like line in space is a {\it necessary} condition for the presence of
finite energy vortex lines.

\section{The Frenet Equation}
\label{sect3}

%

%
Proteins are string-like objects. Thus, to understand proteins we need to develop
the formalism of strings in $\mathbb R^3$, at least to some extent.
 
\vskip 0.2cm
The geometry of a
class $\mathcal C^3$ differentiable string
$\mathbf x(z)$  in $\mathbb R^3$ is governed by the
Frenet equation, described widely in elementary courses of differential geometry. 
The parameter  $z \in [0,L]$ where $L$ is the length of the string 
in $\mathbb R^3$. We can compute the length from 
\begin{equation}
L \ = \ \int\limits_0^L \! dz \,  || {\mathbf x}_z  ||  
\ = \  \int\limits_0^L \! dz \,  \sqrt{ {\mathbf x}_z \cdot {\mathbf x}_z } 
\ \equiv \ \int\limits_0^L \! dz \,  \sqrt{ g }.
\label{njl}
\end{equation} 
Here we recognise the  static version of the standard Nambu-Goto action, with 
generic parameter $z \in [0,L]$. We 
re-parametrize the string to express it in terms of 
the arc-length $s \in [0,L]$ in the ambient
$\mathbb R^3$, by the change of variables
\[
s(z) = \int\limits_0^z || {\mathbf x}_z (z') || d z'
\]
In the following we use the arc-length parametrisation, exclusively.
We consider a single, open string that 
does not self-cross.
We introduce the unit length tangent vector
\begin{equation}
\mathbf t \ = \  
 \frac{ d \hskip 0.2mm \mathbf x (s)} {ds}  \ \equiv \ {\mathbf x}_s
\label{curve}
\end{equation}
the  unit length bi-normal vector
\begin{equation}
\mathbf b \ = \ \frac{  {\mathbf x}_s\times  {\mathbf x}_{ss} } { || 
 {\mathbf x}_s \times {\mathbf x}_{ss} || }
\label{bcont}
\end{equation}
and the unit length normal vector,
\begin{equation}
\mathbf n = \mathbf b \times \mathbf t
\label{ncont}
\end{equation} 
The three vectors  $(\mathbf n, \mathbf b, \mathbf t)$ defines the right-handed, orthonormal Frenet frame. 
We may introduce this framing at every point along the string, whenever
\begin{equation}
{\mathbf x}_s\times  {\mathbf x}_{ss} \ \not= \ 0
\label{kcond}
\end{equation}
We proceed, momentarily, by
assuming this to be the case.
The Frenet equation then  computes the frames along the string as follows,
\begin{equation}
\frac{d}{ds}\!\left(
\begin{matrix} 
{\bf n} \\
{\bf b} \\
{\bf t} \end{matrix} \right) \ =  \   \left( \begin{matrix}
0 & \tau & - \kappa  \\ -\tau & 0 & 0 \\ \kappa & 0 & 0 \end{matrix} \right) 
\left(
\begin{matrix} 
{\bf n} \\
{\bf b} \\
{\bf t} \end{matrix} \right) 
\label{DS1}
\end{equation}
Here 
\begin{equation}
\kappa(s) \ = \ \frac{ || {\mathbf x}_s \times {\mathbf x}_{ss} || } { ||  {\mathbf x}_s||^3 }
\label{kg}
\end{equation}
is the curvature of the string on the osculating plane that is spanned by $\mathbf t$ and $\mathbf n$, and
\begin{equation}
\tau(s) \ = \ \frac{ ( {\mathbf x}_s \times  {\mathbf  x}_{ss} ) \cdot { {\mathbf x}_{sss} }} { || {\mathbf x}_s 
\times  {\mathbf x}_{ss} ||^2 }
\label{tau}
\end{equation}
is the torsion that measures how the string deviates from its osculating plane. Both $\kappa(s)$ and 
$\tau(s)$  are  extrinsic geometric quantities  {\it i.e.}  they depend only on the 
shape of the string in the ambient space $\mathbb R^3$. 
Conversely, if we know the curvature and torsion we can construct the string. For this  
we first solve for $\mathbf t(s)$ from the Frenet equation. We then solve for the string
by integration of (\ref{curve}). The solution is unique, modulo a global  translation and 
rotation of the string.  

Finally, we note that both the curvature (\ref{kg}) and the torsion (\ref{tau}) transform as scalars,  
under reparametrisations of the string. For this  we introduce an infinitesimal local diffeomorphism
along the string,  by deforming $s$ as follows
\begin{equation}
s \to   s + \epsilon(s) 
\label{infi}
\end{equation}
Here $\epsilon(s)$ is an arbitrary infinitesimally small function  such that
\[
\epsilon(0) = \epsilon (L) = 0 = \epsilon_s (0)  = \epsilon_s(L)
\]
Under this reparametrization of the string, the curvature and torsion transform as follows
\begin{equation}
\begin{matrix}
\delta \kappa (s) & = &  - \epsilon(s) \,  {\kappa}_s
\\ ~ \\ 
\delta \tau (s) & = & - \epsilon(s) \,  \tau_s \end{matrix}
\label{diffmor}
\end{equation}
which is the way how scalars transform. 
The Lie algebra of diffeomorphisms (\ref{infi})  is the 
classical Virasoro (Witt) algebra.

\section{Frame rotation and abelian Higgs multiplet}
\label{sect4}

In order to relate the  Abelian Higgs Multiplet with 
extrinsic string geometry,  
we observe that the normal and bi-normal vectors do not appear in (\ref{curve}). 
As a consequence  a SO(2) 
rotation around $\mathbf t (s)$,
%
%
%
%
%
%
%
%
%
%
%
{
\footnotesize
\begin{figure}[h]         
\begin{center}            
  \resizebox{7cm}{!}{\includegraphics[]{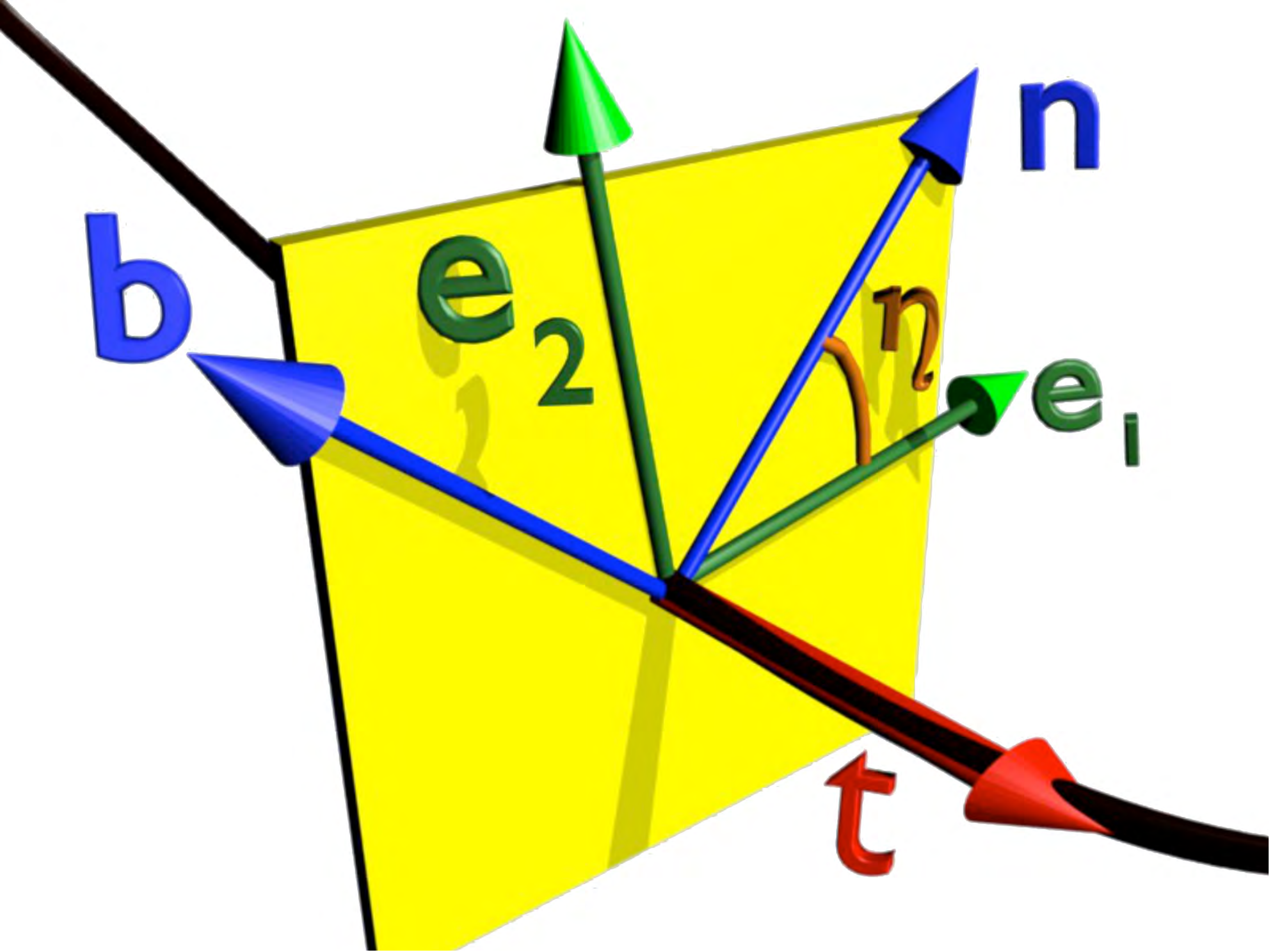}}
\end{center}
\caption {Rotation between the Frenet frames and a generic frame, on the normal plane of the string.}   
\label{fig-19}    
\end{figure}
}
\begin{equation}
\left( \begin{matrix} {\bf n} \\ {\bf b} \end{matrix} \right) \ \to \ \left( \begin{matrix} {{\bf 
e}_1} \\ {\bf e}_2 \end{matrix} \right) \
= \ \left( \begin{matrix} \cos \eta(s) &  \sin \eta(s) \\  - \sin \eta(s) & \cos \eta(s) \end{matrix}\right)
\left( \begin{matrix} {\bf n} \\ {\bf b} \end{matrix} \right).
\label{newframe}
\end{equation}
has no effect on the string.
For  the Frenet equation this rotation gives  
\begin{equation}
\frac{d}{ds} \left( \begin{matrix} {\bf e}_1 \\ {\bf e }_2 \\ {\bf t} \end{matrix}
\right) =
\left( \begin{matrix} 0 & (\tau + \eta_s) & - \kappa \cos \eta \\ 
- (\tau + \eta_s)  & 0 & \kappa \sin \eta \\
\kappa \cos \eta &  - \kappa \sin \eta  & 0 \end{matrix} \right)  
\left( \begin{matrix} {\bf e}_1  \\ {\bf e }_2 \\ {\bf t} \end{matrix}
\right).
\label{contso2}
\end{equation}
We  may utilise the  $\kappa$ dependent terms  in (\ref{contso2})  to promote $\kappa$ into a  complex quantity,  
with modulus that coincides with
the manifestly frame independent geometric curvature (\ref{kg}).
\begin{equation}
\kappa \ \buildrel{\eta}\over\longrightarrow \  \kappa ( \cos \eta + i \sin \eta)  \ \equiv \ \kappa e^{i\eta} 
\label{keta}
\end{equation}
This enables us to  interpret  the transformation  of ($\kappa,\tau$) in (\ref{contso2}) in terms of  
a one-dimensional version of the U(1) gauge
transformation (\ref{u1}): We identify the curvature as  the Higgs field  and the 
torsion as the U(1) gauge field \cite{Niemi-2003},
\begin{equation}
\begin{matrix} \kappa &   \to & \kappa e^{-i\eta} & \equiv  &  \phi  \\ ~ \\
\tau &   \to & \tau + \eta_s & \equiv  &   A_i 
\end{matrix}
\label{ahmulti}
\end{equation}
Note that when we  choose  
\begin{equation}
\eta (s) \ \to \ \eta_B(s) =  - \int_0^s \! \tau (\tilde s) d\tilde s 
\label{partran}
\end{equation}
we arrive at  the {\it unitary}  gauge in terms of the abelian Higgs multiplet. 
This defines
Bishop's  parallel transport framing \cite{Bishop-1975}. The Bishop's frame
vectors ${\bf e}^B_{1,2}$ do not rotate around the tangent vector,
\[
\frac{d}{ds} 
\left( {\bf e}^B_1 + i {\bf e}^B_2 \right)  \ = \ - \kappa \, e^{ - i \eta_B} \mathbf t 
%
\]
Thus, unlike the Frenet 
framing that can not be introduced when the curvature $\kappa(s)$ vanishes,
the Bishop framing can  be introduced and defined in an unambiguous and
continuous manner in that case.
However, it turns out that in the case of proteins which is the subject we are interested in, 
the Bishop frames do not work very well \cite{Hu-2011a}.

\section{The unique string Hamiltonian}
\label{sect7}

The curvature and torsion are the {\it only} available geometric quantities for constructing energy functions of
strings, while (\ref{H1}), ({\ref{chs})  is the  {\it unique} energy of  the Abelian Higgs multiplet
in the Kadanoff-Wilson sense of universality: 

Consider a string, in the limit where the curvature and torsion are slowly varying functions along it.
The shape of the string can not not depend on the  framing, thus its  energy 
can only involve combinations of the curvature and torsion in a manifestly frame independent fashion.   

On the other hand, (\ref{H1}),  ({\ref{chs}) is 
the unique SO(2)$\sim$U(1) invariant 
energy function that involves a complex  Higgs field
and a gauge field. It emerges from general 
arguments and symmetry principles alone, in the limit where the length scale that is 
associated to spatial variations of the field variables becomes very large 
in comparison to other characteristic length scales of the system.

Thus the {\it only} Hamiltonian  that can describe a string and its dynamics in 
the limit of  fields  with slow spatial variations is \cite{Niemi-2003}
\begin{equation}
H \ = \ \int\limits_0^L ds \, \left \{ \, |(\partial_s + i e \,{ \tau})  { \kappa} |^2 + \lambda\, (|
{\kappa} |^2 - m^2)^2 \, \right \}
\ + \ a \! \int\limits_0^L ds \, { \tau}
\label{enes}
\end{equation}
We  have here included the one dimensional version of  the Chern-Simons term (\ref{chs}).
It introduces net helicity along  the
string, breaking the $\mathbb Z_2$ symmetry between strings that are  twisted in
the right-handed and  left-handed sense.

In (\ref{enes}) both  $\kappa$ and  $\tau$ are expressed 
in a generic, arbitrary framing of the string.
The corresponding  gauge invariant variables (\ref{asu})
are the  curvature (\ref{kg}) and torsion (\ref{tau}) that characterise the
extrinsic string geometry. In terms of these gauge invariant variables, which we  from now on denote
by ($\kappa,\tau$)  the Hamiltonian (\ref{enes}) is 
\begin{equation}
H \ = \ \int\limits_0^L ds \, \left \{ \, (\partial_s { \kappa})^2  + e^2 {\kappa}^2 {\tau}^2 + 
\lambda\, ({\kappa}^2 - m^2)^2 \, \right \}
\ + \ a \int\limits_0^L ds \, \tau
\label{enes2}
\end{equation}
where we have simply followed the steps that gave us (\ref{H2}).
Thus  (\ref{enes2}) is the  unique energy of a  string, in terms of geometrically defined
curvature and torsion, and 
in the limit where the spatial variations of curvature and torsion along the string are small.

\section{Integrable hierarchy}
\label{sect8}

Relations exist  between (\ref{enes2}),  the integrable hierarchy of the 
nonlinear Schr\"odinger (NLS) equation, and the Heisenberg spin chain of ferromagnetism. 
For this we introduce the following
Hasimoto 
variable 
\begin{equation}
\psi(s) \ = \ \kappa(s)\,  e^{i e\int\limits_0^s \, ds' \, \tau(s') }
\label{hasi}
\end{equation}  
that combines the curvature and torsion into a single {\it gauge invariant} complex variable.
In terms of (\ref{hasi}), we obtain the Hamiltonian of the  integrable
nonlinear Schr\"odinger  equation \cite{Faddeev-1987,Ablowitz-2003,Kevrekidis-2009} as follows,
\begin{equation}
\kappa_s^2 + e^2 \kappa^2 \tau^2  + \lambda \kappa^4 \ = \  \bar\psi_s \psi_s + \lambda (\bar\psi \psi)^2 \ = \ 
\mathcal H_3 
\label{H3}
\end{equation}
With the Poisson bracket
\[
\{ \psi(s) , \bar \psi(s')\} \ = \  i \delta(s-s')
\]
the lower level conserved densities in the integrable NLS hierarchy 
are the helicity $\mathcal H_{-2}$, length ({\it i.e.} Nambu-Goto) $\mathcal H_{-1}$, 
number operator $\mathcal H_{1}$ and momentum $\mathcal H_{2}$ 
\begin{equation}
\begin{matrix} 
\mathcal H_{-2} & = & \tau & \\ 
\mathcal H_{-1} & = & L & \\
\mathcal H_{1} & = & \kappa^2  & \hskip 0.2cm  \sim \bar \psi \psi \\
\mathcal H_{2} & = & i \kappa^2 \tau & \hskip 0.2cm  \sim  \, \bar \psi \psi_s \\
\end{matrix}
\label{Hnls}
\end{equation}
The energy (\ref{enes2}) is a combination of $\mathcal H_{-2}$, $\mathcal H_{1}$ and $\mathcal H_{3}$. From the
perspective of the NLS hierarchy, 
the momentum $\mathcal H_{2}$ should also be included for completeness. If higher order corrections are
desired the natural candidate is the mKdV density
\[
\mathcal H_4 \ = \  i \kappa \kappa_{sss} + 2 \kappa \kappa_{ss} \tau +  \kappa^2 \tau^3 - e^2 \kappa_s^2 \tau 
+ \frac{3}{4}  \lambda \kappa^4 \tau
\]
\[ 
\sim \ i \, \bar \psi \psi_{sss} \! +  i \frac{\lambda}{2} \bar\psi \psi ( \bar \psi_s
\psi + 4 \bar \psi \psi_s )
\]
with appears as the next level conserved density in the NLS hierarchy. 

We note that the Heisenberg spin chain is obtained
from $\mathcal H_{1}$ using the Frenet equation.
\[
\int\limits_0^L ds \, \mathcal H_{1}  \  =   \int\limits_0^L ds \, \kappa^2 \ = \  \int\limits_0^L ds \, |\mathbf t_s|^2
\]
Furthermore, the combination  of $\mathcal H_{-1}$ and ${\mathcal H}_1$ yields Polyakov's rigid string action \cite{Polyakov-1986}.
In the context of polymers, it relates to the  Kratky-Porod model \cite{Kratky-1949}.

\section{Strings from solitons}
\label{sect9}

Solitons are the paradigm structural
self-organisers in Nature. Solitons become materialised in diverse scenarios 
\cite{Faddeev-1987,Ablowitz-2003,Kevrekidis-2009,Manton-2004}; 
we have already seen that solitons conduct 
electricity in organic polymers. But  solitons can also transmit data in transoceanic cables,
and solitons can transport chemical energy along proteins.
Solitons explain the Mei\ss ner effect in  superconductivity and solitons model 
dislocations in liquid crystals.  Solitons  are used to describe hadronic particles, 
cosmic strings and magnetic monopoles in  high  energy physics.  

We argue that  solitons describe even  life. We
argue that each of us has some  $10^{20}$ solitons in our body. 
These solitons are the building blocks of folded proteins, they are the 
ingredients of all metabolic processes that make us alive.
 
The NLS equation that we obtain from (\ref{H3}),  
is the paradigm equation that supports 
solitons \cite{Faddeev-1987,Ablowitz-2003,Manton-2004,Kevrekidis-2009}.
Depending on the sign of $\lambda$, the soliton 
is either dark ($\lambda >0$) or bright ($\lambda <0$).
The torsion independent contribution to (\ref{enes2})
\begin{equation}
H \ = \ \int\limits_{-\infty}^\infty ds \, \left \{ \,  \kappa_s^2   + 
\lambda\, (\kappa^2 - m^2)^2 \, \right \}
\label{swave}
\end{equation}
reproduces our previous  instanton equation (\ref{nose2a}) with $Q = \sqrt{2}$; the Hamiltonian (\ref{swave})
supports  the double well soliton {\it a.k.a.}  the paradigm {\it topological} soliton:
When $m^2$ is positive and when $\kappa$ can take both positive and negative values, 
the equation of motion 
\[
\kappa_{ss} =  2 \lambda \kappa (\kappa^2 - m^2)
\]
is solved by - see (\ref{topsol})
\begin{equation}
\kappa(s) \ = \ m \, \tanh \left[ m \sqrt{\lambda} (s-s_0)\right]
\label{soliton}
\end{equation}

We have  concluded that the energy function
\begin{equation}
\mathcal E \ = \ \int ds \, \left\{ \, \kappa_s^2 + \lambda (\kappa^2 - m^2)^2 + \frac{d}{2} \kappa^2 \tau^2
- b \kappa^2 \tau - a\tau + \frac{c}{2} \tau^2 \, \right\}
\label{enenls}
\end{equation}
is the most general one,  with a solid geometric foundation; it is 
a linear combination of {\it all} the densities (\ref{H3}), (\ref{Hnls}). Note that in (\ref{enenls})
we have also included 
the Proca mass; this is the last term.  
Even though the Proca mass does 
not appear in the integrable NLS hierarchy, it does 
have a claim to be gauge invariant  \cite{Hu-2013a,Ioannidou-2014}.  Eventually, we shall present a
topological argument why the Proca mass should be included.

The energy (\ref{enenls}) is quadratic in the torsion. Thus we can eliminate 
$\tau$ using its  equation of motion,
\begin{equation}
\frac{\delta \mathcal E}{\delta \tau} \ = \ d \kappa^2 \tau - b \kappa^2 - a + c\tau \ = \ 0
\label{moneq0}
\end{equation}
This  gives
\begin{equation}
\tau [\kappa] \ = \ \frac{ a + b\kappa^2}{c+d\kappa^2} \ \equiv \ \frac{a}{c}  \, \frac{ 1 +  (b/a) \kappa^2}{1+(d/c)
\kappa^2}
\label{taueq}
\end{equation}
and we obtain the following {\it effective} energy for the curvature,
\begin{equation}
\mathcal E_\kappa \ = \ \int ds \, \left\{ \frac{1}{2} \kappa_s^2 + V[\kappa]  \right\}
\label{modNLS}
\end{equation}
with the equation of motion
\begin{equation}
\frac{\delta \mathcal E_\kappa}{\delta \kappa} \ = \ -\kappa_{ss} +  V_\kappa \ = \ 0
\label{monequ}
\end{equation}
where
\begin{equation}
V[\kappa] \ = \ - \left( \frac{ bc - ad}{d}\right) \, \frac{1}{c+d\kappa^2} \ - \ \left( \frac{b^2 + 8\lambda m^2}{2b} \right)
\, \kappa^2 + \lambda \, \kappa^4
\label{V}
\end{equation}
This is a deformation of the potential in (\ref{swave}), the two share the same large-$\kappa$ asymptotics. 
When we select the parameters  properly, we can expect that (\ref{moneq0}), (\ref{monequ}), (\ref{V}) 
continue to support topological solitons.
But we do not know their explicit profile, in terms of elementary functions. 
In the sequel we shall construct these solitons numerically.  

Once we have constructed the soliton of (\ref{monequ}), we  evaluate $\tau(s)$ from  (\ref{taueq}).
We substitute these profiles in the Frenet equation (\ref{DS1}) and solve for $\mathbf t(s)$. We then integrate
(\ref{curve}) to obtain the string $\mathbf x(s)$ that corresponds to the soliton.

\section{Anomaly in the Frenet frames}
\label{sect10}

When the curvature of a string vanishes, there is an anomaly in the Frenet framing.
It turns out that the origin of this anomaly is also
the {\it raison d'edre} for a topological soliton to reside on a string. 

\vskip 0.2cm
Until now,  we have assumed (\ref{kcond}) so that
the curvature (\ref{kg}) does not vanish. But we have also 
observed that in the case of the Abelian Higgs model  it is natural for
the density $\rho$ to vanish, on the world-sheet of a vortex core. 
Thus, in the context of AHM the vanishing  
of $\rho$ relates to important, physically significant  effects.  
Furthermore, the  explicit soliton profile  (\ref{soliton}) displays 
both positive and negative values, and in particular (\ref{soliton}) 
vanishes when $s=s_0$.  Consequently we should consider the possibility that 
$\kappa $ may vanish, and even become negative, along a string. 

We start by extending the curvature (\ref{kg}) so that it has
both positive and negative values. According to  (\ref{keta})
the negative values of $\kappa$ are related to 
the positive ones by a 
$\eta =  \, \pm\pi $ frame rotation,
\begin{equation}
\kappa \ \buildrel{\eta\, = \, \pm \pi}\over\longrightarrow \ e^{\pm i \pi } \kappa \ = \ - \kappa 
\label{kappaeta}
\end{equation}
We compensate for an
extension of (\ref{kg})  to negative values, by  introducing the 
discrete $\mathbb Z_2$ symmetry \cite{Hu-2011a}
\begin{equation}
\begin{matrix} \kappa & \leftrightarrow & - \kappa \\ 
\eta & \leftrightarrow & \eta \pm \pi
\end{matrix} \ \  \Leftrightarrow \ \ \kappa e^{i\eta} \leftrightarrow \kappa e^{i\eta}
\end{equation}

%
%
%
%
%
%
%
%
%
%
%
{
\footnotesize
\begin{figure}[h]         
\begin{center}            
  \resizebox{8 cm}{!}{\includegraphics[]{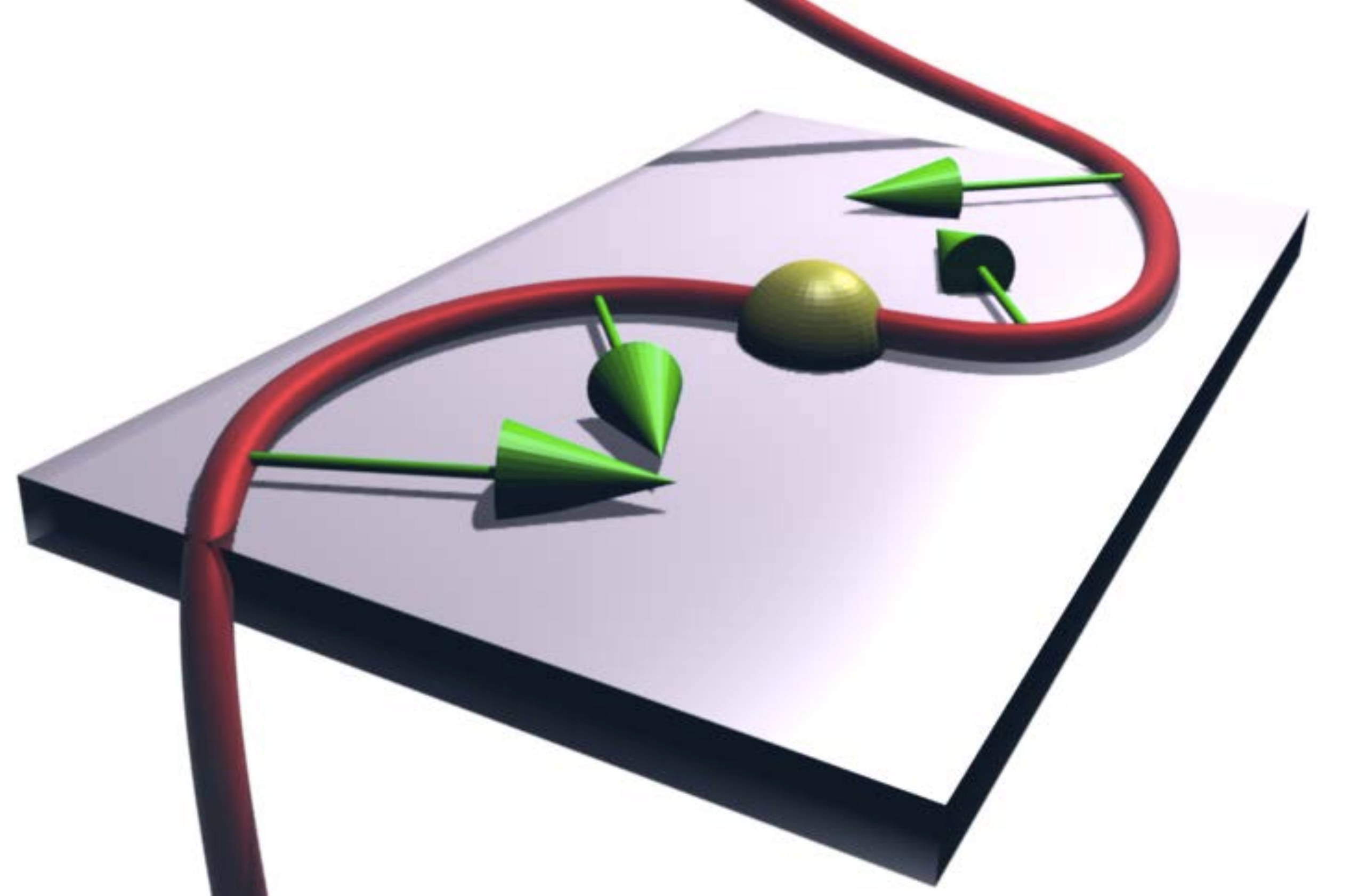}}
\end{center}
\caption {A string with inflection point (ball). At each point the Frenet frame normal vector points towards the center of
    the osculating circle. At the inflection point we have a discontinuity in the direction of the normal vectors: 
    The radius of the osculating  circle diverges and the normal vector are
    reflected in the osculating plane, from one side to the other side of the string. }   
\label{fig-20}    
\end{figure}
}
An (isolated) point where $\kappa(s)$ vanishes is called
an inflection point. Figure \ref{fig-20} shows an example of an inflection point. 
As shown in this Figure, in the limit of a plane curve 
we obtain  a discontinuity in the Frenet frames, when the string  
goes through the inflection point:
The zweibein ($\mathbf n, \mathbf b$) becomes reflected according to
\begin{equation}
(\mathbf n + i \mathbf b)  \ 
\longrightarrow \  - (\mathbf n + i \mathbf b) \ = \  e^{\pm i \pi}(\mathbf n + i \mathbf b)
\label{anom}
\end{equation}
when we have a simple inflection point along the string. 
At the inflection point itself,  the 
Frenet frames can not defined. Thus  we can not unambiguously deduce whether 
we have a jump by $\eta = +\pi$ or by $\eta = -\pi$ at the inflection point.  
We can not conclude whether the Frenet frame vectors  ($\mathbf n, \mathbf b$) become rotated
clockwise or counterclockwise by an angle $\pi$  along the tangent vector. There is a $\mathbb Z_2$  
anomaly in the definition of Frenet framing, due to  inflection points.

To analyse the anomaly,  consider a  string $\mathbf x(s)$ that has 
a simple inflection point when $s=s_0$. Thus $\kappa(s_0)=0$ but $\kappa_s(s_0) \not=0$,
as shown in Figure \ref{fig-20}.
To simplify the  notation we re-define the parameter $s$ so that the inflection point occurs at 
$s_0=0$.  

We can always remove the inflection point by a  generic deformation of the string:
A deformation which is 
restricted to the plane as in Figure  \ref{fig-20} only slides the inflection point along the string 
without removing it. In order to remove the inflection point we need to deform the string off its instantaneous 
tangent plane: The co-dimension of the inflection point in $\mathbb R^3$ is two,  
the inflection point is not generic along a space string.

Consider two different generic deformations,
\begin{equation}
\mathbf x(s) \ \to \ \mathbf x(s) + \delta \mathbf x_{1,2}(s) \ = \ \mathbf x_{1,2}(s)
\label{12curv}
\end{equation}
In the case shown in  Figure \ref{fig-20}, these two deformations 
amount to moving the string either 
slightly up from the plane, or slightly down from the plane, around the inflection point. 
We assume that the deformations are very small,  and compactly supported so that 
\[
\delta \mathbf x_{1,2}(\pm \varepsilon_\pm) = 0
\] 
Here  $\varepsilon_\pm>0$   
are small and determine the parameter values where the deformations 
$ \mathbf x_{1,2}(s)$ bifurcate.

Imagine now a closed string denoted  $\gamma$, that starts from 
$\mathbf x(-\varepsilon_- )$, follows along $\mathbf x_1 $ to $\mathbf x(+\varepsilon_+ )$
and then returns along $\mathbf x_2 $  back to 
$\mathbf x(-\varepsilon_-)$. Introduce the Frenet frame 
normal vectors of $\gamma$; by shifting $\gamma$ slightly 
into the direction of its Frenet frame normals, we obtain 
a second closed string which we call $\tilde\gamma$.  Let  $\mathbf t$ and $\tilde{\mathbf t}$ be the corresponding
tangent vectors.  The Gau\ss~linking number of $\gamma$ and  $\tilde\gamma$ is
\begin{equation}
{\tt Lk} \ = \  \frac{1}{4\pi} \oint\limits_{\gamma} \oint\limits_{\tilde \gamma} ds d\tilde s \,
\frac{\mathbf x - \tilde{\mathbf x} }{|\mathbf x - \tilde{\mathbf x} |^3} \cdot (\mathbf t  \times \tilde {\mathbf t })
\label{gaussln}
\end{equation}
Proceeding along $\mathbf x_{1,2}(s)$ the respective Frenet frames are  continuously 
rotated by $\eta_{1,2}\approx \pm \pi$; in the limit 
where $\delta \mathbf x_{1,2} \to 0$ we would obtain
the discontinuous jump (\ref{anom}). 
By continuity of Frenet framing in the complement 
of inflection points, the linking number has  
values  {\tt Lk}$=\pm 1$ when the $\eta_{1,2}$ 
change in the same direction; we remind  that $\gamma$ proceeds "backwards" 
along $\mathbf  x_2 $. But if the framing along $\mathbf x_1(s)$ and $\mathbf x_2(s)$ rotate
in the opposite directions, we have  {\tt Lk}$=0$.

Accordingly,  the relative sign of $\eta_{1,2}$ depends on the way 
how the inflection point is circumvented: We have a
{\it frame anomaly} in the Frenet framing as $\delta \mathbf x_{1,2}\to 0$,
the value of {\tt Lk} depends on the way  how we define $\delta \mathbf x_{1,2}(s)$.

Finally, we comment on the Bishop framing (\ref{partran}) which from the point of view of the AHM
is like the unitary gauge. Consider our initially planar string and introduce the local deformation 
(\ref{12curv}). Then introduce {\it any} framing that is capable of rotating around the 
tangent vector $\mathbf t(s)$ when we move along the string {\it e.g.} the Frenet framing. Then rotate the string 
any integer number of times within the locally deformed region
so that the framing becomes self-linked and the Gau\ss~ linking number
(\ref{gaussln}) is non-vanishing. In the limit where the support of the local deformation vanishes
the Bishop framing is not capable of observing any residual effect. Indeed,  it is well known
that in the context of the AHM model, unitary gauge can not detect the presence 
of  topological defects such as vortices.

\begin{quote}
{\it Research project: Analyse in detail 
the framing of a string in the presence of an inflection point, using the Bishop's frames 
(\ref{partran}).}
\end{quote}

\section{Perestroika}
\label{pere}

An inflection point together with the corresponding Frenet  frame anomaly can be given an interpretation
in terms of a
string specific bifurcation, which is called {\it inflection point perestroika} 
\cite{Arnold-1990,Arnold-1995,Arnold-1996,Aicardi-2000,Uribe-2004}. It explains
why  a {\it uniquely } defined Frenet framing  across 
the inflection point,  or any other framing that rotates 
around the tangent vector, is not possible:

Consider a segment of a string, along which the torsion $\tau(s)$ 
vanishes. Accordingly the string segment is constrained on a plane, as   in figure 
\ref{fig-20}.  When a string  is constrained on a plane, a 
simple isolated inflection point is generic.  This follows since for a string on plane
the inflection point has co-dimension one. 
Moreover, in the case of a string on plane a single  simple inflection point is a topological
invariant.  It can only be moved around the plane  but not made to disappear; unless it 
escapes the plane which we assume is not the case.  
If we have two simple inflection points along a string on plane, we can bring them together 
to deform the string so that no inflection point remains.  Thus
the inflection point is a {\it mod}(2) topological invariant of a string on plane.

Consider now a generic string in $\mathbb R^3$; a generic string is not constrained on a plane.
The co-dimension of a single simple inflection point is two, thus a generic 
string does not have any inflection points. But along a   
string which moves freely in $\mathbb R^3$,  an isolated simple inflection point appears generically, 
at some point, at some moment, during the motion. 
When this infection point perestroika takes place along the moving string, 
it leaves a trail behind:  The  inflection point perestroika  changes the number of  {\it flattening points},  
which are points along the string where the torsion $\tau(s)$ vanishes \cite{Aicardi-2000,Uribe-2004}. 

At a simple flattening point the curvature $\kappa(s)$ is generically non-vanishing, but 
the torsion $\tau(s)$  changes its sign. Accordingly the inflection point perestroika can only change the   
number of simple flattening points by two. Apparently, it always does \cite{Aicardi-2000,Uribe-2004}.

Unlike the inflection point, a flattening point where $\tau(s)=0$ 
is generic along a static space string. 
Furthermore, unlike a simple inflection point, a single 
simple flattening point that occurs in a one parameter family of strings in $\mathbb R^3$
is a topological invariant. It can not disappear on its own, under local 
deformations that leave the ends of the string intact.
A pair of flattening points can be  combined together, into a single bi-flattening point, which can 
then dissolve. When this  happens, a second string-specific  bifurcation which is 
called {\it bi-flattening perestroika} takes place.

Apparently, inflection point perestroika and bi-flattening perestroika are the only two bifurcations 
where the number of flattening points can change \cite{Uribe-2004}.
The number of simple flattening 
points is  a {\it local} invariant of the string. Besides the 
flattening number, and  the self-linking number in case of a framed string, a generic 
smooth string does not possess any other essential 
local invariants \cite{Aicardi-2000}. The two are also mutually 
independent, even though they often appear together.

For example, one can deduce that the self-linking number of a string increases by one if the torsion is
positive when the string approaches its simple inflection point, and if two simple 
flattening points disappear after the passage of the inflection point. 
Moreover, if the torsion is negative, the self-linking number decreases by one 
when two flattening points disappear after the 
passage \cite{Aicardi-2000}. But when two simple flattening points dissolve
in a bi-flattening perestroika, the self-linking number in general does not change.

\vskip 0.2cm
A bifurcation is the paradigm cause for structural transitions, including phase transitions,  in any
dynamical system. 
Inflection point and bi-flattening perestroika's 
are the only  bifurcations that are string specific. Accordingly these two
perestroika's must have a profound r\^ole 
in determining the physical behaviour and phase structure of string-like configurations.  
In particular, these perestroika's must be 
responsible for any string-specific structural re-organisation that
takes place when the value of the compactness index $\nu$ in (\ref{nuval}) changes.
Since perestroika's relate to the creation and disappearance
of topological solitons such as (\ref{soliton}) along a string,  it is clear that
perestroika's and  topological solitons,  with the ensuing physical effects, 
are commonplace whenever we have a string with an energy function of the 
form (\ref{enenls}).   

\subsection{Example:}  A good example of the interplay between inflection points and flattening points is given
by  (\ref{soliton})  or more generally by a soliton 
solution of (\ref{monequ}), and (\ref{taueq}). For a regular string, the denominator 
of (\ref{taueq}) should not vanish. Thus, in the case of an inflection point the ration ($d/c$) should be positive. 
When ($b/a$) is negative, we have  a symmetric pair of inflection points around the inflection point. Thus starting 
from a one-parameter family of strings $\kappa(s,v)$ with $v$ the parameter, if initially $\kappa(s,v)$ is sufficient 
large and {\it e.g.} positive, and  we are not close to an inflection point, there are no flattening points either. 
When $v$ is varied so that 
the inflection point is approached, a pair of flattening points emerges and remains whenever the
curvature has the profile (\ref{soliton}). 
In particular, we conclude that it is important  to retain the Proca mass term, even a very small one,
as a regulator.

%
%
%
%
%
%
%
%
%
%


\chapter{Discrete Frenet Frames}
\label{sect11}


Proteins are not like continuous differentiable strings. Proteins are like
piecewise linear polygonal strings. Thus, to understand the physical properties of
proteins we need to develop the formalism
of discrete strings. 
Accordingly, we proceed to generalise the Frenet frame formalism to the case of polygonal strings 
that are piecewise linear \cite{Hu-2011a}. 

Let $\mathbf r_i$ with $i=1,...,N$ be the vertices of a piecewise linear discrete string.
At each vertex we introduce the unit tangent vector 
\begin{equation}
\mathbf t_i = \frac{ {\bf r}_{i+1} - {\bf r}_i  }{ |  {\bf r}_{i+1} - {\bf r}_i | }
\label{t}
\end{equation}
the unit binormal vector
\begin{equation}
\mathbf b_i = \frac{ {\mathbf t}_{i-1} - {\mathbf t}_i  }{  |  {\mathbf t}_{i-1} - {\mathbf t}_i  | }
\label{b}
\end{equation}
and the unit normal vector 
\begin{equation}
\mathbf n_i = \mathbf b_i \times \mathbf t_i
\label{n}
\end{equation}
The orthonormal triplet ($\mathbf n_i, \mathbf b_i , \mathbf t_i$) defines a
discrete version of the Frenet  frames  (\ref{curve})-(\ref{ncont}) 
at each position $\mathbf r_i$ along the 
string, as shown in Figure (\ref{fig-21})
%
%
%
%
%
%
%
%
%
%
%
{
\footnotesize
\begin{figure}[h]         
\begin{center}            
  \resizebox{8 cm}{!}{\includegraphics[]{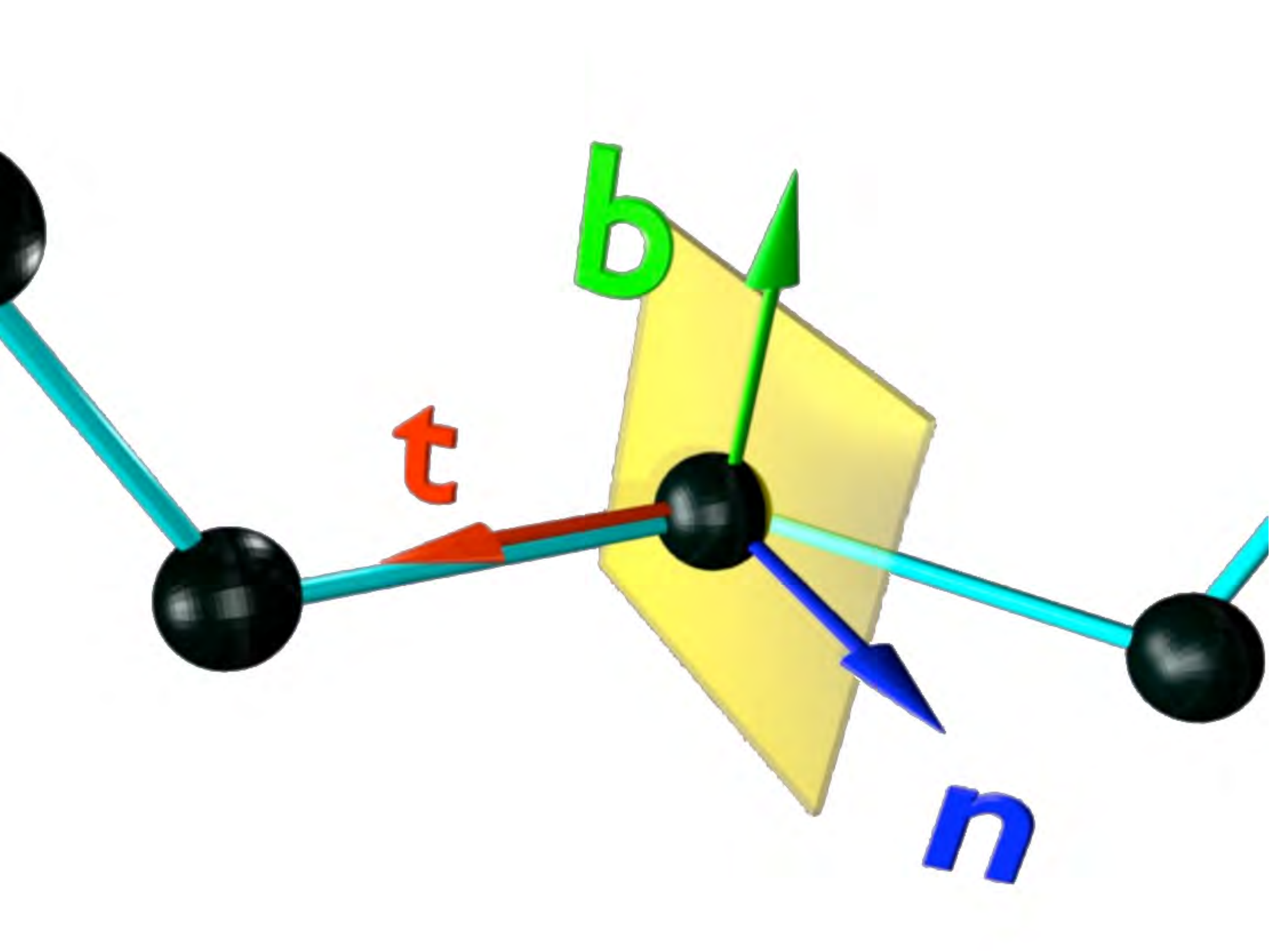}}
\end{center}
\caption {Discrete Frenet frames along a piecewise linear discrete string.   }   
\label{fig-21}    
\end{figure}
}
In lieu of the curvature and torsion, we have the  bond angles and torsion angles, defined as  in
Figure \ref{fig-22}.
%
%
%
%
%
%
%
%
%
%
%
{
\footnotesize
\begin{figure}[h]         
\begin{center}            
  \resizebox{8 cm}{!}{\includegraphics[]{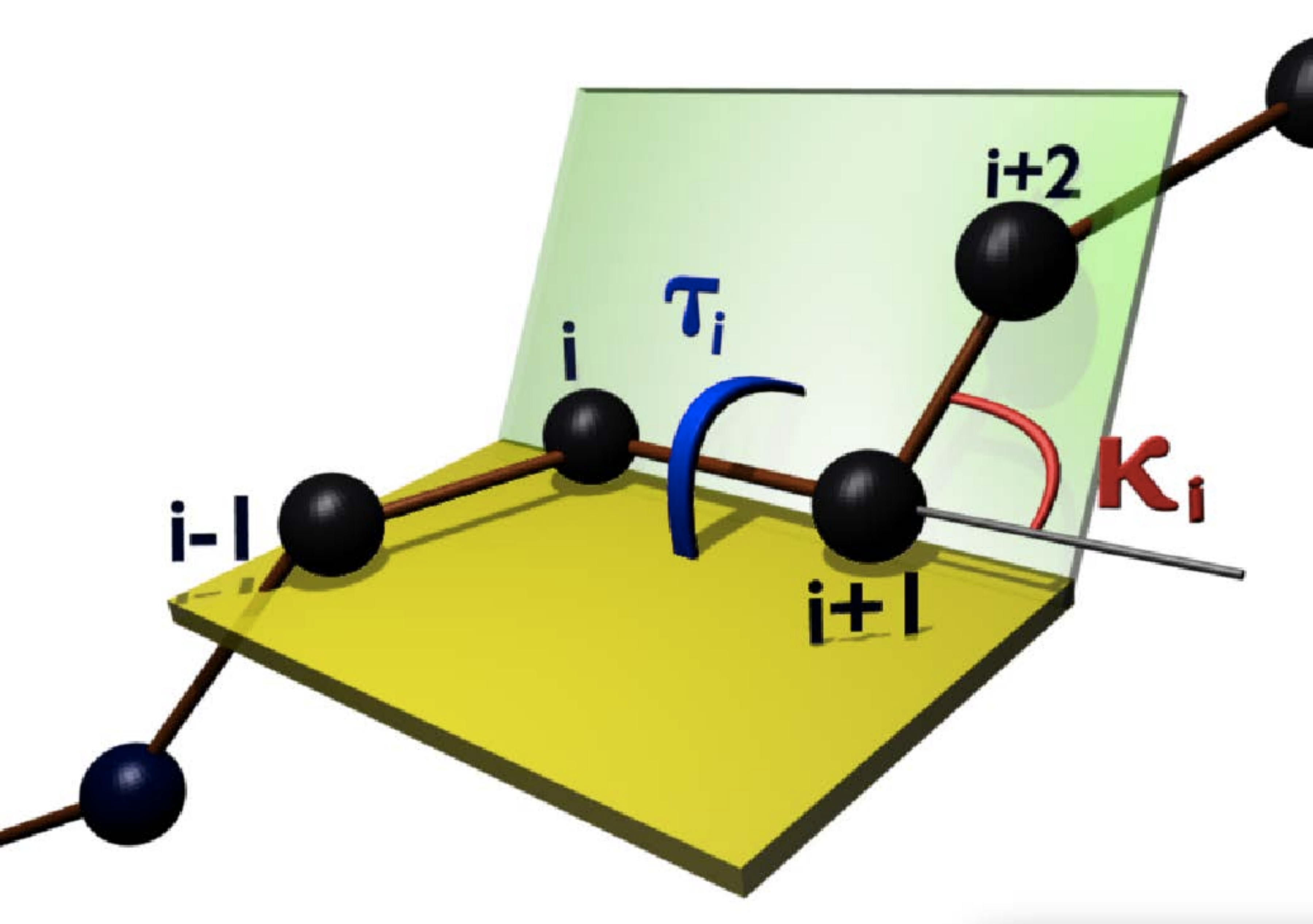}}
\end{center}
\caption {Definition of bond ($\kappa_i$) and torsion ($\tau_i$) angles, along the piecewise linear discrete string.   }   
\label{fig-22}    
\end{figure}
}

When we know the Frenet frames at each vertex, we can compute the values of these angles: The bond angles are
\begin{equation}
\kappa_{i} \ \equiv \ \kappa_{i+1 , i} \ = \ \arccos \left( {\bf t}_{i+1} \cdot {\bf t}_i \right)
\label{bond}
\end{equation}
and the torsion angles are
\begin{equation}
\tau_{i} \ \equiv \ \tau_{i+1,i} \ = \ {\rm sign}\{ \mathbf b_{i-1} \times \mathbf b_i \cdot \mathbf t_i \}
\cdot \arccos\left(  {\bf b}_{i+1} \cdot {\bf b}_i \right) 
\label{tors}
\end{equation}

Conversely, when the values of the bond and torsion angles are all known, 
we can use the discrete Frenet equation
\begin{equation}
\left( \begin{matrix} {\bf n}_{i+1} \\  {\bf b }_{i+1} \\ {\bf t}_{i+1} \end{matrix} \right)
= 
\left( \begin{matrix} \cos\kappa \cos \tau & \cos\kappa \sin\tau & -\sin\kappa \\
-\sin\tau & \cos\tau & 0 \\
\sin\kappa \cos\tau & \sin\kappa \sin\tau & \cos\kappa \end{matrix}\right)_{\hskip -0.1cm i+1 , i}
\left( \begin{matrix} {\bf n}_{i} \\  {\bf b }_{i} \\ {\bf t}_{i} \end{matrix} \right) 
\label{DFE2}
\end{equation}
to  compute  the frame at position $i+i$ 
from the frame at position $i$. Once  all the frames have been constructed,  
the entire string is given by
\begin{equation}
\mathbf r_k = \sum_{i=0}^{k-1} |\mathbf r_{i+1} - \mathbf r_i | \cdot \mathbf t_i
\label{dffe}
\end{equation}
Without any loss of generality we may choose $\mathbf r_0 = 0$, make $\mathbf t_0$ to 
point into the direction of the positive $z$-axis, and let $\mathbf t_1$ lie on the $y$-$z$ plane.

The vectors $\mathbf n_i$ and $\mathbf b_i$ do not appear in (\ref{dffe}). Thus, as in
the case of continuum strings, a discrete string remains intact under frame
rotations of  the ($\mathbf n_i, \mathbf b_i$) zweibein around $\mathbf t_i$.
This local SO(2)  rotation acts on the frames as follows
\begin{equation}
 \left( \begin{matrix}
{\bf n} \\ {\bf b} \\ {\bf t} \end{matrix} \right)_{\!i} \!
\rightarrow  \!  e^{\Delta_i T^3} \left( \begin{matrix}
{\bf n} \\ {\bf b} \\ {\bf t} \end{matrix} \right)_{\! i} =   \left( \begin{matrix}
\cos \Delta_i & \sin \Delta_i & 0 \\
- \sin \Delta_i & \cos \Delta_i & 0 \\ 
0 & 0 & 1  \end{matrix} \right) \left( \begin{matrix}
{\bf n} \\ {\bf b} \\ {\bf t} \end{matrix} \right)_{\! i}
\label{discso2}
\end{equation}
Here $\Delta_i$ is the rotation angle at vertex $i$ and $T^3$ is one of the SO(3) 
generators 
\[
T^1 = \left( \begin{matrix} 0 & 0 & 0 \\
0 & 0 & -1 \\ 0 & 1 & 0 \end{matrix} \right) \ \  T^2 = \left( \begin{matrix} 0 & 0 & 1 \\
0 & 0 & 0 \\ -1 & 0 & 0 \end{matrix} \right) \ \  T^3 = \left( \begin{matrix} 0 & -1 & 0 \\
1 & 0 & 0 \\ 0 & 0 & 0 \end{matrix} \right)
\]
that satisfy the Lie algebra
\[
[T^a , T^b ] = \epsilon^{abc} T^c
\]
Using  these matrices we can write the effect of frame rotation
on the bond and torsion angles as follows
\begin{equation}
\kappa_{i}  \ T^2  \ \to \  e^{\Delta_{i} T^3} ( \kappa_{i} T^2 )\,  e^{-\Delta_{i} T^3}
\label{sok}
\end{equation}
\begin{equation}
\tau_{i}  \ \to \ \tau_{i} + \Delta_{i-1} - \Delta_{i}
\label{sot}
\end{equation}
From the point of view of lattice gauge theories,
the transformation of bond angles is like an adjoint SO(2)$\in$SO(3) 
gauge rotation of a Higgs triplet around the Cartan
generator $T^3$, when the Higgs triplet is  in the
direction of $T^2$. The transformation of torsion angle coincides with that of the 
SO(2) lattice gauge field. Since the $\mathbf t_i$ remain intact under (\ref{discso2}),
the gauge transformation of ($\kappa_i, \tau_i$) has no effect on the geometry of the discrete string. 

{\it A priori}, the fundamental range of the bond angle is  $\kappa_i \in [0,\pi]$ while for the 
torsion angle the range is $\tau_i \in [-\pi, \pi)$. Thus we 
identify ($\kappa_i, \tau_i$) as the canonical 
latitude and longitude angles of a two-sphere $\mathbb S^2$. 
In parallel with the continuum case we find it useful to extend the range
of $\kappa_i$ into negative values $ \kappa_i \in [-\pi,\pi]$ $mod(2\pi)$. 
As in (\ref{kappaeta}) we compensate for this two-fold covering of $\mathbb S^2$ 
by a $\mathbb Z_2$ symmetry which now takes the following form:
\begin{equation}
\begin{matrix}
\ \ \ \ \ \ \ \ \ \kappa_{k} & \to  &  - \ \kappa_{k} \ \ \ \hskip 1.0cm  {\rm for \ \ all} \ \  k \geq i \\
\ \ \ \ \ \ \ \ \ \tau_{i }  & \to &  \hskip -2.5cm \tau_{i} - \pi 
\end{matrix}
\label{dsgau}
\end{equation}
This is a special case of (\ref{sok}), (\ref{sot}), with
\[
\begin{matrix} 
\Delta_{k} = \pi \hskip 1.0cm {\rm for} \ \ k \geq i+1 \\
\Delta_{k} = 0 \hskip 1.0cm {\rm for} \ \ k <  i+1 
\end{matrix}
\]

\section{The C$\alpha$ trace reconstrucion}
\label{sect12}

We have already concluded that the Ramachandran angles are not sufficient for reconstructing
the protein backbones: As shown in Figure (\ref{fig-11}) the reconstructed backbones are not  in the same
universality class with folded proteins. The value of the compactness index $\nu$ is different.  
For a correct reconstruction, we need to utilise all the bond and torsion angles that we have 
defined in Figure \ref{fig-7}. Only for the bond lengths can the average values be used.

We now consider the  protein backbone reconstruction, in terms of virtual C$\alpha$ backbone. 
We identify the vertices in Figure \ref{fig-22} with the C$\alpha$ atoms, so that ($\kappa_i, \tau_i$) are
the virtual C$\alpha$ backbone bond and torsion angles. For the virtual C$\alpha$-C$\alpha$ 
bond lengths we use the average PDB value 
\begin{equation}
 |\mathbf r_{i+1} - \mathbf r_i |  \ \sim \ 3.8 \ \ {\rm \AA}
\label{dist}
\end{equation}
It turns out that, {\it unlike} in the case of the Ramachandran angles, the ensuing 
reconstructed C$\alpha$ backbones reproduce
the original crystallographic structures, with a very high precision; the difference is mostly 
within the range of experimental errors, as measured by the B-factor (\ref{dw}).
%
%
%
%
%
%
%
%
%
%
%
{
\footnotesize
\begin{figure}[h]         
\begin{center}            
  \resizebox{9 cm}{!}{\includegraphics[]{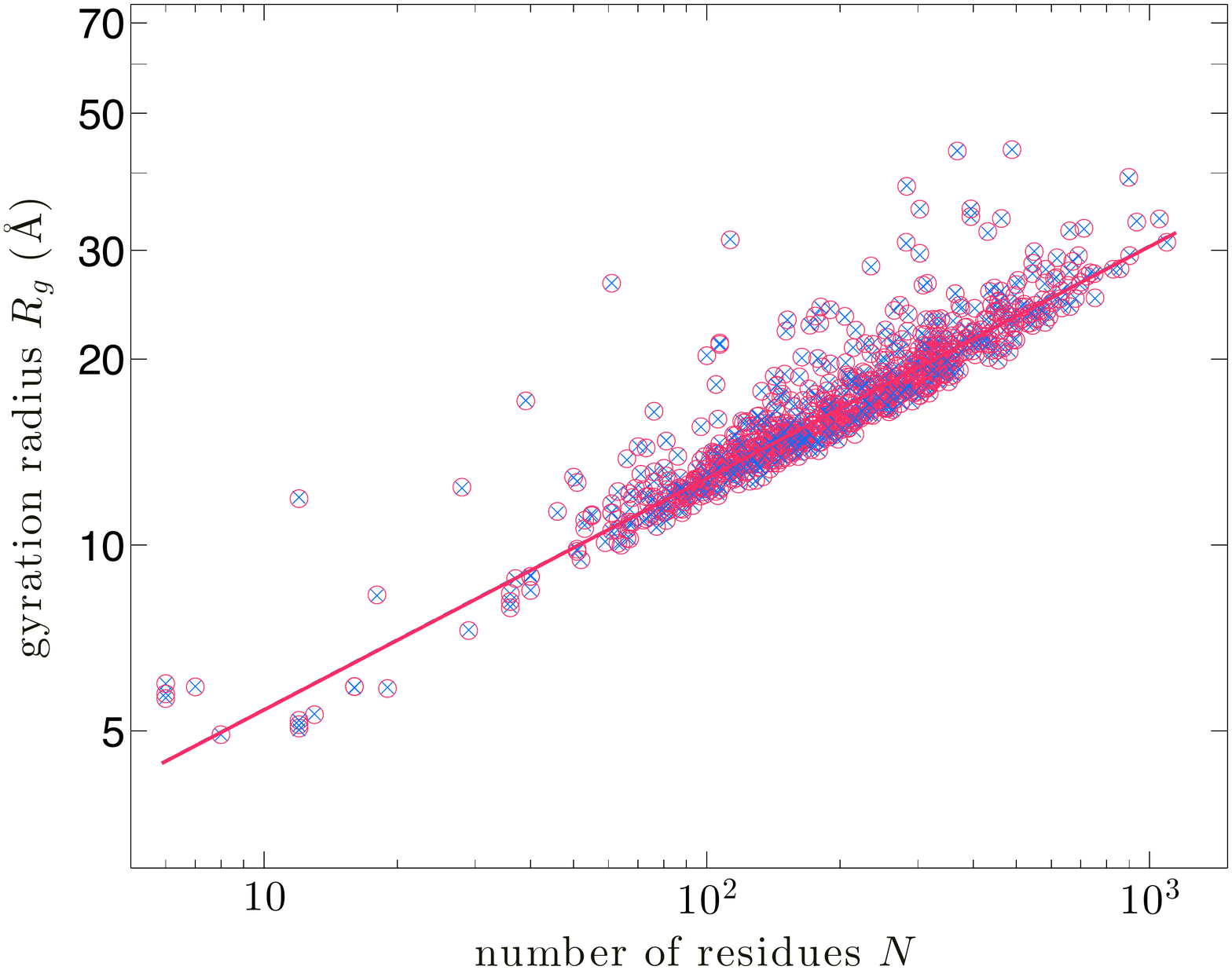}}
\end{center}
\caption {Comparison of the C$\alpha$-C$\alpha$ radius of gyration data between the original PDB structures and those reconstructed in terms of variable virtual bond and torsion angles in combination with optimal C$\alpha$-C$\alpha$ 
virtual bond lengths. The (blue) crosses are the original PDB structures, and the (red) circles are the reconstructed ones. 
The line shows the fits of the radius of gyration. We have no visual difference, between the two cases.  }   
\label{fig-23}    
\end{figure}
}
In Figure \ref{fig-23} we compare the radius of gyration values in our ultra-high resolution
protein structures, for the original PDB structures and those that have been reconstructed using the virtual
C$\alpha$ backbone bond and torsion angles when we use the constant virtual bond length value (\ref{dist}). 
Unlike in the Figure  (\ref{fig-11}), now we observe 
no visual difference. For the reconstructed data, we obtain the relation \cite{Hinsen-2013}
\[
R_g  \ \approx \  2.281 \, N^{0.375} \ \ \ {\rm \AA}
\]
This  is remarkably close to  (\ref{Rg}), the difference is immaterial. Thus we conclude that in the case of crystallographic protein
structures the virtual  C$\alpha$ trace bond and torsion
angles ($\kappa_i, \tau_i$) form a complete set of geometrical local order parameters.

\section{Universal discretised energy} 

The goal is to describe the structure and dynamics of proteins 
{\it beyond } the limitations of an expansion in a small coupling like (\ref{coup}).  
For this we propose to start with an energy 
function where  the virtual C$\alpha$ backbone bond and torsion angles 
appear as local order parameters; recall that these variables form a  complete
set of local order parameters, for backbone reconstruction.   
%
%

Let $F$ be the thermodynamical Helmholtz free energy of a protein. Its  
minimum energy configuration  describes a folded protein, under thermodynamical equilibrium
conditions.  The free energy 
is the sum of the internal energy $U$ and the entropy $S$, at temperature $T$
\begin{equation}
F = U - T S
\label{Helm}
\end{equation}
It is a  function of all the  inter-atomic distances  
\begin{equation}
F = F(r_{\alpha\beta})\ ; \ \ \ r_{\alpha\beta} = |\mathbf r_\alpha - \mathbf r_\beta |
\label{intat}
\end{equation}
where the indices $\alpha,\beta, ... $ extend over all the atoms in the protein system, including those  
of the solvent environment.

We assume that the characteristic length scales over which spatial 
deformations along  the protein backbone around its thermal equilibrium configuration take place,
are large in comparison to the covalent bond lengths;  there are no abrupt wrenches and buckles,
only gradual bends and twists.
We also assume that the C$\alpha$ virtual bond length oscillations 
have a characteristic time scale which is 
short in comparison to the time scale we consider.
The mean value  (\ref{dist}) can then adopted as the universal value of the 
virtual C$\alpha$ bond length, at least in a time averaged sense. The  
completeness of the C$\alpha$  bond and torsion angles
proposes that we consider the response of the  interatomic distances to variations
in these angles,
\[
r_{\alpha\beta} \ = \  r_{\alpha\beta} (\kappa_i, \tau_i)
\]

Suppose that at a local extremum of the free energy, the C$\alpha$ bond and torsion angles have the values
\[
(\kappa_i , \tau_i) \ = \ (\kappa_{i0},  \tau_{i0})
\]
Consider a conformation where the ($\kappa_i, \tau_i$) deviate from  these
extremum values. The deviations are
\begin{equation}
\begin{matrix} 
\Delta \kappa_i & = & \kappa_i - \kappa_{i0} \\
\Delta \tau_i & = & \tau_i - \tau_{i0}
\end{matrix}
\end{equation}
Taylor expand the infrared limit Helmholtz free energy  (\ref{Helm})  around the extremum, 
\[
F\left[ r_{\alpha\beta} =   r_{\alpha\beta} (\kappa_i, \tau_i)\right] \ \equiv \
F(\kappa_i, \tau_i) \ = \ F(\kappa_{i0}, \tau_{i0}) + \sum\limits_k \{ \, 
\frac{\partial F}{\partial \kappa_k}_{|0}\! \Delta \kappa_k 
+ \frac{\partial F}{\partial \tau_k}_{|0}\! \Delta\tau_k \, \} 
\]
\[
+  \sum\limits_{k,l} \{ \, \frac{1}{2}
\frac{\partial^2 F}{\partial \kappa_k \partial \kappa_l }_{|0}\! \Delta \kappa_k  \Delta \kappa_l
+ \frac{\partial^2 F}{\partial \kappa_k \tau_l}_{|0}\! \Delta\kappa_k \Delta \tau_l  +
\frac{1}{2}
\frac{\partial^2 F}{\partial \tau_k \partial \tau_l }_{|0}\! \Delta \tau_k  \Delta \tau_l
\, \} + \mathcal O (\Delta^3)
\]
The first term in the expansion evaluates the free energy at the extremum. Since ($\kappa_{i0}, \tau_{i0}$)
correspond to the extremum, the second term vanishes and we are left with 
the following expansion of the averaged 
free energy,
\[
F(\kappa_i, \tau_i) \ = \ F(\kappa_{i0}, \tau_{i0}) \
\]
\begin{equation}
+  \ \sum\limits_{k,l} \{ \, \frac{1}{2}
\frac{\partial^2 F}{\partial \kappa_k \partial \kappa_l }_{|0}\! \Delta \kappa_k  \Delta \kappa_l
+ \frac{\partial^2 F}{\partial \kappa_k \tau_l}_{|0}\! \Delta\kappa_k \Delta \tau_l  +
\frac{1}{2}
\frac{\partial^2 F}{\partial \tau_k \partial \tau_l }_{|0}\! \Delta \tau_k  \Delta \tau_l
\, \} + \dots 
\label{Fene}
\end{equation}
In the limit where 
the characteristic scale of the extent of spatial deformations around a minimum energy configuration is large
in comparison to a covalent bond length, and the amplitude of these deformations remains small,
we may  {\it re-arrange} 
the expansion (\ref{Fene}) in terms of of the differences in the angles as follows:
First comes local terms. Then comes terms that connect the nearest-neighbors. 
Then comes terms that connect the next-to-nearest neighbours.  And so forth ... 
This re-ordering of the expansion  ensures that we recover the derivative expansion (\ref{weinberg})
in  leading order, when we take the continuum limit where the virtual bond length vanishes. 
Moreover, since the 
free energy  {\it must} remain invariant under the local frame rotations (\ref{sok}),
(\ref{sot}) we conclude 
\cite{Niemi-2003,Danielsson-2010,Chernodub-2010,Molkenthin-2011,Hu-2011,Krokhotin-2011,Krokhotin-2012a,Krokhotin-2013a,Krokhotin-2013b}
that to the leading order the expansion of the free energy  {\it must} coincide  
with a discretisation of the full AHM energy 
({\ref{enenls}):
\begin{equation}
F  = - \sum\limits_{i=1}^{N-1}  2\, \kappa_{i+1} \kappa_{i}  + 
\sum\limits_{i=1}^N
\biggl\{  2 \kappa_{i}^2 + \lambda\, (\kappa_{i}^2 - m^2)^2  
+ \frac{d}{2} \, \kappa_{i}^2 \tau_{i}^2   
- b \, \kappa^2_i \tau_{i}   - a \,  \tau_{i} +  \frac{c}{2}  \tau^2_{i} 
\biggr\} \ + \dots
\label{E1old}
\end{equation}
The corrections include next-to-nearest neighbours couplings and so forth, which are higher order terms
from the point of view of 
our systematic expansion: The approximation (\ref{E1old}) is valid, as long as there are 
no abrupt wrenches and buckles but only long wave length 
gradual bends and twists along 
the backbone. In particular, long range interactions 
are accounted for, as long as they don't introduce any localised 
buckling of the backbone.

In  (\ref{E1old}) $\lambda$, $a$, $b$, $c$, $d$ and $m$ 
depend on the atomic level physical properties and the chemical 
microstructure of the protein and its environment. In principle, these parameters can 
be computed from this knowledge.
 
We note the following:  The free energy (\ref{E1old}) is a deformation of the standard energy function of the discrete 
nonlinear Schr\"odinger  equation (DNLS)  \cite{Faddeev-1987,Ablowitz-2003}. The first sum 
together with the three 
first terms in the second sum is the energy of  the standard  DNLS equation,  
in terms of the discretized Hasimoto variable (\ref{hasi}). 
 The fourth term ($b$) is the conserved momentum, and the fifth term ($a$) is the 
 helicity; these two break the $\mathbb Z_2$ parity symmetry
 and are responsible for right-handed helicity of the C$\alpha$ backbone.
The last ($c$) term is the Proca mass that we again add, as a "regulator". 
Observe in particular the explicit
presence of the non-linear, quartic contribution to the (virtual) bond angle energy. 
This is the familiar double-well potential, shown in Figure  \ref{fig-12}.
Its $\mathbb Z_2$ symmetry 
becomes eventually broken. The breaking of this symmetry
is essential for the emergence  of structure, in the case of proteins. It is the source of {\it Bol'she}
that makes us alive. We note that
this kind of explicit non-linearity is absent in (\ref{hos}).

\vskip 0.2cm
{\bf We summarise:} 
The expression (\ref{E1old}) of the free energy describes the small amplitude fluctuations around the 
local extremum ($\kappa_{i0}, \tau_{i0}$)  in the space of bond and torsion angles. It can be identified as the 
long wavelength (infrared) limit of the full free energy,  in the sense of Kadanoff and Wilson.   To the present 
order of the expansion in powers of ($\kappa_i, \tau_i$), the functional form (\ref{E1old}) is the {\it most general}
long wavelength free energy that is compatible with the principle gauge invariance. This {\it fundamental} 
symmetry principle 
ensures that no physical quantity 
depends on our choice of coordinates (framing) along the backbone.

\begin{quote}
{\it Research project: Develop a method to compute the parameters in (\ref{E1old}) from an all-atom energy function. }
\end{quote}

\section{Discretized solitons}
\label{sect13}

The energy (\ref{E1old}) is a deformation of the integrable energy of the discrete 
nonlinear Schr\"odinger (DNLS) equation \cite{Faddeev-1987,Ablowitz-2003,Kevrekidis-2009}: 
The first term together with the $\lambda$ and $d$ 
dependent terms constitute the (naively) discretized Hamiltonian of the NLS model
in the Hasimoto variable.  The conventional DNLS equation is known to
support solitons. Thus we can try and find soliton solutions of (\ref{E1old}). 

As in (\ref{taueq}) we 
first eliminate the torsion angle, 
\begin{equation}
\tau_i [\kappa] \ = \ \frac{ a + b\kappa_i^2}{c+d\kappa_i^2} \ =  \ a  \frac{ 1 + b\kappa_i^2}{c+d\kappa_i^2}
\ \equiv \ a \hat \tau_i[\kappa]
\label{taueq2}
\end{equation}
For bond angles we then have
\begin{equation}
\kappa_{i+1} = 2\kappa_i - \kappa_{i-1} + \frac{ d V[\kappa]}{d\kappa_i^2} \kappa_i  \ \ \ \ \ (i=1,...,N)
\label{dnlse}
\end{equation}
where we set $\kappa_0 = \kappa_{N+1}=0$, and $V[\kappa]$ is given by (\ref{V}). This equation
is a deformation of the conventional  DNLS equation, and  it is not integrable,  {\it a priori}.
For a numerical solution,  we extend  (\ref{dnlse})  into the following 
iterative equation \cite{Molkenthin-2011}
\begin{equation}
\kappa_i^{(n+1)} \! =  \kappa_i^{(n)} \! - \epsilon \left\{  \kappa_i^{(n)} V'[\kappa_i^{(n)}]  
- (\kappa^{(n)}_{i+1} - 2\kappa^{(n)}_i + \kappa^{(n)}_{i-1})\right\}
\label{ite}
\end{equation}
Here  $\{\kappa_i^{(n)}\}_{i\in N}$ denotes the $n^{th}$ iteration of an initial configuration  
$\{\kappa_i^{(0)}\}_{i\in N}$ and $\epsilon$ is some sufficiently small but otherwise arbitrary 
numerical constant; we often choose $\epsilon = 0.01$ in practical computations. The fixed point
of (\ref{ite}) is clearly independent of the value chosen. But the radius and rate of numerical convergence 
in a simulation  towards the fixed point, 
depends on the value of $\epsilon$:
The fixed point is clearly a solution of (\ref{dnlse}).

Once the numerically constructed fixed point is available, we calculate the corresponding 
torsion angles from (\ref{taueq2}). Then, we obtain the frames from (\ref{DFE2}) and can
proceed to the construction of the 
discrete string, using  (\ref{dffe}).

At the moment we do not know of an analytical expression of the soliton solution 
to the equation (\ref{dnlse}). 
But we have found \cite{Chernodub-2010,Hu-2011,Krokhotin-2012a}
that an {\it excellent} approximative solution can be obtained  by discretizing the topological 
soliton (\ref{soliton}). 
\begin{equation}
\kappa_i \ \approx \   \frac{ 
m_{1}  \cdot e^{ c_{1} ( i-s) } - m_{2} \cdot e^{ - c_{2} ( i-s)}  }
{
e^{ c_{1} ( i-s) } + e^{ - c_{2} ( i-s)} }
\label{An1}
\end{equation}
Here ($c_1, c_2, ,m_{1},m_{2},s$) are parameters. The $m_{1}$ and $m_{2}$ 
specify the asymptotic $\kappa_i$-values of the soliton. Thus, 
these parameters are entirely determined by the character 
of the regular, constant bond and torsion angle structures that are adjacent to the 
soliton. In particular, these parameters are not specific  to the soliton {\it per se}, but to the adjoining  
regular structures.
The parameter $s$ defines the location of the soliton along the 
string.  This leaves us with only two loop specific parameter, the $c_{1}$ and $c_{2}$. 
These parameters quantify the length of the bond angle profile that describes the soliton. 

For the torsion angle, (\ref{taueq2}) involves one parameter ($a$) that we have factored out as the overall relative scale
between the bond angle and torsion angle contributions to the energy; this parameter determines the relative
flexibility of the torsion angles, with respect to the bond angles. Then, there are three additional
parameters ($b/a, c/a, d/a$)  in the remainder $\hat \tau[\kappa]$. Two of these are again determined by the character 
of the regular structures that are adjacent to the soliton. 
As such, these parameters are not specific  to the soliton. The remaining single parameter
specifies the size of the regime where the torsion angle fluctuates.

%


On the regions adjacent to a soliton, 
we have constant values of $(\kappa_i, \tau_i)$.  In the case of a protein, these are the regions 
that correspond to the standard regular secondary structures: In a rough sense, proteins are made of right-handed
$\alpha$-helices, $\beta$-strands and loops; we recommend the reader familiarises herself with these structures,
{\it e.g.} using the resources available at (\ref{uniprot})-(\ref{sbkb}), and in particular at \cite{cath,scop}

For example, the 
standard right-handed $\alpha$-helix is
\begin{equation}
\alpha-{\rm helix:} \ \ \ \ \left\{ \begin{matrix} \kappa \approx \frac{\pi}{2}  \\ \tau \approx 1\end{matrix} \right.
\label{bc1}
\end{equation}
and the standard $\beta$-strand is 
\begin{equation}
\beta-{\rm strand:} \ \ \ \ \left\{ \begin{matrix} \kappa \approx 1 \\ \tau \approx \pi \end{matrix}  \right.
\label{bc2}
\end{equation}
All the other standard  regular secondary structures of proteins such as 3/10 helices, 
left-handed helices {\it etc.} are similarly described by definite constant values of $\kappa_i$ and $\tau_i$.

Protein loops correspond to regions where the values of ($\kappa_i, \tau_i$) 
are variable, protein loops are the soliton proper:  A soliton
is a configuration that interpolates between two regular  structures,  with predetermined constant values of 
($\kappa_i, \tau_i$). Like in Figure (\ref{fig-12}).


\section{Proteins out of thermal equilibrium}
\label{sect17}

When a protein folds towards its native state, it is out of thermal equilibrium. Several studies propose, that
in the case of a small protein the folding  takes place in a manner which is consistent with
Arrhenius' law; we recall that Arrhenius'  law states that the reaction
rate depends exponentially on the ratio of activation energy $E_A$ and temperature,
\[
r  \ \propto  \ \exp\{ - \frac{E_A}{k_B T}\}
\]
On the other hand, in the case of simple spin chains it has been found that the {\it Glauber dynamics}
\cite{Glauber-1963,Bortz-1975}
describes the approach to thermal equilibrium, following Arrhenius's law. Since we
have argued that proteins can be viewed as spin chains, with residues corresponding to the spin
variables, it is natural to try and model the way how a protein folds towards its native state
in terms of Glauber dynamics.
According to Glauber, a
non-equilibrium system approaches thermal equilibrium in a manner which is dictated by
a Markovian  time evolution, defined by 
the probability
distribution   \cite{Glauber-1963,Bortz-1975}
\begin{equation}
\mathcal P = \frac{x}{1+x} \ \  \ \ {\rm with}  \   \ \ \ x =     \exp\{ - \frac{ \Delta E}{k\mathcal T} \}  
\label{P}
\end{equation}
Here $\Delta E$ is the activation energy,  which in a Monte Carlo (MC) simulation is 
the energy difference between consecutive MC time steps, that we compute 
from (\ref{E1old}) in the case of a protein. In addition, in the case of a protein we need to account for
steric constraints: Analysis of PDB structures reveals, that 
the distance between two C$\alpha$ atoms which are {\it not} nearest neighbours
along the backbone, is always larger than (\ref{dist})
\begin{equation}
| \mathbf r_i - \mathbf r_k | > 3.8 \ {\mathrm \AA} \ \ \ {\rm for} \ \ \ |i-k| \geq 2
\label{dist2}
\end{equation}
This condition needs to be included as a requirement, to accept a given
Monte Carlo step during simulation.

The is a clear 
similarity between Arrhenius law and Glauber algorithm.  We also note that the scale of units of
$k \mathcal T$ which appears 
in (\ref{P}) as a temperature factor, should not be directly identified with the  Bolzmannian temperature factor $k_B T$. 
The scale of units depends on the overall scale of the energy function (\ref{E1old}), 
and in particular by our choice of the normalisation factor in the first, nearest neighbor interaction term. 
To determine the unit, we need a renormalisation condition. For this we need 
to perform a proper experimental measurement(s), and compare the predictions  of our model 
to those of the protein that it describes, at that temperature. One suitable renormalisation point could be, 
to try and identify the experimentally measured $\theta$-transition temperature by comparison 
with the properties of our model.

\section{Temperature renormalisation}
\label{4.5}

{\small \it This section is somewhat technical. The details  are not needed in the rest of the lectures. 
We include this section because  we feel that good understanding of temperature renormalisation  of parameters
is relevant to physics of proteins \cite{Krokhotin-2013b}. For example, thus far this has not been really
addressed in any other approach we are aware of. You might find the subject described here to be 
an inspiration for your future research; the exposition is  preliminary and indicative.
}

\vskip 0.2cm
%
%
%
%

In the probability distribution (\ref{P}) 
the nearest-neighbor coupling contribution in (\ref{E1old}) becomes normalised as follows,
\begin{equation}
- \frac{2}{k\mathcal T} \sum\limits_{i=1}^N  \kappa_{i+1} \kappa_i   
\label{norms}
\end{equation}
Thus
the temperature factor $k\mathcal T$ depends on the physical temperature factor $k_B T$ 
in a non-trivial fashion. That is,  we should really write
\begin{equation}
 \frac{2}{k\mathcal T} \ \to \  \frac{J(T)}{k_B T}
\label{JT}
\end{equation}
where $J(T)$ is the strength of the nearest neighbour coupling at Bolzmannian $k_B T $.
Its numerical value depends on the temperature in a manner that is governed by the 
standard renormalisation group equation 
\begin{equation}
T \, \frac{ d J}{dT} = \beta_J (J; \lambda,m,a,b,c,d) \ \sim \ \beta_J(J) + \dots
\label{RGEJ}
\end{equation}
For simplicity  we may assume 
that to leading order the dependence of $\beta_J$ on the other couplings can be ignored.
Note that the parameters and thus their $\beta$-functions, depend
on the properties of the environmental factors such as properties of solvent, the pH of solvent, pressure {\it etc}.

In the low temperature limit we can expand the nearest neighbour coupling as follows, 
\begin{equation}
J(T) \ \approx \  J_0 - J_1 T^\alpha + \dots \ \ \ \ \ \ {\rm as} \ T \to 0
\label{J0}
\end{equation}
Here the value of $J_0 $ is non-vanishing, and
the critical exponent $\alpha$ controls the low temperature behavior of $J(T)$. 
The asymptotic expansion (\ref{J0}) corresponds to a $\beta$-function (\ref{RGEJ}) that in the $T\to 0$ limit
approaches
\[
\beta_J (J) = \alpha( J - J_0) + \dots 
\]
Consequently, at low temperatures we have
\begin{equation}
k{\mathcal T} \ \approx \ \frac{2 k_B}{ J_0} \, T
\label{Tlow}
\end{equation}

In terms of the temperature factor, (\ref{RGEJ}) translates into
\begin{equation}
T  \frac{ d }{dT}\left( \frac{1}{k{\mathcal T}} \right) = - \frac{1}{k {\mathcal T}} + \frac{1}{2 k_B T}\,  
\beta_J\!\left( \frac{2 k_B T}{k{\mathcal T}} \right)
\label{betaT}
\end{equation}
We try to find an approximative  solution in the collapsed phase, when the temperature $T$ is below 
the critical $\theta$-point temperature $T_{\theta}$  where the transition between the collapsed phase and
the random walk phase takes place.  This coincides with the physical temperature value 
that corresponds to the unfolding transition temperature factor value
$k{\mathcal T}_\Theta$.
Let
\[
 \beta_J\!\left( \frac{2 k_B T}{k{\mathcal T}} \right)  \ = \  \frac{2 k_B T}{k{\mathcal T}}  +  F\left( \frac{2 k_B T}{k
 {\mathcal T}} \right)
\]
and define
\[
y = \frac{1}{k{\mathcal T}}  
\]
\[
x = \frac{1}{2k_B T}
\]
The equation (\ref{betaT}) then translates into
\[
\frac{dy}{dx} =  - F(\frac{y}{x})
\]
with the  solution 
\[
\ln ( {c} \, x ) = - \int^{\frac{y}{x}} \frac{du}{F(u) + u }
\]
Here $c$ is an integration constant. We shall assume that 
the leading non-linear corrections are logarithmic; this  is often the case. 
To the leading order we then have
\[
F(u) = (\eta-1) \, u + \alpha  \, u \! \ln u  + \dots 
\]
Note,  that in general there are higher order corrections.
When we re-introduce the original variables and set
\[
\eta = - \alpha \ln J_0
\]
we get for the temperature factor
\begin{equation}
k{\mathcal T }  \ \approx \ \frac{2}{J_0}\,  k_B T \, \exp\{ \frac{J_1}{J_0} T^\alpha \}  
\label{kTt}
\end{equation}
\[
\approx \frac{2}{J_0}\,   k_B tT +  \frac{2J_1}{J_0^2} \, k_B T^{\alpha+1} + ... \ \ \ \ {\rm as } \ \ T \to 0 \ \ \ \ (\alpha >0)
\]
where we have chosen the integration constant so that 
in the low temperature limit we obtain (\ref{Tlow}). 

For the nearest neighbor coupling, (\ref{kTt}) yields
\[
J(T) \ \approx  \ J_0 \exp\{ -\frac{J_1}{J_0} T^\alpha \} 
\]
Thus the coupling between bond angles becomes weak, at an exponential rate, when the temperature 
approaches the transition temperature $T_{\theta}$ between the collapsed phase and the random walk phase.

Similarly, all the other couplings that  are present in (\ref{E1old}) are  temperature dependent.  Each of them
has its own renormalisation group equation. 
For example, the quartic $\kappa_i$ self-coupling 
$\lambda $ in (\ref{E1old}) flows according to a renormalization group 
equation of the  form
\begin{equation}
T \, \frac{ d\lambda  } {dT} \ = \  \beta_\lambda (\lambda) 
\label{betac}
\end{equation}
For simplicity,  we again assume that to the leading order the $\beta_\lambda$ depends only on $\lambda$.

It is natural to interpret $\lambda$ as a measure of the strength 
of hydrogen 
bonds: Structures such as $\alpha$-helices and $\beta$ strands become stable, due to hydrogen bonds.
At the same time,  the value of  $\lambda$ controls the affinity of $\kappa_i$ towards the ground state value of 
the quartic
potential in (\ref{E1old}). 
The hydrogen bonds  are presumed to 
become vanishingly weak, when the protein unfolds. This can take place when the protein approaches 
the transition temperature $T_{\theta} $ which separates 
the collapsed phase from the  random walk phases.  This proposes that  asymptotically, 
\[
\lambda(T)  \  \to  \ \lambda_\theta \ | T - T_{\theta} |^ {\gamma_\lambda} \ \ \ \
\ \ {\rm as}  \ \ T \to T_{\theta}  \ \ {\rm from ~ below}
\]
Here $\gamma_\lambda$ is a critical exponent that controls the way how the strength of (effective)
hydrogen bonds vanish.   More generally, 
we may send $T_\theta \to T_H$, which is the  temperature value where 
hydrogen bonds disappear even between the solvent molecules; in the case of water at atmospheric
conditions  $T_H\approx 100^o\mathrm C$.  In general, we expect that  the  value of $ T_H$ is
higher than that of $T_\theta$.

Above $T>T_{\theta}$, when the hydrogen bonds become vanishingly weak, we 
expect that effectively $\lambda\approx 0$ (or $m\approx 0$) in (\ref{E1old}). 
On the other hand,  we expect that as the temperature decreases the value of $\lambda(T)$ increases,
so that in the low temperature limit  we have
\[
\lambda(T)  \  \to  \ \lambda_0 - \lambda_1 T^{\gamma_0}  + \dots \  \ \ \ \ {\rm as} \ \ T \to 0  
\]
Thus 
\[
\beta_\lambda (\lambda) \ \approx \ \gamma_0 ( \lambda - \lambda_0) + \mathcal O[(\lambda-\lambda_0)^2]
\]
Here $\lambda_0$ should be  close to the value we obtain from PDB, when we compute the parameters in (\ref{E1old}) from
the crystallographic low temperature structure.

\begin{quote}
{\it Research project: Try to numerically evaluate the $\beta$-function (\ref{RGEJ})
in the case of a simple protein, such as villin headpiece
(PDB code 1YRF), using results from detailed experimental measurements.}
\end{quote}

\begin{quote}
{\it Research project: Find out how the parameters in (\ref{hos}), (\ref{ff}) depend on temperature. 
}
\end{quote}

\chapter{Solitons and ordered proteins}
\label{sect16}

Various taxonomy schemes such as CATH and SCOP  \cite{cath,scop}  
have revealed  that  folded proteins are build in a  modular fashion, from a
relatively small number of building blocks.  Despite essentially 
exponential  increase
in the number of new crystallographic protein structures,
novel fold 
topologies are now rarely found and some authors think that most
modular building blocks of proteins are already known 
\cite{Chothia-1992,Skolnick-2009}. This convergence in protein architecture 
demonstrates  that protein folding should be a process which is 
driven by some  kind of  universal structural self-organisation principle.

We know that solitons are the paradigm structural self-organisers. 
Thus we argue  that the soliton solution of the  DNLS equation (\ref{dnlse}), (\ref{taueq2}) 
must be  the universal modular building block from which
folded proteins can be constructed. Indeed, we know that the energy function (\ref{E1old}) is unique,
in the limit where it becomes applicable.  Moreover, 
it has already been shown
that over 92$\%$ of all C$\alpha$-traces of PDB proteins can be
described in terms of no more than  200 different parametrizations 
of the discretised NLS soliton (\ref{An1}),  with a root-mean-square-distance (RMSD) precision 
which is  better than 0.5 \AA~  \cite{Krokhotin-2011}.

Accordingly, we set up to describe  the modular building blocks of proteins in terms of
various parametrizations of the DNLS soliton profile, which is described by the 
equations (\ref{ite}), (\ref{taueq2}), (\ref{DFE2}) and (\ref{dffe}).

\section{ $\Huge \bf \Huge \lambda$-repressor as a multi-soliton}
\label{sect13}

In order to identify the soliton structure of a given protein, we start by computing  
the C$\alpha$ virtual backbone bond and torsion angles from the PDB data. 
We {\it initially}  fix the $\mathbb Z_2$ 
gauge in (\ref{dsgau}) so that all the bond angles take positive values $\kappa_i \in [0,\pi]$.  
A generic protein profile consists of a set of $\kappa_i$ with values 
that are typically between  $\kappa_i \approx 1$ and $\kappa_i \approx \pi/2$; 
the upper bound can be estimated using steric constraints.  
The torsion angle values $\tau_i$ are commonly much more unsettled, and their values
extend more widely  over the entire range $\tau \in [-\pi,+\pi]$. 

As an example we consider the  $\lambda$-repressor which is a protein that controls 
the lysogenic-to-lytic transition  in bacteriophage $\lambda$ infected {\it E. coli} cell. The transition
between the lysogenic and lytic phases  in  bacteriophage $\lambda$ infected {\it E. coli}
is the paradigm example of genetic switch mechanism,  which has been described  in numerous 
molecular biology textbooks and review articles. See for example  \cite{Ptashne-2004,Gottesman-2004}.  
The interplay between the lysogeny maintaining $\lambda$-repressor protein and  the regulator protein that 
controls the transition to the lytic state is a simple model for 
more complex regulatory networks, including those that can  lead to cancer in humans.  

The $\lambda$-repressor structure we consider has 
PDB code 1LMB.  It is a homo-dimer with 92 residues in each of  
the two monomers. It maintains the lysogenic state by binding to DNA with a 
helix-turn-helix motif which is  located between  the  residue sites  33-51. 
The $\lambda$-repressor is a fast-folding
protein:  In \cite{Lindorff-2011} an 80-residue long 
mutant of the $\lambda$-repressor was studied in all-atom simulation.

In figure \ref{fig-24} (left column) we show  the ($\kappa_i,\tau_i$) spectrum of 1LMB, with the convention ({\it i.e.}
$\mathbb  Z_2$ gauge fixing) that $\kappa_i$ is positive. We display the segments 
between residues 19-82. We note that this spectrum is fairly typical, for the  PDB structures we have analysed.
%
%
%
%
%
%
%
%
%
%
%
{
\footnotesize
\begin{figure}[h]         
\begin{center}            
  \resizebox{7 cm}{!}{\includegraphics[]{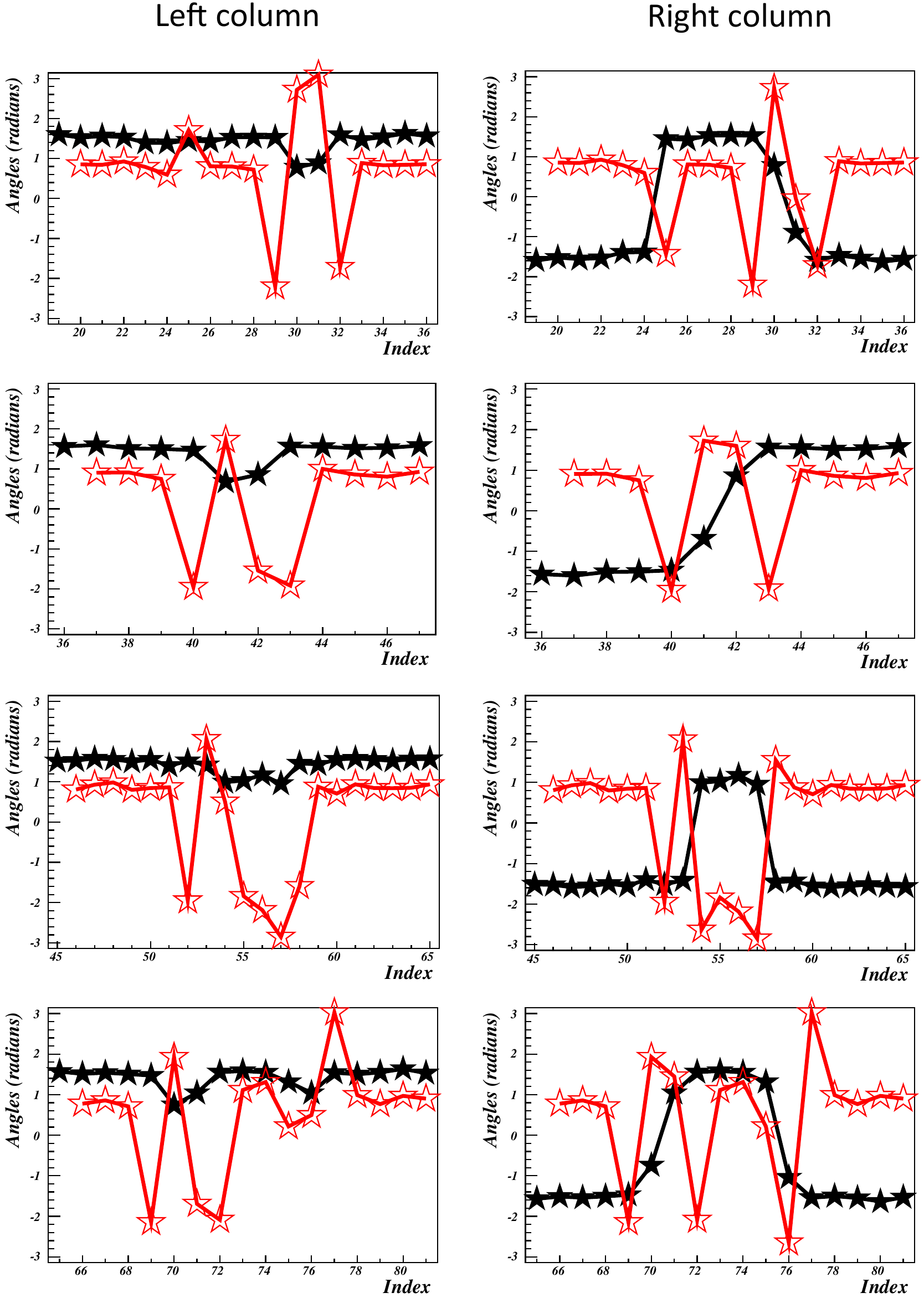}}
\end{center}
\caption {The bond  angle ($\kappa$) and torsion angle ($\tau$) spectrum of $\lambda$-repressor 1LMB, with
indexing that follows PDB.  The left column shows the spectrum in the $\mathbb Z_2$ gauge where all $\kappa_i > 0$
and the right column shows the spectrum after we have implemented the $Z_2$ gauge transformations that identify the
solitons.}   
\label{fig-24}    
\end{figure}
}

The arguments in Section \ref{pere} suggest that in analysing the PDB data, one should initially focus to 
flattening points 
{\it i.e.} points where  $\tau_i$ changes its sign. The flattening points should be located
near a putative  inflection point where a soliton is situated  and perestroika takes place. Accordingly, we 
perform in the spectrum of Figure \ref{fig-24} left column the $\mathbb Z_2$ 
gauge transformations (\ref{dsgau}) in the vicinity of the apparent 
flattening points, to identify the putative multi-soliton profile of $\kappa_i$. 
We observe four regions with an irregular $\tau_i$
profile; these are shown in the  Figures.  By a judicious choice of $\mathbb Z_2$ gauge transformations 
in these regions, we identify 
seven different soliton profiles in $\kappa_i$.  
These profiles are shown in the right column of Figure \ref{fig-24}. 
Each of them is remarkably similar to (\ref{An1}), like a soliton that interpolates between two
regular structures such as (\ref{bc1}), (\ref{bc2}); see also Figure \ref{fig-12}. Moreover,
these soliton profiles are clearly accompanied by putative flattening points, as expected from 
the structure of (\ref{taueq2});  Note that $\tau_i$ is 
multivalued, {\it mod}($2\pi$). Thus the large fluctuations in the values of  $\tau_i$ are deceitful. Once we 
account  for the multivaluedness, the profile of $\tau_i$ is actually quite 
regular. The flexible multi-valuedness of $\tau_i$ is in full accordance with the 
observed, much higher flexibility of the torsion angles in relation to the bond angles, 
which is known to occur in
proteins.

On the basis of  the general considerations in Section \ref{pere}, 
we argue that at least in the  case of 1LMB
the process of folding from a regular unfolded configuration with no solitons 
to the
biologically active natively folded configuration with its solitons is driven by  
inflection and flattening point perestroika's. The initial configuration with no solitons can be chosen to coincide
with the minimum of the second sum in (\ref{E1old}). It could also be {\it e.g.} 
 a uniform  right-handed  $\alpha$-helix (\ref{bc1}), or $\beta$-strand (\ref{bc2}),
or a polyproline-II type conformation. When the protein folds it proceeds from this initial configuration towards
the final configuration, thru successive perestroika's {\it i.e.}
bifurcations. These perestroika's deform the C$\alpha$ backbone, creating  DNLS-like solitons along it and thus
causing the backbone to enter
the space-filling $\nu \sim 1/3$ collapsed phase.

We proceed to determine the parameters in the energy function (\ref{E1old}),
that describes the seven  soliton profiles  in Figure \ref{fig-24} (right column). 
We first train the energy function so that it describes the seven  solitons in
terms of a solution of (\ref{dnlse}), (\ref{taueq2}),  individually. 
The training is performed  by
demanding that the fixed point of the iterative 
equation (\ref{ite}) models the ensuing C$\alpha$ 
backbone structure  as a soliton solution,
with a prescribed precision.  

We have developed 
a program {\it GaugeIT} that implements the $\mathbb Z_2$ gauge transformations
to identify the background, and we have developed a program {\it PropoUI} that trains the energy so that
it has an extremum which models the background in terms of solitons. 
These programs are described at 
\begin{equation}
{\tt http://www.folding-protein.org}
\label{propro}
\end{equation}
In the case of a protein for which the PDB structure is determined with an ultra-high resolution,
typically below 1.0 \AA ngstr\"om, {\it PropoUI} routinely constructs
a solution that describes  a soliton along the C$\alpha$ backbone with a precision which is comparable to
the accuracy of the experimentally measured crystallographic structure; 
recall that  the accuracy of the experimental PDB structure is
estimated by the B-factors using the    
Debye-Waller relation (\ref{dw}). 

%
%
%
%
%
%
%
%
%
%
%
{
\footnotesize
\begin{figure}[h!]         
\begin{center}            
  \resizebox{7 cm}{!}{\includegraphics[]{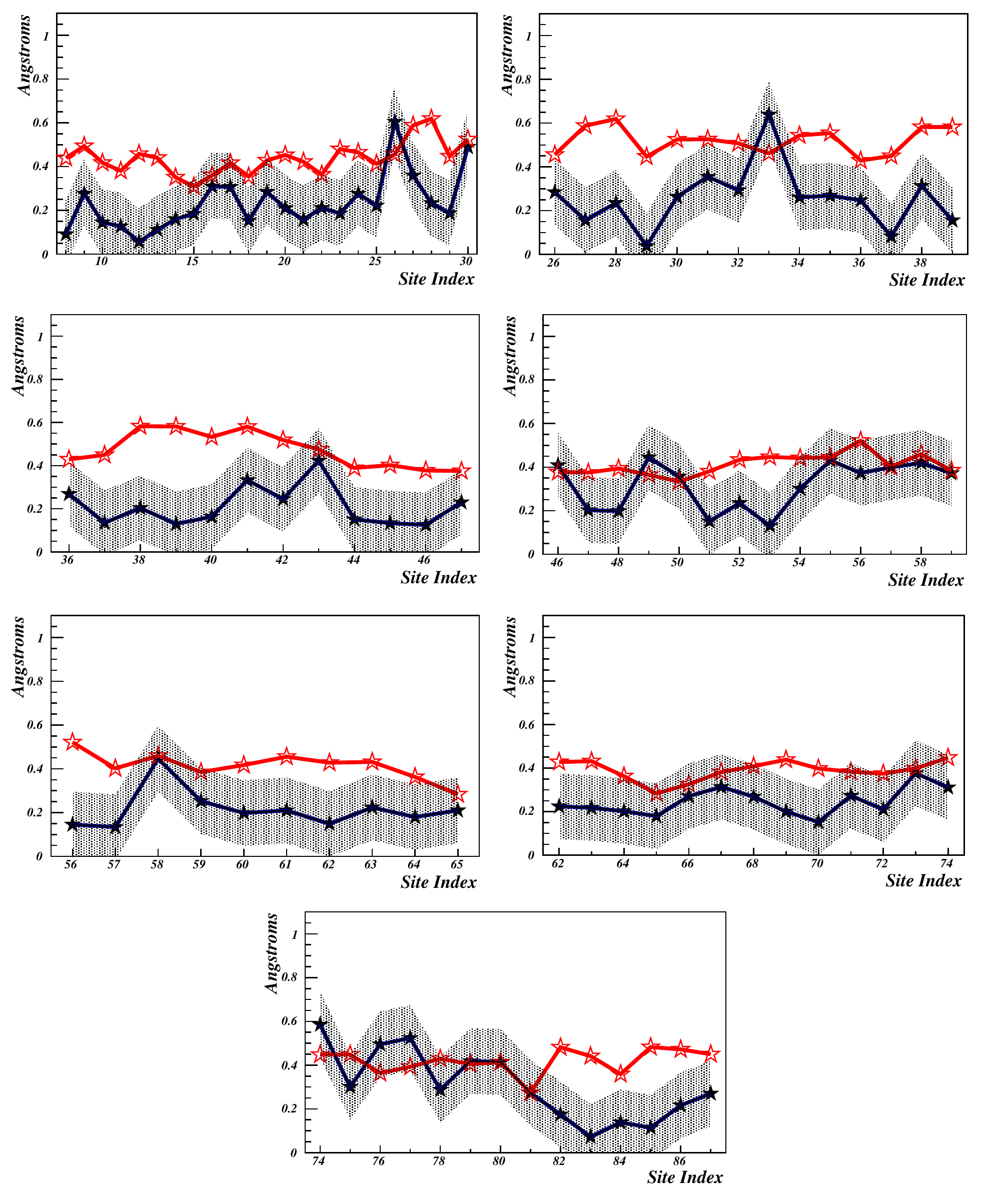}}
\end{center}
\caption {The distance between the PDB backbone of the first 1LMB chain and its  seven solitons. 
    The black line denotes the distance between the PDB structure and the corresponding soliton. 
    The grey area around the black line describes the lower bound estimate of 
    15 pico meter (quantum mechanical) zero point fluctuation 
    distance around each soliton, obtained from Figure (\ref{fig-4}). The red line 
    denotes the Debye-Waller fluctuation distance (\ref{dw}). }   
\label{fig-25}    
\end{figure}
}
In figure  \ref{fig-25} we compare the distance between the
C$\alpha$ backbone, and  the seven individual soliton solutions for 1LMB. 
The B-factor fluctuation distance in the 
experimental structure 1LMB, evacuated from the Debye-Waller relation,  is also displayed. 
As shown in the Figure, the solitons describe the loops with a precision that is fully
comparable with the experimental uncertainties. The grey zone around the soliton profile denotes our
best estimate for the extent of quantum mechanical zero-point fluctuations; according to Figure (\ref{fig-4})
there are practically no Debye-Waller values less than 15 pico meters, which we have chosen here 
as the zero point fluctuation
distance, in the Figure. 
 \vskip 0.2cm

 The next step is to combine the individual solitons into a single multi-soliton solution of the pertinent
 energy function (\ref{E1old}). This can be performed using the program {\it ProproUI}. But instead of
 proceeding with 1LMB, we prefer to consider a somewhat more complex example, the myoglobin
 where techniques such as all-atom MD have thus far failed to produce results.

\section{Structure of myoglobin}
\label{sectmyst}

Myoglobin \cite{Alberts-2014} is the primary oxygen-carrying protein in the muscle cells of mammals. 
Myoglobin is closely related to haemoglobin, which is the oxygen binding protein in blood. Myoglobin
gives meat  its red color; the more red, the more myoglobin.  
Myoglobin also  allows organisms to hold their breath, 
for a  period of time. Diving mammals 
such as whales and seals have a high myoglobin concentration in their muscles.

Myoglobin is the first protein to have its three-dimensional structure determined by x-ray crystallography.  Subsequently myoglobin has remained among the most actively studied
proteins. But theoretically, the investigation of myoglobin {\it e.g.} in an all-atom molecular dynamics simulation
remains a formidable task: The experimentally measured folding time 
from a random chain to the natively folded state  is around 2.5 seconds \cite{Jennings-1993}.
At the same time, the fastest ever built
special purpose MD supercomputer {\it Anton} can produce at most a few microseconds of
{\it in vitro} folding trajectory per day {\it in silico}, in the case of proteins that are much shorter and simpler than
myoglobin. Accordingly it would take something like a  million or so days to reproduce a {\it single} 
myoglobin folding trajectory {\it in silico},
at all-atom level, even with {\it Anton.}  A good 
convergence of Newton's iteration with the energy 
function (\ref{hos}), (\ref{ff}) over such a long time scale  would be truly amazing.

We have already noted that all-atom MD simulation is conceptually a weak coupling 
expansion. As such, it is appropriate for describing 
phenomena over  very short time periods only: The 
time ratio (\ref{coup}) is the small, iterative expansion parameter. In the case of long 
time trajectories, such a weak coupling expansion can not be expected to be very effective, not
even convergent. 
Alternative approximative methods need  to be introduced, to model myoglobin and the large majority of proteins 
that  fold much slower than the  microsecond-to-millisecond scale. 

From the perspective of quantum field theory this means that we need a non-perturbative approach. 
For example, in QCD we do not expect that standard perturbation theory is capable of describing hadrons. 
On the other hand, lattice
QCD is designed for modelling hadrons.  But it can hardly describe the scattering of quarks and gluons.

Indeed,  we have argued that in the case of proteins, 
the energy function (\ref{E1old}) is an example of a non-perturbative approach. It
avoids  altogether the need to introduce a weak coupling expansion parameter, such 
as the time step ratio  (\ref{coup}).  Instead  the C$\alpha$ geometry is modelled  
in terms of small variations in the angular variables,  around their equilibrium conformations. 
Since the approximation does not engage time directly,  it becomes in principle 
possible to describe folding phenomena that can be very difficult, even impossible, 
to model in terms of conventional and presently available all-atom MD.  

We proceed to  apply the energy function (\ref{E1old}) to investigate detailed properties of myoglobin. 
In this Section we analyse the static structure  and in the next Section we consider 
out-of-equilibrium dynamics. 

We first construct the multi-soliton solution that models the myoglobin backbone. We 
use the crystallographic  PDB structure 1ABS as a decoy, to train the energy function.  This structure has been measured 
at very low liquid helium temperatures $\sim$ 20 K.  Thus the thermal B-factor fluctuations (\ref{dw}) are small. 

\begin{quote}
We propose that at this point, the reader downloads  the PDB structure 1ABS, to make  it easier  to 
follow details of our  analysis.  We propose to use the Java interface provided at the PDB 
site  (\ref{pdb}) for visualisation of the backbone and side-chain atoms. We also recommend  the 
analysis tools available on the website of
{\it Molprobity} (\ref{molprobity}) which we shall refer to in the following. We shall  provide the 
values for all the parameters in the  energy function (\ref{E1old})  which the
reader can use as input in the programs {\it ProproUI} and {\it GaugeIT}, that are described at our website (\ref{propro}).
This enables a detailed analysis of the multi-soliton that describes 1ABS, and should help the reader to start his/her independent research.
\end{quote}

We proceed to the construction of the multi-soliton: 
1ABS has154 amino acids, and the PDB index runs over $i=0 \ ,... \ ,153$. Conventionally
one identifies  the structure as a bundle of  eight $\alpha$-helices (A,B,C,...,H) which are
separated by 7 loops as shown in Figure \ref{fig-26}.
%
%
%
%
%
%
%
%
%
%
%
{
\footnotesize
\begin{figure}[h]         
\begin{center}            
  \resizebox{6 cm}{!}{\includegraphics[]{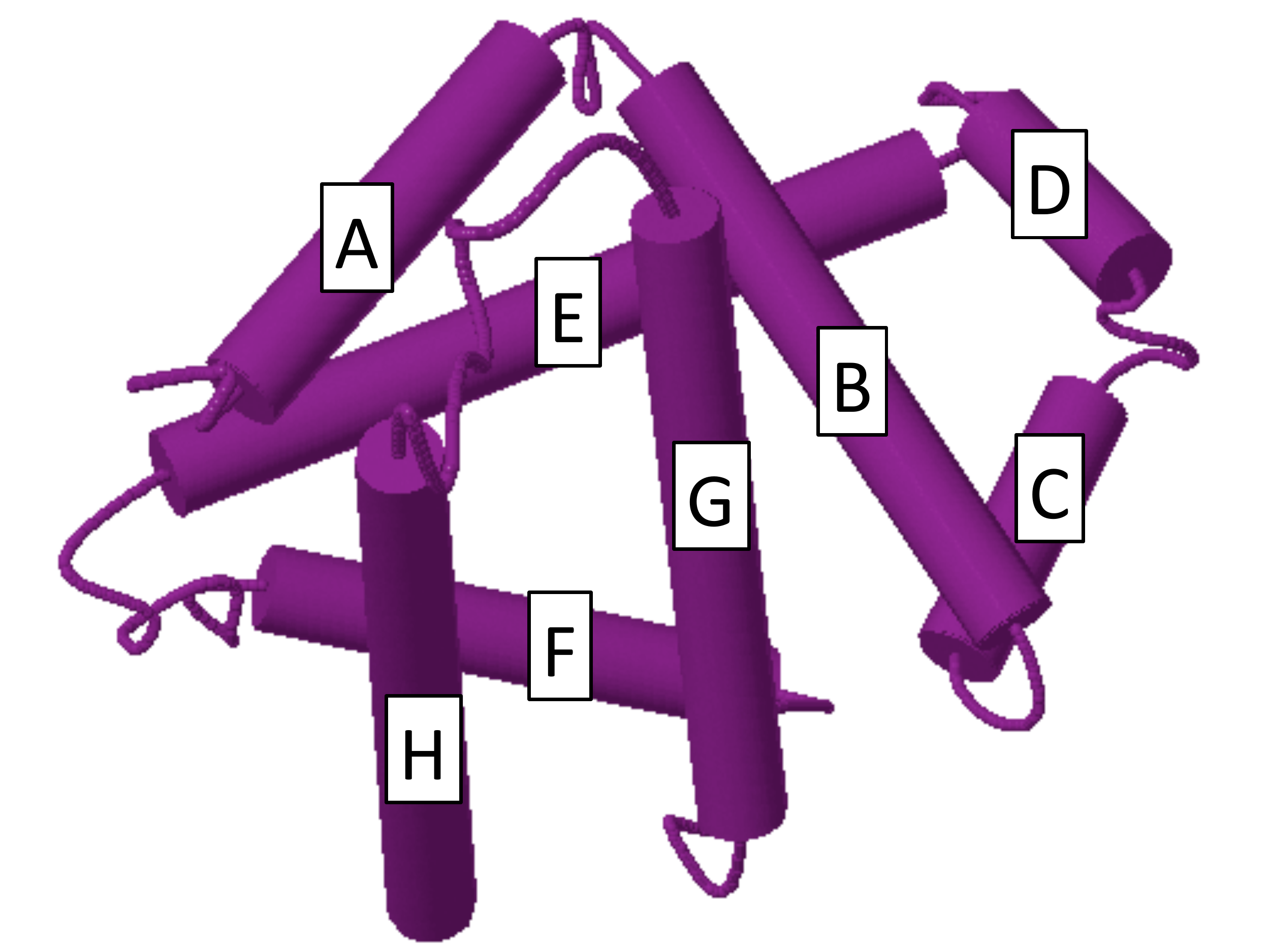}}
\end{center}
\caption {Myoglobin has 8  $\alpha$-helices that are named A,B,C, ... , H}   
\label{fig-26}    
\end{figure}
}

Here we shall limit the construction of the multi-soliton to the sites with PDB index between
N=8-149. That is, we include all the named helices but we do not include the flexible tails at the ends of the backbone. 
These tails could be included, but without much additional insight to the issues that we address.

In Figure (\ref{fig-27}) (top) we show the backbone bond and torsion angle spectrum with the convention that
all $\kappa_i$ are positive.  In Figure (\ref{fig-27}) (bottom) we show the spectrum after we 
have implemented the $\mathbb Z_2$ transformations (\ref{dsgau}) to putatively identify the 
multi-soliton profile; we use the program packages {\it ProproUI} and {\it GaugeIT}  
that are described at our website (\ref{propro}) in our analysis.
%
%
%
%
%
%
%
%
%
%
%
{
\footnotesize
\begin{figure}[h]         
\begin{center}            
  \resizebox{7 cm}{!}{\includegraphics[]{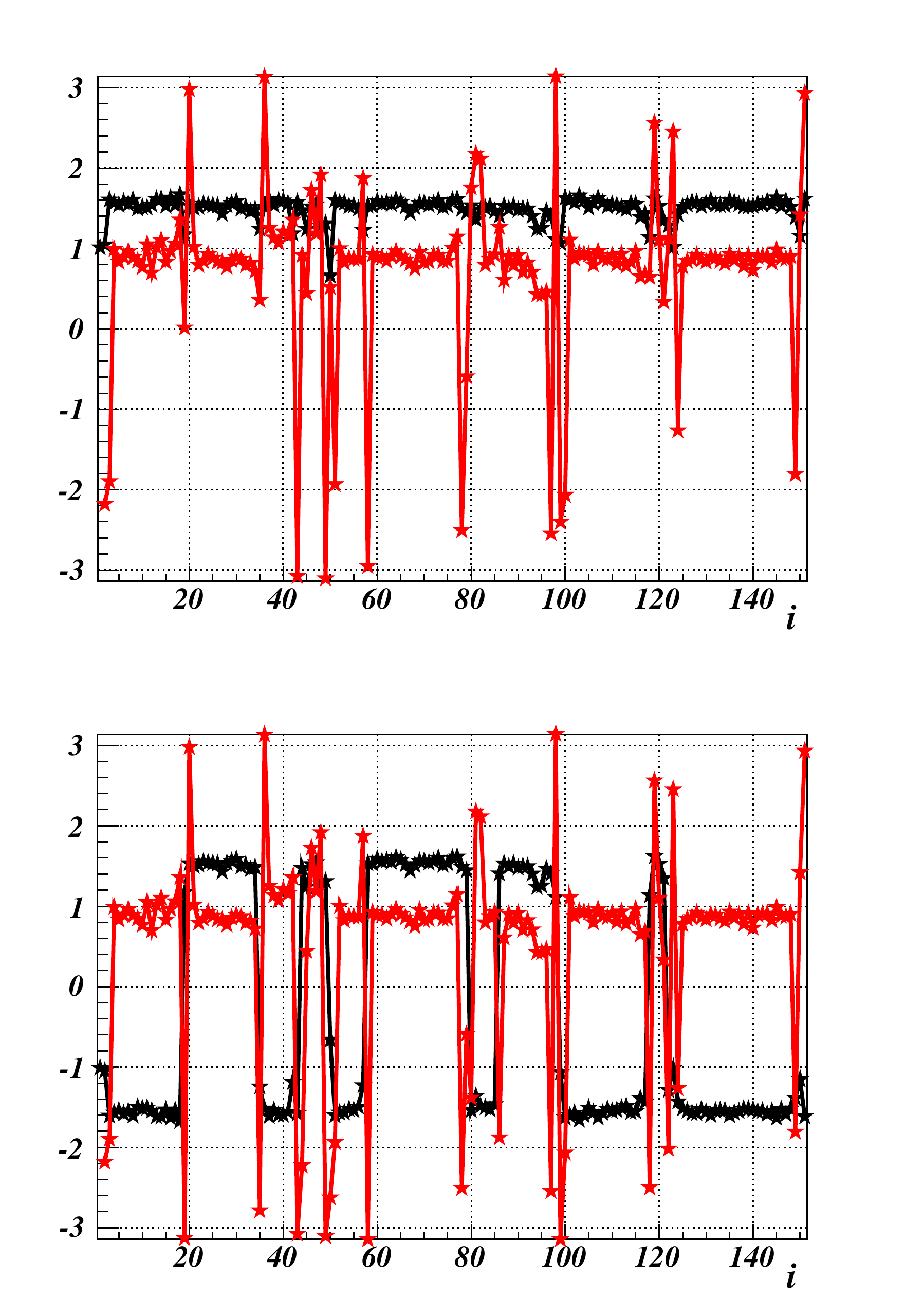}}
\end{center}
\caption {Top: The $\kappa_i$ (black) and $\tau_i$ (red) profiles of 1ABS 
using the standard differential geometric convention that bond angles are positive. Bottom: 
The soliton structure becomes visible in the $\kappa_i$  profile once we implement the transformations (\ref{dsgau}).}   
\label{fig-27}    
\end{figure}
}

We remind that both  the top and bottom of Figure (\ref{fig-27})  correspond to the same intrinsic backbone geometry.
The $\mathbb Z_2$ transformation is a symmetry of the discrete string which is obtained by solving the discrete
Frenet equation. 

 We  conclude from the $\kappa_i$ profile of  the bottom Figure (\ref{fig-27}) that 
the myoglobin backbone has eleven helices 
that are separated by ten single soliton loops. The number of loops and helices is more or less
unambiguously determined by the number of inflection points which we identify visually in Figure (\ref{fig-27}), in the
manner explained in Section \ref{pere}. 
The PDB sites of the 
ten individual soliton profiles that we use for our construction, are identified in Table \ref{loops}.
We  {\it emphasise}   that our geometry based identification of the loops and helices along the 
1ABS backbone does not necessarily need to coincide with the conventional one used {\it e.g.} in crystallography,
which is based on inspection of hydrogen bonds. In particular, according to the conventional classification,
the soliton pair 3 and 4, the soliton pair 6 and 7, and the soliton pair 9 and 10, are all 
interpreted as a single loop.
{
\begin{table}[htb]
\caption{The solitons along the 1ABS  C$\alpha$-backbone,
with indexing starting from the N terminus. We have left out  the end sites that correspond
to monotonous helices, and the N and C termini segments. The type  identifies whether the soliton corresponds to a loop 
that connects $\alpha$-helices and (or) $3/10$-helices. }\vskip 0.5cm
\begin{center}
\begin{tabular}{|c|c|c|c|c|c|}  
\hline 
soliton 
 & 1 & 2  & 3  & 4  & 5
 \tabularnewline 
\hline 
sites &  15-27 & 30-41 & 39-49  & 47-57 & 54-66 
 \tabularnewline \hline
type & $ \alpha$-$\alpha$ & $\alpha $-3/10   & 3/10-3/10 & 3/10-3/10  & 3/10-$\alpha $ 
 \tabularnewline \hline  \hline  
soliton & 6 & 7 & 8 & 9 & 10
 \tabularnewline \hline
site & 72-87 & 83-92 & 94-106 & 110-123 & 121-135 
\tabularnewline \hline
type &  $\alpha$-$\alpha  $ & $\alpha$-$\alpha$ & $\alpha$-$\alpha$ & $\alpha$-$\alpha$ & $\alpha$-$\alpha$
 \tabularnewline  
 \hline 
\end{tabular} 
\end{center}
 \label{loops}
\end{table}
}
From our geometric point of view, the PDB data reveals that in 1ABS, we have four different types of solitons. Those that connect two $\alpha$ helices, those that connect an $\alpha$-helix with a 3/10-helix 
or vice versa, and finally, those that connect two 3/10-helices; see Table \ref{loops}.

In Table \ref{para} we give our parameter values for the multi-soliton
solution. It describes the 1ABS backbone with 0.78 \AA ~ RMSD accuracy. 
{
\begin{table}[htb]
\caption{Parameter values in energy (\ref{E1old}) for the multi-soliton solution that describes 1ABS.  }\vskip 0.5cm
\begin{center}
\begin{tabular}{|c|c|c|c|c|c|c|c|c|}
\hline
soliton &  $\lambda_1$ & $\lambda_2$ & $m_1$ & $m_2$  
\tabularnewline \hline \hline
1  &  9.923 & 2.232 & 1.54097 & 1.54548 
\tabularnewline\hline
2  &  6.48516 & 0.9955 & 1.58013 & 1.54058 
\tabularnewline\hline
3  &  2.05153 & 0.657 & 1.66032 & 1.60224 
\tabularnewline\hline
4  & 0.89676 & 6.74235 & 1.3563& 1.5232 
\tabularnewline\hline
5 & 9.26118 & 0.83376 & 1.55206 & 1.5386 
\tabularnewline\hline
6 &  0.98018 & 2.1337 & 1.45791 & 1.54653 
\tabularnewline\hline
7 & 1.37667 & 3.16891 & 1.47151 & 1.04128 
\tabularnewline\hline
8 &  10.3168 & 4.2801 & 1.18192 & 1.61334 
\tabularnewline\hline
9 &  0.80042 & 1.28973 & 1.5154 & 1.60278 
\tabularnewline\hline
10 & 3.15255 & 0.91475 & 1.55827 & 1.55151 
\tabularnewline\hline\hline
soliton & $a$ & $b$ & $c$ & $d$ 
\tabularnewline \hline \hline
1  & -5.62412 e-08 & -4.13459 e-07 & 1.81044 e-08 & 4.273 e-09
\tabularnewline\hline
2  & -6.25287 e-11& -1.68598 e-05 & 1.47093 e-07 & 2.82807 e-07
\tabularnewline\hline
3  & -9.05135 e-08 & 1.20232 e-06 & 5.10166 e-11 & 5.75389 e-09
\tabularnewline\hline
4  & -2.33413 e-07 & -3.3991 e-07 & 2.36516 e-08 & 7.98841 e-09
\tabularnewline\hline
5 & -9.73035 e-08 & 4.78674 e-07 & 1.03189 e-10 & 4.88194 e-09
\tabularnewline\hline
6 & -7.25906 e-09 & 3.76092 e-09 & 6.82624 e-10 & 1.87212 e-14
\tabularnewline\hline
7 & -1.39052 e-13 & 5.97719 e-13 & 3.77897 e-14 & 5.81911 e-14
\tabularnewline\hline
8 & -1.27193 e-07 & 1.41736 e-06 & 1.07182 e-10 & 1.26295 e-08
\tabularnewline\hline
9 & -2.03487 e-07 & 1.13574 e-06 & 1.46007 e-11 & 7.82707 e-08
\tabularnewline\hline
10 & -1.07811 e-07 & 1.02768 e-06  & 7.49571 e-11 & 7.73639 e-09
\tabularnewline\hline\hline
\end{tabular}
\end{center}
	\label{para}
\end{table}
}
Note that  in those terms in (\ref{E1old}) which  engage the 
torsion angles, the numerical parameter values are consistently 
much smaller than in terms that contain only the bond angles. 
This is in line with the known fact, that in proteins the 
torsion angles  {\it i.e.} dihedrals 
are usually quite flexible  while the bond angles are relatively stiff.

Note also that our energy function has  80 parameters, while there are 153 amino acids in the entire myoglobin
backbone. Thus the energy function (\ref{E1old}) is {\it highly predictive}: The number of free parameters is
even {\it less than the number of amino acids}. This shows that myoglobin displays
structural redundancy, in its  amino acids.

The predictive power of (\ref{E1old}) can alternatively be characterised  as follows: 
When we assume that all the bond lengths have the constant 
value  (\ref{dist}),  we are left with 282 C$\alpha$ angular coordinate values 
in our truncated backbone. These we  need to determine from the properties of the energy function (\ref{E1old}),
in order to construct the backbone from (\ref{DFE2}), (\ref{dffe}).
We have a total of 80 parameters in Table \ref{para},  thus a total of 202 coordinates 
remain to be determined by the multi-soliton solution that minimises the energy  function. 
Therefore, we have  202 unknowns  which are to be {\it predicted} by the model. These predictions  then
directly probe  the physical principles on which (\ref{E1old}) has been built. 

We remind that approaches such as the G\=o model and its variants and various 
elastic network models are descriptive, and
lack this kind of predictive power.
In those models the positions of {\it all} the atoms are assumed to be known {\it a priori}. In addition one needs 
a description how the atoms interact. Thus there are always  {\it more parameters}  than  degrees of freedom, 
and only a description is possible.

%
%
%
%
%
%
%
%
%
%
%
{
\footnotesize
\begin{figure}[htb]         
\begin{center}            
  \resizebox{7 cm}{!}{\includegraphics[]{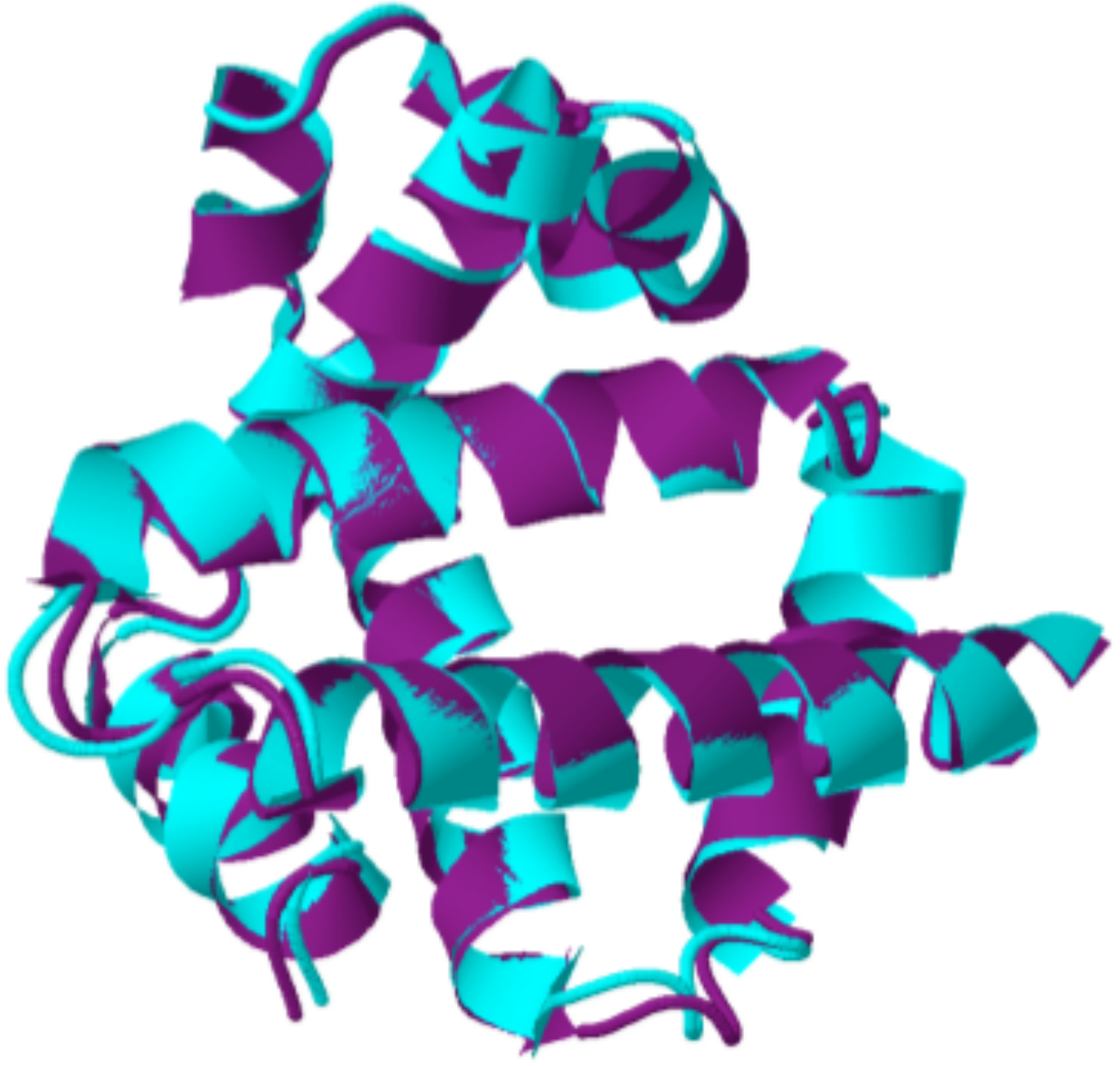}}
\end{center}
\caption {Comparison between the PDB structure 1ABS (dark purple) and the corresponding multi-soliton 
solution (light blue). }   
\label{fig-28}    
\end{figure}
}
In Figure \ref{fig-28} we interlace the 1ABS backbone with the multi-soliton;  the difference is very small.

%
%
%
%
%
%
%
%
%
%
%
%
%
%
%
%
%
%
%
%
%
%
%
%
%
%
%
%
%
%
%
%
%
%
%
%
%
%
{
\footnotesize
\begin{figure}[h]         
\begin{center}            
  \resizebox{8 cm}{!}{\includegraphics[]{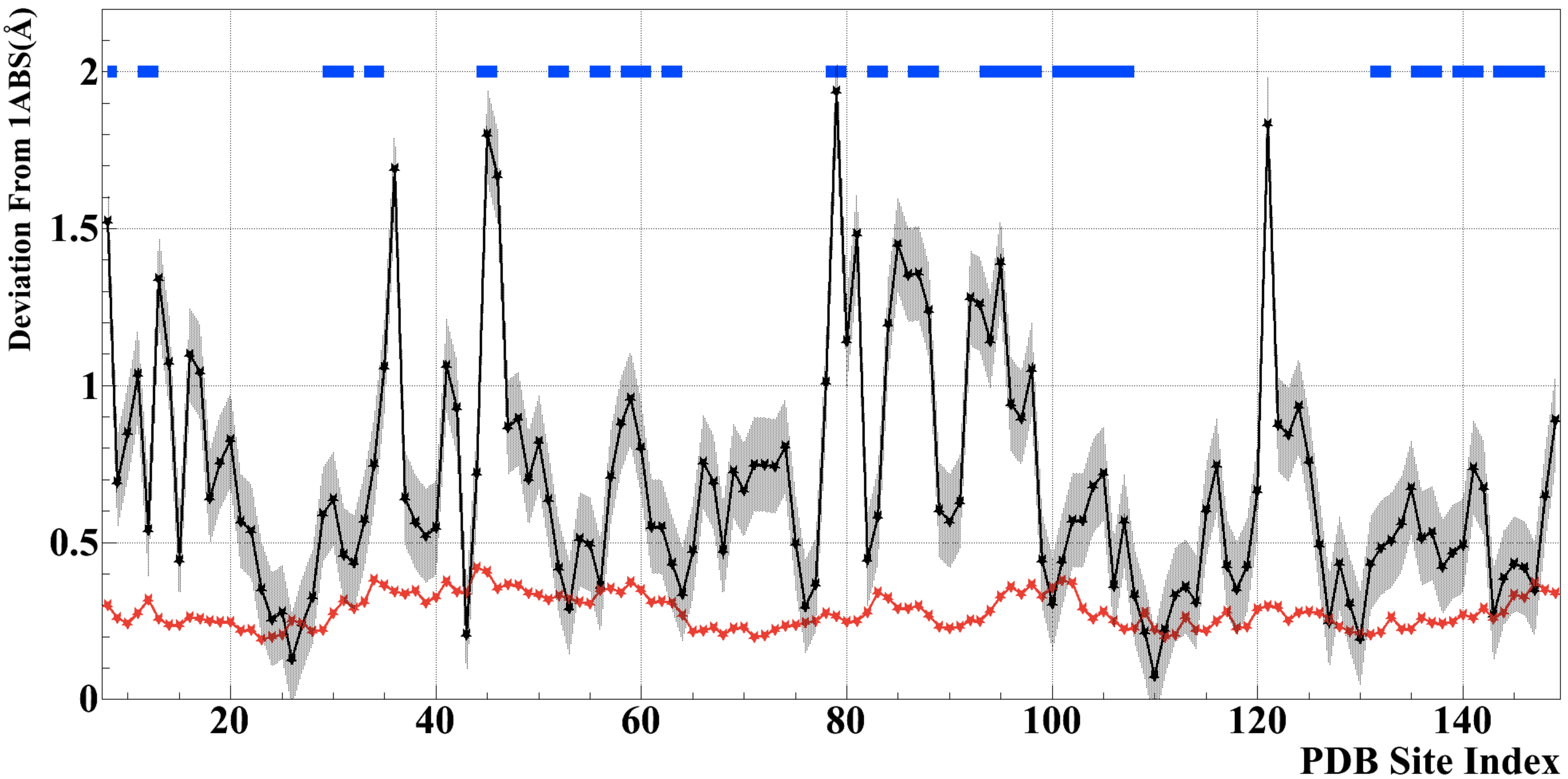}}
\end{center}
\caption {Comparison of the RMSD distance between the 1ABS configuration and the multi-soliton solution, with the Debye-Waller B-factor fluctuation distance around the 1ABS backbone. The blue marking at top, along 2.0 \AA~ line, 
denotes sites where {\it Molprobity} (\ref{molprobity}) detects imperfections.}   
\label{fig-29}    
\end{figure}
}
%
%
%
%
%
In Figure \ref{fig-29} we analyse site-wise the precision of the multi-soliton
configuration with the PDB structure 1ABS. The 15 pico-meter gray-scaled region around the multi-soliton profile 
corresponds again to the regime of zero point fluctuations, that we have deduced from Figure \ref{fig-4}. 
Conceptually, the multi-soliton describes a {\it single} myoglobin
structure in the limit of vanishing temperature.
In particular, as such the multi-soliton does not account for any 
kind of conformational fluctuations that are due to thermal effects, 
lattice imperfections, or any other kind of conformational sub-state effects; we model thermal effects using
the Glauber heath bath (\ref{P}). 
On the other hand, the experimentally measured 1ABS crystal structure should 
not be interpreted as a single static low temperature structure. Instead, it is an average over 
a large number of closely packed crystallographic structures.
A comparison between Figures \ref{fig-25} and \ref{fig-29} shows that in the latter, the distance between the
PDB backbone and the multi-soliton profile is larger than that between the PDB backbone and the individual
solitons in Figure \ref{fig-25}. The Figure \ref{fig-29} describes the {\it single} multi-soliton solution to
the equations (\ref{dnlse}), (\ref{taueq2}) while in Figure \ref{fig-25} we have constructed the individual solitons
by solving (\ref{dnlse}), (\ref{taueq2}) independently for each loop. A similar individual soliton construction in the
case of 1ABS gives profiles that are comparable, even slightly more precise, than those in Figure \ref{fig-25}.
But for energetic studies we need the full multi-soliton with its energy function, we need the local energy minimum of (\ref{E1old})
for the {\it entire} backbone.

We presume that a multi-soliton solution could be constructed with a precision even better than 0.78 \AA~ in
C$\alpha$ RMSD. But the convergence of (\ref{ite}) becomes slow when we use a laptop like MacAir. Thus 
we have simply terminated the process
when we reach the value 0.78 \AA~ which is in any case much better 
than that obtained in any other approach, using any other computer,
to our knowledge.

In Figure \ref{fig-29}  we observe that the distance between the multi-soliton solution
and the C$\alpha$  carbon backbone of 1ABS has its largest values mainly in two 
regimes. These are located {\it roughly} between the sites 35-45,  and between  the sites 79-98;
we propose the reader inspects the structure of 1ABS using the visualisation interface of PDB.
The first regime corresponds to the single soliton that models the loop between helix B and helix C in Figure \ref{fig-26}. 
The second regime corresponds to the location of the helix F.  This helix is part of the "V"-shaped pocket of helices
E and F, where the heme group is located. In particular, the reader can observe that 
helix F includes the proximal histidine at site 93,
which is bonded to the iron ion of the heme.  Note that in addition, our analysis detects an anomaly at around site 121.

In order to understand the origin of the observed deviations  from an ideal  multi-soliton crystal, 
we  check for the presence of potential structural 
disorders in 1ABS using  {\it Molprobity} (\ref{molprobity}); we recommend the reader performs this analysis on the 
{\it Molprobity} web-site.
In Figure \ref{fig-28}, along the top, at the level of the 2.0 \AA ~ line, we have marked with blue those regions where 
according to {\it Molprobity}  we have potential clashes. The  {\it Molprobity} clash score
of 1ABS is 20.32 which puts it in the 10th percentile among structures with comparable resolution,
100  being the best score.   The regions of potential structural clashes correlate with those
regions, where our multi-soliton profile has the largest deviations from the 1ABS backbone.
Except the vicinity of the site 121, which is unproblematic according to {\it Molprobity}.

We first consider the difference between the multi-soliton and the 1ABS backbone 
around the sites 79-98, that was also identified by the {\it Ansatz} as a potentially 
troublesome one. The difference appears to be largely due to a deformation of the
helix F. It could be caused  by a bond between the proximal histidine at site 93 and the oxygen binding
heme iron. This might introduce a strain which modifies the backbone. The effect of the heme is 
not accounted for by our energy function, in the present form. Consequently we propose the
histidine-heme interaction
to be the likely explanation for  the relatively large deviation between our multi-soliton 
profile and the 1ABS backbone, at this position. 

We proceed to consider the difference between the multi-soliton and the 1ABS backbone around the sites 35-45.
These sites are also located very close to the heme. For example, the distance between the C$\alpha$ carbon
at site 45 (Arg) and the hem oxygen 154 is 4.84 \AA, and the C$\alpha$ of Phe at site 43 is even closer to the heme. 
This proximity between the backbone and the heme is reflected in the {\it Molprobity} clash at site 45 (between
C$\delta$ and 154 HEM). We conclude that there could be strain in the backbone structure which is 
due to the heme,  and this could explain the difference between 1ABS backbone and the multi-soliton configuration
in this regime. 

Finally, we note that in Figure \ref{fig-29}, we also have our previously observed anomaly at site 121 (glycine). 
At this point, we have no explanation for the anomaly except that glycine is flexible
and that  this region is on the exterior of the protein. This leaves the hydrophobic 
phenylalanine at nearby site 123 exposed to the solvent. Consequently relatively strong fluctuations 
between several locally conformationally different but energetically degenerated sub-states are possible.

%
%
%
%
%
%
%
%
%
%
%
%
%
%
%
%
%

\section{Dynamical  myoglobin}
\label{sectphase}

%
%
%
%
%
%
%
%
%

The myoglobin stores O$_2$ by binding it to the iron atom, which is 
inside the myoglobin. The oxidisation causes a conversion
from ferrous ion (Fe2+) to ferric ion (Fe3+); the oxidised molecule is called  oxymyoglobin. When the oxygen is
absent,  the molecule is called deoxymyoglobin. We propose the reader finds examples of each from PDB, and  inspects 
the structures using the 3D Java interface.

Numerous detailed experimental investigations have been made both of oxymyoglobin and deoxymyoglobin. But to our
knowledge, the understanding of the oxygen transport mechanism in myoglobin remains incomplete:
We do not yet know exactly, how small non-polar ligands such as O$_2$, CO and NO move 
between the external solvent and the iron containing heme group, which is located inside the myoglobin.  
From the available static crystallographic PDB structures one can not  identify any obvious ligand pathway.
It has been proposed that  the process involves thermally driven large scale conformational motions.   
Collective thermal fluctuations  could open and close gates through which the ligands migrate. Such 
gates are not necessarily visible in the crystallised low temperature structures.   Computational
investigations that model the dynamics of myoglobin might provide a glue how these gates operate.
 
\subsection{Glauber dynamics and phase structure}
\label{subsectphase}

We start our investigation how ligand gates might open and close, by  performing 
heating and cooling simulations of myoglobin with the energy function (\ref{E1old}) in combination with
Glauber dynamics (\ref{P}), (\ref{dist2}) \cite{Krokhotin-2013a}. 
We use the 1ABS multi-soliton we have constructed. The structure
is a carbon-monoxy-myoglobin (MbCO), but with the covalent bond between the CO and iron broken
by photodissociation. In any case, its dynamical properties should serve as a good first-approximation 
model how myoglobin behaves, more generally.

We start our simulations at a low temperature value, from the 
multi-soliton configuration that we have constructed:  A classical soliton solution  is
commonly interpreted as a structure that describes the limit of vanishing 
temperature where thermal fluctuations become very small.

We take 
\[
k{\mathcal T}_L = 10^{-16}
\]
for the numerical value of the low temperature factor, 
in terms of the dimensionless unit
that is determined by our choice of  the overall energy  scale in (\ref{E1old}).  
At this value of the temperature factor  thermal fluctuations are absent in our multi-soliton.
For the numerical high temperature value we choose 
\[
k{\mathcal T}_H =  10^{-13}
\]
By applying the renormalisation group arguments in Section \ref{4.5}  we have related the two temperature factors
$k\mathcal T$ and $k_BT$. Our conversion relation that we shall justify in the sequel, is 
\begin{equation}
k{\mathcal T} = 1.6181 \cdot 10^{-9} k_B \, T \, \exp\{ 0.05506 \, T \}
\label{Ttot}
\end{equation}
We use CGS units so that $k_B = 1.381 \times 10^{-16} \, erg/K$. 

%
%
%
%

Under {\it in vivo} conditions myoglobin always interacts with water. This interaction is 
essential for maintaining the collapsed phase. In our approach we account for the solvent (water) 
implicitly, in terms of the parameter values in (\ref{E1old}). 
In particular, as such our model can only {\it effectively}  take into
account the highly complex phase properties of water at sub-freezing temperatures.   
We do not even try to address the obvious 
complications that appear when the temperature raises above the boiling point of  
water. 

We start the simulation at $k{\mathcal T}_L$. The heating takes place  with an 
exponential rate of increase during 5 million Monte Carlo steps. We model the non-equilibrium response 
using the Glauber protocol (\ref{P}). 
According to (\ref{Ttot}) 
in terms of physical temperature factor  $k_B T$,
the heating process 
should  correspond to an adiabatically slow nearly-linear rate of increase. 

When we arrive at the high temperature $k{\mathcal T}_H = 10^{-13}$ we fully 
thermalise the system by keeping it at this temperature value during 5 million steps. 
We then proceed to cool it back down to $k{\mathcal T}_L$. We use the same rate of cooling as we use for heating,
{\it i.e.} we cool exponentially in $k{\mathcal T}$ during  5 million steps. Each complete 
heating-cooling cycle takes about 3 minutes of wall-clock time when we use  a single 
processor in a standard laptop  (MacAir). Consequently time is no constraint for us and we can collect very,  
very good statistics. In particular,  we have checked that our results and conclusions are quite  insensitive to
the rate of heating and cooling. 

For statistical purposes, we have performed 100 repeated heating and cooling cycles that we have then
analysed in detail; an increase in the number of cycles does not change our conclusions.   
Note that in an all-atom approach a comparable simulation would take over 100.000 years
even with {\it Anton} \cite{Shaw-2008} while for us a few minutes is enough.  

During the simulations, we  follow  the evolution of both the radius of gyration $R_g$ and the RMSD
distance $R_{rmsd}$ between the simulated configuration and the folded 1ABS structure. 
In Figure \ref{fig-30} a) we show the evolution of the radius of gyration, and in Figure \ref{fig-30} b)  
we show the evolution 
of the RMSD distance to 1ABS,   as a function of steps during our 100 repeated 
heating and cooling cycles; Note that in these Figures we have converted the temperature into 
Kelvin scale, using (\ref{Ttot}). 

%
%
%
%
%
%
%
%
%
%
%
{
\footnotesize
\begin{figure}[htb]         
\begin{center}            
  \resizebox{7 cm}{!}{\includegraphics[]{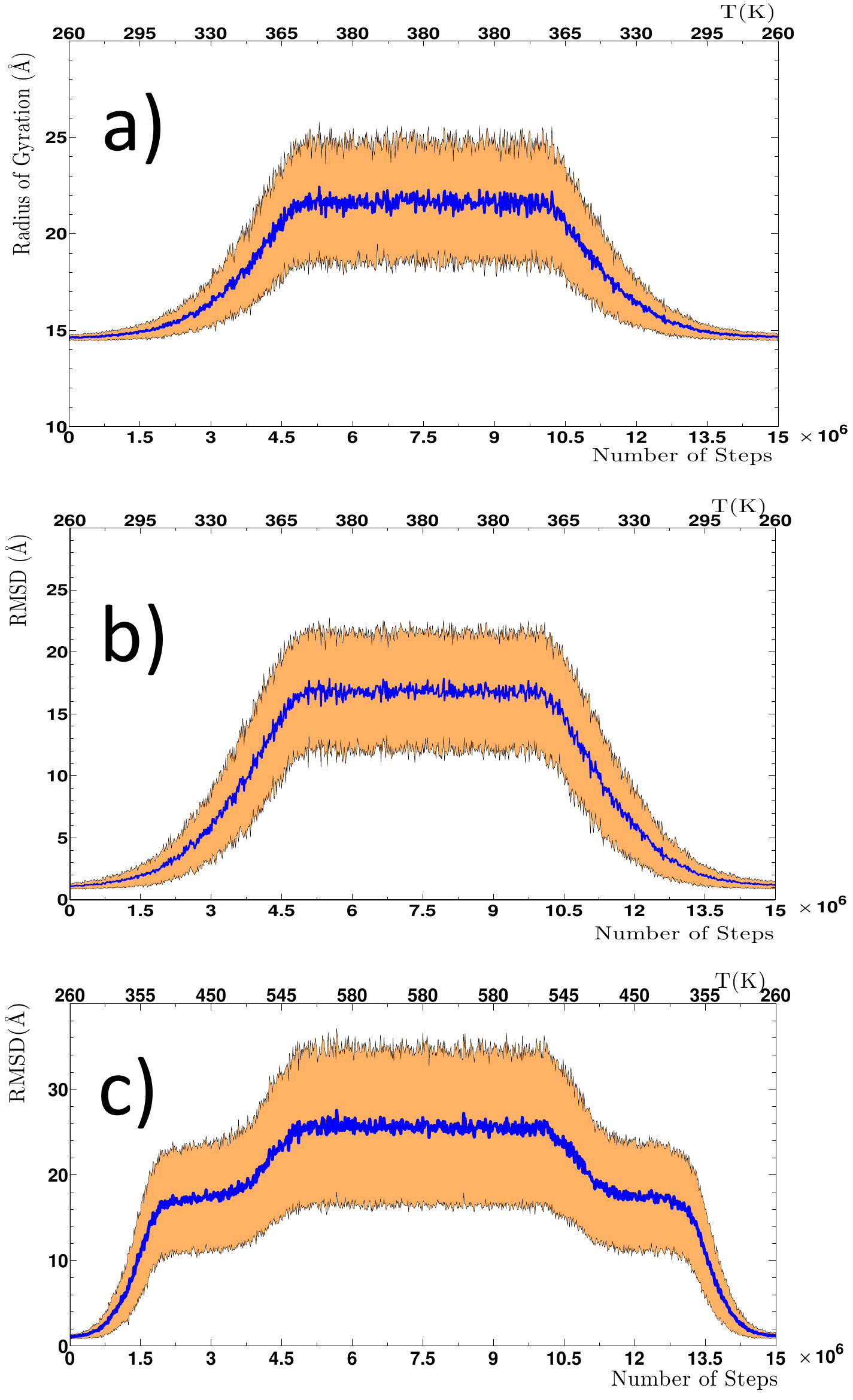}}
\end{center}
\caption {a) The evolution of the radius of gyration during 100 repeated heating and cooling cycles. b) 
The evolution of the RMSD distance between the 1ABS backbone and the simulated configuration during 100 repeated 
heating and cooling cycles. c)
The evolution of the RMSD distance between the 1ABS backbone and the simulated configuration during 100 repeated 
heating and cooling cycles, to very high temperature values. In each Figure the (blue) line denotes the average, and the 
shaded area around it is the extent of one standard deviation fluctuations. 
Along the top axis, we show the temperature in the
Kelvin scale, using the conversion relation (\ref{Ttot}).}   
\label{fig-30}    
\end{figure}
}

We make the
following observations:
At low temperatures, with temperature factors
\[
k{\mathcal T} < 10^{-15}
\]
the radius of gyration is essentially constant $R_g \approx 14.6 $ and only subject to 
very small thermal fluctuations. Between
\[
10^{-15} < k{\mathcal T} < 10^{-14}
\]
we have  a regime when the radius of gyration increases at an accelerating rate in the number of steps. 
The increase in $R_g$ continues until we reach a temperature near $k{\mathcal T}_H$. But for temperatures where
the temperature factor is
in the range
\[
10^{-14} < k{\mathcal T} < k{\mathcal T}_H = 10^{-13}
\]
the rate of increase decelerates so that when we reach the  temperature 
$k{\mathcal T}_H$ we observe no  increase
in $R_g$.  This proposes that we have reached the random walk $\theta$-regime. 
Furthermore, the radius of gyration value 
\[
R_g \ \approx \ 22 \  \ {\rm \AA}
\]
is {\it extremely} close to the experimentally measured value $\sim 23.6$ \AA~ for the molten globule state
of myoglobin. The difference can be entirely attributed to the 12 residues that we have 
excluded (7 from the N-terminal and 5 from the C-terminal) when constructing the multi-soliton.

We have confirmed  that the transition near $k{\mathcal T}_H$ is indeed a
$\theta$-transition between collapsed phase and random walk
phase, by heating the configuration to substantially higher temperature factor values. We have found that above 
this putative $\theta$-transition,
the radius of gyration remains essentially intact  under  temperature variations  all the way to 
\[
k{\mathcal T} = 10^{-8} 
\]
Around this temperature value another transition takes place, presumably 
to the $\nu \sim 3/5$ self-avoiding random walk phase.
In Figure \ref{fig-30} c) we show how the RMS distance between the heated configuration and 
1ABS changes as a function of temperature, during heating and cooling cycles between 
$k{\mathcal T}_L = 10^{-16}$ and $k{\mathcal T} = 10^{-8}$. We observe two clear transitions 
that are consistent with the
transitions between collapsed and random walk phases, and between random walk and self-avoiding random
walk phases according to (\ref{nuval}). %

When we decrease the temperature, the evolution of $R_g$ becomes 
inverted. At the end of the cycle, when the temperature reaches $k{\mathcal T}_L$, 
the configuration returns back to a very close  proximity of the original  folded state; see Figure \ref{fig-30} b). 
This demonstrates the  stability of the 
multi-soliton solution that describes the natively folded myoglobin  
as a local minimum of the energy (\ref{E1old}). 

%
%
%
%
%
%

Figure \ref{fig-31} a)  shows 
the {\it average} values of  $R_g$. These averages are evaluated at several different temperature values, 
over 100 runs.  Both  during the heating period when $0 < x < 7.5 $, and during
the cooling period when $ 7.5 < x < 15 $ where $x$ is the number of MC steps in millions. 
The data can be approximated by a fitting function of the form
\begin{equation}
R_g (x) \ \approx  \  a \cdot \tanh\{ b (x -c)  \} + d
\label{Rg1}
\end{equation}
The parameter values are listed in Table \ref{paraV}.
%
%
%
%
%
%
%
%
%
%
%
{
\footnotesize
\begin{figure}[htb]         
\begin{center}            
  \resizebox{8 cm}{!}{\includegraphics[]{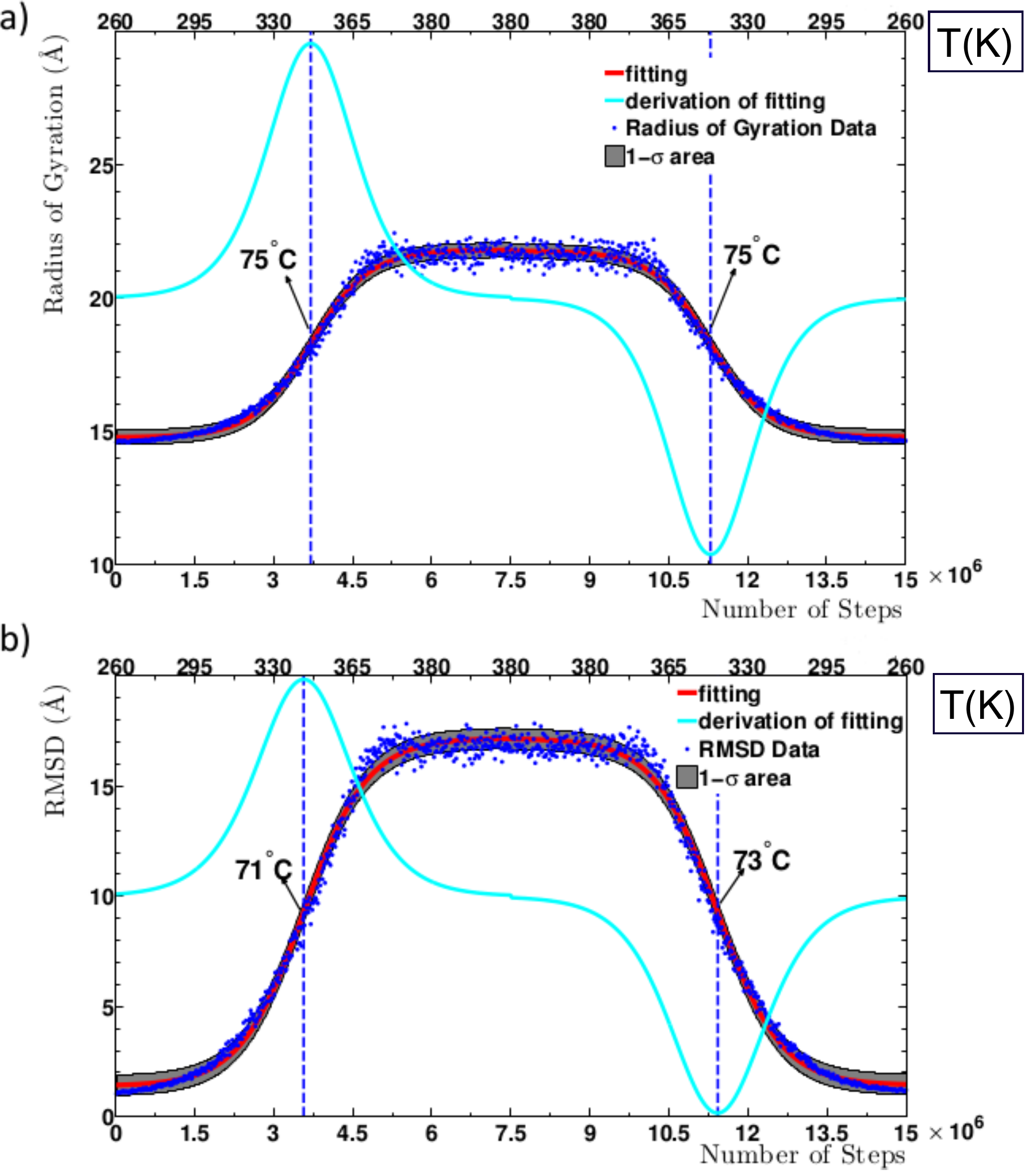}}
\end{center}
\caption {The red line is the fitting of (\ref{Rg1}) to the average values of blue dots which are the numerically computed
values of $R_g$, over the heating and cooling periods. The shaded
area around the red fitting line is one standard deviation estimate. Note: The difference between these three 
is so small that it is barely observable in the Figure. Also shown is the derivative of (\ref{Rg1}) (light blue line). 
Along the top axis, we have converted the temperature into Kelvin scale.}   
\label{fig-31}    
\end{figure}
}
%
%
%
%
%
%
%
{
\begin{table}[htb]
\caption{Parameter values in the fits (\ref{Rg1}), (\ref{rmsd1}) for the two ranges 0 - 7.5 and 7.5 - 15 (in million) of
iteration steps
 }\vskip 0.5cm
 \begin{center}
\begin{tabular}{|ccccc|cccc|}
\multicolumn{1}{c}{}  & \multicolumn{4}{c}{$R_g$} &  \multicolumn{4}{c}{$R_{rmsd}$}  \\
\cline{1-9} 
\multicolumn{1}{|c|}{range}  & \multicolumn{4}{c|}{ \hskip 0.cm a  \hskip 0.cm ~~~~~~~b~ \hskip .cm ~~~~~~c  \hskip .0cm 
~~~~~d }
& 
 \multicolumn{4}{c|}{ \hskip 0.cm a  \hskip 0.cm ~~~~~~~b~ \hskip .cm ~~~~~~c  \hskip .0cm 
~~~~~d}
\\
\cline{1-9}
\multicolumn{1}{|c|}{\hskip 0.1cm 0 - 7.5 \hskip 0.cm }  
& 
\multicolumn{4}{c|}{\hskip -0.cm 3.519~  \hskip 0.cm 0.9047   \hskip 0.cm 3.6855   \hskip 0.cm
~18.29 } 
&  
\multicolumn{4}{c|}{ \hskip 0.cm  7.9 ~  \hskip 0.cm 0.8318  \hskip 0.cm 3.5715  \hskip 0.cm 9.291}  
\\
\multicolumn{1}{|c|}{7.5 - 15\hskip 0.cm}  
& 
\multicolumn{4}{c|}{  \hskip 0.cm -3.486 \hskip 0.cm  0.9193   \hskip 0.cm 11.2965   \hskip 0.cm 18.28 }
&  
\multicolumn{4}{c|}{  \hskip 0.cm -7.872  \hskip 0.cm 0.8327 \hskip 0.cm 11.4255  \hskip 0.cm 9.298 \hskip 0.cm} \\
\cline{1-9} 
\hline
\end{tabular}
\end{center}
\label{paraV}
\end{table}

In Figure \ref{fig-31} a) we display the derivative of (\ref{Rg1}).
We can try and use the maximum of the derivative to identify the 
$\theta$-transition temperature in our model. For this, we assume that 
the experimentally measured \cite{Moriyama-2010,Ochia-2010}
transition 
temperature at $T_c \approx 348 \ K$ is the one that corresponds 
to the $\theta$-transition.   We identify it with the 
maximum of the derivative of $R_g$, to 
conclude that during the heating cycle the $\theta$-transition temperature relates to 
our dimensionless temperature values
as follows,
\begin{equation}
{\mathcal T}_g^{h} \approx 1.63 \cdot 10^{-14}  \approx 348 \ K
\label{renT}
\end{equation}
We use this value to determine one of the two parameters in (\ref{Ttot}).

During the cooling cycle, we find the slightly different
\[
{\mathcal T}_g^{c} \approx 1.71 \cdot 10^{-14}  \approx 349 \ K
\]
We note that an asymmetry between heating and cooling has been observed 
experimentally \cite{Moriyama-2010}.

The RMSD distance between the simulated configuration and the 1ABS backbone depends on temperature 
in a very similar manner. In Figure \ref{fig-31} b) we show 
the comparison between 
simulation, and the corresponding approximation (\ref{Rg1}), 
\begin{equation}
R_{rmsd} \ \approx \ a \cdot \tanh\{ b (x-c) \} + d
\label{rmsd1}
\end{equation}
These parameter values are also listed in Table \ref{paraV} separately
for the heating period $0<x< 7.5$  and for the cooling period $7.5<x<15$.
The Figure \ref{fig-31} b) also 
shows the derivative of $R_{rmsd}(x)$. Like in the case of radius of gyration, we use the maximum of the derivative
to estimate the peak rate of change in the transition temperature. 
During the heating period the increase in $R_{rmsd}$ peaks  at
\[
{\mathcal T}_{rmsd}^{h} \approx 1.35 \cdot 10^{-14} \ \approx \  344 \ K
\]
During the cooling period we find that the peak corresponds to  a slightly higher temperature value,
\[
{\mathcal T}_{rmsd}^{h} \approx 1.45 \ 10^{-14} \ \approx \  346 \ K
\]
These valus are very close to those we observe in the case of $R_g$. 

%
%
%
%
%
%
%
%
%
%
%
%
%
%
%
%

%
%
%
%
%
%
%
%
%
%
%
%
%
%
%
%

\subsection{ Backbone ligand gates}
\label{sect:scale}

We proceed to try and identify potential 
thermally driven backbone ligand gates. We are interested in studying   
how the gates open and close as  the myoglobin is heated and cooled.  
Moreover, thus far we have fixed only one of the two parameters in
(\ref{renT}) using the  $\theta$-transition temperature.
We shall fix the second parameter in the sequel,  
by considering the dynamics of the  backbone ligand gates.


We investigate the shape of the backbone visually, during the heating and cooling.
We find that qualitatively the thermal fluctuations follow  a very similar pattern. 
The backbone  becomes unfolded in more or less the same manner, again and again,
as the temperature becomes increased. The inverse pattern is observed 
during the cooling.

During heating we observe a clear onset of the unfolding transition, 
which we characterise in 
terms of backbone ligand gates. We have identified three major  
gates that we call Gate 1,2 and 3, and we define them as follows:
%
%
%
%
%
%
%
%
%
%
%
{
\footnotesize
\begin{figure}[htb]         
\begin{center}            
  \resizebox{8cm}{!}{\includegraphics[]{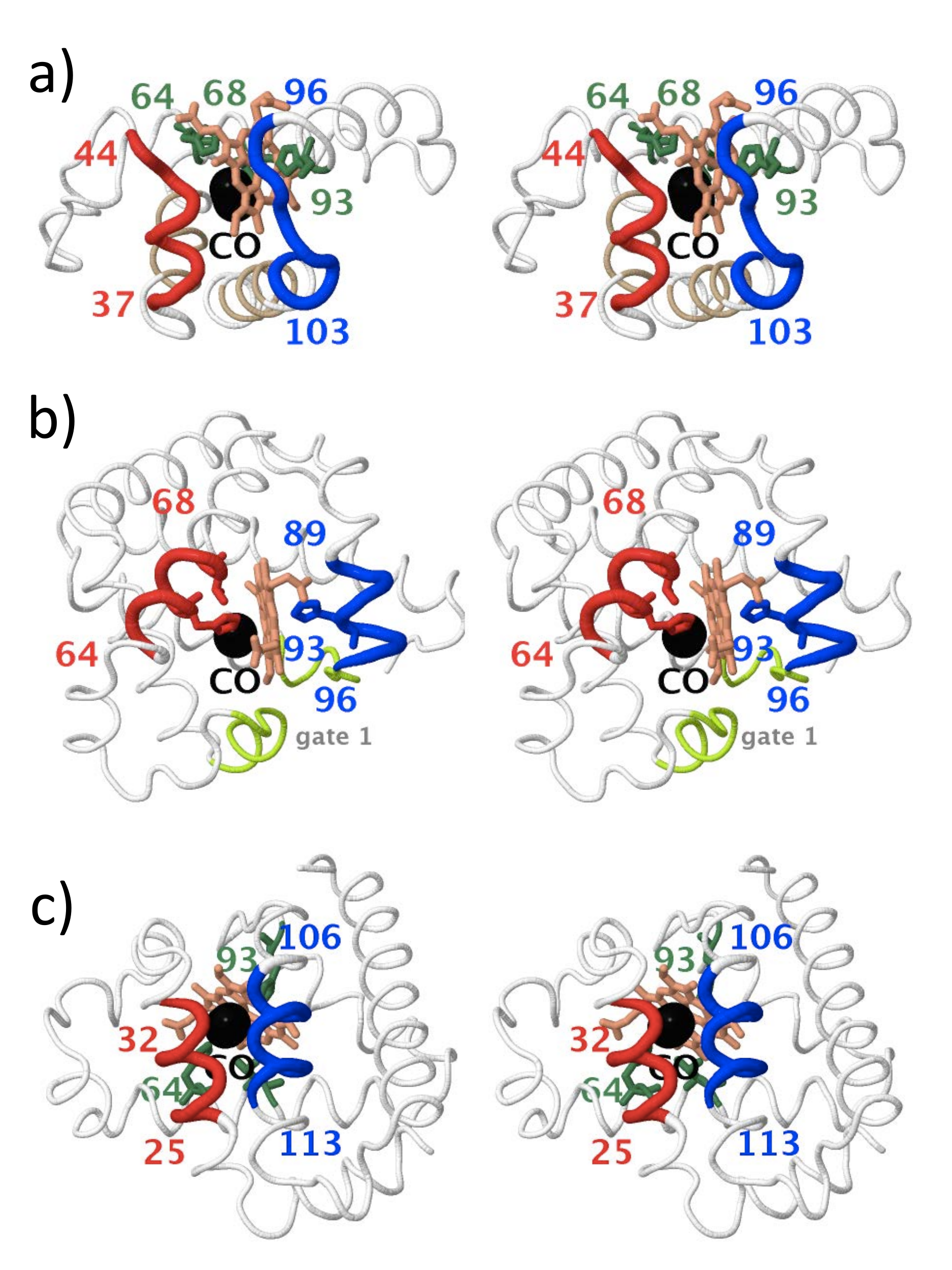}}
\end{center}
\caption {Stereographic cross-eyed view of the ligand Gate 1, 2 and 3 as defined in the text. Figure a) is 
Gate 1 formed by segment  located  between 37 (Pro) and 44 (Asp), and segment that starts at 96 (Lys) 
and ends at 103 (Tyr). Figure b) is Gate 2, between segment that starts from PDB site 61 (Leu) and ends at PDB site 68 (Val), and segment that starts at 89 (Leu) and ends at 96 (Lys). Figure c) is Gate 3, between segment starting from
site 25 (Gly) and ending  at  32 (Leu), and segment that starts at 106 (Phe) and ends at 113 (His).
We also show the location of the heme (orange), the proximal histidine (93), the valine (68), the distal histidine (64) (all green) and the CO (black ellipsoid) of 1ABS.
}   
\label{fig-32}    
\end{figure}
}

The  Gate 1 shown in Figure \ref{fig-32} a)
is defined as the area between the segment  that starts 
from PDB site 37 (Pro) and ends at PDB site 44 (Asp), and the segment  that
starts at 96 (Lys) and ends at 103 (Tyr). The opening of this gate takes place as
the distance between
the two segments increases. The open gate exposes the heme to the solvent.
Figure \ref{fig-32} a) shows the location of this gate along the 1ABS backbone.

The  Gate 2 is located between the helical structures E and F, as shown in \ref{fig-32} b).
This gate extends over the entire length of both helices E and F. Thus, in order to compare it 
with Gate 1 that is composed of segments with only eight residues, we select two opposing
segments along helices E and F,  each with eight amino acids. The first segment, located in the 
helical structure E, starts with site 61 (Leu) and ends with site 68 (Val). The second segment, located 
in the helical structure F opposite to the first segment, starts with site 89 (Leu)  and ends with site 96 (Lys).
We have intentionally selected these two segments to be far from the loop that connects the helices
E and F.  This is because in our simulations, we have observed that the amplitudes 
of the thermal fluctuations in the segment distances tend to increase, 
the further away the segment is located from the connecting loop: The opening and closing of the gate resembles the 
opening and closing of scissors,
with blades formed by helices E and F that are  connected by  the loop between these two helices. 
Note that the first segment along helix E, includes both the distal histidine at site 64 and the valine at the end site 68. This valine
is also inside the heme pocket, and it is presumed to have an important role in CO {\it vs.} O$_2$ discrimination. 
Similarly, the opposite segment in the helical structure F includes the proximal histidine at site 93. 

Finally, the Gate 3 which  is shown in \ref{fig-32} c)  
is  located between the helical structures B and G. Again, in order to compare this 
relatively long gate with Gate 1 
we select two opposing segments,  each with  eight amino acids.
The segment in the helical structure B starts at  site 25 (Gly) and 
ends at site 32 (Leu). The segment in helix G starts at
site 106 (Phe)  and ends at site 113 (His). 

During the heating and cooling cycle of the myoglobin, we follow the size of the three gates. We do  this
by  computing the distance $d_i \ (i=1,2,3)$ between the respective segments, as a functions of temperature. 
We define the distance $d_i$ between the two segments  for each of the three gates,  as follows:
\begin{equation}
d_i = \sqrt{ \  \sum\limits_{n=1}^8 \ (\mathbf x_n - \mathbf y_n )^2 \ \ }
\label{di}
\end{equation}
Here $\mathbf x_n$ are the eight coordinates in the first segment, and $\mathbf y_n$ are the corresponding 
coordinates in the second segment, along Gate  $i=1,2,3$. Note that the two segments in each 
of the three gates, are spatially oriented in an anti-parallel manner with respect to PDB indexing.
Consequently,  in computing (\ref{di}), we invert the indexing in one of the two segments  
with respect to the PDB indexing. 

We start by investigating the temperature dependence of 
the three gates, using the experimental data which is available in  
PDB.  For this we compute the following three gate ratios, from PDB data:
\begin{equation}
\frac{\rm Gate 1}{\rm Gate 2}  = \frac{d_1}{d_2} \ \ \ \ \&  \ \ \ \  \ \frac{\rm Gate 3}{\rm Gate 2}  = \frac{d_3}{d_2}
\ \ \ \ \&  \ \ \ \  \ \frac{\rm Gate 3}{\rm Gate 1}  = \frac{d_3}{d_1}
\label{gr}
\end{equation}
We use all the presently available myoglobin structures in PDB, that have been measured with 
resolution 2.0 \AA~ or better. The results are   
shown  Figures \ref{fig-33}. 
In each Figure,  
we observe  substantial fluctuations in the gate ratios,  in case of PDB data that has been 
taken at around $100 \, K$. But this  reflects only the fact that the majority of PDB data 
has been collected at this temperature value.
Overall, we conclude that the gate ratios show no temperature dependence for $T< 300 \,K$. 
%
%
%
%
%
%
%
%
%
%
%
{
\footnotesize
\begin{figure}[htb]         
\begin{center}            
  \resizebox{7.5cm}{!}{\includegraphics[]{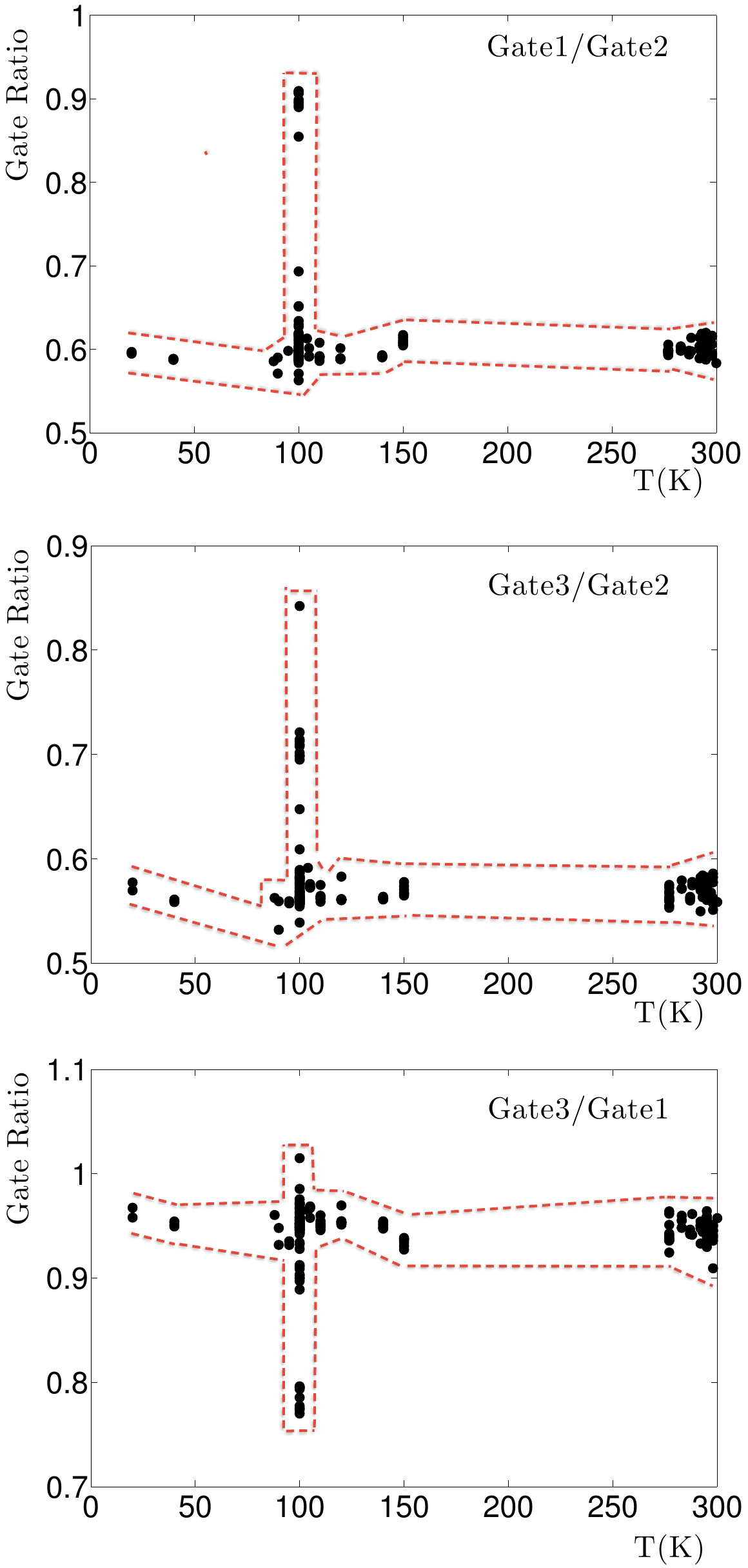}}
\end{center}
\caption {The three Gate ratio (\ref{gr}). The dotted (red) lines are intended to guide eye only. }   
\label{fig-33}    
\end{figure}
}

We proceed to the computation of  the temperature dependence of the gate ratios using our 1ABS multi-soliton with
the energy function (\ref{E1old}). The results are shown in Figure \ref{fig-34} a), b) and c).
 We have found that the Gate 3 is the first gate to open as the temperature increases,
and the last one to close as temperature decreases. The Gate 2 is the one to open last, and the one to close first. 
In the low temperature limit the Gate 3 is about half the size of the Gate 2. But its size exceeds that of the Gate 2 in the
segment separation distance (\ref{di}) at temperature
\[
k{\mathcal T}^{c}_{23} \approx 10^{-14} \ \sim  \ 340 \, K
\]
The transition is very rapid,  in line with the general results of reference \cite{Chernodub-2010b}:  
When the temperature reaches the $\theta$-transition 
value $\sim 348 \, K$,  the Gate 3 is about twice as large as the Gate 2. 

The Gate 1 also opens much faster than Gate 2, but slower than Gate 3. It also closes slower than Gate 2, but faster than 
Gate 3. In the low temperature limit the Gate 1 is about half as wide as Gate 2. But it becomes wider than Gate 2
when the temperature reaches a value
\[
k{\mathcal T}^{c}_{12} \approx 10^{-14} 
\]
However, the Gate 1 does not become quite as wide as Gate 3. 
This is shown in Figure \ref{fig-34} a).
%
%
%
%
%
%
%
%
%
%
%
{
\footnotesize
\begin{figure}[htb]         
\begin{center}            
  \resizebox{7.2cm}{!}{\includegraphics[]{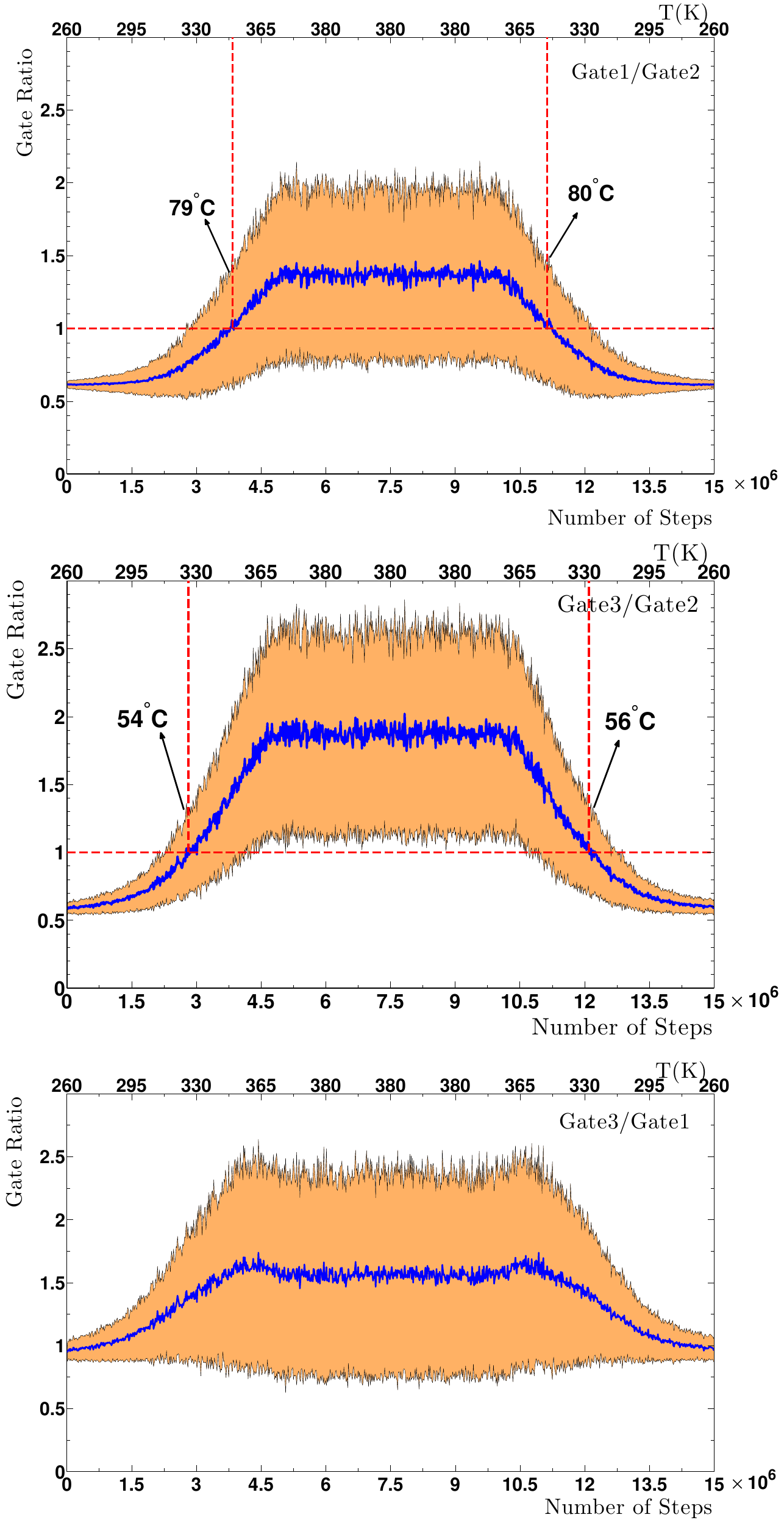}}
\end{center}
\caption {The temperature dependence of the three Gate ratios (\ref{gr}) during our heating and cooling cycles.}   
\label{fig-34}    
\end{figure}
}

We are now in a position to determine the second parameter in (\ref{kTt})  to arrive at  (\ref{Ttot}); 
we remind that one of the two parameters is already determined, in (\ref{renT}).
For this we proceed as follows:
When we compare Figures \ref{fig-33}  we conclude that experimentally, the gate ratios do not display any observable
temperature dependence whenever $T<300 \, K$. Consequently, the lowest possible value of the temperature factor 
$k\mathcal T$ where
Figures \ref{fig-34}  can display any change in the gate ratios, 
should correspond to a temperature which is above $300\, K$.  
When we read off the lowest possible $k\mathcal T$ value where we have an observable 
effect in Figures \ref{fig-34}, we conclude that, necessarily,
\[
k {\mathcal T} \approx 10^{-15} >   k_B \, 300 \, K 
\]
This gives a lower bound. When we adopt this lower bound 
value as our estimate for the gate opening temperature we obtain
the second parameter value in (\ref{Ttot}). 

In reality the actual gate opening temperature can be higher, but at the moment
experimental basis for choosing a higher value is lacking: The single presently existing 
NMR data of myoglobin in PDB is 1MYF. It has been
measured at the slightly higher temperature value of $308 \, $K. But the quality of data 
does not enable us to  improve our estimate.  

We note that a higher gate opening temperature has no qualitative effect to our conclusions, 
quantitatively  the differences are minor. The only effect would be
 a sharpening of the $\theta$-transition
onset.

\vskip 0.2cm
We conclude with a summary of the consequences that  results in Figures 
\ref{fig-34}  might have to ligand migration: We have found that 
to the extent backbone thermal fluctuations play a r\^ole in ligand
migration, the Gate 3 between the helical structures B and G can be very important.  This gate opens very much 
like a baseball glove, as we increase temperature.   The Gate 1 might also play a r\^ole, but probably a lesser one 
than Gate 3. On the other hand, the V-shaped Gate 2 between helices E and F seems to be quite sturdy, 
it does not seem to open as much as the other two gates. The presence of the distal and proximal histidines in
Gate 2 and their attractive interactions with heme, might have an additional stabilising effect that is not accounted by our model. 
Consequently we do not see how the thermal backbone fluctuations that take place in the Gate 2, could 
play a major r\^ole in ligand migration. At least, to the extent that  backbone fluctuations are relevant.  
 
A recent THz time-scale spectroscopy experiment has detected collective thermal fluctuations in the protein,
that might relate to our theoretical proposals  of  ligand gate dynamics \cite{Woods-2014}. However, more detailed
experiments need to be performed.

\chapter{Intrinsically disordered proteins}
\label{sect16a}

The crystallographic protein structures in PDB are {\it ordered} proteins. An ordered 
protein has an essentially  unique  native  fold which can be determined by x-ray crystallography. 
But most proteins are not ordered, most proteins 
do not seem to have an essentially unique native fold. Instead, the low energy landscape  of most proteins
seems to comprise 
several states which are energetically degenerate but conformationally
disparate.  With local energy minima that are 
separated from each other by very 
low free energy barriers. We call them  {\it intrinsically disordered } proteins. Normally 
these proteins can not be crystallised, 
and structural data is in short supply.  
Our aim is to extend the methods that we have developed, to describe the properties of such
proteins. For this we  start by developing some formalism.

But please keep in mind  that there is a grey zone between ordered and intrinsically disordered proteins:
A skilfull crystallographer might be able to produce a crystal out of a protein that others consider
hopeless.

\section{Order vs. disorder}
\label{sectphase2}

%
%
%
%
%
%
%
%
%

When an ordered protein becomes cooled down to low temperatures 
it should assume an essentially unique native fold, that  corresponds to a minimum 
of the low temperature thermodynamic (Helmholtz) free energy. More specifically,  
in the case of a protein with an ordered native fold, a 
cooling should produce a highly localized statistical distribution of
structurally closely related conformational substates. When taken together, this ensemble 
constitutes the folded native state at low temperatures.  
But if a protein is  intrinsically disordered, instead we expect that the low temperature limit 
produces a {\it scattered} statistical distribution of structurally disparate but
energetically comparable ensembles of conformational substates.  Moreover, these different substates should be 
separated from each other only by relatively low energy barriers.  
The  unstructured, disordered character of the protein is a consequence of a 
motion around this scattered landscape:  The protein swings and sways back and 
forth, quite freely,  over the low energy barriers that separate the various energetically 
degenerate but structurally disparate conformations.

We may think that the state space of a  
ordered protein with an essentially  unique native  fold, consists of a
set  of {\it snapshot} states $|\mathfrak s \!> $ that form a {\it tightly} localized and essentially Gau\ss ian distribution  
around an average state $|\, {\mathfrak s} >_{ave}$. When taken together, the set of these
snapshots determine the low temperature folded native state 
as an ensemble of conformational substates. 
In fact, we  expect that for an ordered protein, the extend of conformational variations 
around the average state $|\, {\mathfrak s} >_{ave}$ can be estimated from 
the crystallographic Debye-Waller B-factor.

However,  in the case of  an intrinsically disordered protein 
the low temperature set of 
snapshot states $ |\mathfrak s \!> $ exhibits a {\it disperse} statistical distribution. 
We have no single tightly localised  peak,
with a clearly identifiable average value,  in the statistical distribution of snapshots.
Instead, the statistical distribution of the snapshot 
states is {\it scattered}. The snapshots become
apportioned between several structurally disperse but  energetically 
degenerate conformational substates.  

Of particular interest  are those energetically degenerate substates that are 
separated from each other by relatively low energy barriers. The 
unstructured and unsettled character of an intrinsically 
disordered  protein is  a consequence of thermally induced fluctuations
that move the configuration around the structurally diverse and energetically degenerate landscape: 
The protein swings and sways back and forth between the disparate snapshot states 
$|\mathfrak s \!>$.  Due to very low energy barriers, this dynamics persists even at very low temperatures, 
where quantum mechanical tunneling transitions eventually take over the thermally induced ones. Thus
the unstructured character can persists even at {\it very} low temperatures.

We interpret the ensemble of snapshot states in terms of a Hartree state $|\Phi\!> $ which is a
linear combination of  the form
\begin{equation}
|\Phi\!> \ \simeq \  \sum \hskip -0.5cm \int \limits_i  \, p_i |\mathfrak s_i \!>
\label{HF}
\end{equation}
Here the index set $i$ can have both discrete and continuum portions,  including  various small and large amplitude
collective coordinates.  
We envision that the state space which is spanned by $|\mathfrak s _i\!>$  
can be endowed with a norm (metric) which enables us 
to select and orthonormalize the set of {\it eigen-conformations} (snapshot structures)  
$|\mathfrak s_i \!>$. The detailed construction of a norm in the space of string-like structures 
will not be addressed here.  

When $|\mathfrak s_i \!>$ are normalised, the coefficient $p_i$ determines the 
probability weight for the ensuing eigen-conformation $|\mathfrak s_i \!>$ to contribute in the Hartree state.
In particular, the  Hartree state (\ref{HF}) is a {\it mixed} state and not a pure state,
when it describes an intrinsically disordered protein; the conformational entropy  is non-vanishing,
\[
S = - \sum \hskip -0.5cm \int \limits_i  \, p_i \log p_i \ > \ 0
\]
Note that when we have one single 
value $p_k \approx 1$, and the other $p_i$ with  $i \not=k$  are vanishingly small $p_i\approx 0$, the Hartree
state $|\Phi>$ reduces to a pure state and describes an ordered native fold.

The eigen-conformations  $|\mathfrak s_i \!>$ are time independent,  akin
states in the quantum mechanical 
Heisenberg picture. But the $p_i$ can be time dependent quantities,  they then describe the time evolution of $|\Phi\!>$. 
For sufficiently
long time scales, larger than the characteristic thermal tunneling time between different 
eigen-conformations  $|\mathfrak s_i \!>$,  
the  time dependence of the $p_i$ is governed by a Liouville equation
\begin{equation}
\frac{ d \hat \rho}{dt} \ = \ \frac{\partial \hat \rho}{\partial t} + \{ \hat \rho , H \} \ \equiv \ \mathcal L \, \hat \rho
\label{liouv}
\end{equation}
where 
\[
\hat \rho \ = \ \sum\limits_i  \, p_i |\mathfrak s_i \!><\! \mathfrak s_i |  
\]
is the density matrix. The second term in (\ref{liouv}) is the Poisson bracket with  the (total) semiclassical
Hamiltonian $H$; in a quantum mechanical version the density matrix and 
the Hamiltonian are replaced by the corresponding Heisenberg operators. We note that
for a non-equilibrium system, the total operator
$\mathcal L$ becomes the (semiclassical) Lindblad superoperator.

For a system at or very near a thermodynamical equilibrium,  the $|\mathfrak s_i \!>$ 
concur with the extrema configurations of the low temperature limit of Helmholz free energy.  
The probabilities  $p_i$  are evaluated from corresponding values of the free 
energy,  and (\ref{HF}) acquires the form
\begin{equation}
|\Phi>  \ \simeq \ \frac{1}{Z} \sum \hskip -0.5cm \int  \limits_{ \{ \mathfrak s \} } e^{- \beta E(\mathfrak s ) } |\mathfrak s\!>
\label{thequi}
\end{equation}
where 
\[
Z \ = 
\  \sum \hskip -0.5cm \int \limits_{ \{ \mathfrak s \} } e^{- \beta E( \mathfrak s ) }
\]
and $\beta = 1/kT$ is  the inverse temperature.  For completeness, we note that 
for an extremum  conformation $|\mathfrak s_i \!>$  that is not a local minimum of the 
free energy, a Maslov index contribution needs to be included.

\section{hIAPP and type-II diabetes}
\label{sectphase}

We now proceed to consider an example of an intrinsically disordered protein with very extensive and important 
biological, medical and pharmaceutical ramifications.

\vskip 0.2cm
The human islet amyloid polypeptide (hIAPP),  also known as amylin, 
is a widely studied 37 amino acid polypeptide hormone; for extensive reviews we refer to    
\cite{Westermark-2011,Pillay-2013}. 
It is processed in pancreatic $\beta$-cells from an 89 residue precursor protein, 
by a protease cleavage in combination of post-translational modifications. 
The secretion of hIAPP  responses to meals, and  the peptide co-operates with 
insulin to regulate blood glucose levels.
But hIAPP can also form pancreatic amyloid 
deposits, and
their formation and buildup correlates strongly
with the depletion of islet $\beta$-cells. 
The hIAPP amyloidosis is present in over 90 per cent of the type-II diabetes  patients,
and  the deposits are
considered as the hallmark of the disease in progression. 

It still remains to be fully clarified whether the hIAPP amyloid aggregation is the direct cause 
of apoptosis in the islet $\beta$-cells. Instead, the 
amyloid fibers might only be a consequence of the disease which is ultimately caused by some other 
and yet-to-be-identified agent.  
Among the arguments that support the existence
of a first-hand causal relationship between hIAPP fibrillation and the onset of type-II diabetes, is the observation
that wild-type mice do not develop the disease while  transgenic mice that express hIAPP can 
fall ill with  the disease. 
It has also been observed that a direct contact between the hIAPP amyloid fibrils and
the surfaces of pancreatic islet $\beta$-cells has a toxic effect on the latter.  

There is a real possibility that by understanding
the structural landscape of hIAPP,  in particular how the amyloidosis  
transition takes place, we could gain a major step towards the identification of therapeutic targets and the
development of strategies to combat a potentially deadly disease,
that presently plagues around 5 per cent of  the world's adult population. Indeed,
type-II diabetes is arguably among the most devastating diseases to curse mankind. 
Its annual economic cost has been globally estimated to be in excess of 425 billion 
Euro's, and the number of sufferers is estimated to almost double during the next 20 years.

The structure of aggregated hIAPP fibrils have been studied very extensively \cite{Westermark-2011,Pillay-2013}. 
The fibrils consist of an 
ordered parallel arrangement of monomers, with a zipper-like packing. Apparently the fibril 
formation proceeds by nucleation, with one monomeric hIAPP molecule first assuming a 
hairpin-like structure. This is followed by a piling-up of several monomers, which eventually 
leads to the buildup of amyloid fibrils as the hallmark of the disease in progression. But the 
structure of a full-length monomeric hIAPP, its potentially disease causing dynamical 
conformational state in our pancreatic cells, and why it occasionally may form the 
disease causing hairpin-like structure, all these remain unknown. Moreover, despite the highly 
ordered nature of amyloid fibril aggregates, only very recently have experimental advances 
made it possible to obtain high-resolution models. However, we still largely lack detailed atomic 
level knowledge, needed for drug development. 

The monomeric form of hIAPP is presumed to be an example of an intrinsically disordered protein. 
When biologically active and healthy, it is in an 
unsettled and highly dynamic state. As such it lacks an ordered three dimensional folded 
state that could be studied by conventional x-ray crystallography approaches. Several 
experimental methods which are based e.g. on solution and solid-state NMR and other 
techniques, are currently under development to try and characterise the conformation of 
monomeric hIAPP. But the existing techniques do not yet permit a direct examination of 
the atomic level structure. Detergents such as Sodium dodecyl sulfate (SDS) 
micelles are commonly introduced 
as stabilising agents. 

The detailed atomic level information could in principle be extracted 
by theoretical means with molecular dynamics simulations. However, with explicit water 
presently available computer power can at best cover a dynamical in vitro/in vivo trajectory 
up to around a microsecond per a day {\it in silico}. At the same time amyloid aggregation 
takes hours, even days. Thus the present all-atom computational investigations are
largely dependent on our ability to determine an initial conformation for the simulations. 
Otherwise,  one might only end up simulating the initial condition.

%
%

Due to its  intrinsically disordered character,   the structural data of hIAPP in isolation remains sparse
in PDB. The only presently available  PDB data of hIAPP  consists of two NMR structures and 
one crystallographic  
structure. Both NMR structures 
describe hIAPP  in a complex with SDS micelles, the PDB access codes are  2L86
and 2KB8.  The  sole available crystallographic structure describes 
hIAPP that has been fused with a maltose-binding protein,
the PDB access code is 3G7V. 
We note that these three  structures  are all very different from each other.

The NMR structure 2L86 has been measured at pH of 7.4 {\it i.e.} around
the pH value in the extracellular domain where the hIAPP amyloid 
deposits appear.
On the other hand, the NMR structure 2KB8 has been measured with pH value 
4.6 which is closer to the pH value around 5.5  inside the $\beta$-cell granules of 
pancreas.  We point out that even 
though hIAPP amyloidosis is apparently an
extracellular process,  some  evidence  suggests that the aggregation might have an intracellular origin.
Thus, a thorough investigation of the r\^ole of hIAPP in the onset of type-II diabetes,
should account {\it both} for the extracellular and the
intracellular structural properties of the peptide. In addition, a detailed analysis how hIAPP interacts 
with cell membranes is needed; we note that to some extent, micelles might mimic 
membrane effects.

We conclude, that it has been pointed out, that  hIAPP affects also several other organs
besides pancreas. For example,  hIAPP is known to have binding sites in the brain, where it 
apparently has a  regulatory effect on gastric emptying.  A delayed gastric emptying is commonly diagnosed
in patients with diabetes. But gastroparesis is also a component in a number of other
disorders.  Certainly, the ability of hIAPP to cross the blood-brain barrier and 
affect the central nervous system, relates to its structure. Amyloid fibers can hardly cross the barrier. 
Thus, besides apparently contributing directly to type-II 
diabetes, aggregation should also have a wider  influence on the regulatory activity of hIAPP.

%
%
%
%
%
%
%
%
%
%
%
%
%
%
%

\section{hIAPP as a three-soliton}

In this Section we shall investigate in detail the physical properties of a 28 segment monomer of hIAPP;
we propose that the reader gets access to  the  entry  2L86 in PDB.
The segment we are interested in, consists of the residues 9-36 where several studies have either observed or 
predicted that the amyloid fibril formation starts \cite{Westermark-2011,Pillay-2013}. 
The physical properties of the short N-terminal 
segment that comprises the residues 1-8 is not addressed here. The structure of this segment 
is more involved, due to the disulfide bond that connects the cysteines which are 
located at the residues 2 and 7.  Moreover, it remains to be understood what is the r\^ole, 
if any, of the residues 1-8 in hIAPP aggregation. These residues appear to have a tendency 
towards forming long and stable non-$\beta$-sheet fibers in solution, under the same 
conditions in which hIAPP aggregates into amyloid fibers.

We use the NMR structure 2L86 in PDB as a decoy to train the energy function. We 
construct a multi-soliton configuration as an extremum of the energy function (\ref{E1old}), 
that accurately describes 2L86. Since 2L86 is a composite of hIAPP with SDS micelles, 
we propose the following biological set-up: We consider the structural evolution of an 
isolated hIAPP in the extracellular domain, in a scenario where the polypeptide is initially 
in a direct but residual interaction with the cell membrane. The effect of an initial cell membrane 
interaction is modelled by the effect of SDS micelles in 2L86. Following the construction of 
the multi-soliton configuration, we study the presumed disordered structural landscape of an 
isolated hIAPP; we try and model hIAPP as it enters the extracellular domain. For this 
we subject the multi-soliton to a series of heating and cooling simulations as in the case of myoglobin, 
using the Glauber dynamics. During the heating, we increase the temperature until we detect 
a structural change in the multi-soliton, so that the configuration behaves like a random walker. 
We fully thermalise the configuration at the random walk phase. We then reduce the 
ambient temperature, to cool down the configuration to very low temperature values until 
it freezes into a conformation where no thermal motion prevails. Since an isolated 
hIAPP is intrinsically disordered, instead of a single native fold as in the case of myoglobin 
we expect that the low temperature limit produces a scattered statistical distribution 
of structurally disparate but energetically comparable ensembles of conformational substates
(\ref{HF}).  Moreover, the individual different substates should be separated from each 
other by relatively low energy barriers. The unstructured, disordered character of hIAPP
is then a consequence of a motion around this landscape: It swings and sways back and forth, 
quite freely, over the low energy barriers that separate the various energetically 
degenerate but structurally disparate conformations. We shall find that 
in the case of hIAPP, the heating and cooling procedure which in the case of myoglobin yields
a single low energy state, now produces exactly this kind of structurally 
scattered ensembles of conformations.

In Table \ref{tablehiapp} we have the parameter values,  that we find by  training the energy function (\ref{E1old}) 
to describe 2L86. These parameters values are taken from  \cite{He-2014}.
\begin{table}[tbh]
\begin{center}
\caption{Parameter values for the three-soliton 
configuration that describes 2L86; the soliton 1 cover the PDB segment THR 9 -- ASN 21, the  soliton 2 covers the segment ASN 22 -- ALA 25 and the third soliton covers the segment ILE 26 -- THR 36.  The value of  $a$ is fixed to
$a=-10^{-7}$.}
\vspace{10mm}
\begin{tabular}{|c|ccccccc|}
\hline 
soliton & $q_1$ & $q_2$ & $m_1$ & $m_2$ & $d/a$ & $c/a$ & $b/a$ \\
\hline
 1 & 9.454 & 4.453 & 1.521 & 1.606 &-8.164$\cdot 10^{-2}$ & 
-1.402$\cdot 10^{-3}$ & -2.568 \\
\hline
2 & 2.927 & 2.441 & 1.667 & 1.534 & -4.894$\cdot 10^{-1}$ &  -1.067$\cdot 10^{-3}$ &-19.48 \\
\hline
3 & 1.119 & 8.086 & 1.522 & 1.514 & -3.578$\cdot 10^{-2}$ & -5.907$\cdot 10^{-3}$ & - 1.908
\\
\hline
\end{tabular}
\end{center}
\label{tablehiapp}
\end{table}
%
%
%
%
%
%
%
%

%
%
%
%
%
%
%
%
%
%
%
%
%
%
%
%
%
%
%
{
\footnotesize
\begin{figure}[htb]         
\begin{center}            
  \resizebox{7cm}{!}{\includegraphics[]{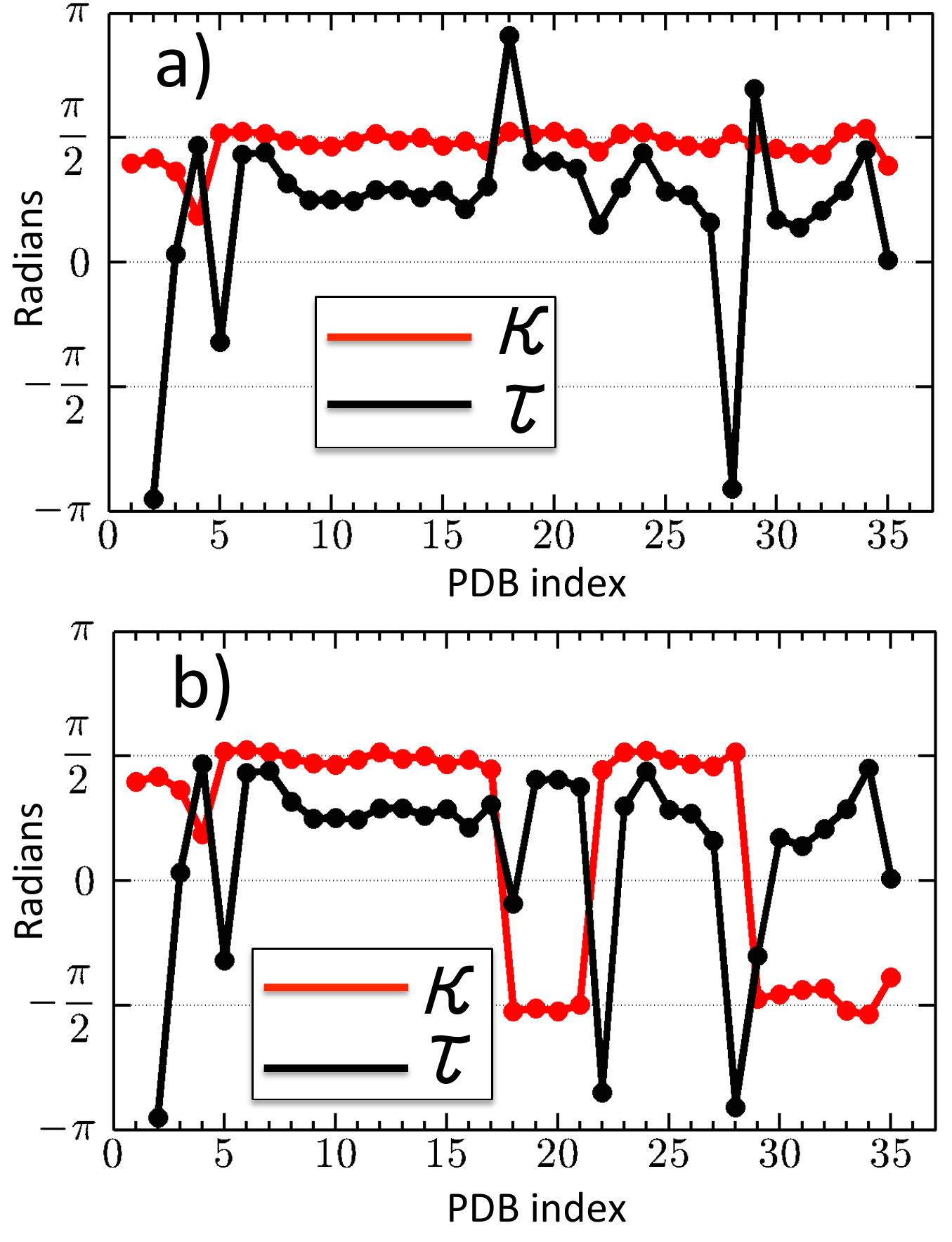}}
\end{center}
\caption {a) The spectrum of the bond and torsion angles of 2L86 (first entry) with the convention that bond angle takes values in $\kappa \in [0,\pi)$. b) The spectrum of the bond and torsion angles that identifies the soliton structures.}   
\label{fig-35}    
\end{figure}
}

%
%
%
%
%
%
%
In Figure \ref{fig-35} a) we show the spectrum of bond and torsion angles for the first  NMR structure of 
2L86, with the convention that the bond angle takes values between $\kappa \in [0,\pi]$. 
In Figure \ref{fig-35} b) we have introduced the $\mathbb Z_2$ symmetry (\ref{dsgau}) to disclose 
three individual solitons along the backbone. The first soliton from the N-terminal
is centered at the site 17. The third soliton is centered at
the site 27. Both of these two solitons correspond to clearly visible  loops  in the three dimensional 
structure,  in the PDB entry 2L86 (see PDB).
The second soliton, centered at site 23, is much less palpable in the 
three dimensional NMR structure. This soliton  appears  more like a bend in an $\alpha$-helical 
structure,  extending from the first soliton to the third soliton.
The $\mathbb Z_2$ transformed ($\kappa,\tau$) profile shown in Figure \ref{fig-35} b)
is the background that we have used in training the energy function (\ref{E1old}).

In Figure \ref{fig-36} we compare the bond and torsion angle spectrum of our three-soliton solution with 
the first NMR structure of 2L86;  the solution  is  obtained  using the program {\it ProPro} 
in (\ref{propro}).
%
%
%
%
%
%
%
%
%
%
%
%
%
%
%
%
%
%
%
%
%
%
%
%
%
{
\footnotesize
\begin{figure}[htb]         
\begin{center}            
  \resizebox{7cm}{!}{\includegraphics[]{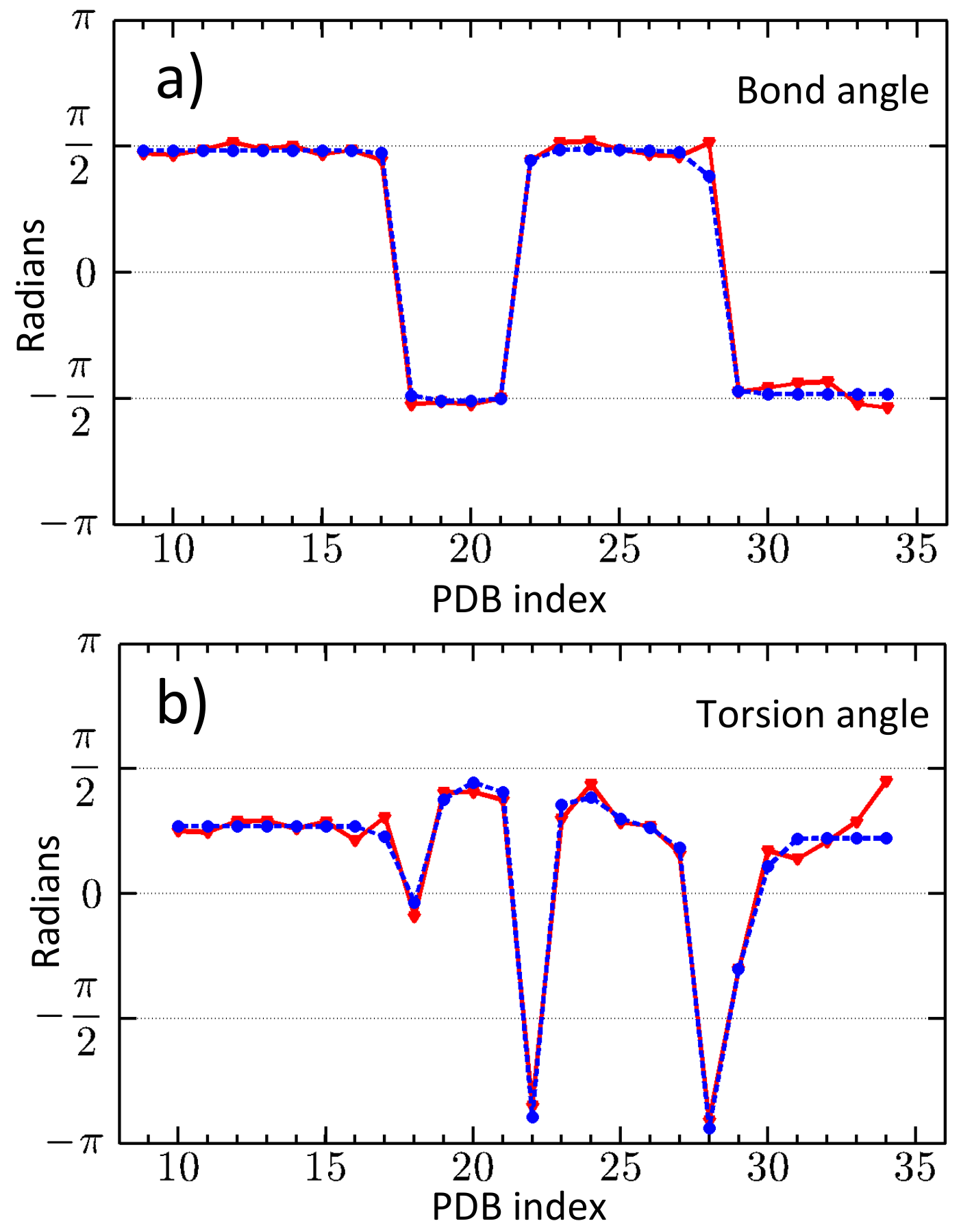}}
\end{center}
\caption {Top: Comparison of the three-soliton bond angle (blue) with the experimental 2L86 bond angle spectrum (red). Bottom: Comparison of the three-soliton 
torsion angle (blue) with the experimental 2L86 torsion angle spectrum (red).}   
\label{fig-36}    
\end{figure}
}
%
%
%
%
%
%
%
The quality of our three-soliton solution is  clearly very good, at the level of the bond and torsion angles.

Figure \ref{fig-37} shows our three-soliton solution, interlaced with the first NMR structure of 2L86.
%
%
%
%
%
%
%
%
%
%
%
%
%
%
%
%
%
%
%
{
\footnotesize
\begin{figure}[htb]         
\begin{center}            
  \resizebox{7.cm}{!}{\includegraphics[]{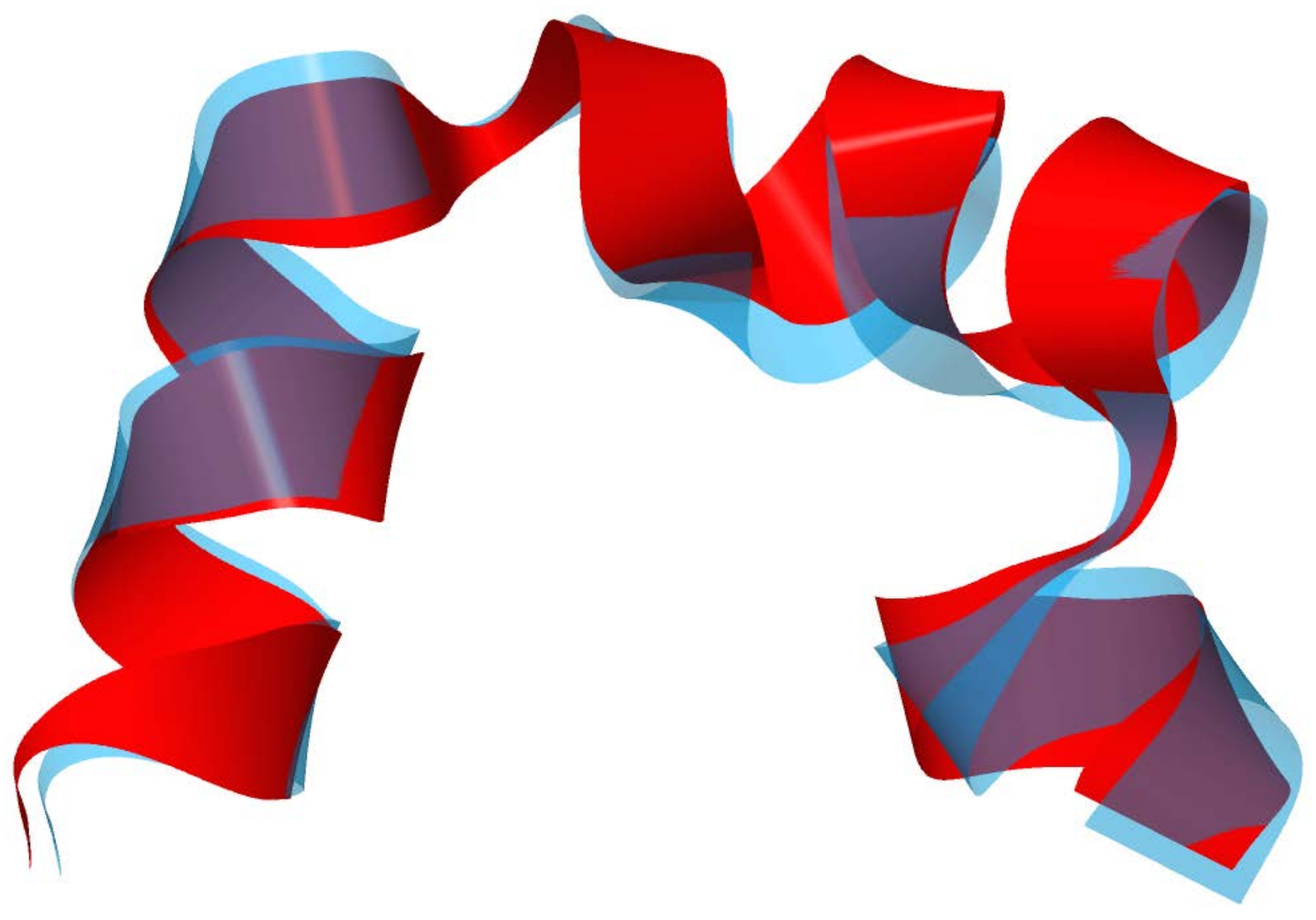}}
\end{center}
\caption {The three-soliton solution (blue) interlaced with the 2L86 experimental structure (red).}   
\label{fig-37}    
\end{figure}
}
%
%
%
%
%
%
The RMSD distance between the experimental 
structure and the three-soliton configuration is 1.17 \AA. This is somewhat large, when compared
to the multi-soliton structures that we have found previously. 
But the resolution of the present experimental NMR structure  is not that good,  
and this is reflected by the somewhat lower
quality of the three-soliton solution, in comparison to the case of high resolution crystallographic structures.

%
%
%
%
%
%
{
\footnotesize
\begin{figure}[htb]         
\begin{center}            
  \resizebox{7.2cm}{!}{\includegraphics[]{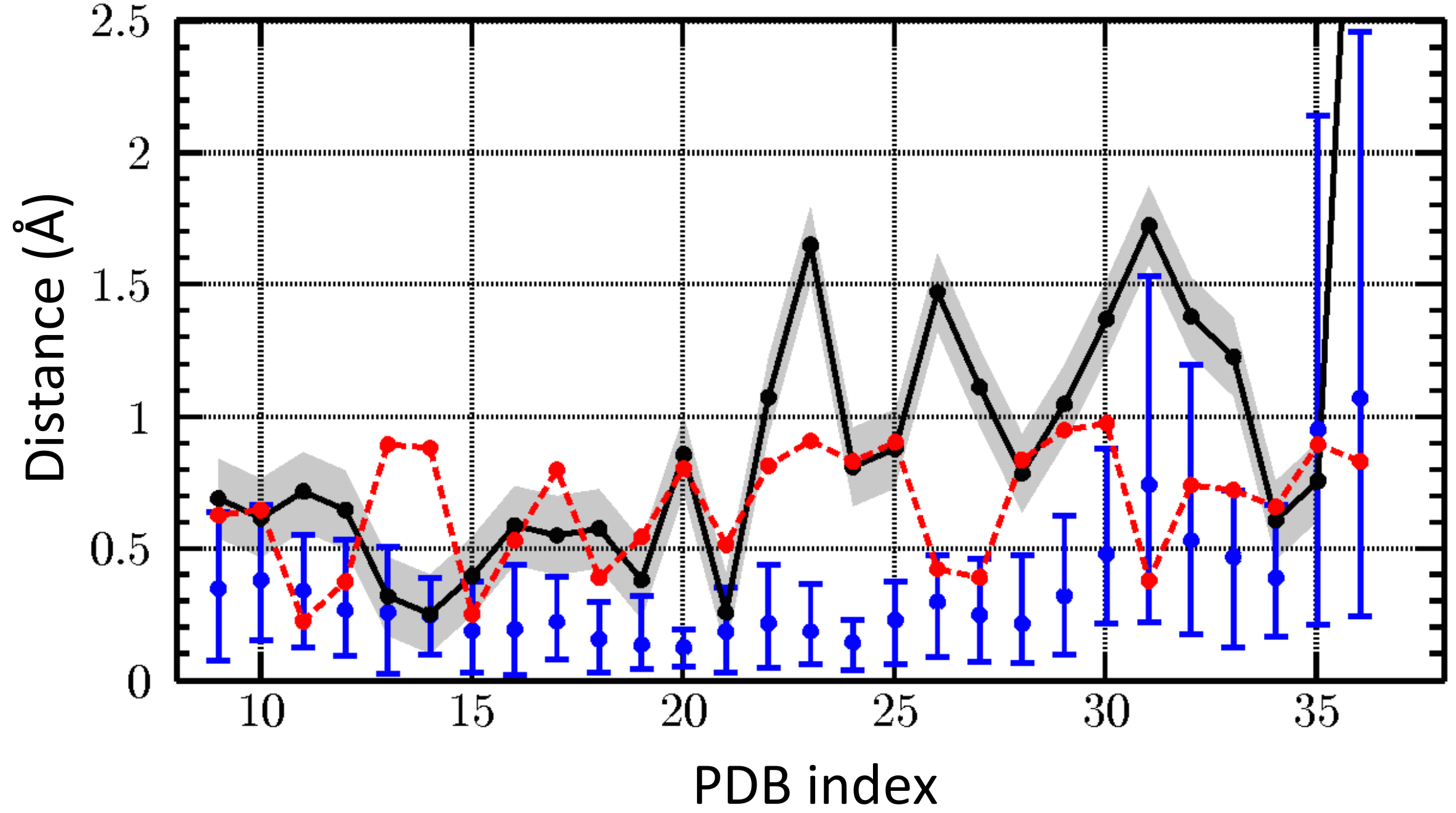}}
\end{center}
\caption {The black line denotes the C$\alpha$ atom distance between the 3-soliton configuration and the model 1 NMR 
configuration 
2L86; the grey region is an estimated 0.15 \AA~ zero-point fluctuation distance from the three-soliton 
configuration. The red line 
denotes the B-factor Debye-Waller fluctuation distance from model 1 of 2L86. The blue-colored points denote the average 
C$\alpha$ distance between the model 1 NMR structure from 
the average of the remaining 19 models on 2L86; the error-bars 
denote the maximal and minimal C$\alpha$ distances.}   
\label{fig-38}    
\end{figure}
}
%
%
%
%
Figure \ref{fig-38} compares the residue-wise C$\alpha$ distances, between the 20 different NMR structures in the
PDB entry 2L86, and our three-soliton solution. For those residues that precede 
the bend-like second soliton which is  centred at site 23, 
the distance between the experimental structures and the numerically constructed
solution is relatively small.  We observe a quantitative change
in the precision of the three-soliton solution, that takes place after site 23.
The distance between the experimental structures 
and the three-soliton solution clearly increases after this residue. 
It could be that this change is due to the SDS micelles,  used in the experimental
set-up to stabilise hIAPP/2L86: SDS is widely used as a 
detergent, to enable NMR structure determination in the case
of proteins with high hydrophobicity.  The mechanism 
of SDS-protein interaction is apparently not yet fully understood. But it is known that the hydrophobic 
tails of SDS molecules interact in particular with the hydrophobic core of a protein. 
These interactions are known to disrupt the native structure to the effect, that
the protein displays an increase in its $\alpha$-helical posture; these additional 
$\alpha$-helical structures tend to be surrounded by SDS micelles.

The residue site 23 of hIAPP is the highly hydrophobic phenylalanine. It is  followed by the 
very flexible glycine at site 24. Thus,  the apparently abnormal
bend which is located  at the site 23 and affects the quality of our three-soliton configuration, could be
due to an interaction between the phenylalanine and the surrounding 
SDS micelles. A high sensitivity of the hIAPP conformation  to the phenylalanine at 
site 23 is well documented.

An analysis of 2L86 structure using {\it Molprobity}  (\ref{molprobity})  suggests  
a propensity towards poor rotamers between the sites 23-36, {\it i.e.}
the region where the quality of our three-soliton solution decreases.

A comparison with the statistically determined 
radius of gyration relation (\ref{Rg})
reveals that for 2L86 the value of $R_g \approx 9.2$ (over residues 
$N=9,...,36$). This is somewhat high.
According to (\ref{Rg}), we expect a value close to $R_g \approx 7.9$ when we set $N=28$. 
The structure of 2L86 should be more compact.

We conclude that most likely the SDS-hIAPP interaction has deformed a loop which, in the absence of micelles,
should be located in the vicinity of the residue number 23. Probably,  the interaction with micelles has converted
this loop into a structure resembling a bend in an $\alpha$-helix. This interaction between hIAPP and SDS
interferes with our construction of the three-soliton configuration, adversely affecting its precision. 

%
%

\section{Heating and cooling  hIAPP}

Following our myoglobin analysis, we proceed to investigate the properties of the three-soliton
model of 2L86 under repeated heating and cooling, using the Glauber algorithm.
%
%
%
%
%
%
%
%
%
{
\footnotesize
\begin{figure}[htb]         
\begin{center}            
  \resizebox{10.2cm}{!}{\includegraphics[]{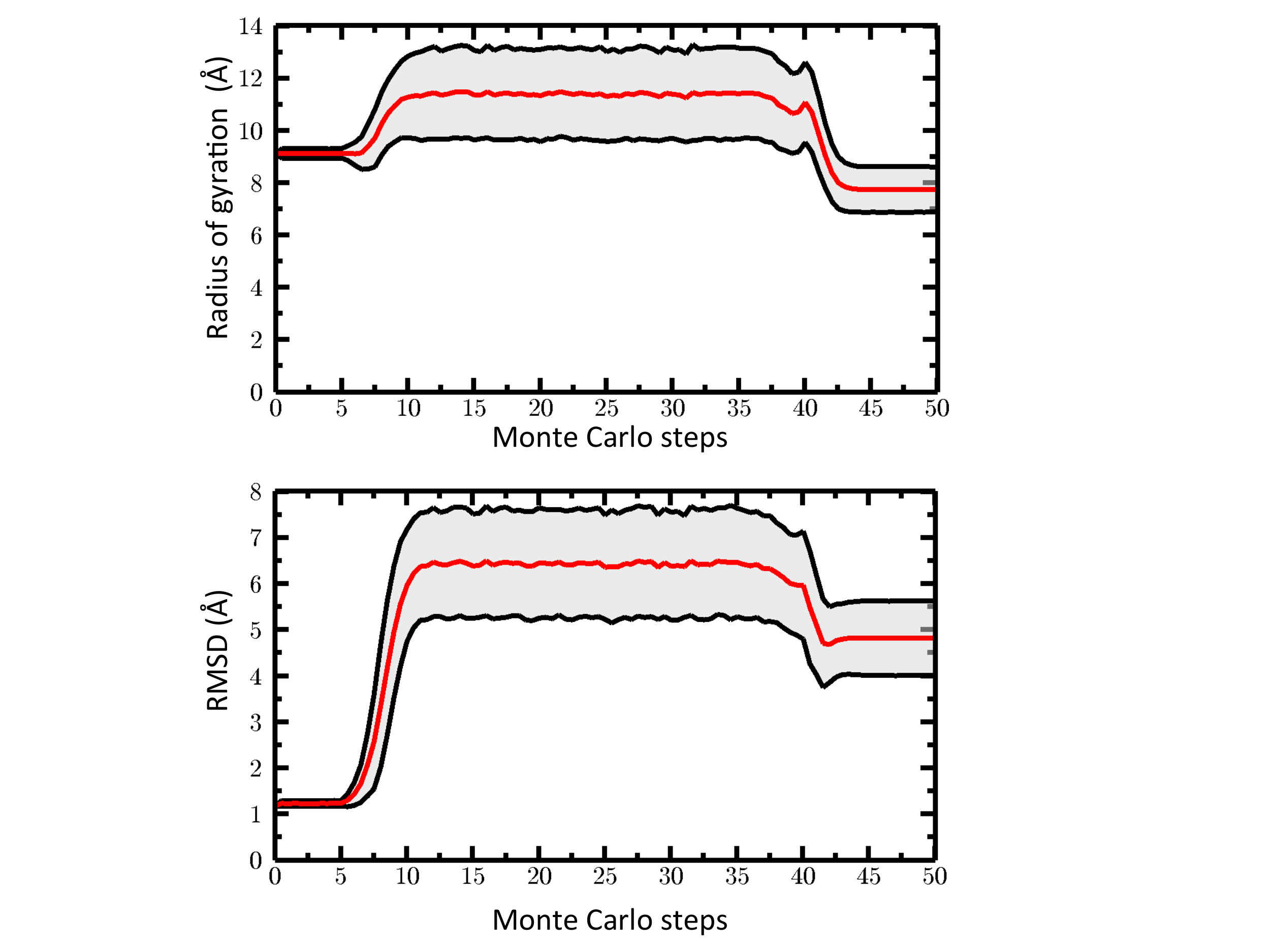}}
\end{center}
\caption {The top figure shows how the radius of gyration of the three-soliton configuration evolves during heating and cooling  
cycle. The bottom figure shows the same for the RMSD distance from the initial configuration. The 
red line is the average value over all configurations, and the grey zone marks the extent of
one standard deviation from the average value. The Monte Carlo steps are displayed in multiplets of  $10^6$.}   
\label{fig-39}    
\end{figure}
}
%
%
%
%
%
%
%

The Figures \ref{fig-39} describe the evolution of the three-soliton configuration during repeated heating and cooling. 
The Figure \ref{fig-39} (top) shows the evolution of the radius of gyration, and the Figure \ref{fig-39} 
(bottom) shows the RMSD distance 
to the PDB structure 2L86. Both the average value and the one standard deviation from the average value are shown.
During the cooling period we observe only one  transition, in both the radius of gyration and the RMSD. 
Thus, based on our previous experience, we are confident that at high temperatures we are in 
the random walk regime. 
The profile of each curve in Figure \ref{fig-39} also shows that the structures are
fully thermalised, both in the high temperature and in the low temperature regimes. 

We observe that the average final value of the radius of gyration 
$R_g \approx 7.8$ is an excellent match
with the prediction  obtained from  (\ref{Rg}). 
In particular, the final configurations are quite different from the initial one:
The RMSD distance between
the initial configuration and the average final configuration is around  $4.8$ \AA.

%
%
%
%
%
%
{
\footnotesize
\begin{figure}[htb]         
\begin{center}            
  \resizebox{8cm}{!}{\includegraphics[]{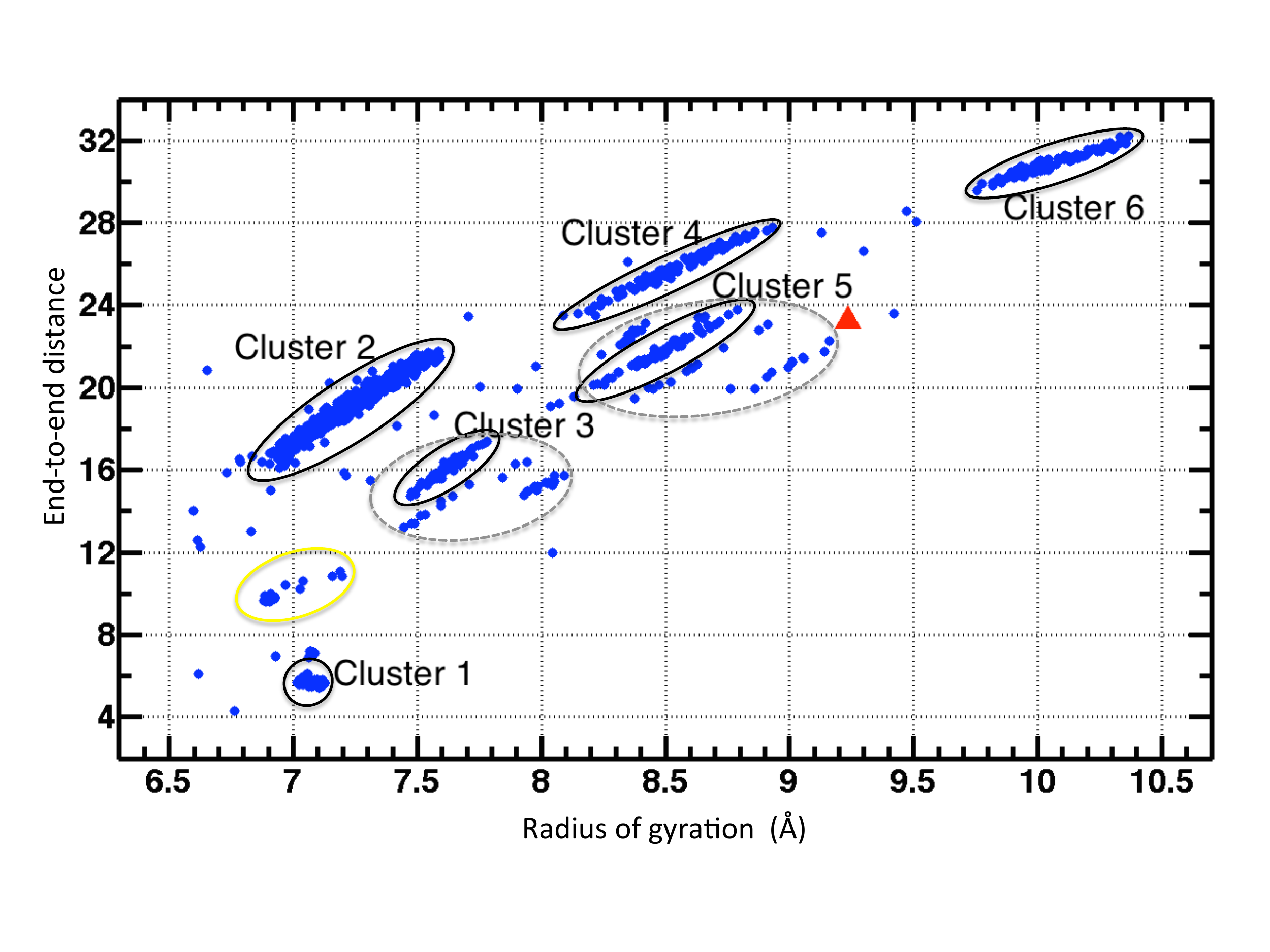}}
\end{center}
\caption {The distribution of all final configurations, in a run with 1.500 full heating and cooling cycles, 
classified in terms of the radius of gyration and end-to-end distance of the final configuration. Each blue 
dot represents a single final configuration. The six major clusters are encircled with a black ellipse; 
a wider grey ellipse around the clusters 3 and 5 includes some nearby scattered states. The red 
triangle identifies the initial configuration, the entry 1 in 2L86. Note also the presence of a 
cluster encircled with yellow between clusters 1 and 2.}   
\label{fig-40}    
\end{figure}
}
%
%
%
%
%
%
%
Figure \ref{fig-40} shows results for a representative simulation with 1.500 complete heating and cooling
cycles; an increase in the number of cycles does not  have a qualitative effect on the result.
The Figure shows the distribution of the final snapshot conformations,  grouped according to their  
radius of gyration versus  end-to-end distance. The final conformations form clusters, and we  identify the six 
major clusters that we observe in our simulations. 
By construction, the clusters correspond to 
local extrema of the energy function we have constructed to model 2L86: The average conformation of  
each cluster can be identified with a particular snapshot state $|\mathfrak s \!>$ in the expansion (\ref{HF}),  (\ref{thequi}). 
Five of the clusters, denoted 2-6 in the Figure, have an apparent spread. This implies that 
the energy has a flat conformational direction
around its extremum.  The clusters 3 and 5 are also somewhat more scattered than the clusters 2, 4 and 6. 
Finally,  the cluster number 1 is a localized one: This cluster corresponds to a sharply localized snapshot state 
$|\mathfrak s \!>$.  Note that the initial conformation, marked with 
red triangle in the Figure \ref{fig-40}, does not appear among the final configurations.  It is apparently 
an unstable extremum of the energy, stabilised by the micelles.

In Figure \ref{fig-41} we display the average conformations in each of the six clusters, interlaced with each other and
the initial 2L86 configuration. In this Figure, the first two C$\alpha$ atoms from the N-terminus are made to coincide. 
We have maximised the alignment of the subsequent C$\alpha$ atoms, to the extent it is possible.  
The Figure reveals the presence of substantial conformational difference between 
the clusters. The totality of the conformations shown in Figure \ref{fig-41} can be given an interpretation in terms of  
the dynamical hIAPP. It is a long time period average picture of the  Hartree state (\ref{HF}),  (\ref{thequi}) that
is a linear combination of the various snapshot conformations $|\mathfrak s_i \!>$ ($i=1,...,6$). 
%
%
%
%
%
%
{
\footnotesize
\begin{figure}[htb]         
\begin{center}            
  \resizebox{7cm}{!}{\includegraphics[]{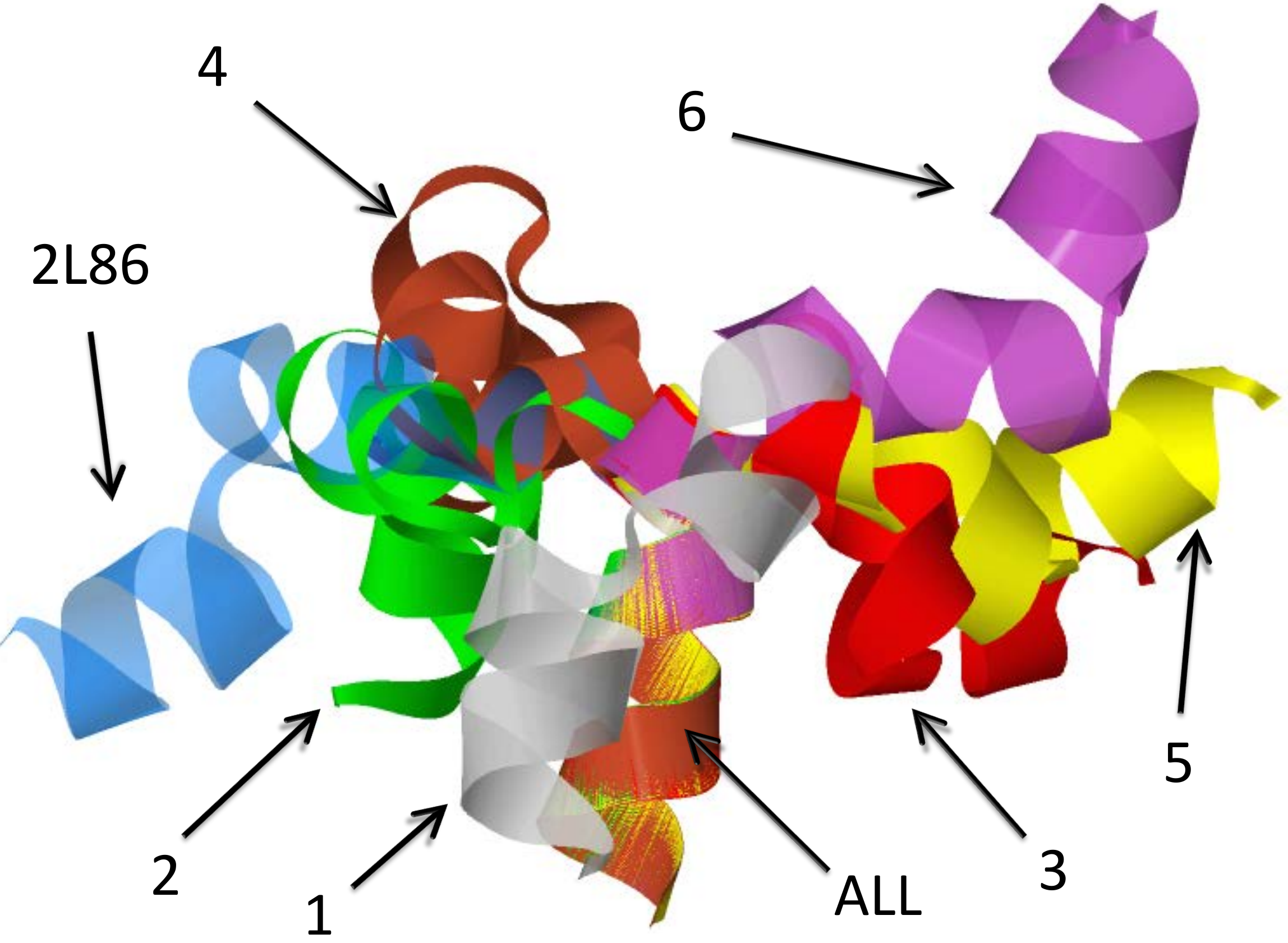}}
\end{center}
\caption {Superposition of all the six major clusters in Figure \ref{fig-40}, interlaced with each other 
and with the PDB entry 2L86.}   
\label{fig-41}    
\end{figure}
}
%
%
%
%
%

%
%
%
%
{
\footnotesize
\begin{figure}[htb]         
\begin{center}            
  \resizebox{8.2cm}{!}{\includegraphics[]{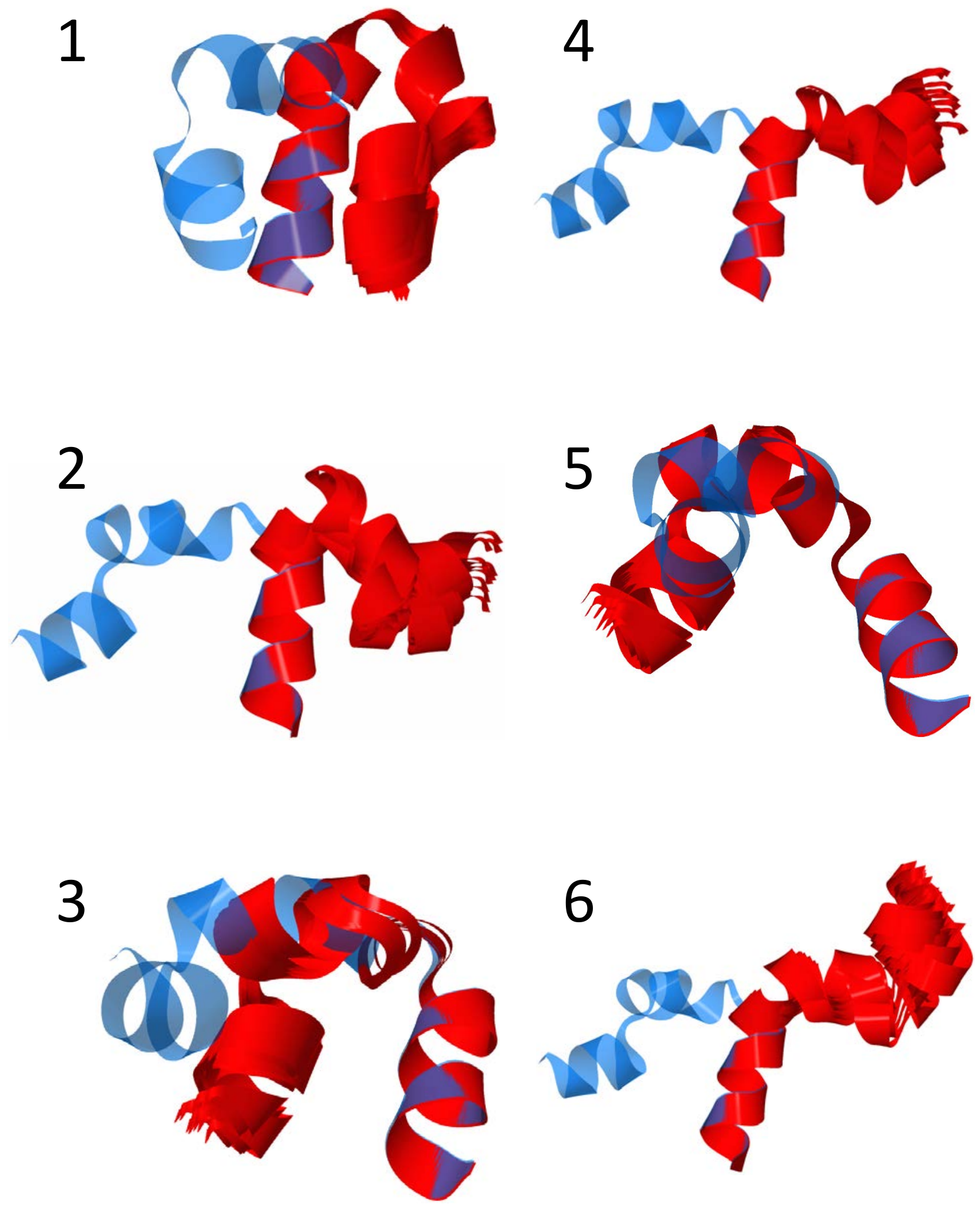}}
\end{center}
\caption {Superposition of ten representative conformations (red) in each of the six clusters, as marked, together with the PDB entry 2L86 (blue).}   
\label{fig-42}    
\end{figure}
}
%
%
%
In Figures \ref{fig-42}
we compare the individual clusters  with the initial 2L86 configuration (in blue). 
In each of these Figures,
we show ten representative entries in each of the clusters (in red), to visualise the extent of conformational fluctuations
within each cluster. We observe that the conformational spread within  each of the six clusters is not very large.

\vskip 0.2cm
In conclusion, we have found that the three-soliton configuration which models the C$\alpha$ 
backbone of the human islet amyloid polypeptide, is quite unsettled:  Its low temperature limit 
comes endowed with six different conformational clusters.  This 
is a marked contrast with the properties of a multi-soliton configuration which models a protein that is 
known to possess a unique folded native state, such as myoglobin.   
The low temperature clustering  of hIAPP is in full accord with the intrinsically disordered character of 
the protein: The different clusters can be viewed as instantaneous snapshot conformations, between which the dynamic 
hIAPP swings and sways in an apparently unsettled manner which is characteristic to an intrinsically disordered protein.  

Only the cluster number one appears different.  This cluster has a much more localised 
conformational distribution than the other five clusters, and the posture  comprises
of two anti-parallel helices. This proposes to us, that the cluster number one is a good candidate to trigger the formation
of hIAPP fibrils and amyloidosis which correlates with diabetes-II.

\vskip 0.2cm
\noindent
{\it My advice to all my students is, be careful with your life-style to keep your hIAPP folds under good control.
}
%
%
%
%
%
%
%
%
%
%
%
%
%
%
%
%

\chapter{Beyond C\Large{$\mathbf \alpha$}}
\label{alpha}

Thus far we have analysed the protein structure and dynamics in
terms of the C$\alpha$ atoms only,  we have argued that the virtual
C$\alpha$ backbone bond and torsion angles form a complete set of
local order parameters to describe the  protein backbone conformation.

The C$\alpha$ atoms  are  in a central r\^ole in x-ray crystallography, where
the experimental determination of protein structure often starts with a 
skeletonisation of the electron  density map: From
Figures \ref{fig-1}, \ref{fig-7} we observe that the C$\alpha$ atoms are located 
centrally. They form the vertices that connect the peptide planes. They
coincide with the branch points between the backbone and the side chain. Thus the C$\alpha$ atoms
are subject to  stringent stereochemical constraints. Accordingly, the first step in
experimental model building is the initial identification of the skeletal C$\alpha$ trace.

The central r\^ole of the C$\alpha$ atoms is widely exploited in structural classification schemes 
like CATH  and SCOP \cite{cath,scop}, in various homology modeling 
techniques \cite{Marti-2000,Schwede-2003,Zhang-2009}
in {\it de novo}  approaches \cite{Dill-2007}, and in the development of 
coarse grained energy functions for folding prediction \cite{Scheraga-2007,Liwo-2014}.
The so-called C$\alpha$-trace problem has been formalised, and it is 
the subject of extensive investigations  \cite{Holm-1991,Lovell-2003,Rotkiewicz-2008}. 
The aim  is to  construct 
an accurate main chain and/or all-atom model 
of the folded protein from the knowledge of the positions of the central C$\alpha$ atoms only.
Both knowledge-based approaches such as  {\it Maxsprout} \cite{maxsprout}  and {\it de novo} methods
like {\it Pulchra} \cite{pulchra} have been developed for this purpose.
In the case of the backbone atoms, various geometric algorithms can be utilised.
For the side chain atoms, most approaches rely either on a statistical
or on a conformer rotamer library in combination with steric constraints,  complemented by an analysis
which is based on diverse scoring functions. For the final fine-tuning of the model, 
all-atom molecular dynamics simulations can be utilised.

The Ramachandran map shown in Figure \ref{fig-10} 
is  used widely both in various analyses of  the protein structures, and as a tool in protein visualisation.  
It describes the statistical distribution of the two dihedral 
angles $\phi$ and $\psi$ that are adjacent to the C$\alpha$ carbons
along the protein backbone. In  the case of side chain atoms, visual analysis methods similar to
the Ramachandran map have been introduced. For example, there is 
the Janin map that can be used to compare observed side chain
dihedrals  in a given protein against their statistical distribution in a manner which is analogous to the 
Ramachandran map. Crystallographic refinement and validation programs like 
{\it Phenix} \cite{phenix},  {\it Refmac} \cite{refmac} and many others,
utilise the statistical data obtained 
from libraries such as Engh and Huber library  \cite{Engh-2001}, that
are built using small molecular 
structures which have been determined with a very high resolution. 
At the level of entire proteins, side chain restraints are commonly derived
from  analysis of high resolution PDB  crystallographic structures \cite{Dunbrack-2002}. 
Backbone independent rotamer libraries that make 
makes no reference to backbone conformation, and both secondary 
structure and backbone dependent rotamer libraries 
have been developed. According to  \cite{Dunbrack-2002} 
the information content in the secondary structure 
dependent libraries and the backbone independent libraries essentially 
coincide, and both  are often used during crystallographic
protein structure model building  and refinement. But for the prediction 
of side-chain conformations  for example 
in the case of  homology modelling and protein design, it is  often an 
advantage to use the more revealing 
backbone dependent rotamer libraries.

\section{"What-you-see-is-what-you-have"}
\label{sectphase}

We shall present a short introductory outline, how the C$\alpha$ Frenet frames can be utilised to
develop new generation visualisation techniques for protein structure analysis, refinement and
validation. Our outline is based on
\cite{Lundgren-2012b,Lundgren-2013b,Peng-2015}.  

Despite the availability of various 3D visualisation
tools such as the Java-based viewer on PDB website, thus far the visualisation of proteins has 
not yet taken {\it full} advantage of modern visualisation techniques. The commonly available 
3D viewers present the protein in the 'laboratory' frame and as such they 
provide mainly an {\it external} geometry based characterisation of the protein structure. 
On the other hand, the method that we describe
is based on internal, {\it co-moving}  framing of the protein backbone: To watch a roller-coaster is not the
same as taking the ride. 
As such our approach 
provides complementary visual information. Starting from the positions of the C$\alpha$ 
atoms,  we aim for a  3D {\it what-you-see-is-what-you-have} type visual map of the 
all-atom structure. 
Indeed, the visualization of a three dimensional discrete framed curve is an important and widely 
studied topic in computer graphics, from the association of ribbons and tubes to the determination 
of camera gaze directions along trajectories.  

In lieu of the backbone dihedral angles that appear as coordinates in the Ramachandran map and
correspond to a toroidal topology, we use the geometry of virtual two-spheres that surround 
each heavy atom. For this we employ the geometric interpretation of the virtual C$\alpha$ backbone
bond and torsion angles in terms of latitude and longitude on the surface of a sphere $\mathbb S^2$.
We shall outline how the approach works in the case of the backbone N and C atoms, and
the side chain C$\beta$ atoms. The approach can be easily extended to visually describe all the higher 
level side chain atoms on the surface of the sphere, level-by-level along 
the backbone and side chains \cite{Lundgren-2012b,Lundgren-2013b,Peng-2015}. 
The outcome is a 3D visual map that 
describes the backbone and side-chain atoms exactly in the 
manner how they are seen by an imaginary, geometrically determined and C$\alpha$ based
miniature observer who roller-coasts along the backbone: At each C$\alpha$ atom the
observer orients herself consistently according to the purely geometrically determined 
C$\alpha$ based discrete Frenet frames.  Thus the visualisation of all the other atoms 
depends only on the C$\alpha$ geometry, there is no reference to the 
other atoms in the initialisation of  the construction. The other atoms - including subsequent 
C$\alpha$ atoms along the backbone chain - are all mapped 
on the surface of a sphere that surrounds the observer, as if these 
atoms were stars in the sky; 
the construction proceeds along the ensuing side chain, until the position of
all heavy atoms have been determined.
This provides a purely geometric and equitable, direct visual information on the statistically 
expected all-atom structure in a given protein, based entirely on the C$\alpha$ trace.  

We start with the bond and torsion angles  (\ref{bond}),   (\ref{tors}) and we choose each  
bond angle to  take values  $\kappa \in [0,\pi]$. We identify the bond angle 
with the latitude angle of a two-sphere which is centered at the C$\alpha$ carbon. 
We orient the sphere so that  the north-pole where $\kappa=0$ is in the direction of $\mathbf t$.  
The torsion angle $\tau\in [-\pi,\pi)$ is the longitudinal angle of the sphere. It is 
defined so that $\tau = 0$ on the great circle that passes both through the north pole 
and through the tip of the normal vector  $\mathbf n$. The longitude angle increases 
towards the counterclockwise direction around the  vector $\mathbf t$.
We find it also useful,  to introduce the stereographic projection 
of the sphere onto the plane.  The standard
stereographic projection  from the south-pole of the sphere to the plane with coordinates ($x,y$) 
is given by
\begin{equation}
x+iy \ \equiv \ r \, e^{i\tau} \ = \ \tan\left( \kappa/2 \right) \, e^{i\tau}
\label{stereo}
\end{equation}
This maps the north-pole where $\kappa=0$ to the origin ($x,y$)$=$($0,0$). The south-pole where 
$\kappa=\pi$ is sent to infinity; see Figure \ref{fig-43}
%
%
%
%
{
\footnotesize
\begin{figure}[htb]         
\begin{center}            
  \resizebox{8.2cm}{!}{\includegraphics[]{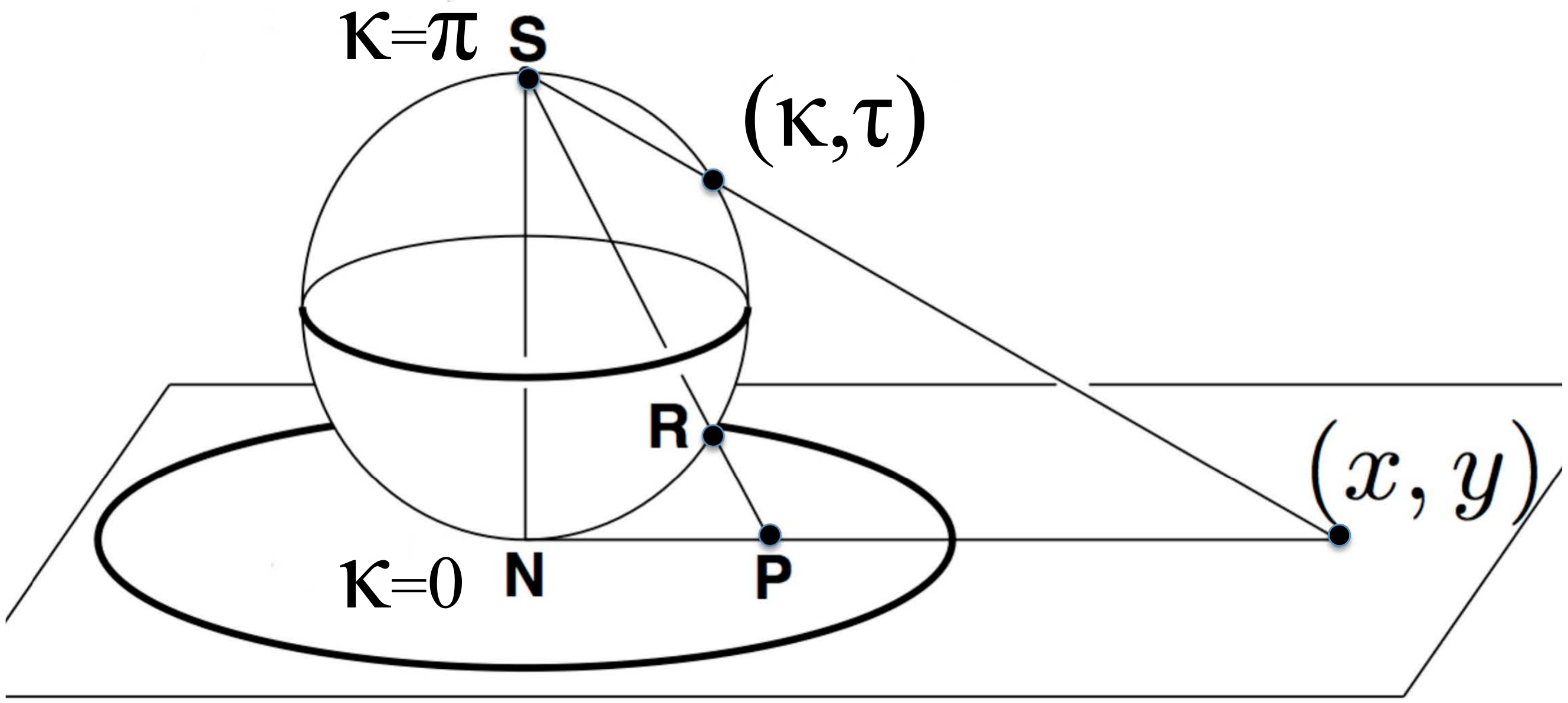}}
\end{center}
\caption {Stereographic projection of two-sphere $\mathbb S^2$ on the plane $\mathbb R^2$, from the southpole.}   
\label{fig-43}    
\end{figure}
}
%
%
%
If need be, the visual effects  of the projection 
can be enhanced by  sending
\[
\kappa \ \to \ f(\kappa) 
\]
where $f(\kappa)$ is a properly chosen function of the latitude angle $\kappa$.

\subsection{The C$\mathbf \alpha$ map}

We first explain how to visually describe the C$\alpha$ trace in terms of the C$\alpha$ 
Frenet frames (\ref{t})-(\ref{n}). Consider the virtual  miniature observer 
who roller-coasts the backbone by moving  between the C$\alpha$ atoms.
At the location of each C$\alpha$ the observer has an orientation that is determined
by the Frenet frames (\ref{t})-(\ref{n}). The base of the $i^{th}$ tangent vector {$\mathbf t_i$ is  
at the position $\mathbf r_{i}$.  The tip of $\mathbf t_i$  
is a point on the surface  of the sphere ($\kappa,\tau$) that surrounds the  observer and points 
towards the north-pole. The vectors $\mathbf n_i$ and $\mathbf b_i$ determine the orientation of the sphere.
These vectors define 
a frame on the normal plane to the backbone trajectory, as shown in Figure \ref{fig-21}. 
The observer maps the various atoms in the protein chain on the  surface of the surrounding 
two-sphere, as if the atoms were stars in the sky.

The map of the  C$\alpha$ backbone is constructed as follows.
The observer first translates the center of the  sphere from the location of the $i^{th}$ C$\alpha$,  
all the way to the location of the $(i+1)^{th}$ C$\alpha$ with {\it no rotation }
of the sphere with respect to the $i^{th}$ Frenet frames.  The observer then 
identifies the direction of $\mathbf t_{i+1}$, {\it i.e.} the 
direction towards the site $\mathbf r_{i+2}$ to which she proceeds from the next C$\alpha$ carbon,
as a point on the surface of the sphere. This determines  the corresponding coordinates ($\kappa_i, \tau_i$).  
After this, the observer re-defines her 
orientation so that it  matches the Frenet framing at the $(i+1)^{th}$ central carbon, and then proceeds
by repeating the construction, exactly in the same manner.
The ensuing map, over the entire backbone, gives an instruction to the observer at each
point $\mathbf r_i$,  how to turn at site $\mathbf r_{i+1}$, to 
reach  the  $(i+2)^{th}$ C$\alpha$ carbon at the point $\mathbf r_{i+2}$.

%
%
%
%
%
%
%
%
%
%
%
%
{
\footnotesize
\begin{figure}[htb]         
\begin{center}            
  \resizebox{8.2cm}{!}{\includegraphics[]{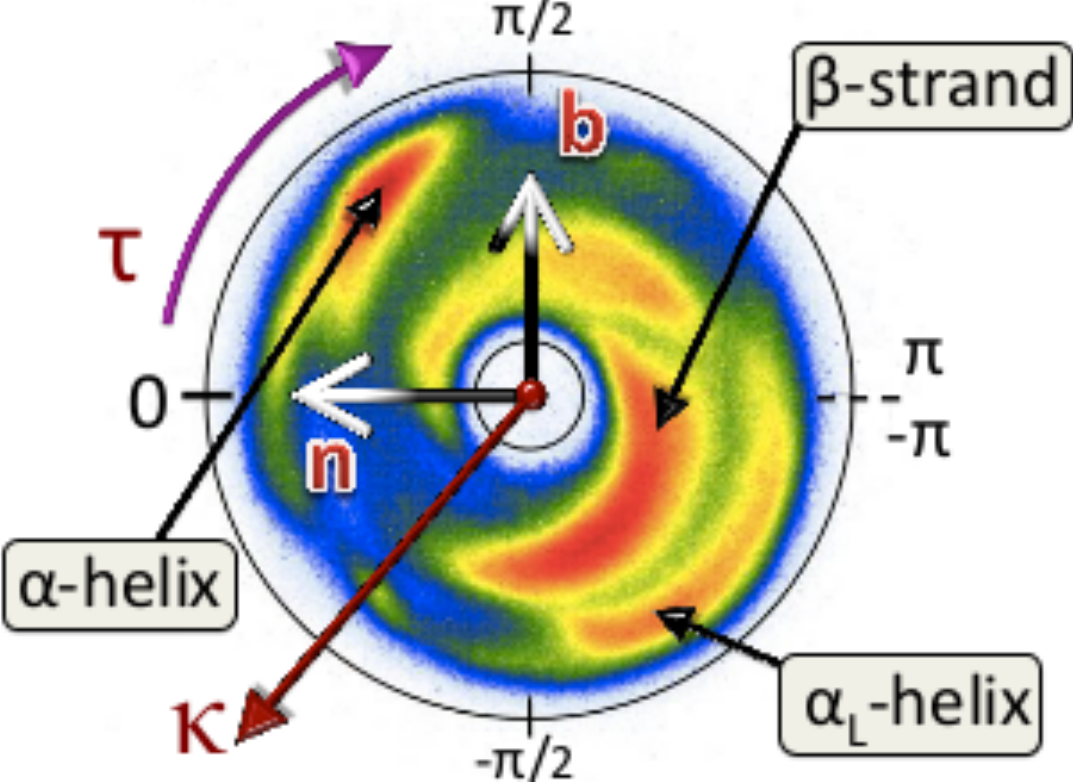}}
\end{center}
\caption {The stereographically projected Frenet frame map of backbone C$\alpha$ atoms, with major 
secondary structures identified. Also shown is the directions of the Frenet frame normal 
vector $\mathbf n$; the vector $\mathbf t$ corresponds to the red circle at the center, and 
it points away from the viewer. The map is constructed using all 
PDB structures that have been measured with better than 2.0 \AA~resolution.}   
\label{fig-44}    
\end{figure}
}
%
%
%
In figure \ref{fig-44}
we show the  C$\alpha$ Frenet frame backbone map. It describes the statistical distribution that we obtain
when we plot all PDB structures which have been 
measured with better than 2.0 \AA~
resolution, and using the stereographic projection (\ref{stereo}).
For our observer, who always fixes her gaze position towards the north-pole 
of the surrounding two-sphere at each C$\alpha$  {\it i.e.}  towards the red dot at the center of the annulus,  
the color intensity in this map 
reveals the probability of the direction at position $\mathbf r_i$, where 
the observer will turns at the next C$\alpha$ carbon, when she moves 
from $\mathbf r_{i+1}$ to $\mathbf r_{i+2}$. In this way, the map is in a direct visual correspondence with
the way how the Frenet frame observer perceives the backbone geometry.  
Note how the probability distribution concentrates within an 
annulus, roughly between the latitude angle values $\kappa \sim1$ and $\kappa \sim 3/2$.
The exterior of the annulus is a sterically excluded region while  the entire interior is in principle sterically 
allowed but very rarely occupied in the case of folded proteins.
In the figure we identify the four major secondary structure regions, according to the PDB classification;
the  $\alpha$-helices, $\beta$-strands, left-handed $\alpha$-helices
and loops.  

\subsection{The  backbone C, N and side chain C$\mathbf \beta$ maps}

Consider our imaginary miniature observer, 
located at the position of a C$\alpha$ atom and oriented according to the discrete
Frenet frames. She  proceeds to observe and record
the backbone heavy atoms N, C and the side-chain C$\beta$ atoms that are covalently bonded to C$\alpha$.
These atoms form the covalently bonded heavy-atom corners of the
C$\alpha$ centered $sp3$-hybridised   tetrahedron.
In figures \ref{fig-45} a)-d)
we show the ensuing density distributions on the surface of the C$\alpha$ centered sphere, the way how they
are seen by the miniature observer.
These figures are constructed from
all  the PDB entries that have been measured using diffraction data with better than 1.0 \AA~resolution.
%
%
%
%
%
%
%
%
{
\footnotesize
\begin{figure}[htb]         
\begin{center}            
  \resizebox{10.2cm}{!}{\includegraphics[]{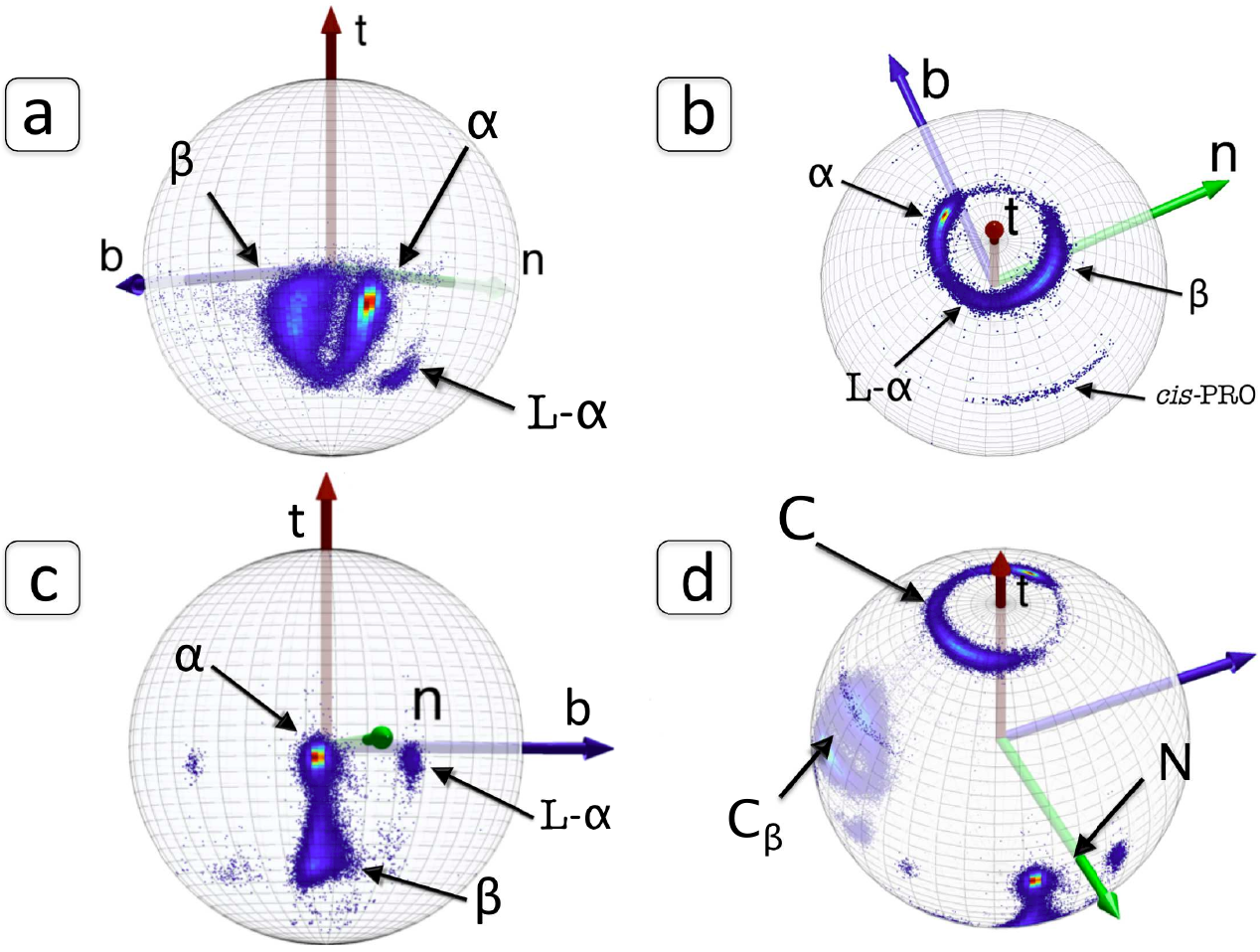}}
\end{center}
\caption {a) Distribution of C$\beta$ atoms in the C$\alpha$ centered Frenet frames in 
PDB structures measured with better than 1.0 \AA~resolution. The three major domains $\alpha$-helices, 
$\beta$-strands and left-handed $\alpha$-helices have been marked. b) Same as a) but for backbone C 
atoms. Note that {\it cis}-prolines are also clearly identifiable. c) Same as a) but for backbone 
N atoms. d) A combination of distributions in a)-c).}   
\label{fig-45}    
\end{figure}
}
%
%
%

As visible in Figures \ref{fig-45} 
in the C$\alpha$ entered Frenet frames the C$\beta$, C and N atoms each oscillate in a manner that depends 
on the local secondary structure. Note that both in the case of C$\beta$ and N, the left-handed $\alpha$ 
region (L-$\alpha$) is distinctly detached from the rest. But in the case of C the L-$\alpha$ 
region is connected with the other regions. On the other hand, in the case of 
N the {\it cis}-prolines form a clearly detached and localized
region, which is not similarly visible in the case of C and C$\beta$.

We now consider the three bond angles
\begin{equation}
\begin{matrix}
\vartheta_{\mathrm N\mathrm C}  & \simeq & \mathrm N- \mathrm C\alpha-\mathrm C \\
\vartheta_{\mathrm N\beta}  & \simeq & \mathrm N-\mathrm C\alpha-\mathrm C\beta \\
\vartheta_{\beta \mathrm C}  & \simeq & \mathrm C\beta-\mathrm C\alpha-\mathrm C \\
\end{matrix}
\label{thetaNCC}
\end{equation}
The $\vartheta_{\mathrm N\mathrm C}$ angle relates to the backbone only, while the definition of the other 
two involves the side chain C$\beta$. In experimental protein structure validation these three 
angles are often presumed to have their ideal values. For example, the 
deviation of the C$\beta$ atom from its ideal position is among the validation criteria used in 
{\it Molprobity} (\ref{molprobity})  that uses it in identifying potential backbone distortions around C$\alpha$. 

In Figure \ref{fig-46} we show the distribution of the three tetrahedral bond angles 
in our 1.0 \AA~ PDB data set. We find that in the case of the two side chain related
angles $ \vartheta_{\mathrm N\beta}$  and $\vartheta_{\beta \mathrm C}$  the 
distribution displays a single peak which is compatible with the ideal values; 
the isolated small peak in Figure \ref{fig-46} b) is due to {\it cis}-prolines. But in the case 
of the backbone specific angle $\vartheta_{\mathrm N\mathrm C} $ we find that this is not the case. The PDB 
data shows  a clear correlation between the $\vartheta_{\mathrm N\mathrm C} $
distribution and the backbone secondary structure.
We remind that $\vartheta_{\mathrm N \mathrm C} $ pertains to the relative orientation of the 
two peptide planes that are connected by
the C$\alpha$. 
%
%
%
%
%
%
%
%
{
\footnotesize
\begin{figure}[htb]         
\begin{center}            
  \resizebox{10.2cm}{!}{\includegraphics[]{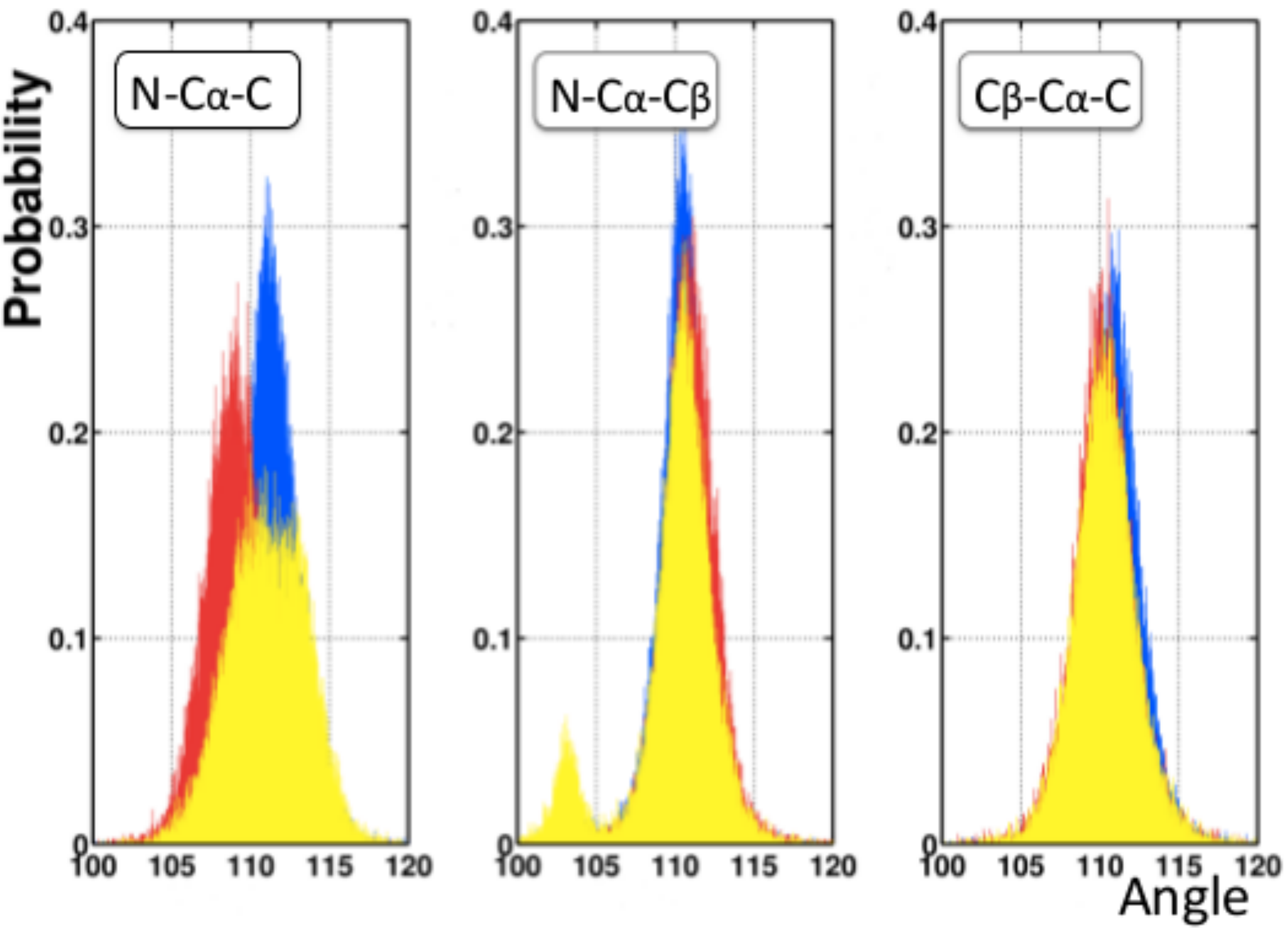}}
\end{center}
\caption {Distribution of the three angles (\ref{thetaNCC})
according to secondary structures. Blue are $\alpha$-helices, red are $\beta$-strands and yellow are loops; 
the small (yellow) peak in N-C$\alpha$-C$\beta$ with angle around 103$^o$ is due to prolines. 
}   
\label{fig-46}    
\end{figure}
}
%
%
%

The two Ramachandran angles ($\phi,\psi$) in Figure \ref{fig-10}  
are directly adjacent to the given C$\alpha$,  and each is specific to a {\it single} peptide plane;  see Figure \ref{fig-7}.
But this is not the case of $\vartheta_{\mathrm N \mathrm C} $  which connects two peptide planes. This angle
contributes to the bending of the backbone. Consequently a systematic secondary 
structure dependence  {\it should} be present  in this angle. 
Due to the very rigid structure of the C$\alpha$ centered $sp3$-hybridized   
covalent tetrahedron one then expects that the secondary structure dependence should also become visible
in the side chain specific $\vartheta_{\mathrm N\beta}$  and $\vartheta_{\beta \mathrm C}$. 
The lack of any observable secondary structure dependence in the 
experimental data of these two angles proposes, that existing validation methods distribute 
the refinement tension entirely to $\vartheta_{\mathrm N\mathrm C} $.

\begin{quote}
{\it Research project: Can you explain in detail, whysecondary structure dependence is absent
in PDB data of $\vartheta_{\mathrm N\beta}$  and $\vartheta_{\beta \mathrm C}$.}
\end{quote}

The construction we have presented, can be continued to visualise all the higher level side chain 
atoms in a protein, beyond C$\beta$ \cite{Lundgren-2012b,Lundgren-2013b,Peng-2015}. It turns 
out that these higher level atoms also display a highly organised, systematic pattern 
akin those shown in Figure \ref{fig-45}.
In particular, the underlying geometry of the C$\alpha$ backbone is clearly visible in these side chain atoms; 
there is a very strong coupling between the C$\alpha$ backbone geometry and the side chain
geometry. Accordingly it becomes possible, in principle, to highly accurately 
determine the all-atom structure of the protein from
the knowledge of the C$\alpha$ atom positions only. This enables one to extend our C$\alpha$ based
approach that builds on (\ref{E1old}),  to construct the full all-atom
structure of the entire protein: The C$\alpha$ positions can be computed from the energy function
(\ref{E1old}), and the remaining atoms are located by s statistical analysis.

\vskip 0.2cm
According to Figure \ref{fig-45} a) in the intrinsic, purely geometric
Frenet frame coordinate system the directions of the C$\beta$ atoms are only subject to small fluctuations. At
the level of C$\beta$, the side chains all point to essentially the same direction. Thus the crystallographic
protein structures are like a spin chain where the side chains are the spin variables, and the collapsed phase 
bears resemblance to  a ferromagnetic phase.

\begin{quote}
{\it Research project: See how far you get using spin chain analogy 
of folded proteins, with side chains as the spin variables.}
\end{quote}

\vskip 2.0cm

{
\footnotesize
\begin{figure}[htb]         
\begin{center}            
  \resizebox{8cm}{!}{\includegraphics[]{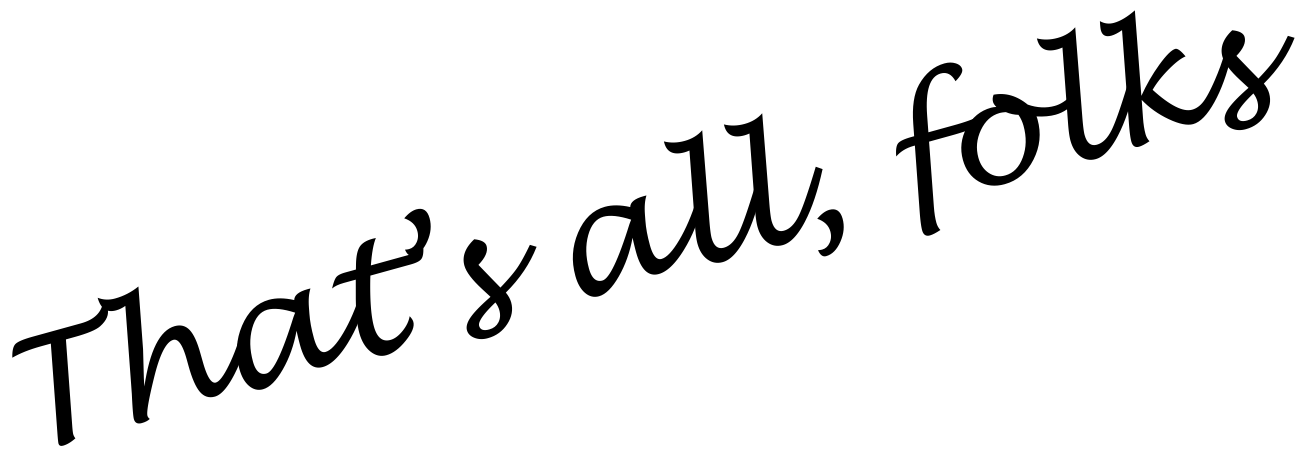}}
\end{center}
\end{figure}
}

%
%
%
%

%
%
%
%
%
%
%
%
%

%
{\footnotesize

}


\thebibliography{0}

\bibitem{Schrodinger-1944} E. Schr\"odinger,
"What is Life?" The Physical
Aspect of the Living Cell {\it Cambridge University Press}
(1948)

\bibitem{Anderson-1972} P. W. Anderson, 
{\it Science} {\bf 177} 393
(1972)



\bibitem{Dill-2007} K. Dill, S.B.  Ozkan, T.R.  Weikl, J.D.  Chodera, V.A.  Voelz, 
{\it Current Opinions in   Structural Biolology}  \textbf{17} 342 (2007)

\bibitem{Dill-2012} K.A. Dill and J.K. MacCallum, 
{\it Science} {\bf 338} 1042
(2012)


\bibitem{Chiti-2006} F. Chiti and C.Dobson, {\it Annual Reviews in Biochemistry} {\bf 75} 333
(2006)


\bibitem{davies-1996} J. Davies, {\it Nature}  {\bf 383} 219 (1996)

\bibitem{walsh-2000} C.T. Walsh, {\it Nature}  {\bf 406} 775 (2000) 

\bibitem{fischbach-2009} M.A. Fischbach, C.T. Walsh, {\it Science} {\bf 325} 1089
(2009)


\bibitem{Alberts-2014} B. Alberts, A. Johnson, J. Lewis, D. Morgan, M. Raff, K. Roberts, P. Walter, 
 Molecular Biology of the Cell 6$^{th}$ Edition {\it Garland Science} (2014)

\bibitem{Frappat-1998}     L. Frappat, P. Sorba, A. Sciarrino, 
{\it Physics Letters } {\bf B250} 214
(1998)

\bibitem{Read-2011} R.J. Read {\it et.al.} 
{\it Structure} {\bf 19}  1395
(2011)

\bibitem{Engh-2001} R.A.  Engh, R. Huber,  in {\it International Tables for Crystallography}, Vol. {\bf F}, 382;  edited by M. G. Rossmann and E. Arnold  
(Kluwer Academic Publishers, Dordrecht, 2001)


\bibitem{Joosten-2009} R. Joosten {\it et.al.}  {\it Journal of Applied Crystallography} {\bf 42}  
376
(2009)
 
 \bibitem{Nature-2009}  See News at {\it Nature} {\bf 459}  1038
 (2009)

\bibitem{Dauter-2014} Z. Dauter, A. Wlodawer, W. Minor, M. Jaskolski, B. Rupp, 
{\it Journal of
the International Union of Crystallography IUCrJ } {\bf  1} 179
(2014)  

\bibitem{Degennes-1979} P.G. De Gennes,  Scaling Concepts in Polymer 
Physics 
{\it Cornell  University Press, Ithaca} (1979)

\bibitem{Schafer-1999} L. Sch\"afer,   Excluded Volume Effects in Polymer So-
lutions, as Explained by the Renormalization Group
{\it Springer Verlag, Berlin} (1999)

\bibitem{Kadanoff-1966} L.P. Kadanoff, 
{\it Physics}  {\bf 2} 263 (1966)

\bibitem{Wilson-1971} K.G. Wilson, 
{\it Physical Review }  {\bf B4} 3174 (1971)

\bibitem{Wilson-1974} K.G. Wilson, J. Kogut, 
{\it Physics Reports} {\bf 12} 75-200 (1974)

\bibitem{Huggins-1941}  M.L. Huggins, 
{\it The Journal of Chemical Physics}
{\bf 9}  440 (1941)


\bibitem{Flory-1942} P.J. Flory, 
{\it The Journal of Chemical Physics}
{\bf  10} 51
(1942)


\bibitem{Li-1995} B. Li, N. Madras, A. Sokal, 
{\it Journal of Statistical Physics}   
 {\bf 80}  661
 (1995).
 


\bibitem{Degennes-1972} P.G. De Gennes, 
{\it Physics  Letters}  {\bf 38A} 339
(1972)


\bibitem{Leguillou-1980}  J.C. LeGuillou, J. Zinn-Justin, 
 {\it Physical  Review}  {\bf B21} 3976 (1980)

 \bibitem{Hinsen-2013} K. Hinsen, S. Hu, G.R. Kneller,  A.J. Niemi, 
{\it The Journal of Chemical Physics} {\bf 139}   124115 (2013) 

\bibitem{Marti-2000} M.A. Marti-Renom,  A.C. Stuart,  A. Fiser,  R. S\'anchez,  F. Melo,  A. Sali, 
{\it Annual Review of Biophysics and Biomolecular Structure} 
{\bf 29} 291
(2000)

\bibitem{Schwede-2003}  T. Schwede, J. Kopp, N. Guex, M.C. Peitsch, 
{\it Nucleic Acids Research}  {\bf 31} 3389 (2003)

\bibitem{Zhang-2009} Y. Zhang, 
{\it Current  Opinions in Structural  Biology}  \textbf{19}  145 (2009)



\bibitem{Chothia-1992} C. Chothia,
{\it Nature} {\bf 357}  543--544 (1992)

\bibitem{Charmm} {\tt http://www.charmm.org/ }

\bibitem{Amber} {\tt http://ambermd.org/ }

\bibitem{Gromacs} {\tt http://www.gromacs.org/ }

\bibitem{Ott-2002} E. Ott, Chaos in Dynamical Systems. {\it Cambridge University Press, New York} (2002)


\bibitem{Korepin-1997} V.E. Korepin, N.M. Bogoliubov, A.G. Izergin, 
Quantum Inverse Scattering Method and Correlation Functions, {\it Cambridge university press, Cambridge} (1997)

\bibitem{Shaw-2008} D. E. Shaw, {\it  et al.}, 
{\it Communications of the ACM}  {\bf 51.7} 91
(2008)

\bibitem{Shaw-2009} D.E. Shaw, {\it et al.},  
{\it High Performance Computing Networking, Storage and Analysis, Proceedings of the Conference} IEEE  (2009)

\bibitem{Lindorff-2011} K. Lindorff-Larsen, S. Piana, R. O. Dror, D.E. Shaw. 
{\it Science} {\bf 334} 517
(2011) 

\bibitem{Gallavotti-2014}  G. Gallavotti,  Nonequilibrium and irreversibility, arXiv:1311.6448v4 [cond-mat.stat-mech] 

\bibitem{Nose-1984} S. Nos\'e,
{\it The Journal of Chemical Physics} {\bf  81} 511
(1984)

\bibitem{Hoover-1985} W.G. Hoover, 
{\it Physical Review}  {\bf A31} 1695  (1985)

\bibitem{Martyna-1992} G.J. Martyna, M.L. Klein, M. Tuckerman, 
{\it The Journal of Chemical Physics} {\bf  97} 2635
(1992)

\bibitem{Liu-2000} Y. Liu, M.E. Tuckerman, 
{\it The Journal of Chemical Physics} {\bf 112} 1685
(2000)

\bibitem{Niemi-2014a} A.J. Niemi 
{\it Theoretical and Mathematical Physics}
{\bf 181} 1235
(2014)

\bibitem{Berendsen-1984} H.J.C. Berendsen,  J. Pl M. Postma, W.F. van Gunsteren, A.R.H.J. DiNola, J.R. Haak. 
{\it The Journal of Chemical Physics} {\bf  81}  
3684
(1984)

\bibitem{Liwo-1997} A. Liwo, S. O{\l}dziej, M.R. Pincus, R.J. Wawak, S. Rackovsky, H.A. Scheraga, 
{\it Journal of Computational  Chemistry} {\bf 18}  849
(1997)

\bibitem{Scheraga-2007} H.A. Scheraga, M.  Khalili, A.  Liwo, 
{\it Annual Reviews in Physical Chemistry} \textbf{58} 57 (2007)

\bibitem{Liwo-2014} A. Liwo, A.J. Niemi, X. Peng, A.K. Sieradzan, A novel coarse-grained description of protein structure and  folding by UNRES force field and  Discrete Nonlinear Schr\"odinger Equation, in {\it 
Frontiers in Computational Chemistry (Bentham Science Publishers)} (2014)

\bibitem{Go-1983}  N. N. G\=o,  
{\it Annual Review of Biophysics and Bioengineering} {\bf 12} 183
(1983) 

\bibitem{Haliloglu-1997}   T. Haliloglu, I. Bahar,  B. Erman, 
{\it 
Physical Review Letters } {\bf 79} 3090
(1997)


\bibitem{Shirakawa-1977} C.K. Chiang, C.R. Fincher Jr, Y.W. Park, A.J. Heeger, H. Shirakawa, E.J. Louis, 
S.C. Gau, A.G. MacDiarmid. 
{\it 
Physical Review Letters} {\bf  39} 1098
(1977)

\bibitem{Jackiw-1981} R. Jackiw, J.R. Schrieffer, 
{\it  Nuclear Physics} {\bf B190}
253
(1981)

\bibitem{Semenoff-1986} A.J. Niemi, G.W. Semenoff, 
{\it Physics Reports} {\bf 135} 99
(1986)

\bibitem{Baskaran-1987} G. Baskaran, Z. Zou,  P. W. Anderson. 
 {\it  Solid state communications} {\bf  63} 973
  (1987) 

\bibitem{Kim-2006} B.J. Kim, H. Koh, E. Rotenberg, S.-J. Oh, H. Eisaki,
N. Motoyama, S. Uchida, T. Tohyama, S. Maekawa, Z.-X. Shen, C. Kim,
{\it Nature Physics} {\bf 2} 397
(2006)

\bibitem{Coleman-1973} S. Coleman, E. Weinberg,
{\it Physical Review} {\bf D7} 1888
(1973)

\bibitem{Chernodub-2008} M.N. Chernodub, L.D. Faddeev, A.J. Niemi, 
{\it   Journal of High Energy Physics} {\bf 12} 014  (2008)

\bibitem{Niemi-2003} A.J. Niemi, 
{\it Physical Review}
{\bf D67} 106004 (2003)

\bibitem{Bishop-1975} R.L. Bishop, 
{\it 
The American Mathematical Monthly}   {\bf  82} 246  (1975)


\bibitem{Hu-2011a} S. Hu, M. Lundgren, A.J. Niemi
{\it Physical Review} {\bf E83} 061908 (2011)

\bibitem{Faddeev-1987} L.D. Faddeev, L.A. Takhtajan, Hamiltonian methods in the theory of solitons {\it
(Springer Verlag, Berlin, 1987)}

\bibitem{Ablowitz-2003} M.J. Ablowitz, B. Prinari, A.D. Trubatch, 
Discrete and Continuous Nonlinear Schr\"odinger Systems {\it 
(London Mat. Soc. Lect. Note Series {\bf 302}, London, 2003)}


\bibitem{Polyakov-1986} A.M. Polyakov, 
{\it Nuclear Physics} {\bf B268} 406 (1986) 

\bibitem{Kratky-1949} O. Kratky, G. Porod,  
{\it Recueil des Travaux Chimiques des PaysBas}  {\bf 68} 
1106
(1949)


\bibitem{Kevrekidis-2009} P.G. Kevrekidis, The Discrete Nonlinear Schr\"odinger Equation: Mathematical Analysis,
Numerical Computations and Physical Perspectives {\it  (Springer-Verlag, Berlin, 2009)}

\bibitem{Manton-2004} N. Manton, P. Sutcliffe,   Topological Solitons {\it  (Cambridge University Press, Cambridge, 2004)}


\bibitem{Hu-2013a} S. Hu, Y. Jiang, A.J. Niemi, 
{\it Physical Review}  105011 (2013)

\bibitem{Ioannidou-2014} T. Ioannidou, Y. Jiang, A.J. Niemi 
{\it Physical Review } {\bf D90} 025012 (2014)

\bibitem{Arnold-1990} V.I. Arnold, Singularities of Caustics and Wave Fronts {\it
(Kluwer Academic Publishers, Dordrechts, 1990)}

\bibitem{Arnold-1995} V.I. Arnold, 
{\it Russian Mathematical Surveys}  {\bf 50} 1 (1995)

\bibitem{Arnold-1996} V.I. Arnold, 
{\it American  Mathematical  Society Translations} {\bf 171} 11 (1996) 

\bibitem{Aicardi-2000} F. Aicardi, 
{\it Functional  Analysis and  Applications} {\bf 34} 79
(2000)

\bibitem{Uribe-2004} R. Uribe-Vargas, 
{\it L'Enseignement Math\'ematique} {\bf 50} 69
(2004)

\bibitem{Danielsson-2010} U.H. Danielsson, M. Lundgren, A.J. Niemi, 
{\it Physical Review} {\bf E82}  021910 (2010)

\bibitem{Chernodub-2010} M.N.  Chernodub, S Hu, A.J. Niemi, 
{\it Physical Review} {\bf E82}  011916 (2010)


\bibitem{Molkenthin-2011} N. Molkenthin, S. Hu, A.J. Niemi, 
{\it Physical Review Letters} {\bf 106} 078102 (2011) 

\bibitem{Hu-2011} S. Hu, A. Krokhotin, A.J. Niemi, X. Peng,  
{\it Physical Review} {\bf  E83}  041907 (2011)

\bibitem{Krokhotin-2011} A. Krokhotin, A.J. Niemi, X. Peng, 
{\it Physical Review} { \bf E85}  031906 (2011)

\bibitem{Krokhotin-2012a}  A.  Krokhotin, M. Lundgren, A.J. Niemi,
{\it Physical Review} {\bf E86}  021923 (2012)

\bibitem{Krokhotin-2013a} A.  Krokhotin,  A.J. Niemi, X. Peng,
{\it The Journal of chemical physics} {\bf 138} 175101 (2013)

\bibitem{Krokhotin-2013b} A.  Krokhotin, M. Lundgren,  A.J. Niemi, X. Peng,
{\it Journal of Physics: Condensed Matter} {\bf  25}  325103 (2013)

\bibitem{cath} {\tt http://www.cathdb.info/ }

\bibitem{scop} {\tt http://scop.mrc-lmb.cam.ac.uk/scop/ }



\bibitem{Glauber-1963} R.J. Glauber, 
{\it Journal of Mathematical Physics}  {\bf 4} 294  (1963)

\bibitem{Bortz-1975}  A.B. Bortz, M.H. Kalos, J.L. Lebowitz, 
{\it Journal of  Computational Physics}  {\bf 17} 10  (1975)

\bibitem{Skolnick-2009} J. Skolnick, A.K. Arakaki, Y.L. Seung, M. Brylinski, 
{\it Proceeding of  National Academy of Sciences USA}   {\bf 106} 15690 (2009)

\bibitem{Ptashne-2004} M. Ptashne,   
A Genetic Switch, Third Edition, Phage Lambda Revisited.  {\it Cold Spring Harbor Laboratory Press, Cold Spring Harbor}
(2004) 

\bibitem{Gottesman-2004} 
M.E. Gottesman,  R.A. Weisberg, 
{\it Microbiology and  Molecular Biology Reviews}  {\bf 68} 796  (2004)

\bibitem{Jennings-1993}  P.A. Jennings, P.E. Wright, 
{\it Science} {\bf 262} 892 (1993)

\bibitem{Moriyama-2010} Y. Moriyama, K. Takeda, 
{\it Journal of Physical Chemistry} {\bf  B114}  2430 (2010)

\bibitem{Ochia-2010} Y. Ochia, Y. Watabane , H. Ozawa, S. Ikegami, N. 
Uchida, S. Watabe, 
{\it Bioscience, Biotechnology, and Biochemistry }  {\bf 74} 1673 (2010)

\bibitem{Chernodub-2010b} M.N. Chernodub, M. Lundgren, A.J. Niemi
{\it Physical Review} {\bf E83} 011126 (2010)

\bibitem{Woods-2014} K.N. Woods, 
{\it Soft Matter} {\bf 10} 4387  (2014)

\bibitem{Westermark-2011} P. Westermark, A. Andersson, G.T. Westermark,
{\it Physiological  Reviews} {\bf 91} 795 (2011) 

\bibitem{Pillay-2013}  K. Pillay, P. Govender, 
{\it BioMed Research International} {\bf 2013} 826706 (2013) 

\bibitem{He-2014} Jianfeng He, Jin Dai, Jing Li, A.J. Niemi,
{\it The
Journal of Chemical Physics} (to appear) 


\bibitem{Holm-1991} L. Holm, C. Sander, {\it Journal of Molecular Biolology}  \textbf{218} 183 (1991)


\bibitem{Lovell-2003} S.C. Lovell, I.W.  Davis, W.B.  Arendall III, P.I.W.  de Bakker, J.M. Word, M.G. 
Prisant, J.S.  Richardson, D.C. Richardson, 
{Proteins}  \textbf{50} 437 (2003)

\bibitem{Rotkiewicz-2008} P. Rotkiewicz, J.  Skolnick,
 {\it Journal of Computational Chemistry}  \textbf{29} 1460 (2008)

\bibitem{maxsprout} {\tt http://www.ebi.ac.uk/Tools/structure/maxsprout/ }

\bibitem{pulchra} {\tt http://cssb.biology.gatech.edu/PULCHRA}

\bibitem{phenix} {\tt http://www.phenix-online.org/}

\bibitem{refmac} { \tt http://www2.mrc-lmb.cam.ac.uk/groups/murshudov/}


\bibitem{Dunbrack-2002}  R.L. Dunbrack Jr.,  
{\it Current  Opinions in Structural Biology} \textbf{12}  431 (2002) 

\bibitem{Lundgren-2012b} M. Lundgren, A.J. Niemi, F. Sha,
{\it Physical Review} {\bf E85}  061909 (2012)

\bibitem{Lundgren-2013b} M. Lundgren, A.J. Niemi, 
{\it Physical Review} {\bf E86} 021904 (2013)

\bibitem{Peng-2015} X. Peng, A. Chenani, S. Hu, Y. Zhou, A.J. Niemi, 
{\it BMC Structural Biology} (to appear)

\endthebibliography
\end{document}